\documentclass[12pt,a4paper,oneside,openright]{report}

\usepackage{moreverb}								
\usepackage{textcomp}								
\usepackage{lmodern}								
\usepackage{helvet}									
\usepackage[utf8]{inputenc}							

\usepackage{enumitem}
\setlist{noitemsep} 

\usepackage{amsmath}								
\usepackage{amsfonts,amssymb,mathrsfs}						

\usepackage{bbold} 

\usepackage{booktabs}
\usepackage{graphicx}								
\usepackage{subfig}									
\numberwithin{equation}{chapter}					
\numberwithin{figure}{chapter}						
\numberwithin{table}{chapter}						

\newcommand\this{\addtocounter{equation}{1}\tag{\theequation}}

\usepackage{lettrine}
\input Romantik.fd
\newcommand*\initfamily{\usefont{U}{Romantik}{xl}{n}}

\usepackage[T1]{fontenc}							
\usepackage[english]{babel}							

\usepackage[final, stretch=10, protrusion=true, tracking=true,spacing=true, kerning=true, expansion=true]{microtype}
\microtypecontext{spacing=nonfrench}

\usepackage{listings}								

\usepackage{chemfig}								

\usepackage{geometry}			
\usepackage{eso-pic}								

\usepackage{float} 									
\usepackage{parskip}								

\usepackage[labelfont=bf, textfont=normal,
			justification=centering, 
			singlelinecheck=false,hypcap=false]{caption} 		

\usepackage{titlesec}		
\titleformat{\chapter}[display]
  {\Huge\bfseries\filcenter}
  {{\fontsize{50pt}{1em}\vspace{-4.2ex}\selectfont \textnormal{\thechapter}}}{1ex}{}[]

\usepackage{fancyhdr}								
\def\layout{1}	
\ifnum\layout=2	
	\fancypagestyle{plain}{			
	\fancyhf{}
	 		
	\fancyfoot[LE,RO]{\thepage}}	
\else			
\fi

\usepackage{tikz}
\usepackage{TikzFeynman}
\usepackage{colortbl}

\arrayrulecolor{blue}
\usepackage{longtable}

\definecolor{marron}{RGB}{60,30,10}
\definecolor{darkblue}{RGB}{0,0,80}
\definecolor{lightblue}{RGB}{80,80,80}
\definecolor{darkgreen}{RGB}{0,80,0}
\definecolor{darkgray}{RGB}{0,80,0}
\definecolor{darkred}{RGB}{80,0,0}

\usepackage{hyperref}								
\hypersetup{bookmarksopen=false,colorlinks,pdfstartview=FitH, citecolor=blue,runcolor=blue,anchorcolor=blue,linkcolor=blue,
			pdftitle = {PM Phd thesis}
   		 	}
    		
\hbadness=\maxdimen
\vbadness=\maxdimen

\begin{document} 

\pagenumbering{roman}
%
%
%
\thispagestyle{empty}
\mbox{}

\thispagestyle{empty}
\begin{center}
	\textbf{\LARGE A Quantum Field Theory for the interaction of\\ pions and rhos} \\[0.5cm]
	
	{ \large \bf{
	Preshin Moodley}}	
	\vfill		
	{ 
	Thesis submitted for the degree of }\\
	{ 
	Doctor of Philosophy in }\\
	{ 
	Theoretical Physics}\\[0.5cm]

	\vfill	
	\begin{figure}[H]
	\centering
	\includegraphics[width=0.7\pdfpagewidth]{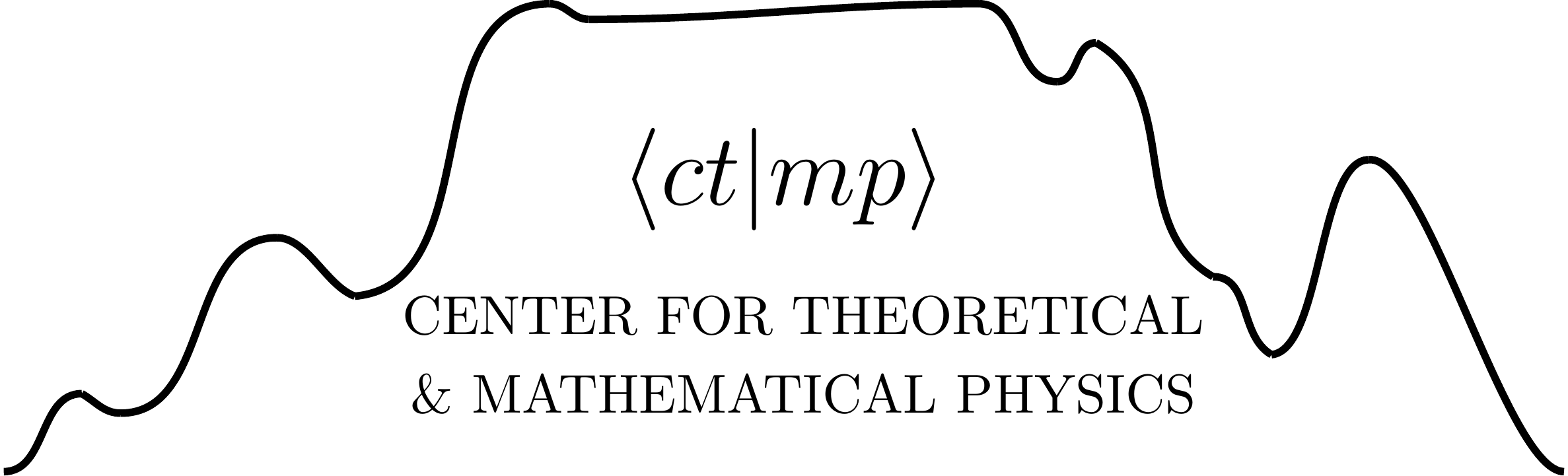} \\	
	\end{figure}	\vspace{5mm}
	
	\vfill	
	\begin{figure}[H]
	\centering
	\includegraphics[width=0.3\pdfpagewidth]{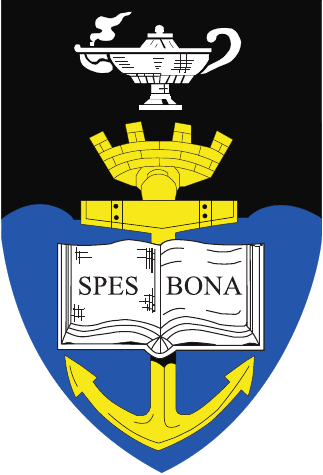} \\	
	\end{figure}	\vspace{5mm}	
	
	\emph{Center for Theoretical \& Mathematical Physics}\\
	Department of Physics \\
	\textsc{University of Cape Town} \\
	Cape Town, South Africa 2017 \\
\end{center}

\newpage
\thispagestyle{plain}

\copyright ~ Preshin Moodley, 2017. \setlength{\parskip}{1cm}

Supervisor: Prof. Cesareo Dominguez, UCT, Department of Physics\\
Co-Supervisor: Prof. Dr. Karl Schilcher, JGU Mainz, Institut für Physik (WA THEP)\\
Co-Supervisor: Prof. Dr. Hubert Spiesberger , JGU Mainz, Institut für Physik (WA THEP)\\
Co-Supervisor: Dr. Gary Tupper, UCT, Department of Physics\\


Ph.D. Thesis\\	
Center for Theoretical \& Mathematical Physics \\
Department of Physics\\
University of Cape Town\\
Cape Town\\
\setlength{\parskip}{0.5cm}

\newpage
A Quantum Field Theory for the interaction of pions and rhos\\
Preshin Moodley\\
Department of Physics\\
University of Cape Town \setlength{\parskip}{0.5cm}

\thispagestyle{plain}			
\setlength{\parskip}{0pt plus 1.0pt}
\section*{Abstract}
We extend the Kroll-Lee-Zumino model in its particle content to include the charged rho vector mesons and the
neutral pion meson. This entailed using the larger SU(2) gauge group. The masses for the vector mesons were generated 
via spontaneous symmetry breaking using the Higgs mechanism. The Lagrangian was then quantized and gauge fixed using the generalized class of $R_\xi$ gauges. Tree scattering lengths were calculated for pion-pion scattering and the values for the $a^0_0$ and $a^2_0$ scattering lengths are found to be comparable with experiment. The one particle irreducible diagrams that contribute to the one loop corrections to the tree scattering lengths are renormalized. 

\newpage			
\thispagestyle{empty}
\mbox{}

\newpage
\thispagestyle{plain}			
\section*{Acknowledgements}
I wish to thank my supervisors who have spurred  me on into the field of particle physics:\\ \\
\textbf{Cesareo Dominguez}, \textit{The Boss}, has been patient for allowing me to learn new techniques and formalism 
for developing this model. His vision of the end result has chartered the path that I had to flesh out. Thank you 
Cesareo for letting me learn slowly so that I could understand all the intricate theory required for this project. We 
both didn't know how big this project would turn out to be at the start, but we made it to the end.  You and Silva 
have invited me into your lives. We have shared many dinners discussing life and science over awesome steaks (and 
sometimes rabbit food) and great wines. I will fondly remember the lessons of cooking great steaks and living life.
\\ \\
\textbf{Karl Schilcher} taught me the elementary aspects of renormalization theory for non Abelian gauge theories and had generously typed up notes for me to read and learn from. Thank you Karl for suggesting alternative approaches to the loop calculations, providing the many references to articles when I needed help, and for insisting on my work being as \textit{über sie gleich} as possible.
\\ \\
\textbf{Hubert Spiesberger} taught me the efficient way of extracting out the Feynman rules from the Lagrangian using 
functional methods and performing loop calculations with the Passarino-Veltman tensor reduction to scalar integrals. 
Thank you Hubert for patiently answering my questions even when they were terribly stupid and I was being very 
stubborn. If you had not shown me the Passarino-Veltman reduction, I would still be stuck doing loop calculations the 
old fashioned way. \textit{``....I have done these calculations more then ten years ago,... and I've forgotten all the 
details, but what remains of that time is experience.''--HS}
\\ \\
\textbf{Gary Tupper}, \textit{Lord of the Strings}, helped me through the journey of learning the non Abelian Higgs 
mechanism, using the 't Hooft-Fujikawa-Lee-Abers gauge fixing function to remove the mixing terms and the quantization 
process using the Faddeev-Popov ghosts. Thank you Dude for answering my constant bombardment of questions, and the 
helpful discussions on gauge theories. We have banished my ignorance to the deeper recesses of my mind...for now.
\\ \\
I have benefited from attending the discussions of gauge theories by Prof. Heribert Weigert for his students, and 
grateful for the invitation to his informal discussion sessions with Prof. Francois Gelis on unitarity cuts, the 
Slavnov-Taylor identities and renormalizability.
\\ \\
I thank Prof. Chris Clarkson and Dr. Obinna Umeh from the Department of Mathematics and Applied Mathematics for 
introducing me to the Mathematica packages used for tensor algebra and answering my endless questions on its usage.
\\ \\
I am grateful to Mawande Lushozi who has been a fellow explorer on this journey. We had many discussions, frustrating 
and rewarding about the workings of gauge theories and the complications of renormalization. My \textit{mfethu} we 
have fought through the fog of mystery surrounding gauge theories, walked through the high halls of multi-loop 
calculations and breath in the rarefied air. 
My office mates Yingwen Zhang, Mirrete Fawzy and Luis Hernandez have kept me entertained and sober of thought. Thank 
you for the awesome experiences, memories and many sushi nights.
\\ \\
Dr. Mehdi Safari and Dr. Iniyan Natarajan read the initial draft chapters and suggested improvements and changes to 
the writing style.
\\ \\
I am grateful to the UCT Physics department which has been my academic home for these past years and has provided me 
with a stimulating environment.
\\ \\
I have been fortunate to meet people who have gone on from being friends to become family, Hari and Verosha, Justin 
and Hilda, Wesley and San, Umut, Martin, Yingwen and Summer, Mawande, Mirrete and Luiz. Thank you for being positive 
influences in my life and keeping me motivated.
\\ \\
And my dear family, who have instilled in me the value of an education. Thank you for letting me pursue my obsession, for the financial support which prevented me from starving and for letting me be me.

%
%
%

\begin{flushright}
\vspace{1.5cm}
\hfill Preshin  Moodley \\
\hfill Cape Town\\
\hfill June 2017
\end{flushright}

\newpage				
\thispagestyle{empty}
\mbox{}

\newpage
\thispagestyle{plain}	
\section*{Declaration of authorship}
I, Preshin Moodley, declare that this thesis titled: ``A Quantum Field Theory for the interaction of pions and rhos'', 
and the work presented in it are my own. I confirm that:
\\
\begin{itemize} 
\item This work was done wholly or mainly while in candidature for a research degree at this University.\\
\item Where any part of this thesis has previously been submitted for a degree or any other qualification at this 
University or any other institution, this has been clearly stated.\\
\item Where I have consulted the published work of others, this is always clearly attributed. \\
\item Where I have quoted from the work of others, the source is always given. With the exception of such quotations, 
this thesis is entirely my own work. \\
\item I have acknowledged all main sources of help. \\
\item Where the thesis is based on work done by myself jointly with others, I have made clear exactly what was done by others and what I 
have contributed myself.\\ \\
\end{itemize}

\noindent Signed: 
\hspace{05 mm}
\\
\rule[0.5em]{10em}{0.5pt} 
 
\noindent Date: 06 June 2017\\
\rule[0.5em]{10em}{0.5pt} 

\newpage
\tableofcontents


\cleardoublepage
\addcontentsline{toc}{chapter}{\listfigurename} 
\listoffigures
\cleardoublepage
\addcontentsline{toc}{chapter}{\listtablename}  
\listoftables

\cleardoublepage
\setcounter{page}{1}
\pagenumbering{arabic}			
\setlength{\parskip}{1.3em}
\linespread{1.3}


\numberwithin{equation}{section}
\chapter{Introduction}

\lettrine[lines=4]{\initfamily\textcolor{darkblue}{S}}{oon} after the success of Quantum Electrodynamics (QED) as a description of the electromagnetic interaction, physicists turned to the investigation of the structure of nucleons and the strong force. Yukawa's conjecture \cite{Yukawa1935} about the existence of a massive particle responsible for mediating the strong interaction was partially successful at explaining the low energy interactions of the nucleons. Numerous new particles were discovered during this period with seemingly no order among them. The quark model was developed by Gell-Man \cite{GELLMANN1964a} and Zweig \cite{Zweig1964a,Zweig1964b} to tame and show some regularity in the particle zoo with hadronic multiplets. Further investigations lead to the developement of the Parton model by Feynman \cite{Feynman1969a,Feynman1969b} which assumed the nucleons were made up of essentially free particles called partons. A nagging question at the time was where were these partons or quarks? The phenomenon of Bjorken scaling \cite{Bjorken1975} incited a search for quantum field theories which exhibited this behaviour. This ended with the discovery of asymptotic freedom in non-Abelian gauge theories by Politzer \cite{Politzer1973}, Wilczek and Gross \cite{GrossWilczek1973} and leading to the formulation of Quantum Chromodynamics (QCD) which is a SU(3)$_c$ non-Abelian gauge with the new isospin degree of freedom identified as color charge. The partons were identified as the quarks and gluons of QCD and were forever bound in colorless unions.

The property of asymptotic freedom was fortunate since it allowed for the analysis of systems of high energy where the strong coupling has a relatively small value and conventional tools like perturbation theory were applicable. New tools needed to be forged to explore low energy systems. The direct approach taken in Lattice QCD has become feasible in recent years with the reduction in the costs of the massive computing power required and the steady increase in computing power. The alternate approach came with the development of models and effective field theories applicable in certain energy regimes. Chiral Perturbation Theory ($\chi$PT) is 
the effective field theory for QCD and has been used for applications up to $\sim 300$ MeV with a failure beyond $\sim 600$ MeV 
\cite{Boyle2009}. The onset of Pertubative QCD has been pushed down to $1$ GeV for some applications \cite{Yeh2002}. There exists 
this gap of ignorance in the interval $\sim\left(0.6, 1\right)$ GeV for which standard analytic tools can not explore. This energy 
region is dominated by the low energy scalar and vector mesons. 

Nambu had suggested that the rho meson could explain the nucleon form factors \cite{Nambu1957}. Sakurai proposed the idea of Vector meson dominance (VMD) where the strong interaction would be 
mediated by vector mesons in a non-Abelian gauge theory \cite{SAKURAI1960,SAKURAI1969}. The idea of VMD is stated generally as: gauge bosons transforms into the lowest energy vector mesons and interact with hadrons through effective vertices.
\begin{figure}[H]
\centering
\includegraphics[width=0.7\linewidth]{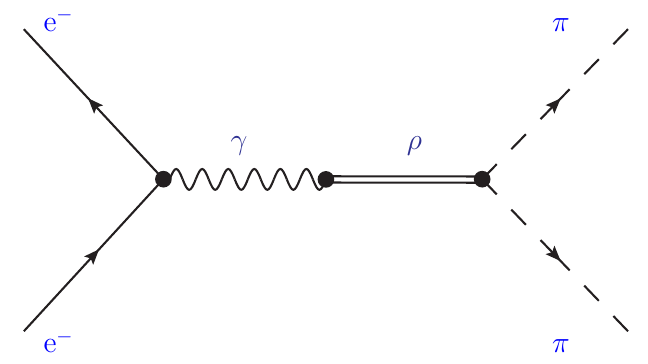}
\caption{Photon transforming into a Rho which decays into pions.}
\label{fig:klz2}
\end{figure}

The photon seems to interact with hadronic matter mainly through the rho vector meson \cite{Bauer1978, engel1996}. Sakurai had suggested interaction terms for VMD but this was not a field theory which could be used for higher order loop analysis. Kroll, Lee and Zumino suggested a model \cite{KLZ1967} involving the charged pions and neutral rho as a candidate for VMD.

\section{The Kroll-Lee-Zumino Model}
\begin{figure}[H]
\centering
\includegraphics[width=0.85\linewidth]{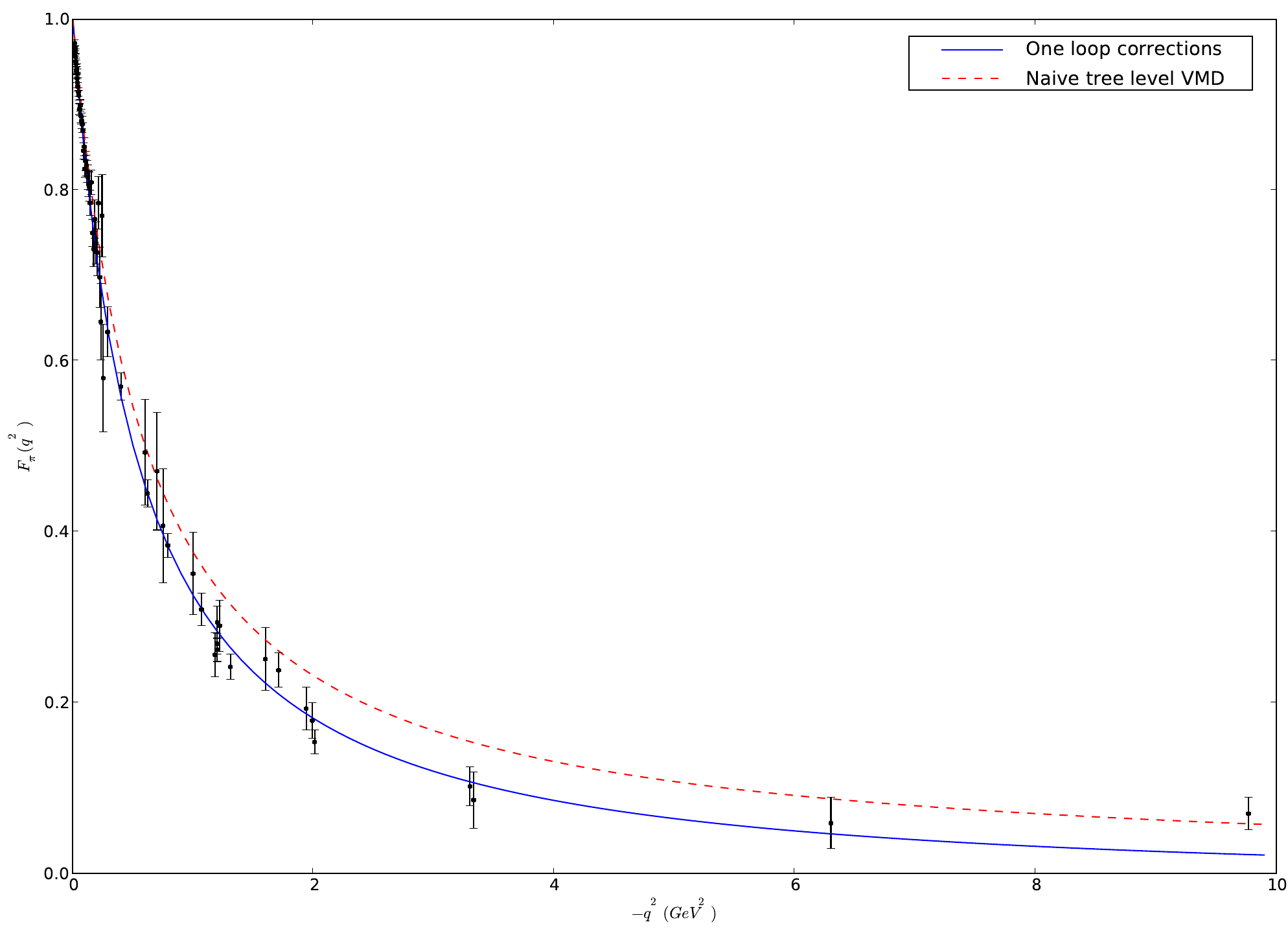}
\caption{Pion form factor for $q^2<0$}
\label{fig:plotall}
\end{figure}

The Kroll-Lee-Zumino Model (KLZ) is a quantum field theory developed to describe the interactions of charged pions and the neutral rho. 
\begin{align}
\mathscr{L} = {\left( {{\partial _\mu }\phi } \right)^* }{\partial ^\mu }\phi  - {m^2}{\phi ^* }\phi  - \frac{1}{4}{F_{\mu \nu }}{F^{\mu \nu }} + \frac{1}{2}{M^2}{A_\mu }{A^\mu } + ie{A^\mu }\left( {{\phi ^* }{\partial _\mu }\phi  - \phi {\partial _\mu }{\phi ^* }} \right) + {e^2}{\phi ^* }\phi {A_\mu }{A^\mu }
\end{align}
with the field strength tensor $F_{\mu\nu}$ defined as
\begin{align}
{F_{\mu \nu }}: = {\partial _\mu }{A_\nu } - {\partial _\nu }{A_\mu }
\end{align}
Here the pions are represented by $\phi$, with $m$ the pion mass. The neutral rho is represented by $A_{\mu}$, with $M$ the rho mass and $e$ represents the strength of the rho-pion coupling which has an approximate value of $e\approx 5.96$. The KLZ is a renormalizable quantum field theory despite the breaking of U(1) gauge symmetry with the presence of the explicit mass term for the vector rho. This is a result of the vector field coupling only to the conserved current \cite{Li2012,GHOSE1972} i.e.
\begin{align}
{J_\mu }&: = {\phi ^* }{\partial _\mu }\phi  - \phi {\partial _\mu }{\phi ^* }
\end{align}
with
\begin{align}
{\partial ^\mu }{J_\mu } &= 0 \label{klzconscurrent}
\end{align}
This can be seen by looking at the propagator of a massive vector field
\begin{align}
i{D_{\mu \nu }} = \frac{{ - i}}{{{k^2} - {M^2} + i\varepsilon }}\left( {{g_{\mu \nu }} - \frac{{{k_\mu }{k_\nu }}}{{{M^2}}}} \right)
\end{align}
which due to the ${k_\mu }{k_\nu }$ term is usually logarithmically divergent. For the case of the KLZ, the above propagator would appear between conserved currents with 
\begin{align}
{k ^\mu }{J_\mu } &= 0
\end{align}
using \eqref{klzconscurrent} in the momentum representation, thus allowing for renormalizability by removing the logarithmically divergent terms. The KLZ being renormalizable makes it an attractive model for analysis, since the renomalizability allows for a systematic calculation of higher order loop corrections without introducing additional parameters into the model. Since the pertubative series expansion parameter is of the form $\dfrac{e}{4\pi}\approx 0.47$,  higher order analysis can be pursued seriously. This can be seen in the application of the KLZ in finite temperature calculations undertaken in \cite{GALE1991}. The KLZ has also been used to analyse the pion form factor by \cite{CADWillers2007}, for space-like transferred momentum, the one loop correction to the form factor is in agreement with data with $\chi^2_F\approx 1.1$ as shown in figure \ref{fig:plotall}.

\begin{figure}[H]
\centering
\includegraphics[width=0.85\linewidth]{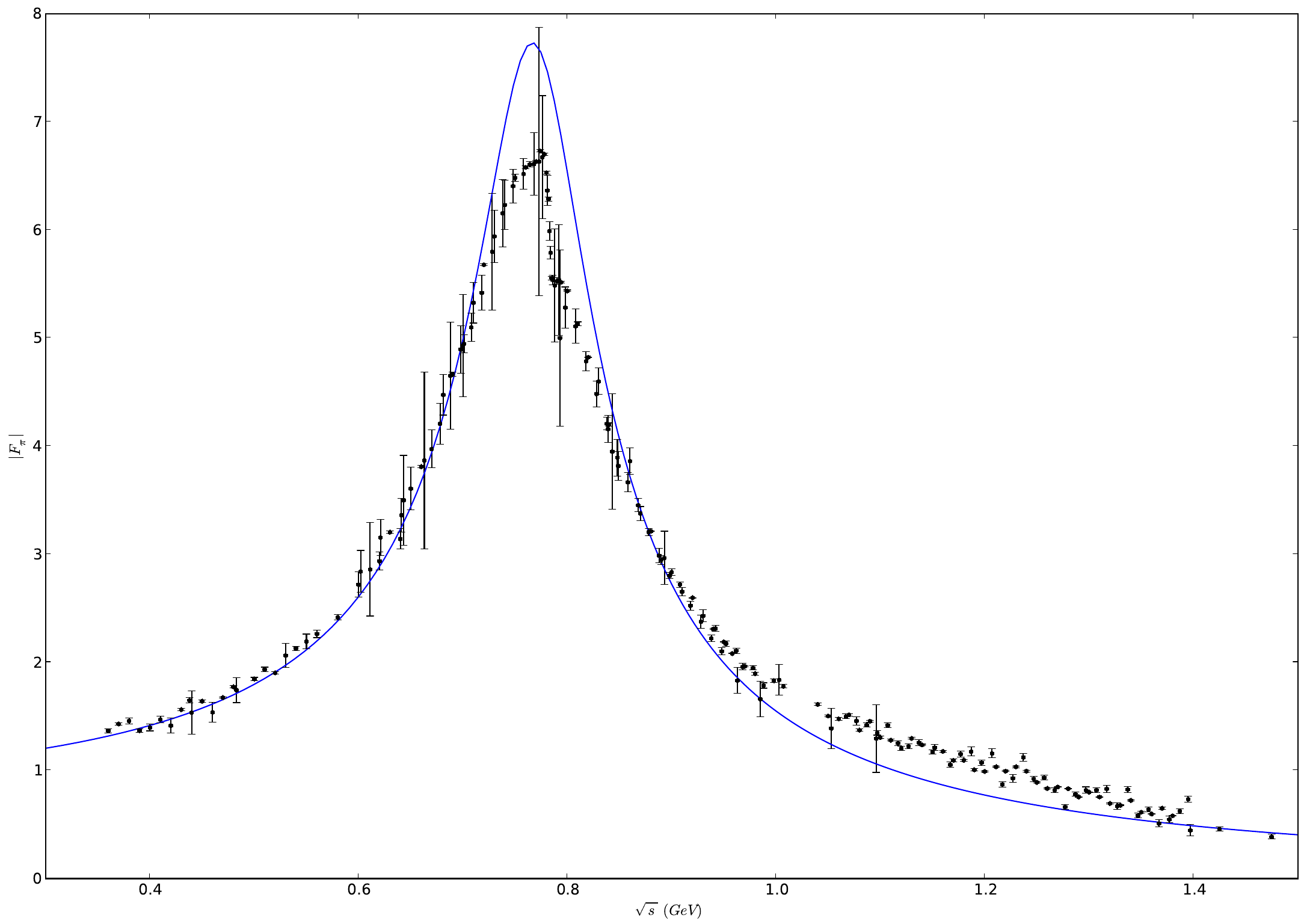}
\caption{Pion form factor for $s=q^2>0$}
\label{fig:plotzerotimelike}
\end{figure}

For time-like transferred momentum the KLZ follows the data closely below the rho peak but starts to deviate from the data beyond the rho peak as can be seen in figure \ref{fig:plotzerotimelike}
This model has also been used to compute the electromagnetic radius of the pion with ${\left\langle {r_\pi ^2} \right\rangle _S} = 0.40~\text{fm}^2$ in \cite{CADWillers2008}.

There are limitations to the KLZ model. The KLZ only provides a partial description of the pion-rho interactions. The particle content is only of the charged pions and neutral rho. There are more pion-rho interactions in nature than are captured by the KLZ, which ignores the neutral pion and charged rhos thus excluding a description of the decay of the charged rhos into pions. To address some of these limitations in the KLZ we shall have to chart a new path forward. The new model must use a larger gauge group to accommodate the full complement of the triplet of pions and rhos. 

As an application of the new model we have chosen to calculate the pion-pion scattering lengths. The pion-pion scattering amplitude has been studied extensively through the Roy equations and Chiral Pertubartion theory \cite{Pelaez2005,Pelaez2006,Pelaez2008,Pelaez2011}. Experimentally, NA48/2 \cite{Batley2006} and DIRAC \cite{Adeva2009} have collected large data sets which allow for the precise extraction of the scattering lengths.  

\section*{Outline}
This document is organized in the following way. In part I, chapter 2, we develop the bare SU(2) model. Chapter 3 is dedicated to Spontaneous symmetry breaking to generate the rho mass. In chapter 4, we quantize the new Lagrangian, introduce the Faddeev-Popov ghosts and fix the gauge. Chapter 5 contains the renormalization transformation and Feynman rules. In Part II, we apply the new model to evaluating the pion-pion scattering lengths. Chapter 6 has the calculations of the tree scattering lengths and results. In chapter 7, we start the program of finding the one loop corrections to the tree scattering lengths. Due to time constraints on a PhD project, the numerical evaluations have been excluded. Finally chapter 8 summarizes with a conclusion.

\part{Generalizing the KLZ}

\chapter{Extending the KLZ}


\section{SU(2) Generalization}

\lettrine[lines=4]{\initfamily\textcolor{darkblue}{W}}{e} will now proceed in generalizing the U(1) KLZ model. What are some simple expectations could we have of such a model? We would like a model describing the interactions of pseudoscalar pions and vector rhos which includes the full triplet of pions $\{\pi^-,\pi^0,\pi^+\}$ and the triplet of rhos $\{\rho^-,\rho^0,\rho^+\}$. We would like it to be renormalizable since it would have some predictive power when considering higher order corrections. We shall make some simplifying assumptions. We shall assume all the particles are point like. We shall ignore the difference between the masses of the charged and neutral pion $(m_{\pi^\pm} - m_{\pi^0} = 4.5936 \pm 0.0005  ~\rm{MeV})$ \cite{Patrignani}. This small difference (small in comparison to their mass) in pion masses is due to the mass difference between the u and d quarks. This approximate flavour symmetry becomes exact in the chiral limit and we have the isospin SU(2) symmetry group. The pions are then in the adjoint representation of SU(2). We also ignore the difference in rho masses $(m_{\rho^\pm} - m_{\rho^0}= 0.7\pm 0.8 ~ \rm{MeV})$ \cite{Patrignani}. 

With these assumptions we can build an effective description of the interactions between the pions and rhos. The rhos will play the role of dynamical gauge
bosons \cite{Bando1985} mediating the interactions between the pions and the rhos themselves. We will let the principle of local gauge invariance guide our construction under which only terms invariant under SU(2) gauge transformations are to be admitted into our Lagrangian.  We begin by writing down a SU(2) globally gauge invariant Lagrangian for scalars representing the pions [$\Phi(x):=\phi^a=\{\phi^1,\phi^2,\phi^3\}$]
\begin{align}
	\mathscr{L} = \frac{1}{2} \left[{\left( {{\partial _\mu }\Phi } \right)^\dag }{\partial ^\mu }\Phi  - b ^2{\Phi ^\dag }\Phi \right]
	\label{lpionf}
\end{align}
where $b$ has dimensions of mass. The Lagrangian \eqref{lpionf} is invariant under the unitary transformation of
\begin{align}
	U&=e^{iT^a \alpha^a} \in \rm{SU(2)} \\
\Phi &\to \Phi' = U\Phi 
\label{pigtrans}
\end{align}
where $T^a$ are the generators of the group and are elements of the Lie algebra $T^a \in \mathfrak{su(2)}$ and $\alpha^a \in \mathbb{R}$ are some constants. The repeated $a$'s imply a sum over $a=\{1,2,3\}$. When we promote the global gauge transformation to a local one with ${\alpha ^a} \to {\alpha ^a}\left( x \right)$, the derivatives in the Lagrangian produce extra terms which breaks the gauge invariance of the Lagrangian. The extra terms are proportional to the generators $T^a$, so to remedy the problem we will introduce an extra field into the partial derivative as 
\begin{align}
{\partial _\mu } \to {D_\mu } := {\partial _\mu } + ie{T^a}A_\mu ^a(x)
\end{align}
Here $D_\mu$ is the gauge covariant derivative, $e$ the gauge coupling which is dimensionless, $[e]=1$ and $A _\mu ^a (x)$ are the gauge fields representing the rhos. The gauge fields are in the adjoint representation of SU(2) so we expanded them using the generators as a basis. The covariant derivative is built to transform such that when we make a unitary transformation on the field $\Phi$, the gauge fields $A_\mu ^a$ must also transform to keep the Lagrangian invariant. This is done by insisting that 
\begin{align}
D{'_\mu }\Phi ' = U{D_\mu }\Phi 
\end{align}
which together with \eqref{pigtrans} gives us the transformation rule for the covariant derivative as
\begin{align}
{D_\mu } \to  D{'_\mu } = U{D_\mu }{U^\dag }
\end{align}
We can infer the transformation rule for the gauge fields from the above condition as
\begin{align*}
D{'_\mu }f &= U{D_\mu }{U^\dag }f\\
\left( {{\partial _\mu } + ieA{'_\mu }} \right)f &= U\left( {{\partial _\mu } + ie{A_\mu }} \right){U^\dag }f\\
 \Rightarrow A{'_\mu } &= UA_\mu{U^\dag } + \frac{1}{{ie}}U{\partial _\mu }{U^\dag } \this
 \label{Atrans}
\end{align*}
with $A{'}_{\mu} =A{'}_\mu^a T^a$ and $f$ was some arbitrary function included to keep track of the derivatives. Now when we make a unitary transformation on the fields and noting that 
\begin{align}
\Phi     \to & \Phi '    = U\Phi \\
{D_\mu } \to & D{'_\mu } = U{D_\mu }{U^\dag }
\end{align}
then the Lagrangian 
\begin{align}
\mathscr{L} = \frac{1}{2}\left[{\left( {{D_\mu }\Phi } \right)^\dag }{D^\mu }\Phi  - b^2{\Phi ^\dag }\Phi\right]
\end{align}
remains invariant. Under the unitary transformation of the scalar field, the gauge field was required to maintain the gauge invariance of 
the Lagrangian. The gauge fields are just passive auxiliary fields at this stage. If the gauge fields are to represent the rhos then they 
must have some dynamic behaviour. So naturally we would like to make this a requirement for the gauge fields. Though since this is a 
nonabelian theory, we must take care when we construct the kinetic term for the gauge fields. For a kinetic term we require a scalar, 
which is Lorentz and gauge invariant, and quadratic in the first derivatives of the field. The commutator serves as 
natural product so we shall construct an operator from the commutator of the covariant derivatives,
\begin{align}
	{{\hat O}_{\mu \nu }}f = \left[ {{D_\mu },{D_\nu }} \right]f
\end{align}
where $f$ is some arbitrary function we included to keep track of the derivatives. Under SU(2) gauge transformations, the operator changes as ${{\hat O}_{\mu \nu }} \to \hat O{'_{\mu \nu }}$
\begin{align*}
\hat O{'_{\mu \nu }}f &= \left[ {D{'_\mu },D{'_\nu }} \right]f \nonumber\\
 &= \left[ {U{D_\mu }{U^\dag },U{D_\nu }{U^\dag }} \right]f \nonumber\\
 &= \left( {U{D_\mu }{D_\nu }{U^\dag } - U{D_\nu }{D_\mu }{U^\dag }} \right)f \nonumber\\
 &= U\left[ {{D_\mu },{D_\nu }} \right]{U^\dag }f \nonumber \\
 &= U{{\hat O}_{\mu \nu }}{U^\dag }f \this
\end{align*}
So we see that the transformed operator is, 
\begin{align}
\hat O{'_{\mu \nu }} = U{{\hat O}_{\mu \nu }}{U^\dag }
\end{align}
We need to ensure that a kinetic term is Lorentz and gauge invariant. We can create a new operator by squaring $O^{\mu\nu}$:
\begin{align*}
\hat O{'_{\mu \nu }}\hat O{'^{\mu \nu }} &= U{{\hat O}_{\mu \nu }}{U^\dag }U{{\hat O}^{\mu \nu }}{U^\dag }\nonumber\\
 &= U{{\hat O}_{\mu \nu }}{{\hat O}^{\mu \nu }}{U^\dag } \this
 \label{O2}
\end{align*}
Looking at (\ref{O2}) we can see that this operator is Lorentz invariant but still dependent on the SU(2) gauge transformation $U$. We also have to deal with (\ref{O2}) not being a scalar, since only scalar functions enter into the action. We have to transform (\ref{O2}) into a scalar while preserving its properties. We can remove the gauge dependence and create a scalar function by taking the trace of (\ref{O2}) as
\begin{align*}
{\rm{Tr}}\left[ {\hat O{'_{\mu \nu }}\hat O{'^{\mu \nu }}} \right] &= {\rm{Tr}}\left[ {U{{\hat O}_{\mu \nu }}{{\hat O}^{\mu \nu }}{U^\dag }} \right] \nonumber\\
 &= {\rm{Tr}}\left[ {{{\hat O}_{\mu \nu }}{{\hat O}^{\mu \nu }}U{U^\dag }} \right]\nonumber\\
 &= {\rm{Tr}}\left[ {{{\hat O}_{\mu \nu }}{{\hat O}^{\mu \nu }}} \right] \this
 \label{O22}
\end{align*}
which is independent of the gauge transformation. Now since this operator \eqref{O22} is the square of the first derivatives of the fields and invariant under Lorentz and gauge transformations, it has the necessary behaviour of a kinetic term of the gauge fields. We will define the nonabelian field strength tensor as 
\begin{align}
	{G_{\mu \nu }} := \frac{1}{ie}{{\hat O}_{\mu \nu }} = \frac{1}{ie}\left[ {{D_\mu },{D_\nu }} \right]
	\label{gmunu}
\end{align}
We can get the explicit form of the tensor by substituting in the covariant derivatives into (\ref{gmunu}), where $f$ is included to keep track of the derivatives.
\begin{align*}
{G_{\mu \nu }}f &= \frac{1}{{ie}}\left[ {{D_\mu },{D_\nu }} \right]f \\
 &= \frac{1}{{ie}}\left[ {{\partial _\mu } + ie{A _\mu },{\partial _\nu } + ie{A_\nu }} \right]f \\
 &= \biggl( {{\partial _\mu }{A _\nu } - {\partial _\nu }{A _\mu } + ie\left[ {{A_\mu },{A_\nu }} \right]} \biggr)f \this
\end{align*}
So the field strength tensor is
\begin{align}
	{G_{\mu \nu }} = {\partial _\mu }{A_\nu } - {\partial _\nu }{A_\mu } + ie\left[ {{A_\mu },{A_\nu }} \right]
\end{align}
We can use the elements of $\rm \mathfrak{su(2)}$ to write ${G_{\mu \nu }}$ as a sum over the generators. Using the Lie algebra
\begin{align}
	\left[ {{T^b},{T^c}} \right] = i{\varepsilon _{abc}}{T^a}
\end{align}
where $\varepsilon _{abc}$ is the Levi-Civita symbol representing the structure constants of the algebra, and $A_\mu=A^a_\mu T^a$, we get the following expression
\begin{align*}
G_{\mu \nu }^a{T^a} &= {\partial _\mu }A _\nu ^a{T^a} - {\partial _\nu }A_\mu ^a{T^a} + ie\left[ {A_\mu ^b{T^b},A_\nu ^c{T^c}} \right] \\
&= \left[ {{\partial _\mu }A_\nu ^a - {\partial _\nu }A_\mu ^a -e{\varepsilon _{abc}} A_\mu ^b A_\nu ^c} \right]{T^a} \this
 \label{gmunu2}
\end{align*}
From (\ref{gmunu2}) we can recognise the familiar abelian field strength tensor in the first two terms in (\ref{gmunu2}). We will define 
\begin{align}
	F_{\mu \nu }^a := {\partial _\mu }A_\nu ^a - {\partial _\nu }A_\mu ^a
\end{align}
then the nonabelian field strength becomes
\begin{align}
	G_{\mu \nu }^a = F_{\mu \nu }^a - e{\varepsilon _{abc}}A_\mu^b A_\nu ^c
\end{align}
We can now define in analogy to (\ref{O22}) a kinetic term for the gauge fields as (the Yang-Mills term)
\begin{align*}
{\mathscr{L}_{YM}} &:=  - \frac{1}{2} \mathrm{Tr} \left[ {{G_{\mu \nu }}{G^{\mu \nu }}} \right] \\
 &=  - \frac{1}{2}G^a_{\mu \nu }G^{b\mu \nu }\mathrm{Tr}\left[ {{T^a}{T^b}} \right] \\
 &=  - \frac{1}{4}G^a_{\mu \nu }G^{a\mu \nu } \\
 &=  - \frac{1}{4}F_{\mu \nu }^a{F^{a\mu \nu }} + \frac{1}{2}e{\varepsilon _{abc}}{A^{b\mu }}{A^{c\nu }}F_{\mu \nu }^a - \frac{1}{4}{e^2}{\varepsilon _{abc}}{\varepsilon _{ade}}A_\mu ^b A_\nu ^c{A^{d\mu }}{A^{e\nu }} \this
 \label{gg}
\end{align*}
where $-\dfrac{1}{2}$ is a factor by convention and we have used the normalization condition 
\begin{align}
\rm{Tr}\left[ {{T^a}{T^b}} \right] = \frac{1}{2}{\delta ^{ab}}
\end{align}
We should pause at this stage and interpret what each term in \eqref{gg} represents. After all we went in search for a kinetic term for the gauge field, but requiring Lorentz and local gauge invariance provided extra terms. So the first term in \eqref{gg} is the pure kinetic term, the second and third terms represent three point and four point interactions among the gauge fields. 
So the full nonabelian Lagrangian is
\begin{align}
 \mathscr{L} &= \frac{1}{2}\left[{\left( {{D_\mu }\Phi } \right)^\dag }{D^\mu }\Phi  - b ^2{\Phi ^\dag }\Phi \right]  - \frac{1}{4}{G^a}_{\mu \nu }{G^a}^{\mu \nu }
\end{align}
Expanding out the covariant derivatives
\begin{align}
\frac{1}{2}{\left( {{D_\mu }\Phi } \right)^\dag }{D^\mu }\Phi = \frac{1}{2}\left\{ {{{\left( {{\partial _\mu }\Phi } \right)}^\dag }{\partial ^\mu }\Phi  + ie\left[ {{{\left( {{\partial _\mu }\Phi } \right)}^\dag }{A^\mu }\Phi  - {\Phi ^\dag }{A^\mu }{\partial _\mu }\Phi } \right] + {e^2}{\Phi ^\dag }{A^\mu }{A_\mu }\Phi } \right\}
\end{align}
we can write out all the terms in components of the fields using the adjoint representation of the generators ${\left( {{T^a}} \right)_{ij}} =  - i{\varepsilon _{aij}}$, so the terms become
\begin{align*}
{\left( {{\partial ^\mu }\Phi } \right)^\dag }{A_\mu }\Phi  &= A_\mu ^a{\left( {{\partial ^\mu }\Phi } \right)^\dag }{T_a}\Phi \\
 &= A_\mu ^a{\left( {{\partial ^\mu }{\phi _i}} \right)^\dag }{\left( {{T^a}} \right)_{ij}}{\phi _j}\\
 &=  - i{\varepsilon _{aij}}A_\mu ^a\left( {{\partial ^\mu }{\phi _i}} \right){\phi _j}\\
\implies {\Phi ^\dag }{A_\mu }\left( {{\partial ^\mu }\Phi } \right) &= {\left[ {{{\left( {{\partial ^\mu }\Phi } \right)}^\dag }{A_\mu }\Phi } \right]^\dag }\\
 &=  - i{\varepsilon _{aij}}A_\mu ^a{\phi _i}\left( {{\partial ^\mu }{\phi _j}} \right)
\end{align*}
with 
\begin{align}
{\left( {{\partial ^\mu }\Phi } \right)^\dag }{A_\mu }\Phi  - {\Phi ^\dag }{A_\mu }\left( {{\partial ^\mu }\Phi } \right) &=  - iA_\mu ^a{\varepsilon _{aij}}\left[ {{\phi _j}\left( {{\partial ^\mu }{\phi _i}} \right) - {\phi _i}\left( {{\partial ^\mu }{\phi _j}} \right)} \right]
\end{align}
and 
\begin{align*}
{\Phi ^\dag }{A^\mu }{A_\mu }\Phi  &= A_\mu ^a{A^{b\mu }}{\phi _i}{\left( {{T^a}} \right)_{ij}}{\left( {{T^b}} \right)_{jk}}{\phi _k}\\
 &=  - {\varepsilon _{aij}}{\varepsilon _{bjk}}{\phi _i}{\phi _k}A_\mu ^a{A^{b\mu }} \this
\end{align*}
So the Lagrangian in component form is
\begin{align*}
\mathscr{L} &= \frac{1}{2}\left( {{\partial _\mu }{\phi _a}{\partial ^\mu }{\phi _a} - {b^2}\phi _a^2} \right) + \frac{1}{2}e{\varepsilon _{aij}}A_\mu ^a\left( {{\phi _j}{\partial ^\mu }{\phi _i} - {\phi _i}{\partial ^\mu }{\phi _j}} \right) - \frac{1}{2}{e^2}{\varepsilon _{aij}}{\varepsilon _{bjk}}{\phi _i}{\phi _k}A_\mu ^a{A^{b\mu }} - \frac{1}{4}F_{\mu \nu }^a{F^{a\mu \nu }} + \\
 &+ \frac{1}{2}e{\varepsilon _{abc}}{A^{b\mu }}{A^{c\nu }}F_{\mu \nu }^a - \frac{1}{4}{e^2}{\varepsilon _{abc}}{\varepsilon _{ade}}A_\mu ^bA_\nu ^c{A^{d\mu }}{A^{e\nu }} \this
\end{align*}

Now there has been an omission from the start, we claimed that under the principle of local gauge invariance all terms which preserve the invariance of the Lagrangian are valid terms for inclusion into the Lagrangian. This allows for the inclusion of a polynomial with an infinite number of terms of the form
\begin{align*}
P(\Phi)&=\sum\limits_{n=1}^{\infty}c_n(\Phi^\dag \Phi)^n \\
&=b^2\Phi^\dag \Phi+ \frac{\lambda_4}{8}(\Phi^\dag \Phi)^2+\sum\limits_{n=3}^{\infty}c_n(\Phi^\dag \Phi)^n \this
\end{align*}
The first term we included from the start with $b$ having dimensions of mass, $[b]=M$, the second term represents a self coupling and $\lambda_4$ is dimensionless, $[\lambda_4]=1$. This quartic coupling term for the pions was left out in the original U(1) KLZ formulation. There is no legitimate reason at this stage to exclude this term so we shall include it in the Lagrangian. For terms greater than $n=3$, the coefficients $c_n$ have dimensions of inverse mass, $[c_n]=M^{-n}$. This poses a problem to our requirement for a renormalizable theory \cite{Bohmbook}. We shall thus ignore all terms for $n\ge 3$. Then finally the Lagrangian is 
\begin{align*}
\mathscr{L}_{\phi A}  &= \frac{1}{2}\left( {{\partial _\mu }{\phi _a}{\partial ^\mu }{\phi _a} - {b^2}\phi _a^2} \right) + \frac{1}{2}e{\varepsilon _{aij}}A_\mu ^a\left( {{\phi _j}{\partial ^\mu }{\phi _i} - {\phi _i}{\partial ^\mu }{\phi _j}} \right) - \frac{1}{2}{e^2}{\varepsilon _{aij}}{\varepsilon _{bjk}}{\phi _i}{\phi _k}A_\mu ^a{A^{b\mu }} - \frac{{{\lambda _4}}}{8}\left( {\phi _a^2\phi _b^2} \right) + \\
 &- \frac{1}{4}F_{\mu \nu }^a{F^{a\mu \nu }} + \frac{1}{2}e{\varepsilon _{abc}}{A^{b\mu }}{A^{c\nu }}F_{\mu \nu }^a - \frac{1}{4}{e^2}{\varepsilon _{abc}}{\varepsilon _{ade}}A_\mu ^bA_\nu ^c{A^{d\mu }}{A^{e\nu }} \this
 \label{mlesslag}
\end{align*}
Let us provide an interpretation for the remaining terms in \eqref{mlesslag}. The first two terms in the parenthesis behave like the kinetic and mass-like\footnote{These are not the couplings nor physical mass identified from experiment. Since we have not renormalized and completed the definition of the theory as yet these terms only have the dimensions of a mass and appearance of couplings.} terms of a scalar field. The third and fourth terms are the three and four point interaction terms between the scalar and vector fields. Note of course the lack of a mass-like term for the gauge field.


\chapter{Spontaneous Symmetry Breaking }


\section{Higgs Mechanism}
\lettrine[lines=4]{\initfamily\textcolor{darkblue}{A}}{s} noted in the last chapter the Lagrangian has a mass term for the pions but none for the rhos. For this quantum field theory to be a good description of reality, it would be wise to endow the particles with the appropriate masses. The term $b^2\Phi ^\dag \Phi$ for the pion was included by hand since the form of the term remained invariant under gauge transformations. We could attempt to do the same for the gauge field by including a term proportional to the form $A^{\mu} A_{\mu}$, but recalling the transformation rule for the gauge field in \eqref{Atrans} shown here below
\begin{align*}
 A{'_\mu } &= UA_\mu{U^\dag } + \frac{1}{{ie}}U{\partial _\mu }{U^\dag }
\end{align*}
We see that the product is
\begin{align}
A{'_\mu }A{'^\mu } = U{A_\mu }{A^\mu }{U^\dag } + \frac{1}{{ie}}\left[ {U{A_\mu }{\partial ^\mu }{U^\dag } + U\left( {{\partial ^\mu }{U^\dag }} \right)U{A_\mu }{U^\dag }} \right] - \frac{1}{{{e^2}}}U\left( {{\partial ^\mu }{U^\dag }} \right)U\left( {{\partial ^\mu }{U^\dag }} \right) 
\end{align}
This does not look very appealing, we could use the identity $U^\dag U =1$
\begin{align}
{\partial ^\mu }\left( {{U^\dag }U} \right) &= 0 \Rightarrow {U^\dag }\left( {{\partial ^\mu }U} \right) =  - \left( {{\partial ^\mu }{U^\dag }} \right)U  \\
{\partial ^\mu }\left( {U{U^\dag }} \right) &= 0 \Rightarrow U\left( {{\partial ^\mu }{U^\dag }} \right) =  - \left( {{\partial ^\mu }U} \right){U^\dag } 
\end{align}
and write the product as
\begin{align}
A{'_\mu }A{'^\mu } = U{A_\mu }{A^\mu }{U^\dag } - \frac{1}{{ie}}U\left[ {{A_\mu }{U^\dag }{\partial ^\mu }U + {U^\dag }\left( {{\partial ^\mu }U} \right){A_\mu }} \right]{U^\dag } + \frac{1}{{{e^2}}}{\partial ^\mu }U{\partial ^\mu }{U^\dag }
\end{align}
We could argue that this expression is not a scalar, and for the case of the Yang-Mills kinetic term it was necessary to take the trace to yield a scalar which also removed the gauge dependence. Unfortunately this does not work for a mass term as
\begin{align*}
\mathrm{Tr}\left[ {A{'_\mu }A{'^\mu }} \right] = \mathrm{Tr}\left[ {{A_\mu }{A^\mu }} \right] - \frac{2}{{ie}}\mathrm{Tr}\left[ {A_\mu {U^\dag }{\partial ^\mu }U} \right] + \frac{1}{{{e^2}}}\mathrm{Tr}\left[ {{\partial ^\mu }U^\dag{\partial ^\mu }{U }} \right] \this
\label{tracegaugemass}
\end{align*}
and \eqref{tracegaugemass} still has remnants of the gauge transformation. So we cannot put in a mass term for the gauge field by hand without breaking gauge invariance. The mass term for the gauge field must be generated dynamically \cite{Quigg2007}. This is done by introducing a new scalar field which has rotational symmetry with respect to some potential. A translation in the field is performed and the previous rotational symmetry becomes hidden. This process is referred to as the Higgs mechanism or spontaneous symmetry breaking. We note that this is not the Higgs mechanism used in the Standard Model to give all the elementary particles their masses. We use a Higgs-like mechanism here only to give mass to the rhos. Generating the vector mass using the Higgs mechanism preserves the renormalizability of the theory \cite{tHooft1971167}.

We introduce a new field $X$ which is a complex doublet that has the gauge transformation
\begin{align}
X  \to X ' = UX
\end{align}
We construct an SU(2) gauge invariant Higgs Lagrangian, with the covariant derivatives acting on the complex doublets
\begin{align}
\mathscr{L} = \left(D_\mu X\right)^\dag \left(D^\mu X\right) - V\left(X,X^\dag\right) - \kappa \left(X^\dagger X \right) \left(\Phi^\dagger\Phi\right)
\end{align}
where the potential $V\left(X,X^\dag\right)$ is 
\begin{align}
V\left(X,X^\dag\right) = \dfrac{\lambda}{8} \left(X^\dagger X \right)^2 - \dfrac{\mu^2}{2}\left(X^\dagger X \right)
\end{align}
which has been tuned for symmetry breaking. The term $\kappa \left(X^\dagger X \right) \left(\Phi^\dagger\Phi\right)$ is present since it is gauge invariant and thus allowed by the principle of local gauge invariance. $\kappa$ is the coupling between the pions and complex doublet and is dimensionless, $[\kappa]=1$. Plotting $V\left(X,X^\dag\right)$ we can see the rotational symmetry present in the potential [under rotations about an axis through $(0,0)$] by the contours which are circles.
\begin{figure}[H]
\centering
\includegraphics[width=0.7\linewidth]{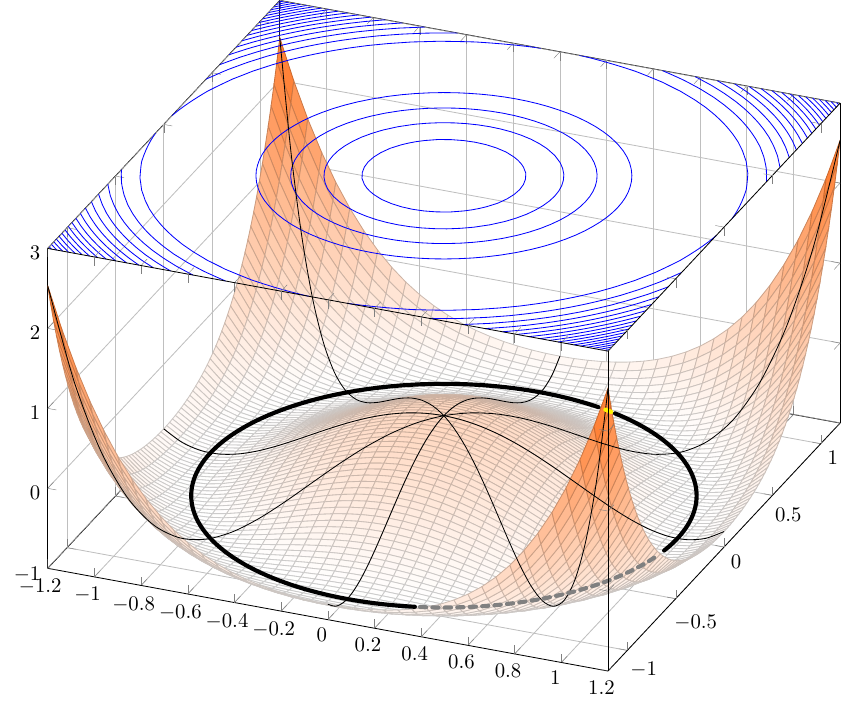}
\caption{Rotationally symmetric symmetry breaking potential.}
\label{fig:higgs-pot}
\end{figure}
We can express the potential in terms of real fields by decomposing the complex double into a doublet of real valued fields. 
\begin{align}
X = \dfrac{1}{\sqrt{2}} \begin{pmatrix}
										x_2 + i x_1 \\
										x_0 - i x_3
							\end{pmatrix}
\end{align}
so
\begin{align}
X^\dagger X 				&= \dfrac{1}{2}x_\alpha x_\alpha \\
\left(X^\dagger X\right)^2 	&= \dfrac{1}{4}x_\alpha x_\alpha x_\beta x_\beta
\end{align}
We note the usage of the repeated Greek index in the above expressions. We only resort to using the Greek index since the sum is over $\{0,1,2,3\}$. The potential in terms of the real fields is
\begin{align}
V\left(x_\alpha\right) = \dfrac{\lambda}{32}x_\alpha x_\alpha x_\beta x_\beta - \dfrac{\mu^2}{4} x_\alpha x_\alpha 
\end{align}
We now search for the minimum (the vacuum) of this potential. 
\begin{align*}
\dfrac{\partial V}{\partial x_\kappa} &= \dfrac{\lambda}{32} \left( 2 x_\alpha \delta_{\alpha \kappa} x^2_\beta + x^2_\alpha \cdot 2x_\beta\delta_{\beta\kappa}  \right) - \dfrac{\mu^2}{4} \cdot 2 x_\alpha \delta_{\alpha \kappa} \\
&= \dfrac{\lambda}{32} \left( 4 x_\kappa x^2_\alpha \right) - \dfrac{\mu^2}{2}  x_\kappa \\
&= x_\kappa \left( \dfrac{\lambda}{8} x^2_\alpha - \dfrac{\mu^2}{2}  \right) \\
&= 0 
\end{align*}
which has solutions
\begin{align}
x_\kappa=0 \qquad\text{OR}\qquad   x^2_\alpha	&=	4\dfrac{\mu^2}{\lambda}
\end{align}
Computing the second derivative to test the above solutions
\begin{align*}
\frac{{{\partial ^2}V}}{{\partial {x_\pi}\partial {x_\kappa}}} = {\delta _{\kappa \pi}}\left( {\frac{\lambda }{8}x_\alpha^2 - \frac{{{\mu ^2}}}{2}} \right) + \frac{\lambda }{8}{x_\kappa}{x_\pi}
\end{align*}
with the results along the axes $\kappa=\pi$
\begin{align*}
{\left. {\dfrac{{{\partial ^2}V}}{{\partial {x_\pi}\partial {x_\kappa}}}} \right|_{{x_\alpha = 0}}} =  - \frac{{{\mu ^2}}}{2} \le 0 ,\qquad
{\left. {\dfrac{{{\partial ^2}V}}{{\partial {x_\pi}\partial {x_\kappa}}}} \right|_{x_\alpha^2 = 4\frac{{{\mu ^2}}}{\lambda }}} = {\mu ^2} \ge 0
\end{align*}
The $x_\kappa=0$ yields a local maximum for the potential and the $x^2_\alpha	=	4\dfrac{\mu^2}{\lambda}$ solution results in the local minimum. We can see these solutions from the plot in figure \ref{fig:higgs-pot}. There is a circle of solutions which are the minima of the potential. We define 
\begin{align}
\nu^2 := \dfrac{\mu^2}{\lambda}\qquad  \chi^2_\alpha 	&=	4\nu^2
\end{align}
We shall pick one of the possible minima with 
\begin{align}
x_1 = x_2 = x_3 = 0 \qquad \text{AND} \qquad x_0=2\nu
\end{align}
The $X$ field has acquired its vacuum expectation value with the above choice of a minimum. The vacuum is located at
\begin{align}
X_\text{M} =  \dfrac{1}{\sqrt{2}} \begin{pmatrix}
0 \\
2\nu
\end{pmatrix}
\label{vevlocation}
\end{align}
Defining a new field $\chi$
\begin{align*}
\chi :=  \dfrac{1}{\sqrt{2}} \begin{pmatrix}
\chi_2 + i \chi_1 \\
\chi_0 - i \chi_3
\end{pmatrix} \this
\end{align*}
we now consider fluctuations about the minimum $X_\text{M}$ by defining
\begin{align*}
 X := \chi + X_M \this
 \label{Xchidef}
\end{align*} 
This is just translating the vacuum to the origin. 
Now we rewrite the Lagrangian in terms of this new translated field. Starting with the potential
\begin{align}
X^\dagger X &= \dfrac{1}{2}\left(\chi_\alpha\chi_\alpha +4\nu \chi_0+4\nu^2 \right)\\
\left(X^\dagger X\right)^2 &= \dfrac{1}{4}\left[ \chi_\alpha\chi_\alpha \chi_\beta\chi_\beta  + 8\nu \chi_0\chi_\alpha \chi_\alpha+16\nu^2\chi^2_0+ 8\nu^2 \chi_\alpha\chi_\alpha+32\nu^3\chi_0+16\nu^4 \right]\nonumber\\
&= \dfrac{1}{4}\chi^2_\alpha \chi^2_\beta + 2\nu\chi_0\chi^2_\alpha + 4\nu^2\chi^2_0 + 2\nu^2 \chi^2_\alpha + 8 \nu^3 \chi_0 + 4\nu^4
\end{align}
So the potential is given by
\begin{align*}
V\left( {{\chi _\alpha}} \right) &= \frac{\lambda }{8}{\left( {{X ^\dag }X } \right)^2} - \frac{{{\mu ^2}}}{2}{X^\dag }X \\
&= \frac{\lambda }{{32}}\chi _\alpha^2\chi _\beta^2 + \frac{{\nu \lambda }}{4}{\chi _0}\chi _\beta^2 + \frac{{{\mu ^2}}}{2}\chi _0^2 - \frac{1}{2}{\mu ^2}{\nu ^2} \this
\end{align*}
The pion-chi interaction becomes
\begin{align*}
\left( {{X ^\dag }X } \right)\left( {{\Phi ^\dag }\Phi } \right) = 2{\nu ^2}\phi _a^2 + 2\nu {\chi _0}\phi _a^2 + \frac{1}{2}\chi _\alpha^2\phi _a^2
\end{align*}
The covariant derivatives expanded out is
\begin{align*}
{\left( {{D_\mu }X } \right)^\dag }{D^\mu }X  
&= {\partial _\mu }{X ^\dag }{\partial ^\mu }X  + ie\left( {{\partial ^\mu }{X ^\dag }{A_\mu }X  - {X ^\dag }{A_\mu }{\partial ^\mu }X } \right) + {e^2}{X ^\dag }{A^\mu }{A_\mu }X \this
\label{higgscovder}
\end{align*}
The kinetic term of the Lagrangian is then
\begin{align*}
{\partial _\mu }{X ^\dag }{\partial ^\mu }X  &= {\partial _\mu }{\chi^\dag }{\partial ^\mu }\chi\\
&= \frac{1}{{\sqrt 2 }}\begin{pmatrix}
	{{\partial _\mu }{\chi _2} - i{\partial _\mu }{\chi _1}}&{{\partial _\mu }{\chi _0} + i{\partial _\mu }{\chi _3}}
	\end{pmatrix}
\frac{1}{{\sqrt 2 }}\begin{pmatrix}
	{{\partial _\mu }{\chi _2} + i{\partial _\mu }{\chi _1}}\\
	{{\partial _\mu }{\chi _0} - i{\partial _\mu }{\chi _3}}
	\end{pmatrix}\\
&= \frac{1}{2}{\partial _\mu }{\chi _\alpha}{\partial ^\mu }{\chi _\alpha} \this
\end{align*}
We note that the gauge field expanded over the generators are
\begin{align*}
{A_\mu } &= A_\mu ^a{T^a}\\
&= \frac{1}{2}	\begin{pmatrix}
	{A_\mu ^3}&{A_\mu ^1 - iA_\mu ^2}\\
	{A_\mu ^1 + iA_\mu ^2}&{ - A_\mu ^3}
				\end{pmatrix} \this\\\\
\implies {A_\mu }{A^\mu } &= \frac{1}{4}\begin{pmatrix}
	{A_\mu ^a{A^{a\mu }}}&0\\
	0&{A_\mu ^a{A^{a\mu }}}
	\end{pmatrix}\\
&= \frac{1}{4}A_\mu ^a{A^{a\mu }}{\mathbb{1}_{2 \times 2}}\this
\end{align*}
The four field coupling term of \eqref{higgscovder} can be expressed using the above expression as
\begin{align*}
{X ^\dag }{A^\mu }{A_\mu }X  &= {X ^\dag }\frac{1}{4}A_\mu ^a{A^{a\mu }}{\mathbb{1}_{2 \times 2}}X \\
&= \frac{1}{2}{\nu ^2}A_\mu ^a{A^{a\mu }} + \frac{1}{2}\nu {\chi _0}A_\mu ^a{A^{a\mu }} + \frac{1}{8}\chi _\alpha^2A_\mu ^a{A^{a\mu }} \this
\end{align*}
The process of picking a vacuum and translating it to the origin, thus hiding the rotational symmetry has generated a mass. The first term in the above equation represents a mass for the gauge field. Lastly we have to deal with the three field coupling term of \eqref{higgscovder}.
\begin{align*}
{\partial ^\mu }{X^\dag }{A_\mu }X &= {\partial ^\mu }{\chi ^\dag }{A_\mu }{X_{\rm{M}}} + {\partial ^\mu }{\chi ^\dag }{A_\mu }\chi \\
 &=  - {\chi ^\dag }{\partial ^\mu }{A_\mu }{X_{\rm{M}}} + {\partial ^\mu }{\chi ^\dag }{A_\mu }\chi \this
\end{align*}
We have integrated by parts (using the integral over the Lagrangian) and discarded the surface term for the first term of the above expression. Substituting the above expression into the three field coupling term
\begin{align*}
{\partial ^\mu }{X^\dag }{A_\mu }X - {X^\dag }{A_\mu }{\partial ^\mu }X &= X_{\rm{M}}^\dag {\partial ^\mu }{A_\mu }\chi  - {\chi ^\dag }{\partial ^\mu }{A_\mu }{X_{\rm{M}}} + {\partial ^\mu }{\chi ^\dag }{A_\mu }\chi  - {\chi ^\dag }{A_\mu }{\partial ^\mu }\chi \\
 &= {\partial ^\mu }A_\mu ^a\left[ {X_{\rm{M}}^\dag {T^a}\chi  - {\chi ^\dag }{T^a}{X_{\rm{M}}}} \right] + {\partial ^\mu }{\chi ^\dag }{A_\mu }\chi  - {\chi ^\dag }{A_\mu }{\partial ^\mu }\chi \this
 \label{threefieldAchichi}
\end{align*}
we see a separation into two types of interaction terms. The last two terms are three field interaction terms but the term in the 
square brackets seems to have two field interaction terms i.e. mixing terms. Let us check
\begin{align*}
{\chi^\dag }{\partial ^\mu }{A_\mu } X_{\rm{M}} &=  \frac{1}{{\sqrt 2 }}\begin{pmatrix}
																			{{\chi _2} - i{\chi _1}}&{{\chi _0} + i{\chi _3}}
																		\end{pmatrix} \frac{1}{2}  {\partial ^\mu } \begin{pmatrix}
																	{A_\mu ^3}&{A_\mu ^1 - iA_\mu ^2}\\
																	{A_\mu ^1 + iA_\mu ^2}&{ - A_\mu ^3}
																	\end{pmatrix} \begin{pmatrix}
																								0\\
																								1
																				  \end{pmatrix}\\
&=   \frac{1}{{2\sqrt 2 }}\left[ {\left( {{\chi _2}{\partial ^\mu }A_\mu ^1 - {\chi _1}{\partial ^\mu }A_\mu ^2 - {\chi _0}{\partial ^\mu }A_\mu ^3} \right) - i\left( {{\chi _2}{\partial ^\mu }A_\mu ^2 + {\chi _1}{\partial ^\mu }A_\mu ^1 + {\chi _3}{\partial ^\mu }A_\mu ^3} \right)} \right]\\
&= \frac{1}{{2\sqrt 2 }}\left[ {-i{\chi _a}{\partial ^\mu }A_\mu ^a + \left( {{\chi _2}{\partial ^\mu }A_\mu ^1 - {\chi _1}{\partial ^\mu }A_\mu ^2 - {\chi _0}{\partial ^\mu }A_\mu ^3} \right)} \right] \this
\label{partmixterms}
\end{align*}

Taking the hermitian conjugate of \eqref{partmixterms} and substituting into the two field interaction terms of \eqref{threefieldAchichi} we get 
%
\begin{align*}
X_{\rm{M}}^\dag {\partial ^\mu }{A_\mu }\chi  - {\chi ^\dag }{\partial ^\mu }{A_\mu }{X_{\rm{M}}}&= \frac{i}{{\sqrt 2 }}{\chi _a}{\partial ^\mu }A_\mu ^a \this
\label{mixingtermsAchi}
\end{align*}
which indeed has only two fields interacting a point. These pesky terms are hard to interpret so we will leave them alone for now and come back to them when we discuss gauge fixing. The remaining term to work on is the three field interaction terms of \eqref{threefieldAchichi}. First the ${A_\mu }\chi$ product
\begin{align*}
{A_\mu }\chi &= \frac{1}{2} \begin{pmatrix}
	{A_\mu ^3}&{A_\mu ^1 - iA_\mu ^2}\\
	{A_\mu ^1 + iA_\mu ^2}&{ - A_\mu ^3}
	\end{pmatrix}\frac{1}{{\sqrt 2 }}\begin{pmatrix}
	{{\chi _2} + i{\chi _1}}\\
	{{\chi _0} - i{\chi _3}}
	\end{pmatrix}\\
&= \frac{1}{{2\sqrt 2 }}\begin{pmatrix}
	{A_\mu ^3{\chi _2} + A_\mu ^1{\chi _0} - A_\mu ^2{\chi _3} + i\left( {A_\mu ^3{\chi _1} - A_\mu ^1{\chi _3} - A_\mu ^2{\chi _0}} \right)}\\
	{A_\mu ^1{\chi _2} + A_\mu ^2{\chi _1} - A_\mu ^3{\chi _0} + i\left( {A_\mu ^1{\chi _1} + A_\mu ^2{\chi _2} + A_\mu ^3{\chi _3}} \right)}
	\end{pmatrix} \this
\end{align*}
then the three field interaction term is
\begin{align*}
&{\partial ^\mu }{\chi^\dag }{A_\mu }\chi \\
&= \frac{1}{{\sqrt 2 }}
 \begin{pmatrix}
		{{\partial ^\mu }{\chi _2} - i{\partial ^\mu }{\chi _1}}&{{\partial ^\mu }{\chi _0} + i{\partial ^\mu }{\chi _3}}
\end{pmatrix}    \frac{1}{{2\sqrt 2 }} 
\begin{pmatrix}
	{A_\mu ^3{\chi _2} + A_\mu ^1{\chi _0} - A_\mu ^2{\chi _3} + i\left( {A_\mu ^3{\chi _1} - A_\mu ^1{\chi _3} - A_\mu ^2{\chi _0}} \right)}\\
	{A_\mu ^1{\chi _2} + A_\mu ^2{\chi _1} - A_\mu ^3{\chi _0} + i\left( {A_\mu ^1{\chi _1} + A_\mu ^2{\chi _2} + A_\mu ^3{\chi _3}} \right)}
\end{pmatrix}\\
&= \frac{1}{4}\bigg\{ \left( {A_\mu ^3{\chi _2} + A_\mu ^1{\chi _0} - A_\mu ^2{\chi _3}} \right){\partial ^\mu }{\chi _2} + \left( {A_\mu ^3{\chi _1} - A_\mu ^1{\chi _3} - A_\mu ^2{\chi _0}} \right){\partial ^\mu }{\chi _1} +\\
&+ i\left[ {{\partial ^\mu }{\chi _2}\left( {A_\mu ^3{\chi _1} - A_\mu ^1{\chi _3} - A_\mu ^2{\chi _0}} \right) - \left( {A_\mu ^3{\chi _2} + A_\mu ^1{\chi _0} - A_\mu ^2{\chi _3}} \right){\partial ^\mu }{\chi _1}} \right] + \\
&+  \left( {A_\mu ^1{\chi _2} + A_\mu ^2{\chi _1} - A_\mu ^3{\chi _0}} \right){\partial ^\mu }{\chi _0} - \left( {A_\mu ^1{\chi _1} + A_\mu ^2{\chi _2} + A_\mu ^3{\chi _3}} \right){\partial ^\mu }{\chi _3} +\\
&+ i\left[ {{\partial ^\mu }{\chi _0}\left( {A_\mu ^1{\chi _1} + A_\mu ^2{\chi _2} + A_\mu ^3{\chi _3}} \right) + \left( {A_\mu ^1{\chi _2} + A_\mu ^2{\chi _1} - A_\mu ^3{\chi _0}} \right){\partial ^\mu }{\chi _3}} \right] \bigg\} \this
\end{align*}
The appearance of the individual $\chi_\alpha$ fields interacting differently is troubling. Taking the hermitian conjugate of the above and substituting into the three field interaction terms of \eqref{threefieldAchichi}
\begin{align*}
{\partial ^\mu }{\chi^\dag }{A_\mu }\chi - {\chi^\dag }{A_\mu }{\partial ^\mu }\chi 
&= \frac{i}{2}\bigg[ \left( {\chi _1}{\partial ^\mu }{\chi _2} - {\chi _2}{\partial ^\mu }{\chi _1} \right)A_\mu ^3 + \left( {\chi _3}{\partial ^\mu }{\chi _0} - {\chi _0}{\partial ^\mu }{\chi _3} \right)A_\mu ^3  + \\
&+\,\,\, \quad  \left( {{\chi _2}{\partial ^\mu }{\chi _3} - {\chi _3}{\partial ^\mu }{\chi _2}} \right)A_\mu ^1 + \left( {{\chi _1}{\partial ^\mu }{\chi _0} - {\chi _0}{\partial ^\mu }{\chi _1}} \right)A_\mu ^1  + \\
&+\,\,\, \quad  \left( {{\chi _2}{\partial ^\mu }{\chi _0} - {\chi _0}{\partial ^\mu }{\chi _2}} \right)A_\mu ^2 + \left( {{\chi _3}{\partial ^\mu }{\chi _1} - {\chi _1}{\partial ^\mu }{\chi _3}} \right)A_\mu ^2 \bigg] \this
\label{needclarity}
\end{align*}
A pattern emerges in the way the individual $\chi_\alpha$ fields interact with the gauge fields. To clarify we define the operator
\begin{align}
A \overleftrightarrow{\partial^\mu} B := A{\partial ^\mu }B - B{\partial ^\mu }A
\end{align}
Rewriting \eqref{needclarity} in terms of the above operator
\begin{align*}
{\partial ^\mu }{\chi^\dag }{A_\mu }\chi - {\chi^\dag }{A_\mu }{\partial ^\mu }\chi &= 
\frac{i}{2}\bigg[ \left( {\chi _1}\overleftrightarrow{\partial^\mu}{\chi _2}\right)A_\mu ^3 + \left( {\chi _3}\overleftrightarrow{\partial^\mu}{\chi _0} \right)A_\mu ^3  + \\
&+\,\,\, \quad  \left( {\chi _2}\overleftrightarrow{\partial^\mu}{\chi _3}\right)A_\mu ^1 + \left( {\chi _1}\overleftrightarrow{\partial^\mu}{\chi _0}\right)A_\mu ^1  + \\
&+\,\,\, \quad  \left( {\chi _2}\overleftrightarrow{\partial^\mu}{\chi _0} \right)A_\mu ^2 + \left( {\chi _3}\overleftrightarrow{\partial^\mu}{\chi _1} \right)A_\mu ^2 \bigg]\\
& = \frac{i}{2}\left[ {\frac{1}{2}{\varepsilon _{abc}}\left( {{\chi _a}\overleftrightarrow{\partial^\mu}{\chi _b}} \right)A_\mu ^c + \left( {{\chi _a}\overleftrightarrow{\partial^\mu}{\chi _0}} \right)A_\mu ^a} \right] \this 
\label{threechiAlevi}
\end{align*}
The pattern for the $\chi_\alpha$ fields interacting allowed us to write the three field interactions compactly in terms of the 
Levi-Civita symbol, with the $\chi_1, \chi_2,\chi_3$ interacting among themselves with the gauge fields and the separate interaction term 
with the $\chi_0$ field. We have all the needed pieces, substituting \eqref{mixingtermsAchi} and \eqref{threechiAlevi} into 
\eqref{threefieldAchichi}
\begin{align*}
{\partial ^\mu }{X^\dag }{A_\mu }X - {X^\dag }{A_\mu }{\partial ^\mu }X &= {\partial ^\mu }A_\mu ^a\left[ {X_{\rm{M}}^\dag {T^a}\chi  - {\chi ^\dag }{T^a}{X_{\rm{M}}}} \right] + {\partial ^\mu }{\chi ^\dag }{A_\mu }\chi  - {\chi ^\dag }{A_\mu }{\partial ^\mu }\chi \\
&={\partial ^\mu }A_\mu ^a\left[ {X_{\rm{M}}^\dag {T^a}\chi  - {\chi ^\dag }{T^a}{X_{\rm{M}}}} \right]+ \frac{i}{2}\left[ {\frac{1}{2}{\varepsilon _{abc}}\left( {{\chi _a}\overleftrightarrow{\partial^\mu}{\chi _b}} \right)A_\mu ^c + \left( {{\chi _a}\overleftrightarrow{\partial^\mu}{\chi _0}} \right)A_\mu ^a} \right] \this
\end{align*}

The covariant derivative in terms of the real $\chi_\alpha$ fields are
\begin{align*}
{\left( {{D_\mu }X } \right)^\dag }{D^\mu }X  &= {\partial _\mu }{\chi ^\dag }{\partial ^\mu }\chi  + ie\left( {{\partial ^\mu }{\chi ^\dag }{A_\mu }\chi  - {\chi ^\dag }{A_\mu }{\partial ^\mu }\chi } \right) + {e^2}{\chi ^\dag }{A^\mu }{A_\mu }\chi \\
&= \frac{1}{2}{\partial _\mu }{\chi _\alpha}{\partial ^\mu }{\chi _\alpha} + ie{\partial ^\mu }A_\mu ^a\left[ {X_{\rm{M}}^\dag {T^a}\chi  - {\chi ^\dag }{T^a}{X_{\rm{M}}}} \right] - \frac{1}{4}e{\varepsilon _{abc}}\left( {{\chi _a}\overleftrightarrow{\partial^\mu}{\chi _b}} \right)A_\mu ^c - \frac{1}{2}e\left( {{\chi _a}\overleftrightarrow{\partial^\mu}{\chi _0}} \right)A_\mu ^a + \\
&+ \frac{1}{2}{e^2}{\nu ^2}A_\mu ^a{A^{a\mu }} 
+ \frac{1}{2}{e^2}\nu {\chi _0}A_\mu ^a{A^{a\mu }} + \frac{1}{8}{e^2}\chi _\alpha^2A_\mu ^a{A^{a\mu }} \this
\end{align*}
Substituting in the covariant derivative, the potential and the $X^\dag X\Phi^\dag \Phi$ interaction terms into the Lagrangian after symmetry breaking
\begin{align*}
\mathscr{L}_H &= {\left( {{D_\mu }X } \right)^\dag }{D^\mu }X  - V\left(X,X^\dag \right) - \kappa \left( {{X^\dag }X } \right)\left( {{\Phi ^\dag }\Phi } \right)\\
&= \frac{1}{2}\left( {{\partial _\mu }{\chi _0}{\partial ^\mu }{\chi _0} - {\mu ^2}\chi _0^2} \right) + \frac{1}{2}{\partial _\mu }{\chi _a}{\partial ^\mu }{\chi _a} + ie{\partial ^\mu }A_\mu ^a\left[ {X_{\rm{M}}^\dag {T^a}\chi  - {\chi ^\dag }{T^a}{X_{\rm{M}}}} \right]  - \frac{1}{4}e{\varepsilon _{abc}}\left( {{\chi _a}\overleftrightarrow{\partial^\mu}{\chi _b}} \right)A_\mu ^c + \\
&- \frac{1}{2}e\left( {{\chi _a}\overleftrightarrow{\partial^\mu}{\chi _0}} \right)A_\mu ^a 
+ \frac{1}{2}{e^2}{\nu ^2}A_\mu ^a{A^{a\mu }} + \frac{1}{2}{e^2}\nu {\chi _0}A_\mu ^a{A^{a\mu }} + \frac{1}{8}{e^2}\chi _\alpha^2A_\mu ^a{A^{a\mu }} - \frac{\lambda }{{32}}\chi _\alpha^2\chi _\beta^2 - \frac{{\nu \lambda }}{4}{\chi _0}\chi _\alpha^2 - \frac{1}{2}\left( {4\kappa {\nu ^2}} \right){\phi^2_a }+ \\
&- 2\kappa \nu {\chi _0}{\phi^2_a } - \frac{1}{2}\kappa \chi _\alpha^2{\phi^2_a }  + \frac{1}{2}{\mu ^2}{\nu ^2} \this
\end{align*}
Due to our choice of the vacuum along the $\chi_0$ direction, the $\chi_0$ field behaves differently from the $\chi_1,\chi_2,\chi_3$ terms. We will separate the $\chi_0$ component out with
\begin{equation}
  \begin{aligned}
    \chi _\alpha^2 &= \chi _0^2 + \chi _a^2 \\
    \chi _\alpha^2\chi _\beta^2 &= \chi _0^4 + 2\chi _0^2\chi _a^2 + \chi _a^2\chi _b^2
  \end{aligned}
  \label{separatehiggsout}
\end{equation}
and make a cosmetic change in relabeling the $\chi_0$ component with ${\chi _0}: = H$ which we shall refer to for convenience as the Higgs field. A reminder again, this is not the Standard Model Higgs field. 
%
\begin{align*}
\mathscr{L}_H &= \frac{1}{2}\left( {{\partial _\mu }H{\partial ^\mu }H - {\mu ^2}{H^2}} \right) + \frac{1}{2}{\partial _\mu }{\chi _a}{\partial ^\mu }{\chi _a} + ie{\partial ^\mu }A_\mu ^a\left[ {X_{\rm{M}}^\dag {T^a}\chi  - {\chi ^\dag }{T^a}{X_{\rm{M}}}} \right] - \frac{1}{4}e{\varepsilon _{abc}}\left( {{\chi _a}\overleftrightarrow{\partial^\mu}{\chi _b}} \right)A_\mu ^c + \\
&- \frac{1}{2}e\left( {{\chi _a}\overleftrightarrow{\partial^\mu}H} \right)A_\mu ^a +
 \frac{1}{2}{e^2}{\nu ^2}A_\mu ^a{A^{a\mu }} + \frac{1}{2}{e^2}\nu HA_\mu ^a{A^{a\mu }} + \frac{1}{8}{e^2}{H^2}A_\mu ^a{A^{a\mu }} + \frac{1}{8}{e^2}\chi _b^2A_\mu ^a{A^{a\mu }} - \frac{\lambda }{{32}}{H^4} - \frac{\lambda }{{16}}{H^2}\chi _a^2 + \\
&- \frac{\lambda }{{32}}\chi _a^2\chi _b^2 - \frac{{\nu \lambda }}{4}{H^3} - \frac{{\nu \lambda }}{4}H\chi _a^2 - \frac{1}{2}\left( {4\kappa {\nu ^2}} \right){\phi^2_a }  - 2\kappa \nu H{\phi^2_a }  - \frac{1}{2}\kappa {H^2}{\phi^2_a }  - \frac{1}{2}\kappa \chi _a^2{\phi^2_a }  + \frac{1}{2}{\mu ^2}{\nu ^2} \this 
\label{higgslagrange}
\end{align*}
Some terms of \eqref{higgslagrange} require some special attention, we point out the entire process of spontaneous symmetry breaking was 
to generate a mass for the gauge field. This was achieved with the generation of the term $\frac{1}{2}{e^2}{\nu ^2}A_\mu ^a{A^{a\mu }}$. 
A contribution to the mass of the pion was generated during the symmetry breaking process, this depended on the coupling $\kappa$ as  
$\frac{1}{2}\left( {4\kappa {\nu ^2}} \right){\phi^2_a }$. A mass term for the Higgs field was also generated during this process with 
the mass term being $\frac{1}{2}{\mu ^2}{H^2}$. The $\chi_a$ are massless and are the three Goldstone fields.  The rotational symmetry is 
still present in the Lagrangian, it is hidden from casual inspection. We could reverse the process and translate the potential back to 
its original configuration. The rotational symmetry would be made explicit again. We noted the appearance of the pesky mixing terms which 
are present in the above Lagrangian. 

\chapter{Quantization }

\section{Faddeev-Popov Ghosts}

\numberwithin{equation}{section}


\lettrine[lines=4]{\initfamily\textcolor{darkblue}{T}}{he} classical Lagrangian is ready for quantization. There is an obvious problem at 
the start. Since this is a gauge theory, there are 
infinitely many field configurations which are related to each other via a gauge transformation which led to equivalent states. This 
leads to an over-counting of the physical states and must be dealt with as it leads to an overall multiplicative divergence. We can 
extract out the contribution from the physically distinct states via a suitable gauge fixing process. This means that the gauge 
transformation partitions the configuration space of gauge fields and sets up an equivalence class for sets of physically equivalent 
gauge fields. This leads to a problem when summing over all field contributions as there are infinitely many physically equivalent 
field configurations leading to a divergence. So we need to count only one member from each partition once and this is done with a gauge 
fixing function which is designed to traverse the configuration space and intersect the set of all physically equivalent gauge fields only once. This is the method pioneered by Feynman and formalized by Faddeev and Popov \cite{FADDEEV1967}. Consider a 
function $g(x)$ and some $w\in \mathbb{R}$ with
\begin{align}
g(x)=w
\end{align}
which has roots at $x_k$ which can be ordered as ${x_1} < {x_2} <  \cdots  < {x_N}$ and $g'\left( {{x_i}} \right) \ne 0$, then 
\begin{align}
\int {dx\delta \left[ {g\left( x \right) - w} \right]} =  \sum\limits_{k = 1}^N {\frac{1}{{\left| {g'\left( {{x_k}} \right)} \right|}}} 
\end{align}
we have the one dimensional identity
\begin{align}
{\left[ {\sum\limits_{k = 1}^N {\frac{1}{{\left| {g'\left( {{x_k}} \right)} \right|}}} } \right]^{ - 1}}\int {dx\delta \left[ {g\left( x \right) - w} \right]}  = 1
\end{align}
We restrict the class of functions $g(x)$ to those which yield only a single root and this simplifies to 
\begin{align}
{\left| {\frac{d}{{dx}}g\left( x \right)} \right|_{x = a}}\int {dx\delta \left[ {g\left( x \right) - w} \right]}  = 1
\end{align}
This can be generalized to the field theoretic version
\begin{align}
\Delta \left[ A \right]\int {\mathcal{D}\mu [U]\delta \left[ {{G^a}\left( {{A^U}} \right) - {w^a}} \right]}  = 1
\label{fpid}
\end{align}
where $\Delta \left[ A \right]$ is a generalization of the derivative of the function $g(x)$, and will later turn out to be a 
Jacobian matrix. $A^U$ indicates the dependence of the gauge field on the transformation $U$. $\mathcal{D}\mu [U]$ is the Haar 
measure \cite{Bohmbook}. The condition placed on $g(x)$ earlier was to select injective mappings as candidates for gauge fixing functions ${G^a}\left( 
{{A^U}} \right)$. This is the problem of the Gribov Ambiguity \cite{ESPOSITO,Ghiotti2007}. For perturbative considerations we will stay within one horizon.

Consider now the path integral for the gauge field
\begin{align*}
Z' &= \int {\mathcal{D}A{e^{iS\left[ A \right]}}} \\
&= \int {\mathcal{D}{A^U}{e^{iS\left[ {{A^U}} \right]}} \cdot 1} \\
&= \int {\mathcal{D}{A^U}{e^{iS\left[ {{A^U}} \right]}}\Delta \left[ {{A^U}} \right]\int {\mathcal{D}\mu [U]\delta \left[ {{G^a}\left( {{A^U}} \right) - {w^a}} \right]} } \this
\end{align*}
We have inserted a $1$ using the identity \eqref{fpid}. We can now make a gauge transformation where ${A^U} \to A$ and noting the Haar measure is invariant under gauge transformations
\begin{align*}
Z' &= \int {\mathcal{D}A{e^{iS\left[ A \right]}}\Delta \left[ A \right]\int {\mathcal{D}\mu [U]\delta \left[ {{G^a}\left( A \right) - {w^a}} \right]} } \\
&= \int {\mathcal{D}\mu [U]} \int {\mathcal{D}A{e^{iS\left[ A \right]}}\Delta \left[ A \right]\delta \left[ {{G^a}\left( A \right) - 
{w^a}\left( x \right)} \right]} 
\end{align*}
We can insert a $1$ in the form of a ratio of Gaussian functionals and absorb the denominator into the measure. 
\begin{align*}
Z' &= \int {\mathcal{D}A{e^{iS\left[ A \right]}}\Delta \left[ A \right]\int {\mathcal{D}\mu [U]\delta \left[ {{G^a}\left( A \right) - 
{w^a}} \right]} } \\
 &= \int {\mathcal{D}\mu [U]} \dfrac{{\int {\mathcal{D}w{e^{ - \frac{i}{{2\xi }}\int {{d^4}x} w_a^2\left( x \right)}}} }}{{\int {\mathcal{D}w{e^{ - \frac{i}{{2\xi }}\int {{d^4}x} w_a^2\left( x \right)}}} }}\int {\mathcal{D}A{e^{iS\left[ A \right]}}\Delta \left[ A \right]\delta \left[ {{G^a}\left( A \right) - {w^a}\left( x \right)} \right]} \\
 &= \int {\mathcal{D}\widetilde \mu [U]} \int {\mathcal{D}w{e^{ - \frac{i}{{2\xi }}\int {{d^4}x} w_a^2\left( x \right)}}} \int {\mathcal{D}A{e^{iS\left[ A \right]}}\Delta \left[ A \right]\delta \left[ {{G^a}\left( A \right) - {w^a}\left( x \right)} \right]} \this
\end{align*}
Since all the fields are independent of the gauge transformation we can factor out the integral over the measure which contributes an overall multiplicative divergence. Note also that $Z'$ can not depend on the parameter $\xi$, since only a $1$ was inserted.
\begin{align}
\frac{{Z'}}{{\int {\mathcal{D}\widetilde \mu [U]} }} = \int {\mathcal{D}A\mathcal{D}w{e^{iS\left[ A \right]}}{e^{ - \frac{i}{{2\xi }}\int {{d^4}x} w_a^2\left( x \right)}}\Delta \left[ A \right]\delta \left[ {{G^a}\left( A \right) - {w^a}\left( x \right)} \right]} 
\end{align}
We can now define the path integral which does not suffer from the over-counting
\begin{align*}
\mathcal{Z} &:= \frac{{Z'}}{{\int {\mathcal{D}\widetilde \mu [U]} }}\\
 &= \int {\mathcal{D}A\mathcal{D}w{e^{iS\left[ A \right]}}{e^{ - \frac{i}{{2\xi }}\int {{d^4}x} w_a^2\left( x \right)}}\Delta \left[ A \right]\delta \left[ {{G^a}\left( A \right) - {w^a}\left( x \right)} \right]} \\
 &= \int {\mathcal{D}A{e^{iS\left[ A \right]}}{e^{ - \frac{i}{{2\xi }}\int {{d^4}x} G_a^2\left( A \right)}}\Delta \left[ A \right]} \this
  \label{pathintegral}
\end{align*}
where we have integrated over the $w$ using the delta functional. We are left with the task of evaluating $\Delta[A]$. Recall the identity
in \eqref{fpid}
\[\Delta \left[ A \right]\int {\mathcal{D}\mu [U]\delta \left[ {{G^a}\left( {{A^U}} \right) - {w^a}} \right]}  = 1\]
with the transformation $U = {e^{i{\alpha ^a}{T^a}}}$, we see that the invariant Haar measure $\mathcal{D}\mu [U] \propto 
\mathcal{D}\alpha$. We can rewrite the identity as
\begin{align}
\Delta \left[ A \right]\int {\mathcal{D}\alpha \delta \left[ {{G^a}\left( {^\alpha A} \right) - {w^a}} \right]}  = 1
\end{align}
We note the change in notation from $A^\alpha$(which looks like the components of a Lorentz vector) to $^\alpha A$(which serves to indicate the dependence of the gauge field on the group parameters $\alpha$). We can make a change of variables 
\begin{align*}
\mathcal{D}G &= \det \left[ {\frac{{\delta {G^b}\left[ {^\alpha A;x} \right]}}{{\delta {\alpha ^c}\left( y \right)}}} 
\right]\mathcal{D}\alpha \\
\implies \mathcal{D}\alpha  &= \mathcal{D}G \cdot \frac{1}{{\det \left[ {\frac{{\delta {G^b}\left[ {^\alpha A;x} \right]}}{{\delta {\alpha ^c}\left( y \right)}}} \right]}} \this
\end{align*}
Substituting in the new measure
\begin{align*}
\Delta \left[ A \right]\int {\mathcal{D}G \cdot \dfrac{1}{{\det \left[ {\frac{{\delta {G^b}\left[ {^\alpha A;x} \right]}}{{\delta {\alpha ^c}\left( y \right)}}} \right]}}\delta \left[ {{G^b}\left( {^\alpha A} \right) - {w^b}} \right]}  &= 1\\
\Delta \left[ A \right] \left. \dfrac{1}{\det  {\left[ {\dfrac{{\delta {G^b}\left[ {^\alpha A;x} \right]}}{{\delta {\alpha ^c}\left( y \right)}}} \right]}} \right|_{ {G^b}\left( {^\alpha A} \right) = {w^b} }  &= 1\\ 
\Delta \left[ A \right] &= \det {\left. {\left[ {\frac{{\delta {G^b}\left[ {^\alpha A;x} \right]}}{{\delta {\alpha ^c}\left( y \right)}}} \right]} \right|_{{G^b}\left( {^\alpha A} \right) = {w^b}}}\\
 &= \det {M^{bc}}\left( {x,y} \right) \this
\end{align*}
We observe that $\Delta[A]$ is the determinant of a matrix of variational derivatives of the gauge fixing function with respect to the group parameters and we defined
\begin{align}
{M^{bc}}\left( {x,y} \right) &:= \frac{{\delta {G^b}\left[ {^\alpha A;x} \right]}}{{\delta {\alpha ^c}\left( y \right)}}
\end{align}
We can choose a class of gauge fixing functions \cite{FujikawaLee,Abers19731}
\begin{align}
{G^b}\left[ {^\alpha A;x} \right] = {\partial ^\mu }A_\mu ^b + \xi ie\left( {X_{\rm{M}}^\dag {T^b}\chi - {\chi^\dag }{T^b}X_{\rm{M}}} \right) 
\label{gffunction}
\end{align}
We shall justify the choice of \eqref{gffunction} in the next section. The derivative of the gauge fixing function with respect to the group parameters is 
\begin{align*}
\frac{{\delta {G^b}\left[ {^\alpha A;x} \right]}}{{\delta {\alpha ^c}\left( y \right)}} &= \int {{d^4}z\left[ {\frac{\delta }{{\delta A_\nu ^d\left( z \right)}}{\partial ^\mu }A_\mu ^b\left( x \right)} \right]\frac{{\delta A_\nu ^d\left( z \right)}}{{\delta {\alpha ^c}\left( y \right)}}}  + \frac{\delta }{{\delta {\alpha ^c}\left( y \right)}}\xi ie\left( {X_{\rm{M}}^\dag {T^b}\chi - {\chi^\dag }{T^b}X_{\rm{M}}} \right) \\
&= \int {{d^4}z\partial _x^\mu {\delta _{bd}}\delta \left( {x - z} \right)g_\mu ^\nu\frac{{\delta A_\nu ^d\left( z \right)}}{{\delta {\alpha ^c}\left( y \right)}}}  + \frac{\delta }{{\delta {\alpha ^c}\left( y \right)}}\xi ie\left( {X_{\rm{M}}^\dag {T^b}\chi - {\chi^\dag }{T^b}X_{\rm{M}}} \right) \this \label{dGFfunc}
\end{align*}

We now turn to finding the variation of the gauge field with respect to the group parameters. Recall that the gauge field transforms as 
\begin{align*}
{{A'}_\mu } &= \frac{1}{{ie}}U{\partial _\mu }{U^\dag } + U{A_\mu }{U^\dag }\\
U &= {e^{i{\alpha ^a}{T^a}}} \this\label{gtrans}
\end{align*}
Since we are looking for the variational derivative we need only concern ourselves with working with the infinitesimal transformation, so for $\delta {\alpha _a} \ll 1$
\begin{align}
U &\simeq 1 + i\delta {\alpha ^a}{T^a} + \mathcal{O}\left( {\delta \alpha _a^2} \right)
\end{align}
The gauge field transforms as 
\begin{align*}
{{A'}_\mu } &\simeq \frac{1}{{ie}}\left( {1 + i\delta {\alpha ^a}{T^a}} \right){\partial _\mu }\left( {1 - i\delta {\alpha ^b}{T^b}} \right) + \left( {1 + i\delta {\alpha ^a}{T^a}} \right){A_\mu }\left( {1 - i\delta {\alpha ^b}{T^b}} \right) + \mathcal{O}\left( {\delta \alpha _a^2} \right)\\
 &=  - \frac{1}{e}\left( {1 + i\delta {\alpha ^a}{T^a}} \right){\partial _\mu }\delta {\alpha ^b}{T^b} + \left( {1 + i\delta {\alpha ^a}{T^a}} \right)\left( {{A_\mu } - i\delta {\alpha ^b}{A_\mu }{T^b}} \right) + \mathcal{O}\left( {\delta \alpha _a^2} \right)\\
 &= {A_\mu } - \frac{1}{e}{\partial _\mu }\delta {\alpha ^a}{T^a} - i\delta {\alpha ^a}A_\mu ^b{T^b}{T^a} + i\delta {\alpha ^a}{T^a}A_\mu ^b{T^b} + \mathcal{O}\left( {\delta \alpha _a^2} \right)\\
 &= {A_\mu } - \frac{1}{e}{\partial _\mu }\delta {\alpha ^a} {\delta ^{ac}} {T^c} + i\delta {\alpha ^a}A_\mu ^b\left[ {{T^a},{T^b}} \right] + \mathcal{O}\left( {\delta \alpha _a^2} \right) \this
\end{align*}
Using the algebra of the group $\left[ {{T^a},{T^b}} \right] = i{\varepsilon _{abc}}{T^c}$.
\begin{align*}
{{A'}_\mu } &= {A_\mu } - \frac{1}{e}{\partial _\mu }\delta {\alpha ^a}{\delta ^{ac}}{T^c} - {\varepsilon _{abc}}{T^c}\delta {\alpha ^a}A_\mu ^b + \mathcal{O}\left( {\delta \alpha _a^2} \right)\\
 &= {A_\mu } - \frac{1}{e}\left[ {{\delta ^{ac}}{\partial _\mu } + e{\varepsilon _{abc}}A_\mu ^b} \right]\delta {\alpha ^a}{T^c} + \mathcal{O}\left( {\delta \alpha _a^2} \right)\\
 &= {A_\mu } - \frac{1}{e}D_\mu ^{ac}\delta {\alpha ^a}{T^c} + \mathcal{O}\left( {\delta \alpha _a^2} \right) \this
\end{align*}
We recognise the covariant derivative $D_\mu ^{ac}$ in the adjoint representation. Expanding the gauge field over the generators
\begin{align}
A^{\prime c}_\mu {T^c} = A_\mu ^c{T^c} - \frac{1}{e}D_\mu ^{ac}\delta {\alpha ^a}{T^c} + \mathcal{O}\left( {\delta \alpha _a^2} \right)
\end{align}
Extracting only the components
\begin{align*}
A^{\prime c}_\mu &= A_\mu ^c - \frac{1}{e}D_\mu ^{ac}\delta {\alpha ^a} + \mathcal{O}\left( {\delta \alpha _a^2} \right) \this\\
\delta A_\nu ^d\left( z \right) &:=  A^{\prime d}_\nu \left( z \right) - A_\nu ^d\left( z \right)\\
 &=  - \frac{1}{e}D_\nu ^{ad}\delta {\alpha ^a}\left( z \right) + \mathcal{O}\left( {\delta \alpha _\nu ^2} \right) \this
\end{align*}
We can now construct the variational derivative, 
\begin{align*}
\dfrac{{\delta A_\nu ^d\left( z \right)}}{{\delta {\alpha ^c}\left( y \right)}} &=  - \frac{1}{e}D_\nu ^{ad}\frac{{\delta {\alpha ^a}\left( z \right)}}{{\delta {\alpha ^c}\left( y \right)}}\\
 &=  - \frac{1}{e}D_\nu ^{ad}{\delta ^{ac}}\delta \left( {z - y} \right)\\
 &=  - \frac{1}{e}D_\nu ^{cd}\delta \left( {z - y} \right) \this \label{dAda}
\end{align*}

Next is the variation of the Higgs and Goldstone fields. Recall the definition of the translated field in \eqref{Xchidef}
from symmetry breaking shown below
\begin{align*}
X: = \chi  + {X_{\rm{M}}}
\end{align*}
Under a gauge transformation this transforms \cite{ChengLi,Weinberg1996QFT2} as
\begin{align}
X' = \chi ' + {X_{\rm{M}}}
\end{align}
with $X' = UX$. So we can deduce the transformation rule for the translated field $\chi'$
\begin{align*}
\chi ' &= UX - {X_{\rm{M}}}\\
 &= U\chi  + \left( {U - 1} \right){X_{\rm{M}}} \this
\end{align*}
For infinitesimal transformations
\begin{align*}
\delta \chi &: = \chi ' - \chi \\
 &= i\delta {\alpha ^c}{T^c}\left( {\chi  + {X_{\rm{M}}}} \right) \this
\end{align*}
The variation of the Higgs and Goldstone fields are 
\begin{align}
\delta \left( {X_{\rm{M}}^\dag {T^b}\chi  - {\chi ^\dag }{T^b}{X_{\rm{M}}}} \right) &= X_{\rm{M}}^\dag {T^b}\delta \chi  - \delta {\chi ^\dag }{T^b}{X_{\rm{M}}} \nonumber\\
 &= i\delta {\alpha ^c}\left[ {X_{\rm{M}}^\dag {T^b}{T^c}\left( {\chi  + {X_{\rm{M}}}} \right) + \left( {{\chi ^\dag } + X_{\rm{M}}^\dag } \right){T^c}{T^b}{X_{\rm{M}}}} \right] \nonumber\\
 &= i\delta {\alpha ^c}\left[ {X_{\rm{M}}^\dag {T^b}{T^c}{X_{\rm{M}}} + X_{\rm{M}}^\dag {T^c}{T^b}{X_{\rm{M}}} + X_{\rm{M}}^\dag {T^b}{T^c}\chi  + {\chi ^\dag }{T^c}{T^b}{X_{\rm{M}}}} \right] \nonumber\\
 &= i\delta {\alpha ^c}\left[ {X_{\rm{M}}^\dag \left\{ {{T^b},{T^c}} \right\}{X_{\rm{M}}} + X_{\rm{M}}^\dag {T^b}{T^c}\chi  + {\chi ^\dag }{T^c}{T^b}{X_{\rm{M}}}} \right]  \label{variationGF} 
\end{align}
We can use ${T^b} = \frac{1}{2}{\sigma ^b}$ and the anti-commutation relations %
\begin{align*}
\left\{ {{T^b},{T^c}} \right\} &= \frac{1}{4}\left\{ {{\sigma ^b},{\sigma ^c}} \right\}\\
 &= \frac{1}{2}{\delta ^{bc}} \this
\end{align*} %
and recalling the location of the vacuum from  \eqref{vevlocation} shown below
\begin{align*}
X_\text{M} =  \dfrac{1}{\sqrt{2}} \begin{pmatrix}
0 \\
2\nu
\end{pmatrix} \implies X_\text{M}^\dag X_\text{M}= 2{\nu ^2}
\end{align*}
the first term of \eqref{variationGF} can be simplified to 
\begin{align}
X_{\rm{M}}^\dag \left\{ {{T^b},{T^c}} \right\}{X_{\rm{M}}} = {\nu ^2}{\delta ^{bc}} \label{xmdagchi}
\end{align}

We now turn to evaluating the last two terms of \eqref{variationGF}. We show two ways to calculate this which serves also as an algebraic check. First, the brute force way is to note the sigma matrices can be parameterized as
\begin{align}
{\sigma ^b} = \begin{pmatrix}
					{{\delta _{b3}}}&{{\delta _{b1}} - i{\delta _{b2}}}\\
					{{\delta _{b1}} + i{\delta _{b2}}}&{ - {\delta _{b3}}}
			  \end{pmatrix}
\end{align}
and the product of the sigma matrices is
\begin{align*}
{\sigma ^c}{\sigma ^b} 
&= \begin{pmatrix}
{{\delta _{ci}}{\delta _{bi}} + i\left( {{\delta _{c1}}{\delta _{b2}} - {\delta _{c2}}{\delta _{b1}}} \right)}&{{\delta _{c3}}{\delta _{b1}} - {\delta _{c1}}{\delta _{b3}} + i\left( {{\delta _{c2}}{\delta _{b3}} - {\delta _{c3}}{\delta _{b2}}} \right)}\\
{{\delta _{c1}}{\delta _{b3}} - {\delta _{c3}}{\delta _{b1}} + i\left( {{\delta _{c2}}{\delta _{b3}} - {\delta _{c3}}{\delta _{b2}}} \right)}&{{\delta _{ci}}{\delta _{bi}} + i\left( {{\delta _{c2}}{\delta _{b1}} - {\delta _{c1}}{\delta _{b2}}} \right)}
\end{pmatrix} \this
\end{align*}
This product of generators acting on the vacuum
\begin{align*}
{T^c}{T^b}X_{\rm{M}}&= \frac{{2\nu }}{{4\sqrt 2 }}{\sigma ^c}{\sigma ^b} \begin{pmatrix}
0\\
1
\end{pmatrix}\\
 &= \frac{\nu }{{2\sqrt 2 }}	\begin{pmatrix}
{{\delta _{ci}}{\delta _{bi}} + i\left( {{\delta _{c1}}{\delta _{b2}} - {\delta _{c2}}{\delta _{b1}}} \right)}&{{\delta _{c3}}{\delta _{b1}} - {\delta _{c1}}{\delta _{b3}} + i\left( {{\delta _{c2}}{\delta _{b3}} - {\delta _{c3}}{\delta _{b2}}} \right)}\\
{{\delta _{c1}}{\delta _{b3}} - {\delta _{c3}}{\delta _{b1}} + i\left( {{\delta _{c2}}{\delta _{b3}} - {\delta _{c3}}{\delta _{b2}}} \right)}&{{\delta _{ci}}{\delta _{bi}} + i\left( {{\delta _{c2}}{\delta _{b1}} - {\delta _{c1}}{\delta _{b2}}} \right)}
\end{pmatrix}	\begin{pmatrix}
0\\
1
\end{pmatrix}\\
 &= \frac{\nu }{{2\sqrt 2 }} \begin{pmatrix}
{{\delta _{c3}}{\delta _{b1}} - {\delta _{c1}}{\delta _{b3}} + i\left( {{\delta _{c2}}{\delta _{b3}} - {\delta _{c3}}{\delta _{b2}}} \right)}\\
{{\delta _{ci}}{\delta _{bi}} + i\left( {{\delta _{c2}}{\delta _{b1}} - {\delta _{c1}}{\delta _{b2}}} \right)}
\end{pmatrix} \this
\end{align*}
Recalling the translated field $\chi$
\begin{align*}
\chi= \frac{1}{{\sqrt 2 }} \begin{pmatrix}
{{\chi _2} + i{\chi _1}}\\
{{\chi _0} - i{\chi _3}}
\end{pmatrix}
\end{align*}
then
\begin{align*}
{\chi^\dag }{T^c}{T^b}{X_{\rm{M}}} &= \frac{1}{{\sqrt 2 }}\left( {\begin{array}{*{20}{c}}
{{\chi _2} - i{\chi _1}}&{{\chi _0} + i{\chi _3}}
\end{array}} \right)\frac{\nu }{{2\sqrt 2 }}\left( {\begin{array}{*{20}{c}}
{{\delta _{c3}}{\delta _{b1}} - {\delta _{c1}}{\delta _{b3}} + i\left( {{\delta _{c2}}{\delta _{b3}} - {\delta _{c3}}{\delta _{b2}}} \right)}\\
{{\delta _{ci}}{\delta _{bi}} + i\left( {{\delta _{c2}}{\delta _{b1}} - {\delta _{c1}}{\delta _{b2}}} \right)}
\end{array}} \right)\\
 &= \frac{\nu }{4} \Big\{ {\chi _2}\left( {{\delta _{c3}}{\delta _{b1}} - {\delta _{c1}}{\delta _{b3}}} \right) + {\chi _1}\left( {{\delta _{c2}}{\delta _{b3}} - {\delta _{c3}}{\delta _{b2}}} \right) + i\left[ {{\chi _2}\left( {{\delta _{c2}}{\delta _{b3}} - {\delta _{c3}}{\delta _{b2}}} \right) - {\chi _1}\left( {{\delta _{c3}}{\delta _{b1}} - {\delta _{c1}}{\delta _{b3}}} \right)} \right] +  \\ 
 &+ {\chi _0}\left( {{\delta _{ci}}{\delta _{bi}}} \right) - {\chi _3}\left( {{\delta _{c2}}{\delta _{b1}} - {\delta _{c1}}{\delta _{b2}}} \right) + i\left[ {{\chi _0}\left( {{\delta _{c2}}{\delta _{b1}} - {\delta _{c1}}{\delta _{b2}}} \right) + {\chi _3}\left( {{\delta _{ci}}{\delta _{bi}}} \right)} \right] \Big\} \this
\end{align*}
Taking the hermitian conjugate of the above expression and substituting into the last two terms of \eqref{variationGF} we obtain
\begin{align}
{\chi^\dag }{T^c}{T^b}{X_{\rm{M}}} + X_{\rm{M}}^\dag {T^b}{T^c}\chi = \frac{1}{2}\nu \left\{ {{\chi _2}\left( {{\delta _{c3}}{\delta _{b1}} - {\delta _{c1}}{\delta _{b3}}} \right) + {\chi _1}\left( {{\delta _{c2}}{\delta _{b3}} - {\delta _{c3}}{\delta _{b2}}} \right) - {\chi _3}\left( {{\delta _{c2}}{\delta _{b1}} - {\delta _{c1}}{\delta _{b2}}} \right) + {\chi _0}\left( {{\delta _{ci}}{\delta _{bi}}} \right)} \right\}
\end{align}
A closer observation of the above expression and we can see the anti-symmetric pattern present in the first three terns. This can be made clearer by using the Levi-Civita symbol
\begin{align*}
{\chi ^\dag }{T^c}{T^b}{X_{\rm{M}}} + X_{\rm{M}}^\dag {T^b}{T^c}\chi  &= \frac{1}{2}\nu \left( {{\chi _a}{\varepsilon _{alm}}{\delta _{cl}}{\delta _{bm}} + {\chi _0}{\delta _{bc}}} \right)\\
 &= \frac{1}{2}\nu \left( {{\chi _0}{\delta _{bc}} - {\chi _a}{\varepsilon _{abc}}} \right) \this
\end{align*}
Secondly, a more elegant way to see this result is to make use of the commutation and anti-commutation relation between the sigma matrices,
\begin{align}
\left[ {{\sigma ^b},{\sigma ^c}} \right] = 2i{\varepsilon _{abc}}{\sigma ^a} ,\quad \left\{ {{\sigma ^b},{\sigma ^c}} \right\} = 2{\delta _{bc}} 
\end{align}
which when summed give an expression for the product of the sigma matrices in terms of the Levi-Civita symbol
\begin{align*}
{\sigma ^b}{\sigma ^c} &= {\delta _{bc}} + i{\varepsilon _{abc}}{\sigma ^a} \\
\implies {T^b}{T^c} &= \frac{1}{4}{\delta _{bc}} + \frac{1}{4}i{\varepsilon _{abc}}{\sigma ^a} \this
\end{align*}
So 
\begin{align}
X_{\rm{M}}^\dag {T^b}{T^c}\chi  = \frac{1}{4}{\delta ^{bc}}X_{\rm{M}}^\dag \chi  + \frac{1}{4}i{\varepsilon _{abc}}X_{\rm{M}}^\dag {\sigma ^a}\chi 
\end{align}
and the pieces (can be obtain using parameterized sigma matrices), are given by
\begin{align*}
X_{\rm{M}}^\dag \chi  &= \frac{1}{{\sqrt 2 }}\left( {\begin{array}{*{20}{c}}
0&{2\nu }
\end{array}} \right)\frac{1}{{\sqrt 2 }}\left( {\begin{array}{*{20}{c}}
{{\chi _2} + i{\chi _1}}\\
{{\chi _0} - i{\chi _3}}
\end{array}} \right) \this\\
 &= \left( {{\chi _0} - i{\chi _3}} \right)\nu \\
X_{\rm{M}}^\dag {\sigma ^a}\chi  &= \frac{1}{{\sqrt 2 }}\left( {\begin{array}{*{20}{c}}
0&{2\nu }
\end{array}} \right)\left( {\begin{array}{*{20}{c}}
{{\delta _{a3}}}&{{\delta _{a1}} - i{\delta _{a2}}}\\
{{\delta _{a1}} + i{\delta _{a2}}}&{ - {\delta _{a3}}}
\end{array}} \right)\frac{1}{{\sqrt 2 }}\left( {\begin{array}{*{20}{c}}
{{\chi _2} + i{\chi _1}}\\
{{\chi _0} - i{\chi _3}}
\end{array}} \right)\\
 &= \nu \left[ {i\left( {{\chi _m}{\delta _{am}}} \right) + {\chi _2}{\delta _{a1}} - {\chi _1}{\delta _{a2}} - {\chi _0}{\delta _{a3}}} \right] \this \label{xmdagsigmachi}
\end{align*}
The result for the $X_{\rm{M}}^\dag {T^b}{T^c}\chi $ product is given by
\begin{align*}
X_{\rm{M}}^\dag {T^b}{T^c}\chi  = \frac{1}{4}{\delta ^{bc}}\left( {{\chi _0} - i{\chi _3}} \right)\nu  + \frac{1}{4}\nu \left[ { - {\chi _a}{\varepsilon _{abc}} + i\left( {{\chi _2}{\varepsilon _{1bc}} - {\chi _1}{\varepsilon _{2bc}} - {\chi _0}{\varepsilon _{3bc}}} \right)} \right]
\end{align*}
Summing the above with its hermitian conjugate yields
\begin{align}
{\chi ^\dag }{T^c}{T^b}{X_{\rm{M}}} + X_{\rm{M}}^\dag {T^b}{T^c}\chi  = \frac{1}{2}\nu \left( {{\chi _0}{\delta ^{bc}} - {\chi _a}{\varepsilon _{abc}}} \right) \label{chidagxm}
\end{align}
which is the same result from the brute force computations. Substituting \eqref{xmdagchi} and \eqref{chidagxm} into \eqref{variationGF}
\begin{align*}
\delta \left( {X_{\rm{M}}^\dag {T^b}\chi  - {\chi ^\dag }{T^b}{X_{\rm{M}}}} \right) &=i\delta {\alpha ^c}\left[ {X_{\rm{M}}^\dag \left\{ {{T^b},{T^c}} \right\}{X_{\rm{M}}} + X_{\rm{M}}^\dag {T^b}{T^c}\chi  + {\chi ^\dag }{T^c}{T^b}{X_{\rm{M}}}} \right]\\
 &= i\delta {\alpha ^c}\left[ {{\nu ^2}{\delta ^{bc}} + \frac{1}{2}\nu \left( {{\chi _0}{\delta ^{bc}} - {\chi _a}{\varepsilon _{abc}}} \right)} \right] \this
 \label{variationGFfin} 
\end{align*}

We have gathered all the ingredients to compute the matrix $M^{bc}$, substituting \eqref{dAda} and \eqref{variationGFfin} into \eqref{dGFfunc}
\begin{align*}
{M^{bc}} &= \frac{{\delta {G^b}\left[ {^\alpha A;x} \right]}}{{\delta {\alpha ^c}\left( y \right)}}\\
 &= \int {{d^4}z\left[ {\partial _x^\mu {\delta _{bd}}\delta \left( {x - z} \right)g_\mu ^\nu \frac{{\delta A_\nu ^d\left( z \right)}}{{\delta {\alpha ^c}\left( y \right)}}} \right]}  + \xi ie\frac{\delta }{{\delta {\alpha ^c}\left( y \right)}}\left( {X_{\rm{M}}^\dag {T^b}\chi  - {\chi ^\dag }{T^b}{X_{\rm{M}}}} \right)\\
 &=  - \frac{1}{e}\partial _x^\mu \int {{d^4}z\left[ {\delta \left( {x - z} \right)D_\mu ^{bc}\delta \left( {z - y} \right)} \right]}  - \xi e\left[ {{\nu ^2}{\delta ^{bc}} + \frac{1}{2}\nu \left( {{\chi _0}{\delta ^{bc}} - {\chi _a}{\varepsilon _{abc}}} \right)} \right]\frac{{\delta {\alpha ^c}\left( x \right)}}{{\delta {\alpha ^c}\left( y \right)}}\\
 &=  - \frac{1}{e}\partial _x^\mu D_\mu ^{bc}\delta \left( {x - y} \right) - \xi e\left[ {{\nu ^2}{\delta ^{bc}} + \frac{1}{2}\nu \left( {{\chi _0}{\delta ^{bc}} - {\chi _a}{\varepsilon _{abc}}} \right)} \right]\delta \left( {x - y} \right)\\
 &= \frac{1}{e}\left[ { - \partial _x^\mu D_\mu ^{bc} - \xi {e^2}{\nu ^2}{\delta ^{bc}} - \frac{1}{2}\xi {e^2}\nu {\chi _0}{\delta ^{bc}} + \frac{1}{2}\xi {e^2}\nu {\chi _a}{\varepsilon _{abc}}} \right]\delta \left( {x - y} \right) \this
\end{align*}

Note that in the third line we have moved the $\partial _x^\mu$ out of the integral since it acts only on the $\delta(x-z)$. The derivative of the covariant derivative in the adjoint representation is 
\begin{align}
\partial _x^\mu D_\mu ^{bc} = \partial _x^\mu \left[ {{\delta ^{bc}}{\partial _{x\mu }} + e{\varepsilon _{bac}}A_\mu ^a} \right]
\end{align}
where the $\partial _x^\mu$ is understood to act on everything to its right. We are still left with the problem of constructing determinant of matrix $M^{bc}$. In its current form the usefulness of the path integral,
\begin{align}
\mathcal{Z}= \int {\mathcal{D}A{e^{iS\left[ A \right]}}{e^{ - \frac{i}{{2\xi }}\int {{d^4}x} G_a^2\left( A \right)}}\Delta \left[ A \right]}
\end{align}
is limited since functional techniques are limited to functionals of Gaussian form and the determinant spoils this form. This can be
remedied by noting that Berezin-type functional integrals \cite{BerezinbookSuperAnalysis} over Grassmann variables results in a determinant in the numerator i.e.
\begin{align*}
\Delta \left[ A \right] &=\det \left[ {{M^{bc}}} \right]\\
 &= \int {\mathcal{D}\bar u \mathcal{D}u{e^{ - ie\int {{d^4}x{d^4}y{{\bar u}^b}\left( x \right){M^{bc}}\left( {x,y} \right){u^c}\left( y \right)} }}} \this
\end{align*}
Where $\bar{u}^a, {u}^a $ are anti-commuting Grassmann fields with
\begin{align}
\left\{ {{u^a},{u^b}} \right\} = \left\{ {{{\bar u}^a},{u^b}} \right\} = \left\{ {{{\bar u}^a},{{\bar u}^b}} \right\} = 0
\end{align}
Since the expressions are long we shall write down the pieces separately
\begin{align*}
- ie\int {d^4}x{d^4}y{{\bar u}^b}\left(x\right){M^{bc}}\left( {x,y} \right){u^c}\left( y \right)
&=  - ie\int {{d^4}x{d^4}y} {{\bar u}^b} \left( x \right)\frac{1}{e} 
\Biggl[  - \partial _x^\mu D_\mu ^{bc} - \xi {e^2}{\nu ^2}{\delta ^{bc}} - \frac{1}{2}\xi {e^2}\nu {\chi _0}{\delta ^{bc}} +\\ 
&+ \frac{1}{2}\xi {e^2}\nu {\chi _a}{\varepsilon _{abc}} \Biggr] \delta \left( {x - y} \right){u^c}\left( y \right)\\
& = i\int {{d^4}x} {{\bar u}^b}\left( x \right)\left[ {\partial _x^\mu D_\mu ^{bc} + \xi {e^2}{\nu ^2}{\delta ^{bc}} + \frac{1}{2}\xi {e^2}\nu {\chi _0}{\delta ^{bc}} - \frac{1}{2}\xi {e^2}\nu {\chi _a}{\varepsilon _{abc}}} \right]{u^c}\left( x \right) \this
\end{align*}
Substituting in the covariant derivative in the adjoint representation for
\begin{align*}
{{\bar u}^b}\left[ {\partial _x^\mu D_\mu ^{bc} + \xi {e^2}{\nu ^2}{\delta ^{bc}}} \right]{u^c} &= {{\bar u}^b}\left[ {{\partial ^\mu }\left( {{\delta ^{bc}}{\partial _\mu } + e{\varepsilon _{bac}}A_\mu ^a} \right) + \xi {e^2}{\nu ^2}{\delta ^{bc}}} \right]{u^c}\\
&= {{\bar u}^a}\left( {{\partial ^\mu }{\partial _\mu } + \xi {e^2}{\nu ^2}} \right){u^a} + e{\varepsilon _{bac}}{{\bar u}^b}{\partial ^\mu }\left( {A_\mu ^a{u^c}} \right)\\
&= {{\bar u}^a}\left( {{\partial ^\mu }{\partial _\mu } + \xi {e^2}{\nu ^2}} \right){u^a} + e{\varepsilon _{abc}}A_\mu ^a\left( {{\partial ^\mu }{{\bar u}^b}} \right){u^c} \this
\end{align*}
Putting all the pieces together for the exponent
\begin{align*}
- ie\int {{d^4}x{d^4}y{{\bar u}^b}\left( x \right){M^{bc}}\left( {x,y} \right){u^c}\left( y \right)} 
 &= i\int {{d^4}x} \Biggl[ 
 {{\bar u}^a}\left( {{\partial ^\mu }{\partial _\mu } + \xi {e^2}{\nu ^2}} \right){u^a} + e{\varepsilon _{abc}}A_\mu ^a\left( {{\partial ^\mu }{{\bar u}^b}} \right){u^c} + \frac{1}{2}\xi {e^2}\nu {\chi _0}{{\bar u}^a}{u^a} + \\
 & - \frac{1}{2}\xi {e^2}\nu {\varepsilon _{abc}}{\chi _a}{{\bar u}^b}{u^c} 
 \Biggr] \this
\end{align*}
The determinant can now be expressed  as 
\begin{align*}
\Delta \left[ A \right] &= \int {\mathcal{D}\bar u \mathcal{D}u{e^{ - ie\int {{d^4}x{d^4}y{{\bar u}^b}\left( x \right){M^{bc}}\left( {x,y} \right){u^c}\left( y \right)} }}} \\
&=\int {\mathcal{D}\bar u \mathcal{D} u{e^{i\int {{d^4}x} \left[ {{{\bar u}^a}\left( {{\partial ^\mu }{\partial _\mu } + \xi {e^2}{\nu ^2}} \right){u^a} + e{\varepsilon _{abc}}A_\mu ^a\left( {{\partial ^\mu }{{\bar u}^b}} \right){u^c} + \frac{1}{2}\xi {e^2}\nu {\chi _0}{{\bar u}^a}{u^a} - \frac{1}{2}\xi {e^2}\nu {\varepsilon _{abc}}{\chi _a}{{\bar u}^b}{u^c}} \right]}}}  \this
\end{align*}

Substituting the determinant into the path integral
\begin{align*}
\mathcal{Z} &= \int {\mathcal{D}A{e^{iS\left[ A \right]}}{e^{ - \frac{i}{{2\xi }}\int {{d^4}x} G_a^2\left( A \right)}}\Delta \left[ A \right]} \\
 &= \int {\mathcal{D}\bar u \mathcal{D} u \mathcal{D} A} {e^{iS\left[ A \right] - \frac{i}{{2\xi }}\int {{d^4}x} G_a^2\left( A \right) + i\int {{d^4}x} \left[ {{{\bar u}^a}\left( {{\partial ^\mu }{\partial _\mu } + \xi {e^2}{\nu ^2}} \right){u^a} + e{\varepsilon _{abc}}A_\mu ^a\left( {{\partial ^\mu }{{\bar u}^b}} \right){u^c} + \frac{1}{2}\xi {e^2}\nu {\chi _0}{{\bar u}^a}{u^a} - \frac{1}{2}\xi {e^2}\nu {\varepsilon _{abc}}{\chi _a}{{\bar u}^b}{u^c}} \right]}}\\
 &= \int {\mathcal{D}\bar u\mathcal{D}u\mathcal{D}A} {e^{i\int {{d^4}x} \left[ {\mathscr{L} - \frac{1}{{2\xi }}G_a^2\left( A \right) + {{\bar u}^a}\left( {{\partial ^\mu }{\partial _\mu } + \xi {e^2}{\nu ^2}} \right){u^a} + e{\varepsilon _{abc}}A_\mu ^a\left( {{\partial ^\mu }{{\bar u}^b}} \right){u^c} + \frac{1}{2}\xi {e^2}\nu {\chi _0}{{\bar u}^a}{u^a} - \frac{1}{2}\xi {e^2}\nu {\varepsilon _{abc}}{\chi _a}{{\bar u}^b}{u^c}} \right]}}\\
 &= \int {\mathcal{D}\bar u\mathcal{D}u\mathcal{D}A} {e^{i\int {{d^4}x{\mathscr{L}_{{\rm{eff}}}}} }} \this
\end{align*}

Where we have defined the effective Lagrangian
\begin{align}
{\mathscr{L}_{{\rm{eff}}}}: = \mathscr{L} - \frac{1}{{2\xi }}G_a^2\left( A \right) + {{\bar u}^a}\left( {{\partial ^\mu }{\partial _\mu } + \xi {e^2}{\nu ^2}} \right){u^a} + e{\varepsilon _{abc}}A_\mu ^a\left( {{\partial ^\mu }{{\bar u}^b}} \right){u^c} + \frac{1}{2}\xi {e^2}\nu {\chi _0}{{\bar u}^a}{u^a} - \frac{1}{2}\xi {e^2}\nu {\varepsilon _{abc}}{\chi _a}{{\bar u}^b}{u^c} \label{leff}
\end{align}

\section{Gauge Fixing Term}
We give a justification of the choice of the gauge fixing function in \eqref{gffunction} shown here:
\begin{align*}
{G^b}\left[ {^\alpha A;x} \right] = {\partial ^\mu }A_\mu ^b + \xi ie\left( {X_{\rm{M}}^\dag {T^b}\chi - {\chi^\dag }{T^b}X_{\rm{M}}} \right) 
\end{align*}
After spontaneous symmetry breaking, some pesky two point mixing terms \eqref{mixingtermsAchi}
\begin{align*}
X_{\rm{M}}^\dag {\partial ^\mu }{A_\mu }\chi  - {\chi ^\dag }{\partial ^\mu }{A_\mu }{X_{\rm{M}}}= \frac{i}{{\sqrt 2 }}{\chi _a}{\partial ^\mu }A_\mu ^a
\end{align*}
were generated which were hard to interpret. These terms survived and were present in the Higgs lagrangian shown below:
\begin{align*}
\mathscr{L}_H &= \frac{1}{2}\left( {{\partial _\mu }H{\partial ^\mu }H - {\mu ^2}{H^2}} \right) + \frac{1}{2}{\partial _\mu }{\chi _a}{\partial ^\mu }{\chi _a} + ie{\partial ^\mu }A_\mu ^a\left[ {X_{\rm{M}}^\dag {T^a}\chi  - {\chi ^\dag }{T^a}{X_{\rm{M}}}} \right] - \frac{1}{4}e{\varepsilon _{abc}}\left( {{\chi _a}\overleftrightarrow{\partial^\mu}{\chi _b}} \right)A_\mu ^c + \\
&- \frac{1}{2}e\left( {{\chi _a}\overleftrightarrow{\partial^\mu}H} \right)A_\mu ^a +
 \frac{1}{2}{e^2}{\nu ^2}A_\mu ^a{A^{a\mu }} + \frac{1}{2}{e^2}\nu HA_\mu ^a{A^{a\mu }} + \frac{1}{8}{e^2}{H^2}A_\mu ^a{A^{a\mu }} + \frac{1}{8}{e^2}\chi _b^2A_\mu ^a{A^{a\mu }} - \frac{\lambda }{{32}}{H^4} - \frac{\lambda }{{16}}{H^2}\chi _a^2 + \\
&- \frac{\lambda }{{32}}\chi _a^2\chi _b^2 - \frac{{\nu \lambda }}{4}{H^3} - \frac{{\nu \lambda }}{4}H\chi _a^2 - \frac{1}{2}\left( {4\kappa {\nu ^2}} \right){\phi^2_a }  - 2\kappa \nu H{\phi^2_a }  - \frac{1}{2}\kappa {H^2}{\phi^2_a }  - \frac{1}{2}\kappa \chi _a^2{\phi^2_a }  + \frac{1}{2}{\mu ^2}{\nu ^2} \this 
\end{align*}
We see that the gauge fixing term,
\begin{align*}
\mathscr{L}_{\rm{GF}} &=  - \frac{1}{{2\xi }}G_a^2\left( A \right)\\
 &=  - \frac{1}{{2\xi }}{\left[ {{\partial ^\mu }A_\mu ^a + \xi ie\left( {X_{\rm{M}}^\dag {T^a}\chi  - {\chi ^\dag }{T^a}{X_{\rm{M}}}} \right)} \right]^2}\\
 &=  - \frac{1}{{2\xi }}{\left( {{\partial ^\mu }A_\mu ^a} \right)^2} - ie{\partial ^\mu }A_\mu ^a\left[ {X_{\rm{M}}^\dag {T^a}\chi  - {\chi ^\dag }{T^a}{X_{\rm{M}}}} \right] + \frac{1}{2}\xi {e^2}{\left( {X_{\rm{M}}^\dag {T^a}\chi  - {\chi ^\dag }{T^a}{X_{\rm{M}}}} \right)^2} \this \label{gaugefixingterm}
\end{align*}
in the effective Lagrangian has been designed specifically to remove the pesky two point mixing terms from symmetry breaking. We now are left to the task of evaluating the last term of \eqref{gaugefixingterm}. Recall the earlier result of \eqref{xmdagsigmachi} of
\begin{align*}
X_{\rm{M}}^\dag {\sigma ^a}\chi = \nu \left[ {i\left( {{\chi _m}{\delta _{am}}} \right) + {\chi _2}{\delta _{a1}} - {\chi _1}{\delta _{a2}} - {\chi _0}{\delta _{a3}}} \right] 
\end{align*}
Taking the hermitian conjugate and finding the difference
\begin{align*}
X_{\rm{M}}^\dag {\sigma ^a}\chi  - \chi {\sigma ^a}X_{\rm{M}} &= 2i\nu {\chi _a} \\
\implies X_{\rm{M}}^\dag {T ^a}\chi  - \chi {T ^a}X_{\rm{M}} &= i\nu {\chi _a} \this
\end{align*}
The gauge fixing terms final form is given by
\begin{align}
\mathscr{L}_{\rm{GF}} =  - \frac{1}{{2\xi }}{\left( {{\partial ^\mu }A_\mu ^a} \right)^2} - ie{\partial ^\mu }A_\mu ^a\left[ {X_{\rm{M}}^\dag {T^a}\chi  - {\chi ^\dag }{T^a}{X_{\rm{M}}}} \right] - \frac{1}{2}\xi {e^2}{\nu ^2}\chi _a^2 \label{gaugefixingtermfinal}
\end{align}

\newpage
\section{Complete Lagrangian}
We make the cosmetic change of relabeling $\chi_0 \to H$ then list the complete Lagrangian
\begin{align*}
{\mathscr{L}_{{\rm{eff}}}}: = \mathscr{L} + {\mathscr{L}_{{\rm{GF}}}} + {{\bar u}^a}\left( {{\partial ^\mu }{\partial _\mu } + \xi {e^2}{\nu ^2}} \right){u^a} + e{\varepsilon _{abc}}A_\mu ^a\left( {{\partial ^\mu }{{\bar u}^b}} \right){u^c} + \frac{1}{2}\xi {e^2}\nu H{{\bar u}^a}{u^a} - \frac{1}{2}\xi {e^2}\nu {\varepsilon _{abc}}{\chi _a}{{\bar u}^b}{u^c} \this
\end{align*}
with 
\begin{align*}
\mathscr{L}: = {\mathscr{L}_{\phi A}} + {\mathscr{L}_H} \this
\end{align*}
the pion-rho Lagrangian
\begin{align*}
\mathscr{L}_{\phi A}  &= \frac{1}{2}\left( {{\partial _\mu }{\phi _a}{\partial ^\mu }{\phi _a} - {b^2}\phi _a^2} \right) + \frac{1}{2}e{\varepsilon _{aij}}A_\mu ^a\left( {{\phi _j}{\partial ^\mu }{\phi _i} - {\phi _i}{\partial ^\mu }{\phi _j}} \right) - \frac{1}{2}{e^2}{\varepsilon _{aij}}{\varepsilon _{bjk}}{\phi _i}{\phi _k}A_\mu ^a{A^{b\mu }} - \frac{{{\lambda _4}}}{8}\left( {\phi _a^2\phi _b^2} \right) + \\
 &- \frac{1}{4}F_{\mu \nu }^a{F^{a\mu \nu }} + \frac{1}{2}e{\varepsilon _{abc}}{A^{b\mu }}{A^{c\nu }}F_{\mu \nu }^a - \frac{1}{4}{e^2}{\varepsilon _{abc}}{\varepsilon _{ade}}A_\mu ^bA_\nu ^c{A^{d\mu }}{A^{e\nu }}  \this
\end{align*}
the Higgs-rho-pion Lagrangian
\begin{align*}
\mathscr{L}_H &= \frac{1}{2}\left( {{\partial _\mu }H{\partial ^\mu }H - {\mu ^2}{H^2}} \right) + \frac{1}{2}{\partial _\mu }{\chi _a}{\partial ^\mu }{\chi _a} + ie{\partial ^\mu }A_\mu ^a\left[ {X_{\rm{M}}^\dag {T^a}\chi  - {\chi ^\dag }{T^a}{X_{\rm{M}}}} \right] - \frac{1}{4}e{\varepsilon _{abc}}\left( {{\chi _a}\overleftrightarrow{\partial^\mu}{\chi _b}} \right)A_\mu ^c + \\
&- \frac{1}{2}e\left( {{\chi _a}\overleftrightarrow{\partial^\mu}H} \right)A_\mu ^a +
 \frac{1}{2}{e^2}{\nu ^2}A_\mu ^a{A^{a\mu }} + \frac{1}{2}{e^2}\nu HA_\mu ^a{A^{a\mu }} + \frac{1}{8}{e^2}{H^2}A_\mu ^a{A^{a\mu }} + \frac{1}{8}{e^2}\chi _b^2A_\mu ^a{A^{a\mu }} - \frac{\lambda }{{32}}{H^4} - \frac{\lambda }{{16}}{H^2}\chi _a^2 + \\
&- \frac{\lambda }{{32}}\chi _a^2\chi _b^2 - \frac{{\nu \lambda }}{4}{H^3} - \frac{{\nu \lambda }}{4}H\chi _a^2 - \frac{1}{2}\left( {4\kappa {\nu ^2}} \right){\phi^2_a }  - 2\kappa \nu H{\phi^2_a }  - \frac{1}{2}\kappa {H^2}{\phi^2_a }  - \frac{1}{2}\kappa \chi _a^2{\phi^2_a }  + \frac{1}{2}{\mu ^2}{\nu ^2} \this
\end{align*}
and the gauge fixing Lagrangian
\begin{align*}
\mathscr{L}_{\rm{GF}} =  - \frac{1}{{2\xi }}{\left( {{\partial ^\mu }A_\mu ^a} \right)^2} - ie{\partial ^\mu }A_\mu ^a\left[ {X_{\rm{M}}^\dag {T^a}\chi  - {\chi ^\dag }{T^a}{X_{\rm{M}}}} \right] - \frac{1}{2}\xi {e^2}{\nu ^2}\chi _a^2  \this
\end{align*}

\chapter{Path to Calculations}

\section{Renormalization Transformation}

\numberwithin{equation}{section}

\lettrine[lines=4]{\initfamily\textcolor{darkblue}{T}}{hus} far we have constructed a Lagrangian which contains bare fields and couplings. Bare in the sense that they are not directly related to experiment. To complete the definition of the theory, we must show how these bare couplings and fields are related to the experimentally measured coupling and give meaning to the fields and couplings. This is done by renormalization \cite{Spiesberger1986,Hioki1986,BardinbookSM}.  Since the complete bare Lagrangian has many terms in it, we shall split up the calculation and deal with the self contained pieces. We begin with the bare pion-rho Lagrangian:
\begin{align*}
\mathscr{L}_{\phi_0 A_0} &= \frac{1}{2}\left( {{\partial _\mu }{\phi _{0a}}{\partial ^\mu }{\phi _{0a}} - m_0^2\phi _{0a}^2} \right) + \frac{1}{2}{e_0}{\varepsilon _{aij}}A_{0\mu }^a\left( {{\phi _{0j}}{\partial ^\mu }{\phi _{0i}} - {\phi _{0i}}{\partial ^\mu }{\phi _{0j}}} \right) - \frac{1}{2}{e_0}^2{\varepsilon _{aij}}{\varepsilon _{bjk}}{\phi _{0i}}{\phi _{0k}}A_{0\mu }^a A_0^{b\mu } + \\
&- \frac{{{\lambda _{{4_0}}}}}{8}\left( {\phi _{0a}^2\phi _{0b}^2} \right) - \frac{1}{4}F_{0\mu \nu }^a F_0^{a\mu \nu } + \frac{1}{2}M_0^2A_{0\mu }^a A_0^{a\mu} - \frac{1}{{2{\xi _0}}}{\left( {{\partial ^\mu }A_{0\mu }^a} \right)^2} + \frac{1}{2}{e_0}{\varepsilon _{abc}} A_0^{b\mu } A_0^{c\nu }F_{0\mu \nu }^a + \\
&- \frac{1}{4}e_0^2{\varepsilon _{abc}}{\varepsilon _{ade}}A_{0\mu }^bA_{0\nu }^c A_0^{d\mu} A_0^{e\nu }
\this \label{pionrholagbare}
\end{align*}
The subscript $0$ is used to indicate bare quantities. We make a redefinition of the fields and couplings in terms of the renormalized fields and couplings:
\begin{equation}
\begin{aligned}
{\phi _{0a}}&: = \phi_a\sqrt {{Z_\phi }} \qquad\qquad\qquad & {m_0}&: = m\sqrt {{Z_m}} \qquad\qquad\qquad & {e_0}&: = e\sqrt {{Z_e}} \\
{M_0}&: = M\sqrt {{Z_M}}		& \frac{1}{{{\xi _0}}}&:= \frac{{\sqrt {{Z_\xi }} }}{\xi }		& {\lambda _{{4_0}}}&:= \lambda_4\sqrt {{Z_{{\lambda _4}}}}
\end{aligned}
\label{renormtrans1}
\end{equation}
Substituting \eqref{renormtrans1} into \eqref{pionrholagbare}
\begin{align*}
{\mathscr{L}_{\phi_0 A_0}} &= \frac{1}{2}\left( {{Z_\phi }{\partial _\mu }{\phi _a}{\partial ^\mu }{\phi _a} - {Z_\phi }{Z_m}{m^2}\phi _a^2} \right) + \frac{1}{2}e{Z_\phi }\sqrt {{Z_e}{Z_A}} {\varepsilon _{aij}}A_\mu ^a\left( {{\phi _j}{\partial ^\mu }{\phi _i} - {\phi _i}{\partial ^\mu }{\phi _j}} \right) + \\
 &- \frac{1}{2}{e^2}{Z_e}{Z_\phi }{Z_A}{\varepsilon _{aij}}{\varepsilon _{bjk}}{\phi _i}{\phi _k}A_\mu ^a{A^{b\mu }} - \frac{{Z_\phi ^2\sqrt {{Z_{{\lambda _4}}}} {\lambda _4}}}{8}\left( {\phi _a^2\phi _b^2} \right) - \frac{1}{4}{Z_A}F_{\mu \nu }^a{F^{a\mu \nu }} + \frac{1}{2}{Z_M}{Z_A}{M^2}A_\mu ^a{A^{a\mu }} + \\
 &- \frac{{{Z_A}\sqrt {{Z_\xi }} }}{{2\xi }}{\left( {{\partial ^\mu }A_\mu ^a} \right)^2} + \frac{1}{2}e\sqrt {{Z_e Z_A^3}} {\varepsilon _{abc}}{A^{b\mu }}{A^{c\nu }}F_{\mu \nu }^a - \frac{1}{4}{e^2}{Z_e}Z_A^2{\varepsilon _{abc}}{\varepsilon _{ade}}A_\mu ^bA_\nu ^c{A^{d\mu }}{A^{e\nu }} \this\label{pionrholagbaresubs}
\end{align*}
The definition of the $Z_i$ factors are given by:
\begin{equation}
\begin{aligned}
{Z_\phi }&:=1+\delta{Z_\phi} \qquad\qquad & {Z_\phi}{Z_m}{m^2}&:={m^2}+\delta{m^2} \qquad\qquad & e{Z_\phi}\sqrt{{Z_e}{Z_A}}&:=e+\delta{Z_{A{\phi^2}}} \\
{e}^2{Z_e}{Z_\phi}{Z_A}&:={e^2}+\delta{Z_{{\phi^2}{A^2}}} & Z_\phi^2\sqrt{{Z_{{\lambda_4}}}}{\lambda _4}&:={\lambda_4}+\delta{Z_{{\phi ^4}}} & {Z_A}&:=1+\delta{Z_A} \\
{Z_M}{Z_A}{M^2}&:= {M^2}+\delta{M^2} & \frac{{{Z_A}\sqrt{{Z_\xi}}}}{\xi}&:=\frac{1}{\xi}+\delta\xi & e\sqrt{{Z_e}Z_A^3} &:= e + \delta {Z_{{A^2}\partial A}} \\
{e^2}{Z_e}Z_A^2 &:= {e^2} + \delta {Z_{{A^4}}}
\end{aligned}
\label{zpionrhofactors}
\end{equation}
and substituting \eqref{zpionrhofactors} into \eqref{pionrholagbaresubs}, the Lagrangian will split into two pieces
\begin{align}
{\mathscr{L}_{\phi_0 A_0}} = {\mathscr{L}_{\phi A}} + {\mathscr{L}_{\phi A}^{\rm{CT}}}
\end{align}
where ${\mathscr{L}_{\phi A}}$ is of a form that resembles the bare Lagrangian ${\mathscr{L}_{\phi_0 A_0}}$
\begin{align*}
{\mathscr{L}_{\phi A}} &= \frac{1}{2}\left( {{\partial _\mu }{\phi _a}{\partial ^\mu }{\phi _a} - {m^2}\phi _a^2} \right) + \frac{1}{2}e{\varepsilon _{aij}}A_\mu ^a\left( {{\phi _j}{\partial ^\mu }{\phi _i} - {\phi _i}{\partial ^\mu }{\phi _j}} \right) - \frac{1}{2}{e^2}{\varepsilon _{aij}}{\varepsilon _{bjk}}{\phi _i}{\phi _k}A_\mu ^a{A^{b\mu }} - \frac{{{\lambda _4}}}{8}\left( {\phi _a^2\phi _b^2} \right) + \\
&- \frac{1}{4}F_{\mu \nu }^a{F^{a\mu \nu }} + \frac{1}{2}{M^2}A_\mu ^a{A^{a\mu }} - \frac{1}{{2\xi }}{\left( {{\partial ^\mu }A_\mu ^a} \right)^2} + \frac{1}{2}e{\varepsilon _{abc}}{A^{b\mu }}{A^{c\nu }}F_{\mu \nu }^a - \frac{1}{4}{e^2}{\varepsilon _{abc}}{\varepsilon _{ade}}A_\mu ^bA_\nu ^c{A^{d\mu }}{A^{e\nu }} \this \label{pionrholagrenorm}
\end{align*}
and the new piece ${\mathscr{L}^{\phi A}_{\rm{CT}}}$ is the counter term Lagrangian
\begin{align*}
{\mathscr{L}_{\phi A}^{\rm{CT}}} &=\frac{1}{2}\left( {\delta {Z_\phi }{\partial _\mu }{\phi _a}{\partial ^\mu }{\phi _a} - \delta {m^2}\phi _a^2} \right) + \frac{1}{2}\delta {Z_e}{\varepsilon _{aij}}A_\mu ^a\left( {{\phi _j}{\partial ^\mu }{\phi _i} - {\phi _i}{\partial ^\mu }{\phi _j}} \right) - \frac{1}{2}\delta {Z_{A{\phi ^2}}}{\varepsilon _{aij}}{\varepsilon _{bjk}}{\phi _i}{\phi _k}A_\mu ^a{A^{b\mu }}+ \\
&- \frac{{\delta {Z_{{\phi ^4}}}}}{8}\left( {\phi _a^2\phi _b^2} \right) - \frac{1}{4}\delta {Z_A}F_{\mu \nu }^a{F^{a\mu \nu }} + \frac{1}{2}\delta {M^2}A_\mu ^a{A^{a\mu }} - \frac{1}{2}\delta \xi {\left( {{\partial ^\mu }A_\mu ^a} \right)^2} + \frac{1}{2}\delta {Z_{{A^2}\partial A}}{\varepsilon _{abc}}{A^{b\mu }}{A^{c\nu }}F_{\mu \nu }^a + \\
&- \frac{1}{4}\delta {Z_{{A^4}}}{\varepsilon _{abc}}{\varepsilon _{ade}}A_\mu ^bA_\nu ^c{A^{d\mu }}{A^{e\nu }} \this \label{pionrholagrenormCT}
\end{align*}
which contains all the $\delta Z_i$ factors. The bare Lagrangian from symmetry breaking is
\begin{align*}
{\mathscr{L}_{H0}} &= \frac{1}{2}\left( {{\partial _\mu }{H_0}{\partial ^\mu }{H_0} - m_{H0}^2{H_0}^2} \right) + \frac{1}{2}\left( {{\partial _\mu }{\chi _{0a}}{\partial ^\mu }{\chi _{0a}} - {\xi _0}{M_0}^2\chi _a^2} \right) - \frac{1}{4}{e_0}{\varepsilon _{abc}}\left( {{\chi _{0a}}\overleftrightarrow{\partial^\mu}{\chi _{0b}}} \right)A_{0\mu }^c + \\
&- \frac{1}{2}{e_0}\left( {{\chi _{0a}}\overleftrightarrow{\partial^\mu}{H_0}} \right)A_{0\mu }^a + \frac{1}{2}{e_0}^2{\nu _0}{H_0}A_{0\mu }^a{A_0}^{a\mu } + \frac{1}{8}{e_0}^2{H_0}^2A_{0\mu }^a{A_0}^{a\mu } + \frac{1}{8}{e_0}^2\chi _{0b}^2A_{0\mu }^a{A_0}^{a\mu } - \frac{{{\lambda _0}}}{{32}}{H_0}^4 + \\
&- \frac{{{\lambda _0}}}{{16}}{H_0}^2\chi _{0a}^2 - \frac{{{\lambda _0}}}{{32}}\chi _{0a}^2\chi _{0b}^2 - \frac{{{\nu _0}{\lambda _0}}}{4}{H_0}^3 - \frac{{{\nu _0}{\lambda _0}}}{4}{H_0}\chi _{0a}^2 - 2{\kappa _0}{\nu _0}{H_0}\phi _{0a}^2 - \frac{1}{2}{\kappa _0}{H_0}^2\phi _{0a}^2 - \frac{1}{2}{\kappa _0}\chi _{0a}^2\phi _{0b}^2 \this \label{higgslagbare}
\end{align*}
Redefining the bare fields and couplings in term of the renormalized fields and couplings
\begin{equation}
\begin{aligned}
{H_0}&: = H\sqrt {{Z_H}} \qquad\qquad\qquad  & m_{H0}&: = {m_H}\sqrt {{Z_{{m_H}}}} \qquad\qquad\qquad  & {\chi _{0a}}&: = {\chi _a}\sqrt {{Z_\chi }} \\
{M_0}&: = M\sqrt {{Z_M}} & {\nu _0}&: = \nu \sqrt {{Z_\nu }} & {\lambda _0}&: = \lambda \sqrt {{Z_\lambda }} \\
{\kappa _0}&: = \kappa \sqrt {{Z_\kappa }} 
\end{aligned}
\label{renormtrans2}
\end{equation}
Reusing some definitions in \eqref{renormtrans1} and \eqref{renormtrans2}, and substituting into \eqref{higgslagbaresubs} gives us
\begin{align*}
{\mathscr{L}_{H0}} &= \frac{1}{2}\left( {{Z_H}{\partial _\mu }H{\partial ^\mu }H - {Z_H}{Z_{{m_H}}}m_H^2{H^2}} \right) + \frac{1}{2}\left( {{Z_\chi }{\partial _\mu }{\chi _a}{\partial ^\mu }{\chi _a} - {\xi _0}{Z_\chi }{Z_M}{M^2}\chi _a^2} \right) + \\
&- \frac{1}{4}e{Z_\chi }\sqrt {{Z_e}{Z_A}} {\varepsilon _{abc}}\left( {{\chi _a}\overleftrightarrow{\partial^\mu}{\chi _b}} \right)A_\mu ^c - \frac{1}{2}e\sqrt {{Z_e}{Z_\chi }{Z_H}{Z_A}} \left( {{\chi _a}\overleftrightarrow{\partial^\mu}H} \right)A_\mu ^a + \frac{1}{2}{e^2}\nu {Z_A}{Z_e}\sqrt {{Z_H}{Z_\nu }} HA_\mu ^a{A^{a\mu }} + \\
&+ \frac{1}{8}{e^2}{Z_e}{Z_H}{Z_A}{H^2}A_\mu ^a{A^{a\mu }} + \frac{1}{8}{e^2}{Z_e}{Z_\chi }{Z_A}\chi _b^2A_\mu ^a{A^{a\mu }} - \frac{\lambda }{{32}}Z_H^2\sqrt {{Z_\lambda }} {H^4} - \frac{\lambda }{{16}}{Z_H}{Z_\chi }\sqrt {{Z_\lambda }} {H^2}\chi _a^2 + \\
&- \frac{\lambda }{{32}}Z_\chi ^2\sqrt {{Z_\lambda }} \chi _a^2\chi _b^2 - \frac{{\nu \lambda }}{4}\sqrt {{Z_\nu }{Z_\lambda }Z_H^3} {H^3} - \frac{{\nu \lambda }}{4}{Z_\chi }\sqrt {{Z_\nu }{Z_\lambda }{Z_H}} H\chi _a^2 - 2\kappa \nu {Z_\phi }\sqrt {{Z_\kappa }{Z_\nu }{Z_H}} H\phi _a^2 + \\
&- \frac{1}{2}\kappa {Z_H}{Z_\phi }\sqrt {{Z_\kappa }} {H^2}\phi _a^2 - \frac{1}{2}\kappa {Z_\chi }{Z_\phi }\sqrt {{Z_\kappa }} \chi _a^2\phi _b^2
\this \label{higgslagbaresubs}
\end{align*}
The definition for the $Z_i$ factors are
\begin{equation}
\begin{aligned}
{Z_H}&:= 1 + \delta{Z_H} & {Z_H}{Z_{{m_H}}}m_H^2&:= m_H^2 + \delta m_H^2 & {Z_\chi }&:= 1 + \delta{Z_\chi } \\
{\xi _0}{Z_\chi }{Z_M}{M^2}&:= \xi {M^2} + \delta M_{\chi\xi} ^2 &
e{Z_\chi }\sqrt {{Z_e}{Z_A}}&:= e + \delta {Z_{A{\chi ^2}}} & e\sqrt {{Z_e}{Z_\chi }{Z_H}{Z_A}}&:= e + \delta {Z_{\chi HA}} \\
{e^2}\nu {Z_A}{Z_e}\sqrt {{Z_H}{Z_\nu }}&:= {e^2}\nu  + \delta {Z_{H{A^2}}} & {e^2}{Z_e}{Z_H}{Z_A}&:= {e^2} + \delta {Z_{{H^2}{A^2}}} & {e^2}{Z_e}{Z_\chi }{Z_A}&:= {e^2} + \delta {Z_{{\chi ^2}{A^2}}} \\
\lambda Z_H^2\sqrt {{Z_\lambda }}&:= \lambda  + \delta {Z_{{H^4}}} & \lambda {Z_H}{Z_\chi }\sqrt {{Z_\lambda }}&:= \lambda  + \delta {Z_{{H^2}{\chi ^2}}} & \lambda Z_\chi ^2\sqrt {{Z_\lambda }}&:= \lambda  + \delta {Z_{{\chi ^4}}} \\
\nu \lambda \sqrt {{Z_\nu }{Z_\lambda }Z_H^3}&:= \nu \lambda  + \delta {Z_{{H^3}}} &
\nu \lambda {Z_\chi }\sqrt {{Z_\nu }{Z_\lambda }{Z_H}}&:= \nu \lambda  + \delta {Z_{H{\chi ^2}}} & \kappa \nu {Z_\phi }\sqrt {{Z_\kappa }{Z_\nu }{Z_H}}&:= \kappa \nu  + \delta {Z_{H{\phi ^2}}} \\
\kappa {Z_H}{Z_\phi }\sqrt {{Z_\kappa }}&:= \kappa  + \delta {Z_{{H^2}{\phi ^2}}} &
\kappa {Z_\chi }{Z_\phi }\sqrt {{Z_\kappa }}&:= \kappa  + \delta {Z_{{\chi ^2}{\phi ^2}}}
\end{aligned}
\label{zhiggsfactors}
\end{equation}
Substituting \eqref{zhiggsfactors} into \eqref{higgslagbaresubs}, the Lagrangian will split into two pieces
\begin{align}
{\mathscr{L}_{H0}} = {\mathscr{L}_{H}} + {\mathscr{L}_{H}^{\rm{CT}}}
\end{align}
where ${\mathscr{L}_{H}}$ is
\begin{align*}
{\mathscr{L}_{H}} &= \frac{1}{2}\left( {{\partial _\mu }{H}{\partial ^\mu }{H} - m_{H}^2{H}^2} \right) + \frac{1}{2}\left( {{\partial _\mu }{\chi _{a}}{\partial ^\mu }{\chi _{a}} - {\xi}{M}^2\chi _a^2} \right) - \frac{1}{4}{e}{\varepsilon _{abc}}\left( {{\chi _{a}}\overleftrightarrow{\partial^\mu}{\chi _{b}}} \right)A_{\mu }^c - \frac{1}{2}{e}\left( {{\chi _{a}}\overleftrightarrow{\partial^\mu}{H}} \right)A_{\mu }^a + \\
&+ \frac{1}{2}{e}^2{\nu}{H}A_{\mu }^a{A}^{a\mu } + \frac{1}{8}{e}^2{H}^2A_{\mu }^a{A}^{a\mu } + \frac{1}{8}{e}^2\chi _{b}^2A_{\mu }^a{A}^{a\mu } - \frac{{{\lambda}}}{{32}}{H}^4 - \frac{{{\lambda}}}{{16}}{H}^2\chi _{a}^2 - \frac{{{\lambda}}}{{32}}\chi _{a}^2\chi _{b}^2 - \frac{{{\nu}{\lambda}}}{4}{H}^3 +\\
&- \frac{{{\nu}{\lambda}}}{4}{H}\chi _{a}^2 - 2{\kappa}{\nu}{H}\phi _{a}^2 - \frac{1}{2}{\kappa}{H}^2\phi _{a}^2 - \frac{1}{2}{\kappa }\chi _{a}^2\phi _{b}^2 \this \label{higgslagrenorm}
\end{align*}
and ${\mathscr{L}_{H}^{\rm{CT}}}$ is the counter term Lagrangian:
\begin{align*}
{\mathscr{L}_{H}^{\rm{CT}}}&= \frac{1}{2}\left( {\delta {Z_H}{\partial _\mu }H{\partial ^\mu }H - \delta m_H^2{H^2}} \right) + \frac{1}{2}\left( {\delta {Z_\chi }{\partial _\mu 
}{\chi _a}{\partial ^\mu }{\chi _a} - \delta M_{\chi\xi} ^2\chi _a^2} \right) - \frac{1}{4}\delta {Z_{A{\chi ^2}}}{\varepsilon _{abc}}\left( {{\chi 
_a}\overleftrightarrow{\partial^\mu}{\chi _b}} \right)A_\mu ^c + \\
&- \frac{1}{2}\delta {Z_{\chi HA}}\left( {{\chi _a}\overleftrightarrow{\partial^\mu}H} \right)A_\mu ^a + \frac{1}{2}\delta {Z_{H{A^2}}}HA_\mu ^a{A^{a\mu 
}} + \frac{1}{8}\delta {Z_{{H^2}{A^2}}}{H^2}A_\mu ^a{A^{a\mu }} + \frac{1}{8}\delta {Z_{{\chi ^2}{A^2}}}\chi _b^2A_\mu ^a{A^{a\mu }} + \\
&- \frac{1}{{32}}\delta {Z_{{H^4}}}{H^4} - \frac{1}{{16}}\delta {Z_{{H^2}{\chi ^2}}}{H^2}\chi _a^2 - \frac{1}{{32}}\delta {Z_{{\chi ^4}}}\chi _a^2\chi 
_b^2 - \frac{1}{4}\delta {Z_{{H^3}}}{H^3} - \frac{1}{4}\delta {Z_{H{\chi ^2}}}H\chi _a^2 - 2\delta {Z_{H{\phi ^2}}}H\phi _a^2 + \\
&- \frac{1}{2}\delta {Z_{{H^2}{\phi ^2}}}{H^2}\phi _a^2 - \frac{1}{2}\delta {Z_{{\chi ^2}{\phi ^2}}}\chi _a^2\phi _b^2 
\this \label{higgslagrenormCT}
\end{align*}
Finally the bare Faddeev-Popov ghost Lagrangian is 
\begin{align*}
{L_{0\rm{FPG}}} &= {{\bar u}_0}^a\left( {{\partial ^\mu }{\partial _\mu } + {\xi _0} M_0^2} \right) u_0^a + {e_0}{\varepsilon _{abc}}A_{0\mu }^a\left( {{\partial ^\mu } \bar u _0^b} \right) u_0^c + \frac{1}{2}{\xi _0} e_0^2{\nu _0}{H_0}{{\bar u}_0}^a u_0^a + \\
&- \frac{1}{2}{\xi _0} e_0^2{\nu _0}{\varepsilon _{abc}}{\chi _{0a}}{{\bar u}_0}^b u_0^c \this \label{fplagbare}
\end{align*}
Only a single field redefinition is required for the bare ghost field:
\begin{align}
u_0^a: = {u^a}\sqrt {{Z_u}} 
\label{renormtrans3}
\end{align}
Substituting the appropriate definitions from \eqref{renormtrans1} and \eqref{renormtrans2}, and \eqref{renormtrans3} into \eqref{fplagbare}
\begin{align*}
{L_{\rm{FPG}}} &= {{\bar u}^a}\left( {{Z_u}{\partial ^\mu }{\partial _\mu } + {\xi _0}{M^2}{Z_u}{Z_M}} \right){u^a} + e{Z_u}\sqrt {{Z_e}{Z_A}} {\varepsilon _{abc}}A_\mu ^a\left( {{\partial ^\mu }{{\bar u}^b}} \right){u^c} + \frac{1}{2}{\xi _0}{e^2}\nu {Z_e}{Z_u}\sqrt {{Z_\nu }{Z_\chi }} H{{\bar u}^a}{u^a} + \\
&- \frac{1}{2}{\xi _0}{e^2}\nu {Z_e}{Z_u}\sqrt {{Z_\nu }{Z_\chi }} {\varepsilon _{abc}}{\chi _a}{{\bar u}^b}{u^c} \this \label{fplagbaresub}
\end{align*}
We define the $Z_i$ factors as
\begin{equation}
\begin{aligned}
{Z_u}&: = 1 + \delta {Z_u}   &{\xi _0}{Z_u }{Z_M}{M^2}&:= \xi {M^2} + \delta M_{u \xi} ^2 & e{Z_u}\sqrt {{Z_e}{Z_A}}&: = e + \delta {Z_{A{u^2}}} \\ {\xi _0}{e^2}\nu {Z_e}{Z_u}\sqrt {{Z_\nu }{Z_\chi }}  &:= \xi{e^2}\nu  + \delta {Z_{H{u^2}}}   &
{\xi _0}{e^2}\nu {Z_e}{Z_u}\sqrt {{Z_\nu }{Z_\chi }}&: = \xi{e^2}\nu  + \delta {Z_{\chi {u^2}}}
\end{aligned}
\label{zfpfactors}
\end{equation}
Substituting \eqref{zfpfactors} into \eqref{fplagbaresub}, the Lagrangian will split into two pieces
\begin{align}
{\mathscr{L}_{0\rm{FPG}}} = {\mathscr{L}_{\rm{FPG}}} + {\mathscr{L}_{\rm{FPG}}^{\rm{CT}}}
\end{align}
where ${\mathscr{L}_{\rm{FPG}}}$ is
\begin{align*}
{\mathscr{L}_{\rm{FPG}}} &= {{\bar u}^a}\left( {{\partial ^\mu }{\partial _\mu } + \xi {M^2}} \right){u^a} + e{\varepsilon _{abc}}A_\mu ^a\left( {{\partial ^\mu }{{\bar u}^b}} \right){u^c} + \frac{1}{2}\xi {e^2}\nu H{{\bar u}^a}{u^a} - \frac{1}{2}\xi {e^2}\nu {\varepsilon _{abc}}{\chi _a}{{\bar u}^b}{u^c}  \this \label{fplagrenorm}
\end{align*}
and ${\mathscr{L}_{\rm{FPG}}^{\rm{CT}}}$ is the counter term Lagrangian:
\begin{align}
{\mathscr{L}_{\rm{FPG}}^{\rm{CT}}}={{\bar u}^a}\left( {\delta {Z_u}{\partial ^\mu }{\partial _\mu } + \delta M_{u\xi} ^2} \right){u^a} + \delta {Z_{A{u^2}}}{\varepsilon _{abc}}A_\mu ^a\left( {{\partial ^\mu }{{\bar u}^b}} \right){u^c} + \frac{1}{2}\delta {Z_{H{u^2}}}H{{\bar u}^a}{u^a}  - \frac{1}{2}\delta {Z_{\chi {u^2}}}{\varepsilon _{abc}}{\chi _a}{{\bar u}^b}{u^c}
\this \label{fplagrenormCT}
\end{align}

The bare field and coupling redefinitions \eqref{renormtrans1}, \eqref{renormtrans2}, \eqref{renormtrans3} together with the $Z_i$ definitions in \eqref{zpionrhofactors}, \eqref{zhiggsfactors} and \eqref{zfpfactors} is the Renormalization Transformation.

\section{Feynman Rules}
The Feynman rule will be computed for the specific example of the rho Green's function/propagator to demonstrate the ideas required to extract out the Feynman rules for this Lagrangian. The Lagrangian for the rho is 
\begin{align}
\mathscr{L}_A =  - \frac{1}{4}F_{\alpha \beta }^c{F^{c\alpha \beta }} + \frac{1}{2}{M^2}A_\alpha ^c{A^{c\alpha }} - \frac{1}{{2\xi }}{\left( {{\partial ^\alpha }A_\alpha ^c} \right)^2}
\end{align}
with the anti-symmetric property of the field strength tensor
\begin{align*}
F_{\alpha \beta }^c  &= {{\partial _\alpha }A_\beta ^c - {\partial _\beta }A_\alpha ^c} \\
&=  - F_{\beta \alpha }^c \this
\end{align*}
We can manipulate the form of the Lagrangian as follows
\begin{align*}
\mathscr{L}_A &=  - \frac{1}{4}\left( {{\partial _\alpha }A_\beta ^c - {\partial _\beta }A_\alpha ^c} \right){F^{c\alpha \beta }} + \frac{1}{2}{M^2}A_\alpha ^c{A^{c\alpha }} - \frac{1}{{2\xi }}{\left( {{\partial ^\alpha }A_\alpha ^c} \right)^2}\\
&=  - \frac{1}{4}\left( {{F^{c\alpha \beta }}{\partial _\alpha }A_\beta ^c - {F^{c\beta \alpha }}{\partial _\alpha }A_\beta ^c} \right) + \frac{1}{2}{M^2}A_\alpha ^c{A^{c\alpha }} - \frac{1}{{2\xi }}{\left( {{\partial ^\alpha }A_\alpha ^c} \right)^2}\\
&=  - \frac{1}{2}{F^{c\alpha \beta }}{\partial _\alpha }A_\beta ^c + \frac{1}{2}{M^2}A_\alpha ^c{A^{c\alpha }} - \frac{1}{{2\xi }}{\left( {{\partial ^\alpha }A_\alpha ^c} \right)^2}\\
&=  - \frac{1}{2}\left[ {{\partial ^\alpha }{A^{c\beta }}{\partial _\alpha }A_\beta ^c - {\partial ^\beta }{A^{c\alpha }}{\partial _\alpha }A_\beta ^c - {M^2}A_\alpha ^c{A^{c\alpha }} + \frac{1}{\xi }{\partial ^\alpha }A_\alpha ^c{\partial ^\beta }A_\beta ^c} \right] \this
\end{align*}
where we have used the anti-symmetric property of the field strength tensor. The action for this Lagrangian is:
\begin{align*}
S &= \int {{d^4}z \mathscr{L}_A} \\
&=  - \frac{1}{2}\int {{d^4}z\left[ {{\partial ^\alpha }{A^{c\beta }}{\partial _\alpha }A_\beta ^c - {\partial ^\beta }{A^{c\alpha }}{\partial _\alpha }A_\beta ^c - {M^2}A_\alpha ^c{A^{c\alpha }} + \frac{1}{\xi }{\partial ^\alpha }A_\alpha ^c{\partial ^\beta }A_\beta ^c} \right]} \\
&=  - \frac{1}{2}\int {{d^4}z\left[ { - A_\nu ^a\left( {\partial _z^2 + {M^2}} \right){A^{a\nu }} + A_\nu ^a{\partial ^\mu }{\partial ^\nu }A_\mu ^a - \frac{1}{\xi }A_\nu ^a{\partial ^\nu }{\partial ^\mu }A_\mu ^a} \right]} \\
&= \frac{1}{2}\int {{d^4}zA_\alpha ^c\left( z \right)\left[ {\left( {\partial _z^2 + {M^2}} \right){g^{\alpha \beta }} - \left( {1 - \frac{1}{\xi }} \right)\partial _z^\alpha \partial _z^\beta } \right]A_\beta ^c\left( z \right)}  \this
\end{align*}
where we have performed integration by parts with the surface terms contributing zero. The two-point function in configuration space is defined as
\begin{align*}
{\Gamma ^{\left( 2 \right)}}\left( {x,y} \right)&: = \frac{{{\delta ^2}}}{{\delta A_\mu ^a\left( x \right)\delta A_\nu ^b\left( y \right)}}iS\\
&= i{\delta ^{ab}}\left[ {\left( {\partial _y^2 + {M^2}} \right){g^{\mu \nu }} - \left( {1 - \frac{1}{\xi }} \right)\partial _y^\mu \partial _y^\nu } \right]\delta \left( {y - x} \right) \this
\end{align*}
and the Fourier transform of the two-point function is
\begin{align*}
{\widetilde \Gamma ^{\left( 2 \right)}}\left( {p,q} \right) &= \int {{d^4}x{d^4}y} {e^{ - i\left( {px + qy} \right)}}{\Gamma ^{\left( 2 \right)}}\left( {x,y} \right)\\
&= i{\delta ^{ab}}\left[ {\left( { - {q^2} + {M^2}} \right){g^{\mu \nu }} + \left( {1 - \frac{1}{\xi }} \right){q^\mu }{q^\nu }} \right]{\left( {2\pi } \right)^4}{\delta ^4}\left( {p + q} \right) \this
\end{align*}
We are now left to the task of finding the Green's function which is the reciprocal of  
\begin{align}
\overline \Gamma  _{ab}^{\mu \nu } \left(k\right) =  - i{\delta ^{ab}}\left[ {\left( { - {k^2} + {M^2}} \right){g^{\mu \nu }} + \left( {1 - \frac{1}{\xi }} \right){k^\mu }{k^\nu }} \right] \label{greensfunctoperator}
\end{align}
where
\begin{align}
\overline \Gamma  _{ab}^{\alpha \beta } \left(k\right) D_{\beta bc}^\mu \left(k\right) &= {\delta _{ac}}{g^{\alpha \mu }} \label{recipricalD}
\end{align}
and $D_{\beta bc}^\mu \left(k\right) $ the Green's function. We can extract out the isospin indices $D_{\beta bc}^\mu \left(k\right) = {\delta _{bc}}D_\beta ^\mu \left(k\right)$ and substituting into \eqref{recipricalD}
\begin{align*}
 - i{\delta ^{ab}}\left[ {\left( { - {k^2} + {M^2}} \right){g^{\alpha \beta }} + \left( {1 - \frac{1}{\xi }} \right){k^\alpha }{k^\beta }} \right] \cdot {\delta _{bc}}D_\beta ^\mu \left(k\right) &= {\delta _{ac}}{g^{\alpha \mu }}\\
\implies \left[ {\left( { - {k^2} + {M^2}} \right){g^{\alpha \beta }} + \left( {1 - \frac{1}{\xi }} \right){k^\alpha }{k^\beta }} \right]D_\beta ^\mu  \left(k\right)&= i{g^{\alpha \mu }} \this \label{rhopropdef}
\end{align*}
We need to find a $D_\beta ^\mu  \left(k\right)$ which satisfies \eqref{rhopropdef}. The most general tensor we can construct is a linear combination of $g_\beta ^\mu$ and ${k^\mu }{k_\beta }$ with 
\begin{align}
D_\beta ^\mu \left(k\right) := Ag_\beta ^{ \mu}  + B{k^\mu }{k_\beta } \label{gendefD}
\end{align}
with $A,B \in  \mathbb{C}$. Substituting \eqref{gendefD} into \eqref{rhopropdef}
\begin{align*}
\left[ {\left( { - {k^2} + {M^2}} \right){g^{\alpha \beta }} + \left( {1 - \frac{1}{\xi }} \right){k^\alpha }{k^\beta }} \right]\left( {Ag_\beta ^\mu  + B{k^\mu }{k_\beta }} \right) &= i{g^{\alpha \mu }}\\
\implies A\left( { - {k^2} + {M^2}} \right){g^{\alpha \mu }} + \left[ {B\left( {{M^2} - \frac{{{k^2}}}{\xi }} \right) + A\left( {1 - \frac{1}{\xi }} \right)} \right]{k^\alpha }{k^\mu } &= i{g^{\alpha \mu }} \this
\end{align*}
Since $g_\beta ^\mu$ and ${k^\mu }{k_\beta }$ are linearly independent we just have to match coefficients with 
\begin{align}
A\left( { - {k^2} + {M^2}} \right) &= i \quad \textrm{and} \quad B\left( {{M^2} - \frac{{{k^2}}}{\xi }} \right) + A\left( {1 - \frac{1}{\xi }} \right) = 0
\end{align}
which has solutions: 
\begin{align}
B &= \dfrac{{A\left( {\xi  - 1} \right)}}{{{k^2} - \xi {M^2}}} \\ 
A &=  - \dfrac{i}{{{k^2} - {M^2}}}
\end{align}
The rho Green's function/propagator is then given by
\begin{align*}
D_{ab}^{\mu \nu } &= {\delta _{ab}}{D^{\mu \nu }}\\
 &=  - \frac{{i{\delta _{ab}}}}{{{k^2} - {M^2} +i\varepsilon}}\left[ {{g^{\mu \nu }} + \frac{{\xi  - 1}}{{{k^2} - \xi {M^2}}}{k^\mu }{k^\nu }} \right] \this
\end{align*}
The counter term Lagrangian for the rho propagator is 
\begin{align}
{\mathscr{L}_{A}^{\rm{CT}}} = - \frac{1}{4}\delta {Z_A}F_{\mu \nu }^a{F^{a\mu \nu }} + \frac{1}{2}\delta {M^2}A_\mu ^a{A^{a\mu }} - \frac{1}{2}\delta \xi {\left( {{\partial ^\mu }A_\mu ^a} \right)^2}
\end{align}
Going through the above procedure again for the counter term Lagrangian but stopping at \eqref{greensfunctoperator}, we get the counter term for the rho two point function
\begin{align}
\overline \Gamma  _{ab\rm{CT}}^{\mu \nu } = i\delta_{ab}\left[ \left(-k^2\delta Z_A +\delta M^2\right)g^{\mu \nu}- \left( \delta\xi-\delta Z_A \right)k^\mu k^\nu \right]
\end{align}
\sloppy
\begin{center}
	\captionof{table}{Feynman rules for this QFT. All momenta are incoming with $\displaystyle\sum\limits_{i}p_i =0$} \label{tabfrule}
	\begin{longtable}[l]{@{}p{2.8in}p{4.3in}@{}}
		
		\toprule[1pt] 
		\multicolumn{1}{l}{{Kinetic(P)/Interaction(I)/Counter(C)}} & \multicolumn{1}{l}{{Feynman Rule}} 
		\\ 
		\midrule
		\endfirsthead
		
		\multicolumn{2}{c}{{Table \ref{tabfrule} -- continued from previous page}} 
		\\
		\\
		\toprule[1pt] 
		\multicolumn{1}{l}{{Kinetic(P)/Interaction(I)/Counter(C)}} & \multicolumn{1}{l}{{Feynman Rule}} 
		\\ 
		\midrule
		\endhead
		
		\midrule[0.1pt]
		\bottomrule[1pt] 
		\endlastfoot

		\\

    	P.1	 $\mathscr{L}=\frac{1}{2}\left( \partial^\mu \phi_a \partial_\mu \phi_a - m^2 \phi_a^2\right)$
    		 &
    		 \begin{minipage}{\linewidth}
    			\begin{tikzpicture}[scale=1]
		    		\draw[\Fphi] (2,2) -- (5,2);
	    			 \fill[fill=black] (7.3,2) node {$\bar\Gamma^{2ab} = \dfrac{i\delta^{ab}}{k^2-m^2+i\epsilon} $};
	    			 \fill[fill=black] (3.5,2.25) node {$k$};
	    			 \fill[fill=black] (2.25,2.25) node {$a$};
 	    			 \fill[fill=black] (4.75,2.25) node {$b$};
    		 	\end{tikzpicture}
    		 \end{minipage}
    		\\ \\
    	\small{P.2} $\mathscr{L}=-\dfrac{(F^a_{\mu\nu})^2}{4}+\dfrac{M^2(A^a_\mu)^2}{2} -\dfrac{(\partial^\mu A^a_\mu )^2}{2\xi} $
    		 &
    		 \begin{minipage}{\linewidth}
    			\begin{tikzpicture}[scale=1]
		    		\draw[\Frho] (2,2) -- (5,2);
	    			 \fill[fill=black] (9.05,2) node {$\Gamma _{ab}^{2\mu \nu } =  \dfrac{{-i{\delta ^{ab}}}}{{{k^2} - {M^2}+i\epsilon}}\left[ {{g^{\mu \nu }} + \dfrac{{\left( {\xi  - 1} \right){k^\mu }{k^\nu }}}{{{k^2} - \xi {M^2}}}} \right]$};
	    			 \fill[fill=black] (3.5,2.25) node {$k$};
	    			 \fill[fill=black] (2.25,2.25) node {$a,\mu $};
	    			 \fill[fill=black] (4.75,2.25) node {$b,\nu $};
    		 	\end{tikzpicture}
    		 \end{minipage}
    		\\ \\
    	P.3	 $\mathscr{L}=\frac{1}{2}\left( \partial^\mu H \partial_\mu H - m_H^2 H^2 \right)$
    		 &
    		 \begin{minipage}{\linewidth}
    			\begin{tikzpicture}[scale=1]
		    		\draw[\Fhiggs] (2,2) -- (5,2);
	    			 \fill[fill=black] (7.2,2) node {$\bar\Gamma^{2} = \dfrac{i}{k^2-m^2_H+i\epsilon} $};
	    			 \fill[fill=black] (3.5,2.25) node {$k$};
    		 	\end{tikzpicture}
    		 \end{minipage}
	    	\\ \\
    	P.4	 $\mathscr{L}=\frac{1}{2}\left( \partial^\mu \chi_a \partial_\mu \chi_a- \xi M^2 \chi_a\chi_a \right)$
    		 &
    		 \begin{minipage}{\linewidth}
    			\begin{tikzpicture}[scale=1]
		    		\draw[\Fchi] (2,2) -- (5,2);
	    			 \fill[fill=black] (7.4,2) node {$\bar\Gamma^{2ab} = \dfrac{i\delta^{ab}}{k^2-\xi M^2+i\epsilon} $};
	    			 \fill[fill=black] (3.5,2.25) node {$k$};
	    			 \fill[fill=black] (2.25,2.25) node {$a$};
	    			 \fill[fill=black] (4.75,2.25) node {$b$};	    			 
    		 	\end{tikzpicture}
    		 \end{minipage}
	    	\\ \\
    	P.5	 $\mathscr{L}=  - {{\bar u}^a}\left( {{\partial ^\mu }{\partial _\mu } + \xi {M^2}} \right){u^a}$
    		 &
    		 \begin{minipage}{\linewidth}
    			\begin{tikzpicture}[scale=1]
		    		\draw[\Fgst] (2,2) -- (5,2);
	    			 \fill[fill=black] (7.5,2) node {$\bar\Gamma^{2ab} = \dfrac{i\delta^{ab}}{k^2-\xi M^2+i\epsilon} $};
	    			 \fill[fill=black] (3.5,2.25) node {$k$};
	    			 \fill[fill=black] (2.25,2.25) node {$a$};
	    			 \fill[fill=black] (4.75,2.25) node {$b$};	    			 
    		 	\end{tikzpicture}
    		 \end{minipage}
	    	\\ \\ \midrule[0.2pt] \\
    	I.1.	 $\mathscr{L} = \frac{1}{2}e{\varepsilon _{aij}}A_\mu ^a\left( {{\phi _i}{\partial ^\mu }{\phi _j} - {\phi _j}{\partial ^\mu }{\phi _i}} \right)$ 
    		&
    		 \begin{minipage}{\linewidth}
    			\begin{tikzpicture}[scale=1]
    			 	 \FThreeVert{2}{2}{\Fphi}{$m$}{\Fphi}{$n$}{\Frho}{$b,\nu$}
	    			 \fill[fill=black] (6.7,2) node {${\bar\Gamma^\nu _{bmn} } = e{\varepsilon _{bmn}}{\left( {{p_1} - {p_2}} \right)^\nu } $};
    		 	\end{tikzpicture}
    		 \end{minipage}
	    	\\ \\     		 
    	I.2.	$\mathscr{L} =  - \frac{1}{2}{e^2}{\varepsilon _{aij}}{\varepsilon _{bjk}}{\phi _i}{\phi _k}A_\mu ^a{A^{b\mu }}$
    		&
    		\begin{minipage}{\linewidth}
    			\begin{tikzpicture}[scale=1]
					 \FFourVert{2}{2}{\Fphi}{$m$}{\Fphi}{$n$}{\Frho}{$c,\nu$}{\Frho}{$d,\alpha$}
    			 	 \fill[fill=black] (7.8,2) node {$\bar \Gamma _{cdmn}^{\alpha \nu } = i{e^2}{g^{\alpha \nu }}\left( {{\varepsilon _{jdn}}{\varepsilon _{jcm}} + {\varepsilon _{jdm}}{\varepsilon _{jcn}}} \right)$ };
	    	 	\end{tikzpicture}
	    	 \end{minipage}	
    		\\ \\
    	I.3.	$\mathscr{L} =  - \frac{\lambda }{{32}}{H^4}$
    		&
    		\begin{minipage}{\linewidth}
    		 	\begin{tikzpicture}[scale=1]
     				\FFourVert{2}{2}{\Fhiggs}{}{\Fhiggs}{}{\Fhiggs}{}{\Fhiggs}{}
     				\fill[fill=black] (5.7,2) node {$\bar \Gamma  =  - \frac{3}{4}i\lambda $ };
     			\end{tikzpicture}
    		\end{minipage}
		\\ \\
		I.4.	$\mathscr{L} =  - \frac{\lambda }{{16}}{H^2}\chi _a^2$
			&
			\begin{minipage}{\linewidth}
    		 	\begin{tikzpicture}[scale=1]
     				\FFourVert{2}{2}{\Fhiggs}{$a$}{\Fhiggs}{$b$}{\Fchi}{$b$}{\Fchi}{$c$}
     				\fill[fill=black] (5.7,2) node {${{\bar \Gamma }^{bc}} =  - \frac{1}{4}i\lambda {\delta ^{bc}}$};
     			\end{tikzpicture}
    		\end{minipage}			
		\\ \\
		I.5.	$\mathscr{L} =  - \frac{\lambda }{{32}}\chi _a^2\chi _b^2$
			&
			\begin{minipage}{\linewidth}
    		 	\begin{tikzpicture}[scale=1]
     				\FFourVert{2}{2}{\Fchi}{$b$}{\Fchi}{$c$}{\Fchi}{$d$}{\Fchi}{$e$}
     				\fill[fill=black] (7.7,2) node {${{\bar \Gamma }^{bcde}} =  - \frac{1}{4}i\lambda \left( {{\delta ^{cd}}{\delta ^{be}} + {\delta ^{bd}}{\delta ^{ce}} + {\delta ^{bc}}{\delta ^{de}}} \right)$};
     			\end{tikzpicture}
    		\end{minipage}
			
		\\ \\
		I.6.	$\mathscr{L} =  - \frac{1}{4}\lambda \nu {H^3}$
			&
			\begin{minipage}{\linewidth}
    		 	\begin{tikzpicture}[scale=1]
     				\FThreeVert{2}{2}{\Fhiggs}{}{\Fhiggs}{}{\Fhiggs}{}
     				\fill[fill=black] (5.7,2) node {$\bar \Gamma  =  - \frac{3}{2}i\lambda \nu $};
     			\end{tikzpicture}
    		\end{minipage}			
			
		\\ \\
		I.7.	$\mathscr{L} =  - \frac{1}{4}\lambda \nu H\chi _a^2$
			&
			\begin{minipage}{\linewidth}
    		 	\begin{tikzpicture}[scale=1]
	    		 	\FThreeVert{2}{2}{\Fchi}{$b$}{\Fchi}{$c$}{\Fhiggs}{}
     				\fill[fill=black] (6.,2) node {${{\bar \Gamma }^{bc}} =  - \frac{1}{2}i\lambda \nu {\delta ^{bc}}$};
     			\end{tikzpicture}
    		\end{minipage}			
			
		\\ \\
		I.8.	$\mathscr{L} =  - \frac{1}{4}e{\varepsilon _{abc}}\left( {{\chi _a}{\partial ^\mu }{\chi _b} - {\chi _b}{\partial ^\mu }{\chi _a}} \right)A_\mu ^c $ 
			&
			\begin{minipage}{\linewidth}
    		 	\begin{tikzpicture}[scale=1]
	    		 	\FThreeVert{2}{2}{\Fchi}{$a$}{\Fchi}{$b$}{\Frho}{$c,\mu$}
     				\fill[fill=black] (6.8,2) node {$\bar \Gamma _{cab}^\nu  = -\frac{1}{2}e{\varepsilon _{cab}}{\left( {{p_1} - {p_2}} \right)^\mu }$};
     			\end{tikzpicture}
    		\end{minipage}			
			
		\\ \\
		I.9.	$\mathscr{L} = - \frac{1}{2}e\left( {{\chi _a}{\partial ^\mu }H - H{\partial ^\mu }{\chi _a}} \right)A_\mu ^a$
			&
			\begin{minipage}{\linewidth}
    		 	\begin{tikzpicture}[scale=1]
     				\FThreeVert{2}{2}{\Fchi}{$b$}{\Fhiggs}{}{\Frho}{$c,\nu$}
     				\fill[fill=black] (6.8,2) node {$\bar \Gamma _{bc}^\nu  = \frac{1}{2}e{\delta ^{bc}}{\left( {{p_1} - {p_2}} \right)^\nu }$};
     			\end{tikzpicture}
    		\end{minipage}				
			
		\\ \\
		I.10. $\mathscr{L} = \frac{1}{2}{e^2}\nu HA_\mu ^a{A^{a\mu }}$
			&
			\begin{minipage}{\linewidth}
    		 	\begin{tikzpicture}[scale=1]
     				\FThreeVert{2}{2}{\Fhiggs}{}{\Frho}{$b,\nu$}{\Frho}{$c,\alpha$}
     				\fill[fill=black] (6.4,2) node {$\bar \Gamma _{bc}^{\alpha \nu } = i{e^2}\nu {\delta ^{bc}}{g^{\alpha \nu }}$};
     			\end{tikzpicture}
    		\end{minipage}			

		\\ \\
		I.11. $\mathscr{L} = \frac{1}{8}{e^2}{H^2}A_\mu ^a{A^{a\mu }}$
			&
			\begin{minipage}{\linewidth}
    		 	\begin{tikzpicture}[scale=1]
     				\FFourVert{2}{2}{\Fhiggs}{}{\Fhiggs}{}{\Frho}{$b,\nu$}{\Frho}{$c,\alpha$}
     				\fill[fill=black] (6.8,2) node {$\bar \Gamma _{bc}^{\alpha \nu } = \frac{1}{2}i{e^2}{\delta ^{bc}}{g^{\alpha \nu }}$};
     			\end{tikzpicture}
    		\end{minipage}			
			
		\\ \\    
		I.12. $\mathscr{L} = \frac{1}{8}{e^2}\chi _b^2A_\mu ^a{A^{a\mu }}$
			&
			\begin{minipage}{\linewidth}
    		 	\begin{tikzpicture}[scale=1]
     				\FFourVert{2}{2}{\Fchi}{$c$}{\Fchi}{$d$}{\Frho}{$e,\nu$}{\Frho}{$f,\alpha$}
     				\fill[fill=black] (7.2,2) node {$\bar \Gamma _{cdef}^{\alpha \nu } = \frac{1}{2}i{e^2}{\delta ^{cd}}{\delta ^{ef}}{g^{\alpha \nu }}$};
     			\end{tikzpicture}
    		\end{minipage}			
			
		\\ \\
		I.13. $\mathscr{L} =  - \frac{1}{2}\kappa {H^2}\phi _n^2$
			&
			\begin{minipage}{\linewidth}
    		 	\begin{tikzpicture}[scale=1]
     				\FFourVert{2}{2}{\Fhiggs}{}{\Fhiggs}{}{\Fphi}{$a$}{\Fphi}{$b$}
     				\fill[fill=black] (6.8,2) node {${{\bar \Gamma }^{ab}} =  - 2i\kappa {\delta ^{ab}}$};
     			\end{tikzpicture}
    		\end{minipage}				
			
		\\ \\
		I.14. $\mathscr{L} =  - \frac{1}{2}\kappa \chi _a^2\phi _n^2$
			&
			\begin{minipage}{\linewidth}
    		 	\begin{tikzpicture}[scale=1]
     				\FFourVert{2}{2}{\Fchi}{$b$}{\Fchi}{$c$}{\Fphi}{$e$}{\Fphi}{$f$}
     				\fill[fill=black] (7.2,2) node {${{\bar \Gamma }^{abef}} =  - 2i\kappa {\delta ^{ab}}{\delta ^{ef}}$};
     			\end{tikzpicture}
    		\end{minipage}				
			
		\\ \\
		I.15. $\mathscr{L} =  - 2\kappa \nu H\phi _n^2$
			&
			\begin{minipage}{\linewidth}
    		 	\begin{tikzpicture}[scale=1]
		   		 	\FThreeVert{2}{2}{\Fhiggs}{}{\Fphi}{$a$}{\Fphi}{$b$}
	    			\fill[fill=black] (6.8,2)  node {${{\bar \Gamma }^{ab}} =  - 4i\kappa \nu {\delta ^{ab}}$};
     			\end{tikzpicture}
    		\end{minipage}			
			
		\\ \\
		I.16. $\mathscr{L} =  - e{\varepsilon _{abc}}\left( {{\partial ^\mu }{{\bar u}_a}} \right){u_b}A_\mu ^c$
			&
			\begin{minipage}{\linewidth}
    		 	\begin{tikzpicture}[scale=1]
		   		 	\FThreeVert{2}{2}{\Fgst}{$f$}{\Fgst}{$g$}{\Frho}{$h,\nu$}
	    			\fill[fill=black] (7.2,2) node {$\bar \Gamma _{fgh}^\nu  =  - e{\varepsilon _{fgh}}{\left( {{p_2} + {p_3}} \right)^\nu }$};
     			\end{tikzpicture}
    		\end{minipage}		
		\\ \\
		I.17. $\mathscr{L} =  \dfrac{1}{2}e {\varepsilon _{abc}} F^a _{\mu\nu}A^{b\mu} A^{c_\nu}$
			&
			\begin{minipage}{\linewidth}
    		 	\begin{tikzpicture}[scale=1]
		   		 	\FThreeVert{2}{2}{\Frho}{$a,\alpha$}{\Frho}{$b,\beta$}{\Frho}{$c,\gamma$}
	    			\fill[fill=black] (5.5,-0.5) node {$\bar \Gamma^{abc}_{\alpha \beta \gamma}= e{\varepsilon _{abc}} \left[
	    			 \left(p_3-p_2\right)_\alpha g_{\beta\gamma} + \left(p_1-p_3\right)_\beta g_{\gamma\alpha}  + \left(p_2-p_1\right)_\gamma g_{\beta\alpha}\right]$};
     			\end{tikzpicture}
    		\end{minipage}	  
		\\ \\
		I.18. $\mathscr{L} =  \frac{1}{2}\xi {e^2}\nu {H}{{\bar u}^b}{u^b}$
			&
			\begin{minipage}{\linewidth}
    		 	\begin{tikzpicture}[scale=1]
		   		 	\FThreeVert{2}{2}{\Fgst}{$a$}{\Fgst}{$c$}{\Fhiggs}{}
	    			\fill[fill=black] (6.8,2) node {${{\bar \Gamma }_{ac}} = \frac{1}{2}i\xi {e^2}\nu {\delta _{ac}}$};
     			\end{tikzpicture}
    		\end{minipage}      		
		\\ \\
		I.19. $\mathscr{L} =   - \frac{1}{2}\xi {e^2}\nu {\varepsilon _{abc}}{\chi _a}{{\bar u}^b}{u^c}$
			&
			\begin{minipage}{\linewidth}
    		 	\begin{tikzpicture}[scale=1]
		   		 	\FThreeVert{2}{2}{\Fgst}{$d$}{\Fgst}{$e$}{\Fchi}{f}
	    			\fill[fill=black] (6.8,2) node {${{\bar \Gamma }_{fed}} =  - \frac{1}{2}i\xi {e^2}\nu {\varepsilon _{fed}}$};
     			\end{tikzpicture}
    		\end{minipage}      		
		\\ \\
		I.20. $\mathscr{L} =  -\dfrac{1}{8}\lambda _4 \phi_a^2 \phi_b^2$
			&
			\begin{minipage}{\linewidth}
    		 	\begin{tikzpicture}[scale=1]
		   		 	\FFourVert{2}{2}{\Fphi}{$a$}{\Fphi}{$b$}{\Fphi}{$c$}{\Fphi}{$d$}
	    			\fill[fill=black] (7.6,2) node {${{\bar \Gamma }_{abcd}} =  - i{\lambda _4}\left( {{\delta _{ab}}{\delta _{cd}} + {\delta _{bc}}{\delta _{da}} + {\delta _{bd}}{\delta _{ac}}} \right)$};
     			\end{tikzpicture}
    		\end{minipage} 
    	\\ \\						
		I.21. $\mathscr{L} =  -\dfrac{1}{4}e^2 \varepsilon _{abc}\varepsilon _{aef} A^b_{\mu}A^c_{\nu}A^{e\mu} A^{f\nu}$
			&
			\begin{minipage}{\linewidth}
    		 	\begin{tikzpicture}[scale=1]
		   		 	\FFourVert{2}{2}{\Frho}{$a,\mu$}{\Frho}{$b,\nu$}{\Frho}{$c,\rho$}{\Frho}{$d,\sigma$}
	    			\fill[fill=black] (7,2) node {$\Gamma _{\mu \nu \pi \sigma }^{abcd} = - i{e^2}\bigl[{\varepsilon _{fab}}{\varepsilon _{fcd}}\left( {{g_{\mu \rho }}{g_{\nu \sigma }} - {g_{\mu \sigma }}{g_{\nu \rho }}} \right) + $ };
		    			\fill[fill=black] (7.5,1.25) node {$ + {\varepsilon _{fac}}{\varepsilon _{fdb}}\left( {{g_{\mu \sigma }}{g_{\rho \nu }} - {g_{\mu \nu }}{g_{\rho \sigma }}} \right) +  $};
		    			\fill[fill=black] (7.5,0.5) node {$ + {\varepsilon _{fad}}{\varepsilon _{fbc}}\left( {{g_{\mu \nu }}{g_{\rho \sigma }} - {g_{\mu \rho }}{g_{\nu \sigma }}} \right)\bigr]  $};
     			\end{tikzpicture}
    		\end{minipage}      		
		\\ \\
    	\\ 
		\midrule[0.2pt] 
		\\
		C.1.\begin{minipage}{\linewidth}
    		 	\begin{tikzpicture}[scale=1]
	    			\fill[fill=black] (0,0) node {$\mathscr{L} = - \dfrac{1}{4}\delta {Z_A}F_{\mu \nu }^a{F^{a\mu \nu }} + \dfrac{1}{2}{\delta {M^2}A_\mu ^a{A^{a\mu }}}$};
    				\fill[fill=black] (0,-1) node {$ - \dfrac{1}{2}{\delta \xi {{\left( {{\partial ^\mu }A_\mu ^a} \right)}^2}}$};
    		   	\end{tikzpicture}
    	 	\end{minipage} 
	    	&			
			\begin{minipage}{\linewidth}
	    	 	\begin{tikzpicture}[scale=1]
					\draw[\Frho] (2,1.5) -- (5,1.5);
	    			 \fill[fill=black] (9.25,1.5) node {$\overline \Gamma  _{ab\rm{CT}}^{\mu \nu }= i{\delta ^{ab}}\bigl[ \left( { {k^2}\delta {Z_A} - \delta {M^2}} \right){g^{\mu \nu }}$};
					 \fill[fill=black] (9.25,0.5) node {$ + \left( {\delta \xi  - \delta {Z_A}}\right){k^\mu }{k^\nu } \bigr]$};
	    			 \fill[fill=black] (3.5,1.95) node {$k$};					 
	    			 \fill[fill=black] (2.25,1.75) node {$a,\mu $};
	    			 \fill[fill=black] (4.75,1.75) node {$b,\nu $};
	    			 \ct{3.5}{1.5}{0.2}
	    		\end{tikzpicture}
	    	\end{minipage} 
		\\ \\
    	C.2.
    	\begin{minipage}{\linewidth}
    	   	\begin{tikzpicture}[scale=1]
    			\fill[fill=black] (-0.2,0) node {${\mathscr{L}} = \frac{1}{2}\left( {{\delta Z_H}{\partial _\mu }H{\partial ^\mu }H - {Z_H}{Z_{{m_H}}}m_H^2{H^2}} \right)$};
    	    \end{tikzpicture}
   	 	\end{minipage} 
	    	&			
			\begin{minipage}{\linewidth}
	    	 	\begin{tikzpicture}[scale=1]
					\draw[\Fhiggs] (2,2) -- (5,2);
	    			 \fill[fill=black] (8.75,2) node {$\overline \Gamma  _{\rm{CT}}= i\left( { {k^2}\delta {Z_H} - \delta {m_H ^2}} \right)$};
	    			 \fill[fill=black] (3.5,2.45) node {$k$};
	    			 \ct{3.5}{2}{0.2}
	    		\end{tikzpicture}
	    	\end{minipage} 
		\\ \\
    	C.3. $\mathscr{L} = \frac{1}{2}\delta {Z_e}{\varepsilon _{aij}}A_\mu ^a\left( {{\phi _j}{\partial ^\mu }{\phi _i} - {\phi _i}{\partial ^\mu }{\phi _j}} \right)$
	    	&			
			\begin{minipage}{\linewidth}
	    	 	\begin{tikzpicture}[scale=1]
			 		\FThreeVertCTS{2}{1}{\Fphi}{$a$}{\Fphi}{$b$}{\Frho}{$c,\mu$}
			 		 \fill[fill=black] (7.25,1) node {$\overline \Gamma  _{ab\rm{CT}}^{\mu }=\delta Z_e{\varepsilon _{cab}}{\left( {{p_1} - {p_2}} \right)^\mu }$};
	    		\end{tikzpicture}
	    	\end{minipage} 
		\\ \\		
    	C.4. $\mathscr{L} = - 2\delta {Z_{H{\phi ^2}}}H\phi _a^2$
	    	&			
			\begin{minipage}{\linewidth}
	    	 	\begin{tikzpicture}[scale=1]
			 		\FThreeVertCTS{2}{1}{\Fphi}{$a$}{\Fphi}{$b$}{\Fhiggs}{}
			 		 \fill[fill=black] (6.85,1) node {$\overline \Gamma  _{ab\rm{CT}}=- 4i\delta {Z_{H{\phi ^2}}}{\delta ^{ab}}$};
	    		\end{tikzpicture}
	    	\end{minipage} 
		\\ \\	
    	C.5. $\mathscr{L} = - \dfrac{1}{8}{\delta {Z_{{\phi ^4}}}}\left( {\phi _a^2\phi _b^2} \right)$
	    	&			
			\begin{minipage}{\linewidth}
	    	 	\begin{tikzpicture}[scale=1]
		 			\FFourPointCounterTermS{2}{2}
			 		\fill[fill=black] (7.25,2) node {$\overline \Gamma  _{abcd\rm{CT}}= - i{{\delta {Z_{{\phi ^4}}}}}\left( {{\delta _{ab}}{\delta _{cd}} + {\delta _{bc}}{\delta _{da}} + {\delta _{bd}}{\delta _{ac}}} \right)$};
	    		\end{tikzpicture}
	    	\end{minipage} 
		\\ \\							
		\\ \\	
    \end{longtable} 	
\end{center}

%
\part{Quantum Corrections}
\chapter{Scattering Lengths at Tree Level}

\section{Useful Formulae}

\numberwithin{equation}{section}

\lettrine[lines=1]{\initfamily\textcolor{darkblue}{C}}{onsider} an on-shell scattering process of 
\begin{align*}
\phi^a\left(p_1\right) + \phi^b\left(p_2\right) \to \phi^c\left(p_3\right) + \phi^d\left(p_4\right)
\end{align*}
with the 4-momentum conservation equation
\begin{align*}
p_1^\mu +p_2^\mu = p_3^\mu + p_4^\mu
\end{align*}
Since these particle are on-shell $p_i^2=m^2$. We can define the Mandlestam invariants:
\begin{equation}
\begin{aligned}
  s &:= {\left( {{p_1} + {p_2}} \right)^2} = {\left( {{p_3} + {p_4}} \right)^2} \\ 
  t &:= {\left( {{p_1} - {p_3}} \right)^2} = {\left( {{p_4} - {p_2}} \right)^2} \\ 
  u &:= {\left( {{p_1} - {p_4}} \right)^2} = {\left( {{p_3} - {p_2}} \right)^2} 
\end{aligned}
\label{stumandle}
\end{equation}
We can rewrite the dot products $p_1 \cdot p_2$ and $p_3 \cdot p_4$ using the $s$ Mandlestam variable,
\begin{align}
 s 	&= p_1^2 + 2{p_1} \cdot {p_2} + p_2^2 \nonumber\\ 
	&= 2m ^2 + 2{p_1} \cdot {p_2} \nonumber\\ 
 \Rightarrow{p_1} \cdot {p_2} &= \frac{{s - 2m ^2}}{2} \\ 
 s 	&= p_3^2 + 2{p_3} \cdot {p_4} + p_4^2 \nonumber\\ 
 \Rightarrow {p_3} \cdot {p_4} &= {p_1} \cdot {p_2} = \frac{{s - 2m^2}}{2}
\end{align}
A summary of all the dot products are listed below %
\begin{equation}
\begin{aligned}
 {p_3} \cdot {p_4} &= {p_1} \cdot {p_2} = \frac{{s - 2m ^2}}{2} \\
 {p_2} \cdot {p_4} &= {p_1} \cdot {p_3} = \frac{{2m ^2 - t}}{2} \\ 
 {p_2} \cdot {p_3} &= {p_1} \cdot {p_4} = \frac{{2m ^2 - u}}{2}
\end{aligned}
\label{dotprods}
\end{equation}

Consider now the case of the elastic pion scattering process in the center of mass frame, with the incoming three momentum $\underline{q}$ and the outgoing three momentum $\underline{q'}$ with
\begin{align}
  \underline q &\ne \underline q ' \\ 
 q:=|\underline q| &= |\underline q '|
\end{align}
where $q$ is defined as the magnitude $\underline q$. For the $t$ channel process,
\begin{figure}[H]
\centering
\includegraphics[width=0.7\linewidth]{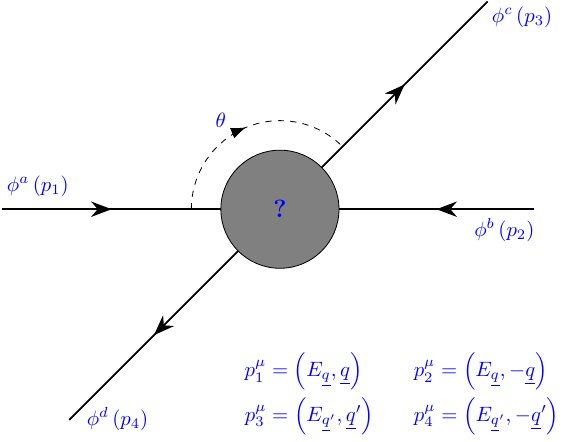}
\caption{$t$ channel scattering process}
\label{fig:tchannel}
\end{figure}
Let the 3-momenta for $\phi^a$ and $\phi^c$  be $\underline{q}$ and $\underline{q}'$ respectively, such that
\begin{align}
 p_1^\mu &= \left( {{E_{{\underline q}}},\underline q } \right) = \left( {\sqrt {m ^2 + |\underline q {|^2}} ,\underline q } \right) \\ 
 {p_3^\mu} &= \left( {{E_{{\underline q'}}},\underline {q'} } \right) = \left( {\sqrt {m ^2 + |\underline {q'} {|^2}} ,\underline {q'} } \right) = \left( {\sqrt {m^2 + |\underline q {|^2}} ,\underline {q'} } \right) 
\end{align}
The product of $p_1 \cdot p_3$ is
\begin{align}
 {p_1} \cdot {p_3} &= {\left( {\sqrt {m ^2 + |\underline q {|^2}} } \right)^2} - \underline q  \cdot \underline {q'}  \nonumber\\ 
  &= m ^2 + |\underline q {|^2} - |\underline q ||\underline {q'} |\cos \theta  \nonumber\\ 
  &= m ^2 + {q^2}\left( {1 - \cos \theta } \right) 
  \label{p1p3}
\end{align}
Substituting ${p_1} \cdot {p_3}$ from \eqref{dotprods} into \eqref{p1p3} the $t$ Mandlestam variable is now given by
\begin{align}
 t &= 2m ^2 - 2{p_1} \cdot {p_3} \nonumber\\ 
   &= 2m ^2 - 2\left[ {m^2 + {q^2}\left( {1 - \cos \theta } \right)} \right] \nonumber\\ 
   &=  - 2{q^2}\left( {1 - \cos \theta } \right) 
\end{align}
Consider the $u$ channel scattering process. 
\begin{figure}[H]
\centering
\includegraphics[width=0.7\linewidth]{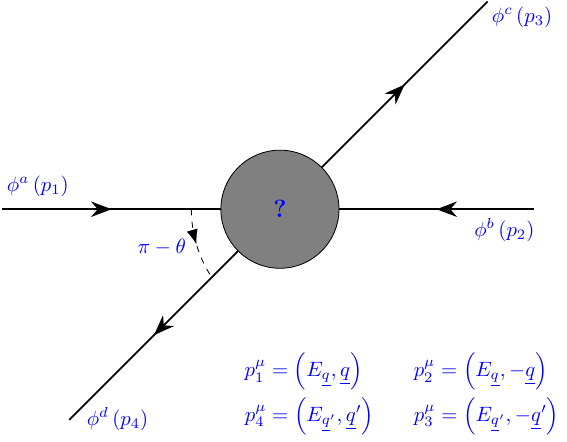}
\caption{$u$ channel scattering process}
\label{fig:uchannel}
\end{figure}
Let the 3-momenta for $\phi^a$ and $\phi^d$  be $\underline{q}$ and $\underline{q}'$ respectively, such that
\begin{align}
 {p_1} &= \left( {{E_{{\underline q}}},\underline q } \right) = \left( {\sqrt {m ^2 + |\underline q {|^2}} ,\underline q } \right) \\ 
 {p_4} &= \left( {{E_{{\underline q'}}},\underline {q'} } \right) = \left( {\sqrt {m ^2 + |\underline {q'} {|^2}} ,\underline {q'} } \right) = \left( {\sqrt {m^2 + |\underline q {|^2}} ,\underline {q'} } \right)
\end{align}
The product of $p_1 \cdot p_4$ is
\begin{align}
 {p_1} \cdot {p_4} &= m^2 + {q^2}\left( {1 + \cos \theta } \right) \label{p1p4}
\end{align}
Substituting ${p_1} \cdot {p_4}$ from \eqref{dotprods} into \eqref{p1p4} the $u$ Mandlestam variable is now given by
\begin{align}
 u =  - 2{q^2}\left( {1 + \cos \theta } \right) 
\end{align}
The three Mandlestam variables are related to each with
\begin{align}
s + t + u = \sum\limits_{j = 1}^4 {p_j^2}  = 4m^2 \label{mandlestamsum}
\end{align}
We can use \eqref{mandlestamsum} to find the $s$ Mandlestam variable from $t$ and $u$.
\begin{align}
 s &= 4m^2 - t - u \nonumber\\ 
  &= 4m^2 + 2{q^2}\left( {1 - \cos \theta } \right) + 2{q^2}\left( {1 + \cos \theta } \right) \nonumber\\ 
  &= 4m^2 + 4{q^2} 
\end{align}
We shall define the cosine of the scattering angle and the ratio $R$
\begin{align}
z &:= \cos \theta	\label{zcostheta} \\
R &:= \frac{q^2}{m^2} \label{Rqsqtomsq}
\end{align}
Then the necessary formulae for the following calculations in terms of $z$ and $R$ are
\begin{align}
 s &= 4{m^2}\left(1+R\right) \label{mandletams}\\ 
 t &=  - 2{m^2}\left( {1 - z} \right)R \label{mandletamt}\\ 
 u &=  - 2{m^2}\left( {1 + z} \right)R \label{mandletamu} \\ 
 s - t &= 2m^2\left[2 + \left(3 - z\right)R\right] \\ 
 s - u &= 2m^2\left[2 + \left(3 + z\right)R\right]\\ 
 t - u &= 4{m^2}Rz
\end{align}

\section{Tree Scattering Amplitudes}
We have now reached the stage to calculate the pion scattering lengths. A list of the required tree diagrams are generated from the Feynman rules for pion-pion scattering and are shown below. 

\begin{tikzpicture}[scale=1]
	
	\FScatteringS{0}{0}{\Fphi}{\Frho} 
	\fill[fill=black] (5,0) node {$+$};		
	\FScatteringT{7.25}{0}{\Fphi}{\Frho}
	\fill[fill=black] (9.5,0) node {$+$};		
	\FScatteringU{11.75}{0}{\Fphi}{\Frho}
	\fill[fill=black] (14.25,0) node {$+$};		

	\FScatteringS{0}{-7}{\Fphi}{\Fhiggs} 
	\fill[fill=black] (5,-7) node {$+$};		
	\FScatteringT{7.25}{-7}{\Fphi}{\Fhiggs}
	\fill[fill=black] (9.5,-7) node {$+$};		
	\FScatteringU{11.75}{-7}{\Fphi}{\Fhiggs}
	\fill[fill=black] (14.25,-7) node {$+$};			
 	\FFourScattering{1.25}{-13}
\end{tikzpicture}

For the $s$ channel scattering process mediated by the rho:
\begin{figure}[H]
\centering
	\begin{tikzpicture}[scale=1]
		\FScatteringS{5}{0}{\Fphi}{\Frho} 
	\end{tikzpicture}
\end{figure}
With the definitions for the verticies
\begin{align*}
{\Gamma _1} &= e{\varepsilon _{fab}}{\left( {{p_1} - {p_2}} \right)^\alpha } & {\Gamma _2} &= e{\varepsilon _{hdc}}{\left( {{p_3} - {p_4}} \right)^\beta }& D_{\alpha \beta }^{fh}\left( p \right) &=  - \frac{{i{\delta _{fh}}{g_{\alpha \beta }}}}{{{p^2} - {M^2}}}
\end{align*}
The amplitude is given by:
\begin{align*}
M_s^A &= {\Gamma _2} \cdot D_{\alpha \beta }^{fh}\left( p \right) \cdot {\Gamma _1}\\
 &=  - i\frac{{{e^2}}}{{{p^2} - {M^2}}}{\varepsilon _{fab}}{\varepsilon _{fdc}}\left( {{p_1} - {p_2}} \right) \cdot \left( {{p_3} - {p_4}} \right)\\
 &= {S_A}\left( {{\delta _{ad}}{\delta _{bc}} - {\delta _{ac}}{\delta _{bd}}} \right) \this
\end{align*}
with 
\begin{align*}
{S_A}&:=  - i\frac{{{e^2}}}{{s - {M^2}}}\left( {{p_1} - {p_2}} \right) \cdot \left( {{p_3} - {p_4}} \right) \\
 &=  - i\frac{{{e^2}}}{{s - {M^2}}}\left( {t - u} \right)\\
 &=  - 4i{e^2}{m^2}\frac{{zR}}{{4{m^2}\left( {1 + R} \right) - {M^2}}} \this
\end{align*}

For the $t$ channel scattering process mediated by the rho:
\begin{figure}[H]
\centering
	\begin{tikzpicture}[scale=1]
		\FScatteringT{5}{0}{\Fphi}{\Frho} 
	\end{tikzpicture}
\end{figure}
With the definitions for the verticies
\begin{align*}
{\Gamma _1} &=  - e{\varepsilon _{fca}}{\left( {{p_1} + {p_3}} \right)^\alpha } & {\Gamma _2} &= e{\varepsilon _{hbd}}{\left( {{p_2} + {p_4}} \right)^\beta } & D_{\alpha \beta }^{fh}\left( p \right) &=  - \frac{{i{\delta _{fh}}{g_{\alpha \beta }}}}{{{p^2} - {M^2}}}
\end{align*}
The amplitude is given by:
\begin{align*}
M_t^A &= {\Gamma _2} \cdot D_{\alpha \beta }^{fh}\left( p \right) \cdot {\Gamma _1}\\
 &= i\frac{{{e^2}}}{{{p^2} - {M^2}}}\left( {{p_1} + {p_3}} \right) \cdot \left( {{p_2} + {p_4}} \right)\left( {{\delta _{bc}}{\delta _{ad}} - {\delta _{ab}}{\delta _{cd}}} \right)\\
 &= {T_A}\left( {{\delta _{ab}}{\delta _{cd}} - {\delta _{bc}}{\delta _{ad}}} \right) \this
\end{align*}
with 
\begin{align*}
{T_A} &:=  - i\frac{{{e^2}}}{{t - {M^2}}}\left( {{p_1} + {p_3}} \right) \cdot \left( {{p_2} + {p_4}} \right)\\
 &=  - i\frac{{{e^2}}}{{t - {M^2}}}\left( {s - u} \right)\\
 &= 2i{e^2}{m^2}\frac{{2 + \left( {3 + z} \right)R}}{{2{m^2}\left( {1 - z} \right)R + {M^2}}} \this
\end{align*}

For the $u$ channel scattering process mediated by the rho:
\begin{figure}[H]
\centering
	\begin{tikzpicture}[scale=1]
		\FScatteringU{5}{0}{\Fphi}{\Frho} 
	\end{tikzpicture}
\end{figure}
With the definitions for the verticies
\begin{align*}
{\Gamma _1} &=  - e{\varepsilon _{fda}}{\left( {{p_1} + {p_4}} \right)^\alpha } & {\Gamma _2} &= e{\varepsilon _{hbc}}{\left( {{p_2} + {p_3}} \right)^\beta } & D_{\alpha \beta }^{fh}\left( p \right) &=  - \frac{{i{\delta _{fh}}{g_{\alpha \beta }}}}{{{p^2} - {M^2}}}
\end{align*}
The amplitude is given by:
\begin{align*}
M_u^A &= {\Gamma _2} \cdot D_{\alpha \beta }^{fh}\left( p \right) \cdot {\Gamma _1}\\
 &= i\frac{{{e^2}}}{{{p^2} - {M^2}}}{\varepsilon _{fda}}{\varepsilon _{fbc}}\left( {{p_1} + {p_4}} \right) \cdot \left( {{p_2} + {p_3}} \right)\\
 &= i\frac{{{e^2}}}{{{p^2} - {M^2}}}\left( {{p_1} + {p_4}} \right) \cdot \left( {{p_2} + {p_3}} \right)\left( {{\delta _{ac}}{\delta _{bd}} - {\delta _{ab}}{\delta _{cd}}} \right)\\
 &= {U_A}\left( {{\delta _{ab}}{\delta _{cd}} - {\delta _{ac}}{\delta _{bd}}} \right) \this
\end{align*}
with 
\begin{align*}
{U_A}&:=  - i\frac{{{e^2}}}{{u - {M^2}}}\left( {{p_1} + {p_4}} \right) \cdot \left( {{p_2} + {p_3}} \right) \\
 &=  - i\frac{{{e^2}}}{{u - {M^2}}}\left( {s - t} \right)\\
 &= 2i{e^2}{m^2}\frac{{2 + \left( {3 - z} \right)R}}{{2{m^2}\left( {1 + z} \right)R + {M^2}}} \this
\end{align*}

The sum of the amplitudes due to the rho mediator
\begin{align*}
T_A^{ab,cd} &= {S_A}\left( {{\delta _{ad}}{\delta _{bc}} - {\delta _{ac}}{\delta _{bd}}} \right) + {T_A}\left( {{\delta _{ab}}{\delta _{cd}} - {\delta _{bc}}{\delta _{ad}}} \right) + {U_A}\left( {{\delta _{ab}}{\delta _{cd}} - {\delta _{ac}}{\delta _{bd}}} \right)\\
 &= \left( {{T_A} + {U_A}} \right){\delta _{ab}}{\delta _{cd}} - \left( {{S_A} + {U_A}} \right){\delta _{ac}}{\delta _{bd}} + \left( {{S_A} - {T_A}} \right){\delta _{ad}}{\delta _{bc}} \this
\end{align*}

For the $s$ channel scattering process mediated by the Higgs:
\begin{figure}[H]
\centering
	\begin{tikzpicture}[scale=1]
		\FScatteringS{5}{0}{\Fphi}{\Fhiggs} 
	\end{tikzpicture}
\end{figure}
With the definitions for the verticies
\begin{align*}
{\Gamma _1} &=  - 4i\kappa \nu {\delta _{ab}} & {\Gamma _2} &=  - 4i\kappa \nu {\delta _{cd}} & D\left( p \right) &= \frac{i}{{{p^2} - m_H^2}}
\end{align*}
The amplitude is given by:
\begin{align*}
M_s^H &= {\Gamma _2} \cdot D\left( p \right) \cdot {\Gamma _1}\\
 &=  - i\frac{{16{\kappa ^2}{\nu ^2}}}{{{p^2} - m_H^2}}{\delta _{ab}}{\delta _{cd}}\\
 &= {S_H}{\delta _{ab}}{\delta _{cd}} \this
\end{align*}
with 
\begin{align}
{S_H}: =  - i\frac{{16{\kappa ^2}{\nu ^2}}}{{s - m_H^2}} 
\end{align}

For the $t$ channel scattering process mediated by the Higgs:
\begin{figure}[H]
\centering
	\begin{tikzpicture}[scale=1]
		\FScatteringT{5}{0}{\Fphi}{\Fhiggs} 
	\end{tikzpicture}
\end{figure}
With the definitions for the verticies
\begin{align*}
{\Gamma _1} &=  - 4i\kappa \nu {\delta _{ac}} & {\Gamma _2} &=  - 4i\kappa \nu {\delta _{bd}} & D\left( p \right) &= \frac{i}{{{p^2} - m_H^2}}
\end{align*}
The amplitude is given by:
\begin{align*}
M_t^H &= {\Gamma _2} \cdot D\left( p \right) \cdot {\Gamma _1}\\
 &=  - i\frac{{16{\kappa ^2}{\nu ^2}}}{{{p^2} - m_H^2}}{\delta _{ac}}{\delta _{bd}}\\
 &= {T_H}{\delta _{ac}}{\delta _{bd}} \this
\end{align*}
with 
\begin{align}
{T_H}: =  - i\frac{{16{\kappa ^2}{\nu ^2}}}{{t - m_H^2}}
\end{align}

For the $u$ channel scattering process mediated by the Higgs:
\begin{figure}[H]
\centering
	\begin{tikzpicture}[scale=1]
		\FScatteringU{5}{0}{\Fphi}{\Fhiggs} 
	\end{tikzpicture}
\end{figure}
With the definitions for the verticies
\begin{align*}
{\Gamma _1} &=  - 4i\kappa \nu {\delta _{ad}} & {\Gamma _2} &=  - 4i\kappa \nu {\delta _{bc}} &D\left( p \right) &= \frac{i}{{{p^2} - m_H^2}}
\end{align*}
The amplitude is given by:
\begin{align*}
M_u^H &= {\Gamma _2} \cdot D\left( p \right) \cdot {\Gamma _1}\\
 &=  - i\frac{{16{\kappa ^2}{\nu ^2}}}{{{p^2} - m_H^2}}{\delta _{ad}}{\delta _{bc}}\\
 &= {U_H}{\delta _{ad}}{\delta _{bc}} \this
\end{align*}
with 
\begin{align}
{U_H}: =  - i\frac{{16{\kappa ^2}{\nu ^2}}}{{u - m_H^2}}
\end{align}

The sum of the amplitudes due to the Higgs mediator
\begin{align}
T_H^{ab,cd} = {S_H}{\delta _{ab}}{\delta _{cd}} + {T_H}{\delta _{ac}}{\delta _{bd}} + {U_H}{\delta _{ad}}{\delta _{bc}}
\end{align}

For the four pion scattering process:
\begin{figure}[H]
\centering
	\begin{tikzpicture}[scale=1]
		\FFourScattering{5}{0}
	\end{tikzpicture}
\end{figure}
With the definitions for the verticies
\begin{align*}
{S_\lambda }&: =  - i{\lambda _4} & \Gamma &= {S_\lambda }\left( {{\delta _{ab}}{\delta _{cd}} + {\delta _{ad}}{\delta _{bc}} + {\delta _{ac}}{\delta _{bd}}} \right)\\
\implies {M^\phi } &= \Gamma 
\end{align*}
\begin{align}
T_\lambda^{ab,cd} ={S_\lambda }\left( {{\delta _{ab}}{\delta _{cd}} + {\delta _{ad}}{\delta _{bc}} + {\delta _{ac}}{\delta _{bd}}} \right)
\end{align}

\section{Isospin Amplitudes}
The most general form of the scattering amplitude is
\begin{align}
{M^{ab,cd}} = F\left( {s,t,u} \right){\delta _{ab}}{\delta _{cd}} + G\left( {s,t,u} \right){\delta _{ac}}{\delta _{bd}} + H\left( {s,t,u} \right){\delta _{ad}}{\delta _{bc}} \label{GeneralAplitude}
\end{align}
The amplitude can be decomposed over an isospin invariant basis 
\begin{align}
{M^{ab,cd}} = \sum\limits_{m = 0}^2 {{T^m}P_m^{abcd}} \label{IsospinAmpDef}
\end{align}
where $T^m$ are the isospin amplitudes and the basis vectors \cite{Bjorken1965} are
\begin{align}
P_0^{abcd}&: = \frac{1}{3}{\delta _{ab}}{\delta _{cd}}\\
P_1^{abcd}&: = \frac{1}{2}\left( {{\delta _{ac}}{\delta _{bd}} - {\delta _{ad}}{\delta _{bc}}} \right)\\
P_2^{abcd}&: = \frac{1}{2}\left( {{\delta _{ac}}{\delta _{bd}} + {\delta _{ad}}{\delta _{bc}}} \right) - \frac{1}{3}{\delta _{ab}}{\delta _{cd}}
\end{align}
Expanding the amplitude \eqref{IsospinAmpDef}
\begin{align}
{M^{ab,cd}} = \frac{1}{3}\left( {{T^0} - {T^2}} \right){\delta _{ab}}{\delta _{cd}} + \frac{1}{2}\left( {{T^1} + {T^2}} \right){\delta _{ac}}{\delta _{bd}} - \frac{1}{2}\left( {{T^1} - {T^2}} \right){\delta _{ad}}{\delta _{bc}}
\label{AmplitudeInIsospinBasis}
\end{align}
and identifying coefficients between \eqref{GeneralAplitude} and \eqref{AmplitudeInIsospinBasis}, we obtain a system of linear equations
\begin{align}
\frac{1}{3}\left( {{T^0} - {T^2}} \right) &= F\left( {s,t,u} \right)\\
\frac{1}{2}\left( {{T^1} + {T^2}} \right) &= G\left( {s,t,u} \right)\\
 - \frac{1}{2}\left( {{T^1} - {T^2}} \right) &= H\left( {s,t,u} \right)
\end{align}
which has the solution,
\begin{align}
{T^0} &= 3F\left( {s,t,u} \right) + G\left( {s,t,u} \right) + H\left( {s,t,u} \right)\\
{T_1} &= G\left( {s,t,u} \right) - H\left( {s,t,u} \right)\\
{T_2} &= G\left( {s,t,u} \right) + H\left( {s,t,u} \right)
\end{align}
%
The isospin amplitudes for rho mediation are given by:
\begin{align*}
T_A^0&:= 3\left( {{T_A} + {U_A}} \right) - \left( {{S_A} + {U_A}} \right) + \left( {{S_A} - {T_A}} \right)\\
 &= 2\left( {{T_A} + {U_A}} \right)\\
 &=4i{e^2}{m^2}\left[ {\frac{{2 + \left( {3 + z} \right)R}}{{2{m^2}\left( {1 - z} \right)R + {M^2}}} + \frac{{2 + \left( {3 - z} \right)R}}{{2{m^2}\left( {1 + z} \right)R + {M^2}}}} \right] \this
\end{align*}
\begin{align*}
T_A^1 &: =  - \left( {{S_A} + {U_A}} \right) - \left( {{S_A} - {T_A}} \right)\\
 &=  - 2{S_A} + {T_A} - {U_A}\\
 &= 2i{e^2}{m^2}\left[ {\frac{{4zR}}{{4{m^2}\left( {1 + R} \right) - {M^2}}} + \frac{{2 + \left( {3 + z} \right)R}}{{2{m^2}\left( {1 - z} \right)R + {M^2}}} - \frac{{2 + \left( {3 - z} \right)R}}{{2{m^2}\left( {1 + z} \right)R + {M^2}}}} \right] \this
\end{align*}
\begin{align*}
T_A^2&: =  - \left( {{S_A} + {U_A}} \right) + \left( {{S_A} - {T_A}} \right)\\
 &=  - \left( {{T_A} + {U_A}} \right)\\
 &=  - \frac{1}{2}T_A^0 \this
\end{align*}
The isospin amplitudes for Higgs mediator are given by:
\begin{align*}
T_H^0 &:= 3{S_H} + {T_H} + {U_H}\\
 &=  - 3i\frac{{16{\kappa ^2}{\nu ^2}}}{{s - m_H^2}} - i\frac{{16{\kappa ^2}{\nu ^2}}}{{t - m_H^2}} - i\frac{{16{\kappa ^2}{\nu ^2}}}{{u - m_H^2}}\\
 &=  16i{\kappa ^2}{\nu ^2}\left[ -{\frac{3}{{4{m^2}\left( {1 + R} \right) - m_H^2}} + \frac{1}{{2{m^2}\left( {1 - z} \right)R + m_H^2}} + \frac{1}{{2{m^2}\left( {1 + z} \right)R + m_H^2}}} \right] \this
\end{align*}
\begin{align*}
T_H^1 &:= {T_H} - {U_H}\\
 &=  - i\frac{{16{\kappa ^2}{\nu ^2}}}{{t - m_H^2}} + i\frac{{16{\kappa ^2}{\nu ^2}}}{{u - m_H^2}}\\
 &= 16i{\kappa ^2}{\nu ^2}\left[ {\frac{1}{{2{m^2}\left( {1 - z} \right)R + m_H^2}} - \frac{1}{{2{m^2}\left( {1 + z} \right)R + m_H^2}}} \right] \this
\end{align*}
\begin{align*}
T_H^2 &:= {T_H} + {U_H}\\
 &=  - i\frac{{16{\kappa ^2}{\nu ^2}}}{{t - m_H^2}} - i\frac{{16{\kappa ^2}{\nu ^2}}}{{u - m_H^2}}\\
 &= 16i{\kappa ^2}{\nu ^2}\left[ {\frac{1}{{2{m^2}\left( {1 - z} \right)R + m_H^2}} + \frac{1}{{2{m^2}\left( {1 + z} \right)R + m_H^2}}} \right] \this
\end{align*}
The isospin amplitudes for the four pion vertex are given by:
\begin{align}
T_\lambda ^0 &:=  - 5i{\lambda _4} & T_\lambda ^1 &: =0 &T_\lambda ^2 &: =  - 2i{\lambda _4}
\end{align}
The sum of the amplitudes of the Higgs and 4 pion tree diagrams are:
\begin{align*}
T_{H\lambda }^0 &= T_H^0 + T_\lambda ^0\\
 &= 16i{\kappa ^2}{\nu ^2}\left[ { - \frac{3}{{4{m^2}\left( {1 + R} \right) - m_H^2}} + \frac{1}{{2{m^2}\left( {1 - z} \right)R + m_H^2}} + \frac{1}{{2{m^2}\left( {1 + z} \right)R + m_H^2}}} \right] - 5i{\lambda _4} \this
\end{align*}
\begin{align*}
T_{H\lambda }^1 &= T_H^1 + T_\lambda ^1\\
 &= 16i{\kappa ^2}{\nu ^2}\left[ {\frac{1}{{2{m^2}\left( {1 - z} \right)R + m_H^2}} - \frac{1}{{2{m^2}\left( {1 + z} \right)R + m_H^2}}} \right] \this
\end{align*}
\begin{align*}
T_{H\lambda }^2 &= T_H^2 + T_\lambda ^2\\
 &= 16i{\kappa ^2}{\nu ^2}\left[ {\frac{1}{{2{m^2}\left( {1 - z} \right)R + m_H^2}} + \frac{1}{{2{m^2}\left( {1 + z} \right)R + m_H^2}}} \right] - 2i{\lambda _4} \this
\end{align*}

\section{Scattering Lengths}
Scattering lengths are computed from the coefficients of the partial wave scattering amplitude with the partial wave scattering 
amplitude obtained from the projection of the isospin amplitudes over the Legendre polynomials. We begin with the isospin amplitude 
${T^I}\left( {{q^2},z} \right)$ expressed as a sum over partial scattering amplitudes $T_m^I\left( q^2 \right)$
\begin{align}
{T^I}\left( {{q^2},z} \right) = 32\pi \sum\limits_{m = 0}^\infty  {\left( {2m + 1} \right){P_m}\left( z \right)T_{m}^I\left( {{q^2}} \right)} 
\end{align} 
where ${P_m}\left( z \right)$ is the Legendre polynomials indexed by $m$ with $z = \cos \theta$  as defined in \eqref{zcostheta}. Using the orthogonality of the Legendre polynomials on the inner product:
\begin{align}
\int\limits_{ - 1}^1 {{P_m}\left( z \right){P_n}\left( z \right)dz}  = \frac{2}{{2m + 1}}{\delta _{mn}}
\end{align}
the partial scattering amplitudes can be extracted by multiplying ${T^I}\left( {{q^2},z} \right)$ by the Legendre polynomials and integrating:
\begin{align*}
\int\limits_{ - 1}^1 {{P_n}\left( z \right){T^I}\left( {{q^2},z} \right)dz} &= 32\pi \sum\limits_{m = 0}^\infty  {\left( {2m + 1} \right)\int\limits_{ - 1}^1 {{P_m}\left( z \right){P_n}\left( z \right)dz}  \cdot T_m^I\left( {{q^2}} \right)} \\
 &= 32\pi \sum\limits_{m = 0}^\infty  {\left( {2m + 1} \right)\frac{2}{{2m + 1}}{\delta _{mn}} \cdot T_m^I\left( {{q^2}} \right)} \\
 &= 64\pi T_n^I\left( {{q^2}} \right)\\ \\
\implies T_n^I\left( {{q^2}} \right) &= \frac{1}{{64\pi }}\int\limits_{ - 1}^1 {{P_n}\left( z \right){T^I}\left( {{q^2},z} \right)dz} \this
\end{align*}

A Maclaurin series expansion can be made in terms of $q^2$ \cite{Colangelo2001125} as 
\begin{align}
T_n^I &= i{\left( {\frac{{{q^2}}}{{m_\pi ^2}}} \right)^n}\left[ {a_n^I + b_n^I\left( {\frac{{{q^2}}}{{m_\pi ^2}}} \right) +  \ldots } \right] \nonumber \\
&= i R^n \left[ {a_n^I + b_n^I R +  \ldots } \right]
\end{align}
where $R=\dfrac{q^2}{m^2}$ which was defined in \eqref{Rqsqtomsq} and $a_n^I$ and $b_n^I$ are the scattering lengths. We shall project the isospin amplitudes using the first three Legendre polynomials which are 
\begin{align}
 {P_0}\left( z \right) &= 1 & {P_1}\left( z \right) &= z  & {P_2}\left( z \right) &= \frac{1}{2}\left( {3{z^2} - 1} \right)
\end{align}
The first two scattering lengths are calculated below. For $T^0$
\begin{align*}
{T^0}\left( R \right) &= T_A^0 + T_{H\lambda }^0\\
 &= 4i{e^2}{m^2}\left[ {\frac{{2 + \left( {3 + z} \right)R}}{{2{m^2}\left( {1 - z} \right)R + {M^2}}} + \frac{{2 + \left( {3 - z} \right)R}}{{2{m^2}\left( {1 + z} \right)R + {M^2}}}} \right] + \\
 &+ 16i{\kappa ^2}{\nu ^2}\left[ { - \frac{3}{{4{m^2}\left( {1 + R} \right) - m_H^2}} + \frac{1}{{2{m^2}\left( {1 - z} \right)R + m_H^2}} + \frac{1}{{2{m^2}\left( {1 + z} \right)R + m_H^2}}} \right] \this
\end{align*}
The partial wave amplitude $T_0^0$ is 
\begin{align*}
T_0^0\left( R \right) &= \frac{1}{{64\pi }}\int\limits_{ - 1}^1 {{P_0}\left( z \right){T^0}\left( {R,z} \right)dz} \\
 &= a_0^0 + b_0^0R \this
\end{align*}
Setting $R=0$ gives us
\begin{align*}
a_0^0 &= {\left. {T_0^0\left( R \right)} \right|_{R = 0}}\\
 &= \frac{{{e^2}}}{{2\pi }}\frac{{{m^2}}}{{{M^2}}} + \frac{{{\kappa ^2}{\nu ^2}}}{\pi }\left[ {\frac{1}{{m_H^2}} - \frac{3}{{2\left( {4{m^2} - m_H^2} \right)}}} \right] - \frac{{5{\lambda _4}}}{{32\pi }} \this
\end{align*}
and differentiating with respect to $R$ and then setting to zero
\begin{align*}
b_0^0 &= {\left. {\frac{{dT_0^0\left( R \right)}}{{dR}}} \right|_{R = 0}}\\
 &= \frac{{{e^2}}}{{4\pi }}\frac{{{m^2}}}{{{M^2}}}\left( {3 - 4\frac{{{m^2}}}{{{M^2}}}} \right) + \frac{{2{m^2}{\kappa ^2}{\nu ^2}}}{\pi }\left[ {\frac{3}{{{{\left( {4{m^2} - m_H^2} \right)}^2}}} - \frac{1}{{m_H^4}}} \right] \this
\end{align*}
The above calculations can be repeated for $T^1, T^2$ and the three Legendre polynomials $P_0(z),P_1(z)$ and $P_2(z)$. 

\section{Tree Scattering Lengths Results}
A summary of the results for the scattering lengths are listed below:
\begin{align*}
a_0^0 &= \frac{{{e^2}}}{{2\pi }}\frac{{{m^2}}}{{{M^2}}} + \frac{{{\kappa ^2}{\nu ^2}}}{\pi }\left[ {\frac{1}{{m_H^2}} - \frac{3}{{2\left( {4{m^2} - m_H^2} \right)}}} \right] - \frac{{5{\lambda _4}}}{{32\pi }} \\
b_0^0 &= \frac{{{e^2}}}{{4\pi }}\frac{{{m^2}}}{{{M^2}}}\left( {3 - 4\frac{{{m^2}}}{{{M^2}}}} \right) + \frac{{2{m^2}{\kappa ^2}{\nu ^2}}}{\pi }\left[ {\frac{3}{{{{\left( {4{m^2} - m_H^2} \right)}^2}}} - \frac{1}{{m_H^4}}} \right]\\
a_1^0 &= 0 \\ b_1^0 &= 0\\ 
a_2^0 &= \frac{{2{e^2}}}{{15\pi }}\frac{{{m^4}}}{{{M^4}}}\left( {1 + 4\frac{{{m^2}}}{{{M^2}}}} \right) + \frac{{16}}{{15\pi }}\frac{{{m^4}{\kappa ^2}{\nu ^2}}}{{m_H^6}} \\ 
b_2^0 &= \frac{{4{e^2}}}{{5\pi }}\frac{{{m^6}}}{{{M^6}}}\left( {1 - 12\frac{{{m^2}}}{{{M^2}}}} \right) - \frac{{96}}{{5\pi }}\frac{{{m^8}{\kappa ^2}{\nu ^2}}}{{m_H^8}}\\ \\ 
a_0^1 &= 0 \\ b_0^1 &= 0\\ 
a_1^1 &= \frac{{{e^2}}}{{24\pi }}\frac{{16{m^6} - 3{m^2}{M^4}}}{{4{m^2}{M^4} - {M^6}}} + \frac{2}{{3\pi }}\frac{{{m^2}{\kappa ^2}{\nu ^2}}}{{m_H^4}} \\
b_1^1 &= \frac{{{e^2}}}{{6\pi }}\dfrac{{{m^4}\left( { - \dfrac{64}{3}\dfrac{m^6}{M^6} + 16\dfrac{m^4}{M^4} - 4\dfrac{m^2}{M^2} + 1} \right)}}{{{M^4}{{\left( 1- 4\dfrac{m^2}{M^2} \right)}^2}}} - \frac{8}{{3\pi }}\frac{{{m^4}{\kappa ^2}{\nu ^2}}}{{m_H^6}}\\
a_2^1 &= 0 \\ b_2^1 &= 0\\
\\ \\
a_0^2 &=  - \frac{{{e^2}}}{{8\pi }}\frac{{{m^2}}}{{{M^2}}} + \frac{{{\kappa ^2}{\nu ^2}}}{{2\pi m_H^2}} - \frac{{{\lambda _4}}}{{32\pi }} \\ b_0^2 &=  - \frac{{{e^2}}}{{48\pi }}\frac{{{m^2}}}{{{M^2}}}\left( {3 - 4\frac{{{m^2}}}{{{M^2}}}} \right) - \frac{1}{{3\pi }}\frac{{{m^2}{\kappa ^2}{\nu ^2}}}{{m_H^4}}\\
a_1^2 &= 0 \\ b_1^2 &= 0\\
a_2^2 &=  - \frac{{{e^2}}}{{30\pi }}\frac{{{m^4}}}{{{M^4}}}\left( {1 + 4\frac{{{m^2}}}{{{M^2}}}} \right) + \frac{8}{{15\pi }}\frac{{{m^4}{\kappa ^2}{\nu ^2}}}{{m_H^6}} \\ b_2^2 &=  - \frac{{{e^2}}}{{15\pi }}\frac{{{m^6}}}{{{M^6}}}\left( {1 - 12\frac{{{m^2}}}{{{M^2}}}} \right) - \frac{{16}}{{5\pi }}\frac{{{m^6}{\kappa ^2}{\nu ^2}}}{{m_H^8}}
\end{align*}

For the experimental input we shall use the average masses of the charged and uncharged pions and rhos:
\begin{align}
m &=0.1372734 \pm  0.0000007 ~\rm{GeV}& M &=0.77649 \pm 0.00034~\rm{GeV}
\end{align}
which are taken from \cite{Patrignani}. The rho-pion-pion coupling,
\begin{align}
e &= 5.96\pm 0.20
\end{align}
is taken from \cite{CADLushozi2015}. The value of the vacuum expectation value can be inferred from the definition of the mass of the rho
\begin{align*}
M&:=e\nu \\
\implies \nu &=\frac{M}{e}=0.130 \pm 0.004 ~\rm{GeV} \this
\end{align*}
The pion decay constant and its ratio in the chiral limit,
\begin{align*}
F_\pi &\approx 0.093 ~\rm{GeV} & \frac{F_\pi}{F} &=1.0627\pm0.0028
\end{align*}
is taken from \cite{Borsanyi2013}. The four pion coupling is taken from \cite{Scherer2003}:
\begin{align*}
\lambda_4 &:= \left(\frac{m}{F}\right)^2 \\
&=2.45074\pm 0.1568 \this
\end{align*}
The mass of the symmetry breaking field $m_H$ is taken to be the mass of the $f_0(500)$:
\begin{align}
m_H &:= m_{\rm{f_0}}=0.450 \pm 0.016 ~\rm{GeV}
\end{align}
This value is taken from \cite{Pelaez2016}. Using the above values and $a^0_0$ and $b^0_0$ we can infer an average value for $\kappa$ of:
\begin{align}
\kappa = 1.31 \pm 0.03
\end{align}
The values for the scattering lengths are tabulated below with NABKLZ referring to the model developed in this thesis.:

\begin{table}[H]
	\begin{center}
		\begin{tabular}{@{}cccccccc@{}}
			
			\toprule
Lengths&Weinberg$^d$&$\chi$PT$(1^{st}\mathcal{O})^a$&$\chi$PT$(2^{nd}\mathcal{O})^a$&NABKLZ$^\dagger$&Colangelo$^b$&Bijnens$^c$&Exp$^{abc}$\\
			\midrule
			\midrule
$a^0_0$ &$0.20$  	&$0.16$	 &$0.20$				&$0.21$ 					&$0.220$				&$0.219$& $0.220\pm0.005$ \\
$b^0_0$ &			&$0.18$	 &$0.26$				&$0.30$ 					&$0.276$				&$0.279$& $0.25\pm0.03$\\

$a^0_1$ &   	   	& 		 & 						&$0$						& 						& &\\
$b^0_1$ &		   	& 		 & 						&$0$						& 						& &\\

$a^0_2\times 10^{3}$ & 			&$0$&$2$&$2.06$		& $1.75$	&$2.2$  & $1.7\pm 3$ \\
$b^0_2\times 10^{4}$ &		    &   &   & $-5.23$ 	& $-3.55$   &$-3.2$ &\\
			
			\midrule 
			
$a^1_0$ &	 	   		&        &   					&$0$						& 			& & \\
$b^1_0$ &	 	   		&        &   					&$0$						&  			& &\\

$a^1_1$ &			  	&$0.030$ &$0.036$				&$0.0528$				&$0.0379$	&$0.0378$ &$0.038 \pm 0.002 $\\
$b^1_1$ &				&$0$	 &$0.043$ 				&$0.0053$				&$0.0057$	&$0.0059$ &\\
	
$a^1_2$ &		   		&  	     &  					&$0$						& 						& &\\
$b^1_2$ &		   		&  	     &  					&$0$						& 						& &\\
			
			\midrule
			
$a^2_0$ & $-0.06$			 &$-0.045$&$-0.041$		&$-0.0456$				&$-0.0444$		&$-0.0420$& $-0.044\pm 0.001$ \\
$b^2_0$ &			 &  	  &  			&$-0.0225$				&$-0.0803$		&$-0.0756$&$-0.082\pm0.008$\\

$a^2_1$ &		   	 & 		  & 			&$0$					&				& &\\
$b^2_1$ &		   	 & 		  & 			&$0$					&				& &\\

$a^2_2\times 10^{4}$ & 			  	&$0$ 	&$3.5$ 		& $-2.03$	& $1.70$ 	&$2.90$		&$1.3\pm 3$ \\
$b^2_2\times 10^{4}$ &				&$-8.9$ &$-7$ 	& $-0.53$	& $-3.26$	&$-3.60$	&$-8.2$\\

			\bottomrule
		\end{tabular}
	\end{center}
	\medskip
		\raggedright
		{\footnotesize
			\hskip 0.2cm $^\dagger$ Tree results  \\
			\hskip 0.2cm $^a$ Results taken from \cite{donoghue2014dynamics}\\
			\hskip 0.2cm $^b$ Results taken from \cite{Colangelo2001125} \\
			\hskip 0.2cm $^c$ Results taken from \cite{Bijnens2000}\\
			\hskip 0.2cm $^d$ Results taken from \cite{Weinberg1966}\\
		}
	\caption{Summary of predicted values and experimental data of the scattering length.}
	\label{treeresultstable}
\end{table}

\begin{figure} 
\centering
\includegraphics[width=0.9\linewidth]{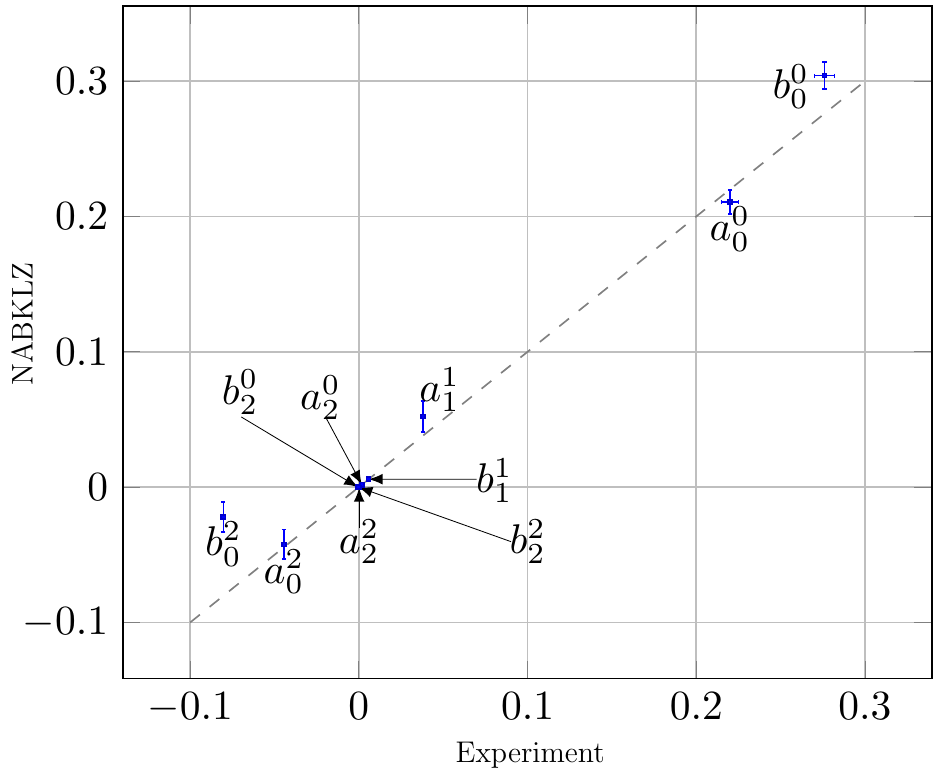}
\caption{Predicted vs Experimental values}
\label{fig:testdataplotted2}
\end{figure}

A parity plot can be made to show the predicted versus experimental values for the scattering lengths. The dashed line in figure \ref{fig:testdataplotted2} is the reference line of predicted values equal to experimental values.

\chapter{One Loop Corrections}
\lettrine[lines=4]{\initfamily\textcolor{darkblue}{T}}{he} tree results for the scattering lengths can be improved upon by including in the next order correction terms from the pertubative series. This can be done systematically by first computing the one particle irreducible (1PI) diagrams for the self energies and verticies. These 1PI diagrams then serve as the building blocks for the higher order analysis \cite{DasbookQFT}. We shall list the topologies that contribute to the self energies and verticies. In total there are $\sim 85$ diagrams.

\section{One Loop Topologies}
Diagrams that contribute to the self energy of the rho. \\ \\
\begin{minipage}{\linewidth}
\begin{tikzpicture}[scale=0.95]
	\Fbubble{0}{2}{\Frho}{\Fphi}{\Fphi}{\Frho};
	\Ftadpole{4.5}{2}{\Frho}{\Fphi};
	\Fbubble{9}{2}{\Frho}{\Frho}{\Frho}{\Frho};
	\Ftadpole{13.5}{2}{\Frho}{\Frho};

	\Fbubble{0}{-1}{\Frho}{\Fchi}{\Fchi}{\Frho};
	\Ftadpole{4.5}{-1}{\Frho}{\Fchi};
	\Fbubble{9}{-1}{\Frho}{\Frho}{\Fhiggs}{\Frho};
	\Ftadpole{13.5}{-1}{\Frho}{\Fhiggs};
	
	\Fbubble{0}{-4}{\Frho}{\Fchi}{\Fhiggs}{\Frho};
	\Fbubble{4.5}{-4}{\Frho}{\Fgst}{\Fgst}{\Frho};
	\FPropCounterTerm{9}{-4}{\Frho};
\end{tikzpicture}
\end{minipage}
\newpage
Diagrams that contribute to the self energy of the Higgs. \\ \\

\begin{minipage}{\linewidth}
\begin{tikzpicture}[scale=1]
	\Fbubble{0}{2}{\Fhiggs}{\Fphi}{\Fphi}{\Fhiggs};
	\Ftadpole{4.5}{2}{\Fhiggs}{\Fphi};
	\Fbubble{9}{2}{\Fhiggs}{\Frho}{\Frho}{\Fhiggs};
	\Ftadpole{13.5}{2}{\Fhiggs}{\Frho};

	\Fbubble{0}{-1}{\Fhiggs}{\Fchi}{\Fchi}{\Fhiggs};
	\Ftadpole{4.5}{-1}{\Fhiggs}{\Fchi};
	\Fbubble{9}{-1}{\Fhiggs}{\Fhiggs}{\Fhiggs}{\Fhiggs};
	\Ftadpole{13.5}{-1}{\Fhiggs}{\Fhiggs};

	\Fbubble{0}{-4}{\Fhiggs}{\Frho}{\Fchi}{\Fhiggs};
	\FPropCounterTerm{4.5}{-4}{\Fhiggs};
\end{tikzpicture}
\end{minipage}
\\ \\
The lollipop diagrams for the rho and Higgs self energies will be included separately later during the analysis. 
Diagrams that contribute to the vertex of the $\phi\phi\rho$. \\ 
\begin{minipage}{\linewidth}
\begin{tikzpicture}[scale=1]
	\Ftriangle{2}{2}{\Fphi}{\Frho}{\Fphi}{\Frho};
	\Ftriangle{6.5}{2}{\Frho}{\Fphi}{\Frho}{\Frho};
	\Ftriangle{11}{2}{\Fphi}{\Fhiggs}{\Fphi}{\Frho};
	\Ftriangle{15.5}{2}{\Fhiggs}{\Fphi}{\Frho}{\Frho};
	
	\Fbubbletoprop{1}{-2.5}{\Frho}{\Frho};
	\Fbubbletoprop{5.5}{-2.5}{\Fphi}{\Frho};
	\Fbubbletoprop{10}{-2.5}{\Fhiggs}{\Frho};
	\FtriangleSeagullT{15}{-2.5}{\Frho}
	
	\FtriangleSeagullB{2}{-7}{\Frho}
	\FThreePointCounterTerm{6.25}{-7}{\Frho};
\end{tikzpicture}
\end{minipage}
\\ \\
Diagrams that contribute to the vertex of the $\phi\phi H$. \\ \\
\begin{minipage}{\linewidth}
\begin{tikzpicture}[scale=1]
	\Ftriangle{2}{2}{\Fphi}{\Frho}{\Fphi}{\Fhiggs};
	\Ftriangle{6.5}{2}{\Frho}{\Fphi}{\Frho}{\Fhiggs};
	\Ftriangle{11}{2}{\Fphi}{\Fhiggs}{\Fphi}{\Fhiggs};
	\Ftriangle{15.5}{2}{\Fhiggs}{\Fphi}{\Fhiggs}{\Fhiggs};
	
	\Fbubbletoprop{1}{-2.5}{\Frho}{\Fhiggs};
	\Fbubbletoprop{5.5}{-2.5}{\Fphi}{\Fhiggs};
	\Fbubbletoprop{10}{-2.5}{\Fhiggs}{\Fhiggs};
	\FtriangleSeagullT{15}{-2.5}{\Fhiggs}
	
	\FtriangleSeagullB{2}{-7}{\Fhiggs}
	\FThreePointCounterTerm{5.5}{-7}{\Fhiggs};
\end{tikzpicture}
\end{minipage}
\\ \\
Diagrams that contribute to the $\phi\phi$ scattering via a loop. \\\\
\begin{minipage}{\linewidth}
\begin{tikzpicture}[scale=1]
	\FFourtoFour{0.5}{2}{\Frho};
	\FFourtoFour{5}{2}{\Fhiggs};
	\FFourtoFour{9.5}{2}{\Fchi};
	\FFourtoFour{14}{2}{\Fphi};
\end{tikzpicture}
\end{minipage}
\\ \\ \\
\begin{minipage}{\linewidth}
\begin{tikzpicture}[scale=1]
	\FThreetoFourtriangle{2}{2}{\Fphi}{\Frho}{\Fphi};
	\FThreetoFourtriangle{6.5}{2}{\Fhiggs}{\Fphi}{\Fhiggs};
	\FThreetoFourtriangle{11}{2}{\Frho}{\Fphi}{\Frho};
	\FThreetoFourtriangle{15.5}{2}{\Fphi}{\Fhiggs}{\Fphi};
\end{tikzpicture}
\end{minipage}
\\ \\ \\
\begin{minipage}{\linewidth}
\begin{tikzpicture}[scale=1]
	\Fbox{1}{1}{\Fphi}{\Frho}{\Fphi}{\Frho};
	\Fbox{5.5}{1}{\Frho}{\Fphi}{\Frho}{\Fphi};
	\Fbox{10}{1}{\Fhiggs}{\Fphi}{\Fhiggs}{\Fphi};
	\Fbox{14.5}{1}{\Fphi}{\Fhiggs}{\Fphi}{\Fhiggs};
	\FFourPointCounterTerm{1.75}{-2.25};
\end{tikzpicture}
\end{minipage}
\\\\
\section{Dimensional Regularization}
To parameterize the divergences that result in higher order loop calculations, we shall use dimensional regularization \cite{tHooft1972189}. This is done by analytically continuing the spacetime dimension from $n=4$ dimensions to $n=4-2\varepsilon$ dimensions \cite{CollinsbookRenorm}. Making this change in the spacetime dimension will capture the divergences in terms of the form $\dfrac{1}{\varepsilon}$ and $\dfrac{p^2}{\varepsilon}$ which can be appropriately absorbed by the counter terms. A further consequence of changing the spacetime dimension to $n=4-2\varepsilon$ is that the couplings are not dimensionless. This can be remedied by explicitly taking out the extra mass dimension. The action in $n$ dimensions is 
\begin{align}
S = \int {{d^n}x \mathscr{L}} 
\end{align}
Since the action is dimensionless
\begin{align}
\left[ S \right] = 1 \implies \left[ {{d^n}x} \right] = {\left[ {dx} \right]^n} = {M^{ - n}} \text{ and }\left[ \mathscr{L} \right] = {M^n}
\end{align}
Using the dimension of the Lagrangian we can determine the dimensions of the fields:
\begin{align}
\left[ A \right] = \left[ \phi  \right] = \left[ \chi  \right] = \left[ H \right] = M^{\frac{n-2}{2}}
\end{align} 
The dimensionful couplings can now be worked out to be 
\begin{align}
\left[ e \right] = {M^{\frac{{4 - n}}{2}}} ,\quad \left[ {{\lambda _4}} \right] = {M^{4 - n}} ,\quad \left[ \nu  \right] = {M^{\frac{{n - 4}}{2}}} ,\quad \left[ \lambda  \right] = {M^{4 - n}} ,\quad \left[ \kappa  \right] = {M^{4 - n}}
\end{align}
We shall define a parameter $\mu$ which is referred to as the renormalization scale with dimensions of mass, $\left[\mu\right] = M$ and define dimensionless couplings in terms of the renormalization scale as 
\begin{align}
e \sim {\mu ^{\frac{{4 - n}}{2}}}e ,\quad {\lambda _4} \sim {\mu ^{4 - n}}{\lambda _4} ,\quad \nu  \sim {\mu ^{\frac{{n - 4}}{2}}}\nu ,\quad \lambda  \sim {\mu ^{4 - n}}\lambda ,\quad \kappa  \sim {\mu ^{4 - n}}\kappa 
\end{align}
We will suppress the explicit appearance of the scale during the calculations but absorb it into the definition of the Passarino-Veltman scalar functions.

\section{Passarino-Veltman Reduction}
An efficient way to compute the one loop diagrams is to use 
the Passarino-Veltman reduction technique \cite{Passarino1979151}. This technique allows us to compute a Feynman 
diagram in terms of a few master scalar integrals \cite{tHooft1979365,Denner1991}, which in principle only have to be 
computed once. The Passarino-Veltman reduction technique is implemented by using the denominators of the propagators 
to write all the scalar products between external momenta and the loop momentum. Practically this means expressing 
the 
numerators of the Feynman amplitude in terms of the denominator, leaving only scalar integrals to be evaluated. This is demonstrated 
below for the case of a vector self energy diagram. 
\\\\
\begin{minipage}{\linewidth}
\centering
\begin{tikzpicture}[scale=1]
	\Fbubble{0}{2}{\Frho}{\Fphi}{\Fphi}{\Frho};
\end{tikzpicture}
\end{minipage}
The Feynman verticies are:
\begin{align*}
{\Gamma _1} = e{\varepsilon _{adc}}{\left( {2k + p} \right)^\mu } \qquad {\Gamma _2} = e{\varepsilon _{bef}}{\left( {2k + p} \right)^\nu }
\end{align*}
with the self energy contribution due the pion bubble:
\begin{align*}
 - i\pi _{ab1}^{\mu \nu } &= \int {\frac{{{d^n}k}}{{{{\left( {2\pi } \right)}^n}}}{\Gamma _1} \cdot {D^{df}}\left( k \right) \cdot {\Gamma _2} \cdot {D^{ce}}\left( {k + p} \right)} \\
 &= \int {\frac{{{d^n}k}}{{{{\left( {2\pi } \right)}^n}}}} \frac{{e{\varepsilon _{adc}}{{\left( {2k + p} \right)}^\mu } \cdot {\delta ^{df}} \cdot e{\varepsilon _{bef}}{{\left( {2k + p} \right)}^\nu }{\delta ^{ce}}}}{{\left( {{k^2} - {m^2}} \right)\left[ {{{\left( {k + p} \right)}^2} - {m^2}} \right]}}\\
 &= {e^2}{\varepsilon _{cda}}{\varepsilon _{cdb}}\int {\frac{{{d^n}k}}{{{{\left( {2\pi } \right)}^n}}}} \frac{{{{\left( {2k + p} \right)}^\mu }{{\left( {2k + p} \right)}^\nu }}}{{\left( {{k^2} - {m^2}} \right)\left[ {{{\left( {k + p} \right)}^2} - {m^2}} \right]}}\\
 &= {e^2}{\varepsilon _{cda}}{\varepsilon _{cdb}}I_1^{\mu \nu } \this
\end{align*}
Using ${\varepsilon _{cda}}{\varepsilon _{cdb}} = 2{\delta _{ab}}$
\begin{align}
 - i\pi _{ab1}^{\mu \nu } = 2{e^2}{\delta _{ab}}I_1^{\mu \nu }
\end{align}

We can make some definitions for aiding in the calculation:
\begin{align*}
N_1^{\mu \nu }&:= {\left( {2k + p} \right)^\mu }{\left( {2k + p} \right)^\nu }\\
 &= 4{k^\mu }{k^\nu } + 2\left( {{k^\mu }{p^\nu } + {p^\mu }{k^\nu }} \right) + {p^\mu }{p^\nu } \this\\
{D_1}&:= {k^2} - {m^2} \this\\
{D_2}&:= {\left( {k + p} \right)^2} - {m^2}\\
 &= {k^2} + 2k \cdot p + {p^2} - {m^2}\\
 &= {D_1} + 2k \cdot p + {p^2} \this
\end{align*}
So we have the relations
\begin{align}
{k^2} &= {D_1} + {m^2}\\
2k \cdot p &= {D_2} - {D_1} - {p^2}
\end{align}
The scalar product have now been expressed in terms of the denominators. We can split the integral over the transverse and longitudinal parts
\begin{align*}
I_1^{\mu \nu }&:= \int {\frac{{{d^n}k}}{{{{\left( {2\pi } \right)}^n}}}} \frac{{N_1^{\mu \nu }}}{{{D_1}{D_2}}}\\
 &= {f_{T1}}P_T^{\mu \nu } + {f_{L1}}P_L^{\mu \nu } \this
\end{align*}
where the projectors are defined as
\begin{align}
P_L^{\mu \nu }&:=\frac{p^\mu p^\nu}{p^2} \\
P_T^{\mu \nu }&:= g^{\mu\nu} - \frac{p^\mu p^\nu}{p^2}
\end{align}
The coefficients can be extracted using the transverse and longitudinal projectors.
\begin{align}
{P_{L\mu \nu }}I_1^{\mu \nu } &= {f_{L1}}\\
{P_{T\mu \nu }}I_1^{\mu \nu } &= \left( {n - 1} \right){f_{T1}}
\end{align}
The projectors acting on the numerator $N_1^{\mu \nu }$ yield
\begin{align*}
{P_{L\mu \nu }}N_1^{\mu \nu } &= \frac{1}{{{p^2}}}{\left( {2k \cdot p + {p^2}} \right)^2}\\
 &= \frac{1}{{{p^2}}}\left( {D_2^2 - 2{D_1}{D_2} + D_1^2} \right) \this\\
\end{align*}
and
\begin{align*}
{P_{T\mu \nu }}N_1^{\mu \nu } &= {g_{\mu \nu }}N_1^{\mu \nu } - {P_{L\mu \nu }}N_1^{\mu \nu }\\
{g_{\mu \nu }}N_1^{\mu \nu } &= 4{k^2} + 2k \cdot p + {p^2}\\
 &= 2{D_1} + 2{D_2} + 4{m^2} - {p^2} \this
\end{align*}
The necessary ingredients have been assembled to extract the longitudinal coefficient
\begin{align*}
{f_{L1}} &= \int {\frac{{{d^n}k}}{{{{\left( {2\pi } \right)}^n}}}} \frac{{{P_{L\mu \nu }}N_1^{\mu \nu }}}{{{D_1}{D_2}}}\\
 &= \frac{1}{{{p^2}}}\int {\frac{{{d^n}k}}{{{{\left( {2\pi } \right)}^n}}}} \frac{{D_2^2 - 2{D_1}{D_2} + D_1^2}}{{{D_1}{D_2}}}\\
 &= \frac{1}{{{p^2}}}\int {\frac{{{d^n}k}}{{{{\left( {2\pi } \right)}^n}}}\left( {\frac{{{D_2}}}{{{D_1}}} + \frac{{{D_1}}}{{{D_2}}} - 2} \right)} \\
 &= \frac{1}{{{p^2}}}\int {\frac{{{d^n}k}}{{{{\left( {2\pi } \right)}^n}}}\left( {\frac{{{D_2}}}{{{D_1}}} + \frac{{{D_1}}}{{{D_2}}}} \right)} \this
\end{align*}
Since there are no massless particles in this field theory, we can safely make use of the Veltman conjecture by setting all dimensionless integrals to zero i.e.
\begin{align}
\int \frac{{{d^n}k}}{{{{\left( {2\pi } \right)}^n}}} \, k^{2\alpha} = 0
\end{align}
for $\alpha \in \mathbb{C}$. We are left with the task of evaluating 
\begin{align}
\int {\frac{{{d^n}k}}{{{{\left( {2\pi } \right)}^n}}}\frac{{{D_1}}}{{{D_2}}}}  = \int {\frac{{{d^n}k}}{{{{\left( {2\pi } \right)}^n}}}\frac{{{k^2} - {m^2}}}{{{{\left( {k + p} \right)}^2} - {m^2}}}} 
\end{align}
Making a change of variables
${l^\mu } =  - {\left( {k + p} \right)^\mu } \Rightarrow {k^\mu } =  - {\left( {l + p} \right)^\mu } \Rightarrow \displaystyle  \int {\dfrac{{{d^n}k}}{{{{\left( {2\pi } \right)}^n}}}}  = \displaystyle \int {\dfrac{{{d^n}l}}{{{{\left( {2\pi } \right)}^n}}}}$. Substituting in the new measure
\begin{align*}
\int {\frac{{{d^n}k}}{{{{\left( {2\pi } \right)}^n}}}\frac{{{D_1}}}{{{D_2}}}}  &= \int {\frac{{{d^n}l}}{{{{\left( {2\pi } \right)}^n}}}\frac{{{{\left( {l + p} \right)}^2} - {m^2}}}{{{l^2} - {m^2}}}} \\
 &= \int {\frac{{{d^n}k}}{{{{\left( {2\pi } \right)}^n}}}\frac{{{D_2}}}{{{D_1}}}} \\
 &= \int {\frac{{{d^n}k}}{{{{\left( {2\pi } \right)}^n}}}\left( {1 + \frac{{2k \cdot p}}{{{k^2} - {m^2}}} + \frac{{{p^2}}}{{{k^2} - {m^2}}}} \right)} \\
 &= {p^2}{A_0}\left( m \right) \this
\end{align*}
where we have used the Passarino-Veltman scalar function $A_0(m)$ (see Appendix \ref{ScalarFunctions}). So the longitudinal coefficient is
\begin{align*}
{f_{L1}} &= \frac{1}{{{p^2}}}\int {\frac{{{d^n}k}}{{{{\left( {2\pi } \right)}^n}}}\left( {\frac{{{D_2}}}{{{D_1}}} + \frac{{{D_1}}}{{{D_2}}}} \right)} \\
 &= 2{A_0}\left( m \right) \this
\end{align*}
The transverse coefficient can be evaluated as
\begin{align*}
\left( {n - 1} \right){f_{T1}} &= {P_{T\mu \nu }}I_1^{\mu \nu }\\
 &= \int {\frac{{{d^n}k}}{{{{\left( {2\pi } \right)}^n}}}} \frac{{{P_{T\mu \nu }}N_1^{\mu \nu }}}{{{D_1}{D_2}}}\\
 &= \int {\frac{{{d^n}k}}{{{{\left( {2\pi } \right)}^n}}}} \left( {\frac{{{g_{\mu \nu }}N_1^{\mu \nu }}}{{{D_1}{D_2}}} - \frac{{{P_{L\mu \nu }}N_1^{\mu \nu }}}{{{D_1}{D_2}}}} \right)\\
 &= \int {\frac{{{d^n}k}}{{{{\left( {2\pi } \right)}^n}}}} \frac{{2{D_1} + 2{D_2} + 4{m^2} - {p^2}}}{{{D_1}{D_2}}} - \int {\frac{{{d^n}k}}{{{{\left( {2\pi } \right)}^n}}}} \frac{{{P_{L\mu \nu }}N_1^{\mu \nu }}}{{{D_1}{D_2}}}\\
 &= \int {\frac{{{d^n}k}}{{{{\left( {2\pi } \right)}^n}}}} \left[ {\frac{2}{{{D_1}}} + \frac{2}{{{D_2}}} + \frac{{4{m^2} - {p^2}}}{{{D_1}{D_2}}}} \right] - {f_{L1}}\\
 &= 4{A_0}\left( m \right) + \left( {4{m^2} - {p^2}} \right){B_0}\left( {m;p,m} \right) - 2{A_0}\left( m \right)\\
 &= 2{A_0}\left( m \right) + \left( {4{m^2} - {p^2}} \right){B_0}\left( {m;p,m} \right)\\\\
\implies{f_{T1}} &= \frac{1}{{n - 1}}\left[ {2{A_0}\left( m \right) + \left( {4{m^2} - {p^2}} \right){B_0}\left( {m;p,m} \right)} \right] \this
\end{align*}
So the integral $I_1^{\mu \nu }$ is
\begin{align}
I_1^{\mu \nu } &= \frac{1}{{n - 1}}\left[ {2{A_0}\left( m \right) + \left( {4{m^2} - {p^2}} \right){B_0}\left( {m;p,m} \right)} \right]P_T^{\mu \nu } + 2{A_0}\left( m \right)P_L^{\mu \nu }
\end{align}
with the self energy due to a pion bubble. 
\begin{align*}
 - i\pi _{ab1}^{\mu \nu } &= 2{e^2}{\delta _{ab}}I_1^{\mu \nu }\\
 &= \frac{{2{e^2}{\delta _{ab}}}}{{n - 1}}\left[ {2{A_0}\left( m \right) + \left( {4{m^2} - {p^2}} \right){B_0}\left( {m;p,m} \right)} \right]P_T^{\mu \nu } + 4{e^2}{\delta _{ab}}{A_0}\left( m \right)P_L^{\mu \nu } \this
\end{align*}

\section{Summary of one loop self energies}
Appendix \ref{selfenergies} contains the details for the complete one loop corrections for the self energies. We shall 
summarize the results below:
\subsection{Rho Self Energy}
\begin{minipage}{\linewidth}
\centering
\begin{tikzpicture}[scale=1]
	\Fbubble{0}{2}{\Frho}{\Fphi}{\Fphi}{\Frho};
\end{tikzpicture}
\end{minipage}
\begin{align}
 - i\pi _{ab1}^{\mu \nu } = \frac{{2{e^2}{\delta _{ab}}}}{{n - 1}}\left[ {2{A_0}\left( m \right) + \left( {4{m^2} - {p^2}} \right){B_0}\left( {m;p,m} \right)} \right]P_T^{\mu \nu } + 4{e^2}{\delta _{ab}}{A_0}\left( m \right)P_L^{\mu \nu }
 \label{rhoselfEfirst}
\end{align}

\begin{minipage}{\linewidth}
\centering
\begin{tikzpicture}[scale=1]
	\Ftadpole{4.5}{2}{\Frho}{\Fphi};
\end{tikzpicture}
\end{minipage}
\begin{align}
 - i\pi _{ab2}^{\mu \nu } &=  - 4{e^2}{\delta _{ab}}{A_0}\left( m \right)P_T^{\mu \nu } - 4{e^2}{\delta _{ab}}{A_0}\left( m \right)P_L^{\mu \nu }
\end{align}

\begin{minipage}{\linewidth}
\centering
\begin{tikzpicture}[scale=1]
	\Fbubble{9}{-1}{\Frho}{\Frho}{\Fhiggs}{\Frho};
\end{tikzpicture}
\end{minipage}
\begin{align}
 - i\pi _{ab3}^{\mu \nu } =  - {e^2}{M^2}{\delta _{ab}}{B_0}\left( {M;p,{m_H}} \right)P_T^{\mu \nu } - {e^2}{M^2}{\delta _{ab}}{B_0}\left( {M;p,{m_H}} \right)P_L^{\mu \nu }
\end{align}

\begin{minipage}{\linewidth}
\centering
\begin{tikzpicture}[scale=1]
	\Ftadpole{13.5}{-1}{\Frho}{\Fhiggs};
\end{tikzpicture}
\end{minipage}
\begin{align}
 - i\pi _{ab4}^{\mu \nu } &=  - \frac{1}{2}{e^2}{\delta _{ab}}{A_0}\left( {{m_H}} \right)P_T^{\mu \nu } - \frac{1}{2}{e^2}{\delta _{ab}}{A_0}\left( {{m_H}} \right)P_L^{\mu \nu }
\end{align}

\begin{minipage}{\linewidth}
\centering
\begin{tikzpicture}[scale=1]
	\Fbubble{9}{-1}{\Frho}{\Fchi}{\Fhiggs}{\Frho};
\end{tikzpicture}
\end{minipage}
\begin{align*}
 - i\pi _{ab5}^{\mu \nu }&=  - \frac{{{e^2}{\delta _{ab}}}}{{4\left( {n - 1} \right)}} \bigg\{ {A_0}\left( M \right) + {A_0}\left( {{m_H}} \right) + \left( {2m_H^2 + 2{M^2} - {p^2}} \right){B_0}\left( {M;p,{m_H}} \right) +\\ &- \frac{{m_H^2 - {M^2}}}{{{p^2}}}\left[ {{A_0}\left( M \right) - {A_0}\left( {{m_H}} \right) + \left( {m_H^2 - {M^2}} \right){B_0}\left( {M;p,{m_H}} \right)} \right] \bigg\} P_T^{\mu \nu } + \\
 &- \frac{1}{4}{e^2}{\delta _{ab}}\left\{ {{A_0}\left( M \right) + {A_0}\left( {{m_H}} \right) + \frac{{m_H^2 - {M^2}}}{{{p^2}}}\left[ { - {A_0}\left( {{m_H}} \right) + {A_0}\left( M \right) + \left( {m_H^2 - {M^2}} \right){B_0}\left( {M;p,{m_H}} \right)} \right]} \right\}P_L^{\mu \nu } \this
\end{align*}

\begin{minipage}{\linewidth}
\centering
\begin{tikzpicture}[scale=1]
	\Fbubble{9}{-1}{\Frho}{\Fchi}{\Fchi}{\Frho};
\end{tikzpicture}
\end{minipage}
\begin{align}
 - i\pi _{ab8}^{\mu \nu } &=   \frac{{{e^2}{\delta _{ab}}}}{{2\left( {n - 1} \right)}}\left[ {2{A_0}\left( M \right) + \left( {4{M^2} - {p^2}} \right){B_0}\left( {M;p,M} \right)} \right]P_T^{\mu \nu } + {e^2}{\delta _{ab}}{A_0}\left( M \right)P_L^{\mu \nu }
\end{align}

\begin{minipage}{\linewidth}
\centering
\begin{tikzpicture}[scale=1]
	\Ftadpole{13.5}{-1}{\Frho}{\Fchi};
\end{tikzpicture}
\end{minipage}
\begin{align}
 - i\pi _{ab6}^{\mu \nu } &=  - \frac{3}{2}{e^2}{\delta ^{ab}}{A_0}\left( M \right)P_T^{\mu \nu } - \frac{3}{2}{e^2}{\delta ^{ab}}{A_0}\left( M \right)P_L^{\mu \nu }
\end{align}

\begin{minipage}{\linewidth}
\centering
\begin{tikzpicture}[scale=1]
	\Fbubble{9}{-1}{\Frho}{\Fgst}{\Fgst}{\Frho};
\end{tikzpicture}
\end{minipage}
\begin{align}
- i\pi _{ab7}^{\mu \nu } &= \frac{{2{e^2}{\delta _{ab}}}}{{n - 1}}\left[ {\frac{1}{2}{A_0}\left( M \right) + \left( {{M^2} - \frac{1}{4}{p^2}} \right){B_0}\left( {M;p,M} \right)} \right]P_T^{\mu \nu } + 2{e^2}{\delta _{ab}}\left[ {\frac{1}{2}{A_0}\left( M \right) + \frac{1}{4}{p^2}{B_0}\left( {M;p,M} \right)} \right]P_L^{\mu \nu }
\end{align}

\begin{minipage}{\linewidth}
\centering
\begin{tikzpicture}[scale=1]
	\Fbubble{9}{-1}{\Frho}{\Frho}{\Frho}{\Frho};
\end{tikzpicture}
\end{minipage}
\begin{align*}
 - i\pi _{ab9}^{\mu \nu } &=   2{e^2}{\delta _{ab}}\left\{ {\frac{{4n - 5}}{{n - 1}}{A_0}\left( M \right) + \frac{1}{{2\left( {n - 1} \right)}}\left[ 4{\left( {3n - 4} \right){M^2} + \left( {6n - 5} \right){p^2}} \right]B\left( {M;p,M} \right)} \right\}P_T^{\mu \nu } + \\
 &+ 2{e^2}{\delta _{ab}}\left[ {\left( {2n - 1} \right){A_0}\left( M \right) + \frac{1}{2}\left( {4{M^2} - {p^2}} \right)B\left( {M;p,M} \right)} \right]P_L^{\mu \nu } \this
\end{align*}

\begin{minipage}{\linewidth}
\centering
\begin{tikzpicture}[scale=1]
	\Ftadpole{13.5}{-1}{\Frho}{\Frho};
\end{tikzpicture}
\end{minipage}
\begin{align}
 - i\pi _{ab10}^{\mu \nu } &=  - 4{e^2}{\delta _{ab}}\left( {n - 1} \right){A_0}\left( M \right)P_T^{\mu \nu } - 4{e^2}{\delta _{ab}}\left( {n - 1} \right){A_0}\left( M \right)P_L^{\mu \nu }
\end{align}

\begin{minipage}{\linewidth}
\centering
\begin{tikzpicture}[scale=1]
	\FFlollipop{9}{-5.5}{\Frho}{\Fhiggs}{fill=gray!70};
\end{tikzpicture}
\end{minipage}
\begin{align*}
 - i\pi _{ab11}^{\mu \nu } &= \frac{3}{2}{M^2}{\delta _{ab}}\left\{ {\lambda \frac{{{A_0}\left( {{m_H}} \right)}}{{m_H^2}} + \left[ {\lambda  + {e^2}\left( {n - 1} \right)} \right]\frac{{{A_0}\left( M \right)}}{{m_H^2}} + 8\kappa \frac{{{A_0}\left( m \right)}}{{m_H^2}}} \right\}P_T^{\mu \nu } + \\
 &+ \frac{3}{2}{M^2}{\delta _{ab}}\left\{ {\lambda \frac{{{A_0}\left( {{m_H}} \right)}}{{m_H^2}} + \left[ {\lambda  + {e^2}\left( {n - 1} \right)} \right]\frac{{{A_0}\left( M \right)}}{{m_H^2}} + 8\kappa \frac{{{A_0}\left( m \right)}}{{m_H^2}}} \right\}P_L^{\mu \nu } 
 \this  \label{rhoselfElast}
\end{align*}

\subsection{Higgs Self Energy}
\begin{minipage}{\linewidth}
\centering
\begin{tikzpicture}[scale=1]
	\Fbubble{0}{2}{\Fhiggs}{\Fphi}{\Fphi}{\Fhiggs};
\end{tikzpicture}
\end{minipage}
\begin{align}
 - i\pi _1^H = 48{\kappa ^2}{\nu ^2}{B_0}\left( {m;p,m} \right) \label{higgsselfEfirst}
\end{align}

\begin{minipage}{\linewidth}
\centering
\begin{tikzpicture}[scale=1]
	\Ftadpole{4.5}{2}{\Fhiggs}{\Fphi};
\end{tikzpicture}
\end{minipage}
\begin{align}
 - i\pi _2^H = 6\kappa {A_0}\left( m \right)
\end{align}

\begin{minipage}{\linewidth}
\centering
\begin{tikzpicture}[scale=1]
	\Fbubble{9}{-1}{\Fhiggs}{\Frho}{\Frho}{\Fhiggs};
\end{tikzpicture}
\end{minipage}
\begin{align}
 - i\pi _3^H = 3{e^4}{\nu ^2}n{B_0}\left( {M;p,M} \right)
\end{align}

\begin{minipage}{\linewidth}
\centering
\begin{tikzpicture}[scale=1]
	\Ftadpole{13.5}{-1}{\Fhiggs}{\Frho};
\end{tikzpicture}
\end{minipage}
\begin{align}
 - i\pi _4^H = \frac{3}{2}{e^2}n{A_0}\left( M \right)
\end{align}

\begin{minipage}{\linewidth}
\centering
\begin{tikzpicture}[scale=1]
	\Fbubble{9}{-1}{\Fhiggs}{\Fchi}{\Fchi}{\Fhiggs};
\end{tikzpicture}
\end{minipage}
\begin{align}
 - i\pi _5^H = \frac{3}{4}{\lambda ^2}{\nu ^2}{B_0}\left( {M;p,M} \right)
\end{align}

\begin{minipage}{\linewidth}
\centering
\begin{tikzpicture}[scale=1]
	\Ftadpole{13.5}{-1}{\Fhiggs}{\Fchi};
\end{tikzpicture}
\end{minipage}
\begin{align}
 - i\pi _6^H = \frac{3}{4}\lambda {A_0}\left( M \right)
\end{align}

\begin{minipage}{\linewidth}
\centering
\begin{tikzpicture}[scale=1]
	\Fbubble{9}{-1}{\Fhiggs}{\Fhiggs}{\Fhiggs}{\Fhiggs};
\end{tikzpicture}
\end{minipage}
\begin{align}
 - i\pi _7^H = \frac{9}{4}{\lambda ^2}{\nu ^2}{B_0}\left( {{m_H};p,{m_H}} \right)
\end{align}

\begin{minipage}{\linewidth}
\centering
\begin{tikzpicture}[scale=1]
	\Ftadpole{13.5}{-1}{\Fhiggs}{\Fhiggs};
\end{tikzpicture}
\end{minipage}

\begin{align}
 - i\pi _8^H = \frac{3}{2}\lambda {A_0}\left( {{m_H}} \right)
\end{align}

\begin{minipage}{\linewidth}
\centering
\begin{tikzpicture}[scale=1]
	\Fbubble{9}{-1}{\Fhiggs}{\Frho}{\Fchi}{\Fhiggs};
\end{tikzpicture}
\end{minipage}
\begin{align}
 - i\pi _9^H =  - \frac{3}{4}{e^2}\left[ {{A_0}\left( M \right) + \left( {{M^2} + 2{p^2}} \right){B_0}\left( {M;p,M} \right)} \right]
\end{align}

\begin{minipage}{\linewidth}
\centering
\begin{tikzpicture}[scale=1]
	\Fbubble{9}{-1}{\Fhiggs}{\Fgst}{\Fgst}{\Fhiggs};
\end{tikzpicture}
\end{minipage}
\begin{align}
 - i\pi _{10}^H = \frac{3}{4}{e^2}{M^2}{B_0}\left( {M;p,M} \right)
\end{align}

\begin{minipage}{\linewidth}
\centering
\begin{tikzpicture}[scale=1]
	\FFlollipop{9}{-5.5}{\Fhiggs}{\Fhiggs}{fill=gray!70};
\end{tikzpicture}
\end{minipage}
\begin{align}
 - i\pi _{11}^H =  - \frac{9}{4}\lambda {\nu ^2}\left\{ {\lambda \frac{{{A_0}\left( {{m_H}} \right)}}{{m_H^2}} + \left[ {\lambda  + {e^2}\left( {n - 1} \right)} \right]\frac{{{A_0}\left( M \right)}}{{m_H^2}} + 8\kappa \frac{{{A_0}\left( m \right)}}{{m_H^2}}} \right\} \label{higgsselfElast}
\end{align}

\section{Summary of one loop vertices}
We shall use these definitions for the calculations of the one loop corrections to the three point vertices:
\begin{align*}
{p^\mu } &= p_1^\mu  + p_2^\mu \\
p \cdot \left( {{p_1} - {p_2}} \right) &= \left( {{p_1} + {p_2}} \right) \cdot \left( {{p_1} - {p_2}} \right)\\
&= 0\\
	{p_1} \cdot {p_2} &= \frac{1}{2}\left[ {{{\left( {{p_1} + {p_2}} \right)}^2} - p_1^2 - p_2^2} \right]\\
	&= \frac{1}{2}{p^2} - {m^2}\\
	p \cdot {p_1} &= \left( {{p_1} + {p_2}} \right) \cdot {p_1} \\
	&= \frac{1}{2}{p^2} 	\\
	p \cdot {p_2} &= \left( {{p_1} + {p_2}} \right) \cdot {p_2}\\
	&= \frac{1}{2}{p^2}
\end{align*}
\begin{align*}
\begin{array}{*{80}{c}}
{p \cdot {p_1} = \left( {{p_1} + {p_2}} \right) \cdot {p_1}}&{p \cdot {p_2} = \left( {{p_1} + {p_2}} \right) \cdot {p_2}}\\
{ = \frac{1}{2}{p^2}}&{= \frac{1}{2}{p^2}}
\end{array}
\end{align*}
where $\{a,p_1\}, \{b,p_2\}$ are incoming momenta of the pions and $\{c,p^{\mu}\}$ is the momentum of the outgoing rho.

\subsection{Pion-Pion-Rho Vertex}

\begin{minipage}{\linewidth}
\centering
\begin{tikzpicture}[scale=1]
	\Ftriangle{2}{0}{\Fphi}{\Frho}{\Fphi}{\Frho};
\end{tikzpicture}
\end{minipage}
\begin{align*}
V_1^\mu  &=  - i{e^3}{\varepsilon _{cab}}\frac{{{{\left( {{p_1} - {p_2}} \right)}^\mu }}}{{4{m^2} - {p^2}}} \Biggl[\frac{1}{{{m^2}}}\left( {4{m^2} - {p^2}} \right)\left[ {{A_0}\left( M \right) - {A_0}\left( m \right)} \right]\\
 &+ \frac{1}{{{m^2}}}\left[ {16{m^4} - 6{m^2}\left( {{M^2} + {p^2}} \right) + {M^2}{p^2}} \right]{B_0}\left( {m,m,M} \right) - 2\left( {4{m^2} - {M^2} - 2{p^2}} \right){B_0}\left( {m,p,m} \right) + \\
 &+\left( {4{m^2} - {M^2} - 2{p^2}} \right)\left( {4{m^2} - 2{M^2} - {p^2}} \right){C_0}\left( {m;{p_2},M;p,m} \right)\Biggr] \this \label{pionpionrhovertfirst}
\end{align*}

\begin{minipage}{\linewidth}
\centering
\begin{tikzpicture}[scale=1]
	\Ftriangle{6.5}{2}{\Frho}{\Fphi}{\Frho}{\Frho};
\end{tikzpicture}
\end{minipage}
\begin{align*}
V_{2}^\mu &= 2i{e^3}{\varepsilon _{cab}}\frac{\left( {{p_1} - {p_2}} \right)^\mu }{{4{m^2} - {p^2}}} \Biggl[\left( {1 - \frac{{{p^2}}}{{4{m^2}}}} \right)\left[ {{A_0}\left( m \right) - {A_0}\left( M \right)} \right] + \left( { - 4{m^2} + 2{M^2} + \frac{{3{p^2}}}{2} - \frac{{{M^2}{p^2}}}{{4{m^2}}}} \right){B_0}\left( {m;m,M} \right) + \\
  &- \left( {{M^2} + \frac{{{p^2}}}{2}} \right){B_0}\left( {M;p,M} \right) + \left( { - 4{m^2}{M^2} + {M^4} - 2{m^2}{p^2} + {M^2}{p^2} + \frac{1}{4}{p^4}} \right){C_0}\left( {m;{p_1},M; - {p_2},M} \right) \Biggr] \this
\end{align*}

\begin{minipage}{\linewidth}
\centering
\begin{tikzpicture}[scale=1]
	\Ftriangle{11}{2}{\Fphi}{\Fhiggs}{\Fphi}{\Frho};
\end{tikzpicture}
\end{minipage}
\begin{align*}
V_3^\mu  &= \frac{{32i{\kappa ^2}{\nu ^2}e{\varepsilon _{cab}}{{\left( {{p_1} - {p_2}} \right)}^\mu }}}{{4{m^2} - {p^2}}}\left[ {{B_0}\left( {m;m,{m_H}} \right) - {B_0}\left( {m;p,m} \right) + \frac{1}{2}\left( {4{m^2} - {p^2} - 2m_H^2} \right){C_0}\left( {{m_H};{p_1},m; - {p_2},m} \right)} \right] \this
\end{align*}

\begin{minipage}{\linewidth}
\centering
\begin{tikzpicture}[scale=1]
	\Ftriangle{15.5}{2}{\Fhiggs}{\Fphi}{\Frho}{\Frho};
\end{tikzpicture}
\end{minipage}
\begin{align*}
V_4^\mu  &=  - 2i\kappa {\nu ^2}{e^3}{\varepsilon _{cab}}\frac{{{{\left( {{p_1} - {p_2}} \right)}^\mu }}}{{4{m^2} - {p^2}}} \Biggl[{B_0}\left( {m,m,M} \right) + {B_0}\left( {m,m,{m_H}} \right) - 2{B_0}\left( {M,p,{m_H}} \right) + \\
& + \left( { - 8{m^2} + {M^2} + m_H^2 + {p^2}} \right){C_0}\left( {m;{p_1},M;- {p_2},{m_H}} \right) \Biggr] + \\
 &- 2i\kappa {\nu ^2}{e^3}{\varepsilon _{cab}}\frac{{{{\left( {{p_1} + {p_2}} \right)}^\mu }}}{{{p^2}}}\Biggl[{B_0}\left( {m,m,{m_H}} \right) - {B_0}\left( {m,m,M} \right) + \\
+&\left( {{M^2} - m_H^2 - {p^2}} \right){C_0}\left( {m; {p_1},M;- {p_2},{m_H}} \right) \Biggr] \this
\end{align*}

\begin{minipage}{\linewidth}
\centering
\begin{tikzpicture}[scale=1]
	\Ftriangle{15.5}{2}{\Frho}{\Fphi}{\Fhiggs}{\Frho};
\end{tikzpicture}
\end{minipage}
%
\begin{align*}
V_5^\mu &=  - 2i\kappa {\nu ^2}{e^3}{\varepsilon _{cab}}\frac{{{{\left( {{p_1} - {p_2}} \right)}^\mu }}}{{4{m^2} - {p^2}}} \Biggl[{B_0}\left( {m,m,M} \right) + {B_0}\left( {m,m,{m_H}} \right) - 2{B_0}\left( {M,p,{m_H}} \right) + \\
& + \left( { - 8{m^2} + {M^2} + m_H^2 + {p^2}} \right){C_0}\left( {m;{p_1},M;- {p_2},{m_H}} \right) \Biggr] + \\
 &- 2i\kappa {\nu ^2}{e^3}{\varepsilon _{cab}}\frac{{{{\left( {{p_1} + {p_2}} \right)}^\mu }}}{{{p^2}}} \Biggl[ - {B_0}\left( {m,m,{m_H}} \right) + {B_0}\left( {m,m,M} \right) + \\
&+\left( { - {M^2} + m_H^2 + {p^2}} \right){C_0}\left( {m;{p_1},M;- {p_2},{m_H}} \right) \Biggr] \this
\end{align*}

\begin{minipage}{\linewidth}
\centering
\begin{tikzpicture}[scale=1]
	\Fbubbletoprop{1}{-2.5}{\Frho}{\Frho};
\end{tikzpicture}
\end{minipage}
\begin{align}
{V_6} = 0
\end{align}

\begin{minipage}{\linewidth}
\centering
\begin{tikzpicture}[scale=1]
	\Fbubbletoprop{1}{-2.5}{\Fphi}{\Frho};
\end{tikzpicture}
\end{minipage}
\begin{align}
{V_7} = 0
\end{align}

\begin{minipage}{\linewidth}
\centering
\begin{tikzpicture}[scale=1]
	\Fbubbletoprop{1}{-2.5}{\Fchi}{\Frho};
\end{tikzpicture}
\end{minipage}
\begin{align}
{V_8} &= 0
\end{align}

\begin{minipage}{\linewidth}
\centering
\begin{tikzpicture}[scale=1]
	\FtriangleSeagullT{15}{-2.5}{\Frho}
\end{tikzpicture}
\end{minipage}

\begin{align*}
{V_9}&=  - 3i{e^3}{\varepsilon _{cab}}\frac{{p_1^\mu }}{{2{m^2}}}\left[ {{A_0}\left( m \right) - {A_0}\left( M \right) + \left( {{M^2} - 3{m^2}} \right){B_0}\left( {m;m,M} \right)} \right] \this
\end{align*}

\begin{minipage}{\linewidth}
\centering
\begin{tikzpicture}[scale=1]
	\FtriangleSeagullB{2}{-7}{\Frho}
\end{tikzpicture}
\end{minipage}

\begin{align*}
{V_{10}} &= 3i{e^3}{\varepsilon _{cab}}\frac{{p_2^\mu }}{{2{m^2}}}\left[ {{A_0}\left( m \right) - {A_0}\left( M \right) + \left( {{M^2} - 3{m^2}} \right){B_0}\left( {m;m,M} \right)} \right] \this \label{pionpionrhovertlast}
\end{align*}

\begin{align}
{V_9} + {V_{10}} =  - 3i{e^3}{\varepsilon _{cab}}\frac{1}{{2{m^2}}}\left[ {{A_0}\left( m \right) - {A_0}\left( M \right) + \left( {{M^2} - 3{m^2}} \right){B_0}\left( {m;m,M} \right)} \right]\left( {p_1^\mu  - p_2^\mu } \right) 
\end{align}

\subsection{Pion-Pion-Higgs Vertex}
\begin{minipage}{\linewidth}
\centering
\begin{tikzpicture}[scale=1]
	\Ftriangle{2}{2}{\Fphi}{\Frho}{\Fphi}{\Fhiggs};
\end{tikzpicture}
\end{minipage}
\begin{align}
V_1^H &=  - 8\kappa \nu {e^2}{\delta _{ab}}\left[ {2{B_0}\left( {m;m,M} \right) - {B_0}\left( {m;p,m} \right) + \left( {4{m^2} - {M^2} - 2{p^2}} \right){C_0}\left( {M;{p_1},m; - {p_2},m} \right)} \right] \label{pionpionhiggsvertfirst}
\end{align}

\begin{minipage}{\linewidth}
\centering
\begin{tikzpicture}[scale=1]
	\Ftriangle{6.5}{2}{\Frho}{\Fphi}{\Frho}{\Fhiggs};
\end{tikzpicture}
\end{minipage}
\begin{align}
V_2^H &=  - 2{e^4}\nu {\delta _{ab}}\left[ {2{B_0}\left( {M;p,M} \right) - {B_0}\left( {m;m,M} \right) + \left( {4{m^2} - {M^2} - \frac{1}{2}{p^2}} \right){C_0}\left( {m;{p_1},M; - {p_2},M} \right)} \right]
\end{align}

\begin{minipage}{\linewidth}
\centering
\begin{tikzpicture}[scale=1]
	\Ftriangle{11}{2}{\Fphi}{\Fhiggs}{\Fphi}{\Fhiggs};
\end{tikzpicture}
\end{minipage}
\begin{align}
V_3^H &= 64{\kappa ^3}{\nu ^3}{\delta _{ab}}{C_0}\left( {m;{p_2},{m_H};p,m} \right)
\end{align}

\begin{minipage}{\linewidth}
\centering
\begin{tikzpicture}[scale=1]
	\Ftriangle{15.5}{2}{\Fhiggs}{\Fphi}{\Fhiggs}{\Fhiggs};
\end{tikzpicture}
\end{minipage}
\begin{align}
V_4^H &= 24{\kappa ^2}{\nu ^3}\lambda {\delta _{ab}}{C_0}\left( {m;{p_1},{m_H}; - {p_2},{m_H}} \right)
\end{align}

\begin{minipage}{\linewidth}
\centering
\begin{tikzpicture}[scale=1]
	\Fbubbletoprop{1}{-2.5}{\Frho}{\Fhiggs};
\end{tikzpicture}
\end{minipage}
\begin{align}
V_5^H &= 4n{e^4}\nu {\delta _{ab}}{B_0}\left( {M;p,M} \right)
\end{align}

\begin{minipage}{\linewidth}
\centering
\begin{tikzpicture}[scale=1]
	\Fbubbletoprop{5.5}{-2.5}{\Fphi}{\Fhiggs};
\end{tikzpicture}
\end{minipage}
\begin{align}
V_6^H &= 20\kappa \nu {\lambda _4}{\delta _{ab}}{B_0}\left( {m;p,m} \right)
\end{align}

\begin{minipage}{\linewidth}
\centering
\begin{tikzpicture}[scale=1]
	\Fbubbletoprop{10}{-2.5}{\Fhiggs}{\Fhiggs};
\end{tikzpicture}
\end{minipage}
\begin{align}
V_7^H &= 3\kappa \lambda \nu {\delta _{ab}}{B_0}\left( {{m_H};p,{m_H}} \right)
\end{align}

\begin{minipage}{\linewidth}
\centering
\begin{tikzpicture}[scale=1]
	\Fbubbletoprop{10}{-2.5}{\Fchi}{\Fhiggs};
\end{tikzpicture}
\end{minipage}
\begin{align}
V_8^H &= 3\kappa \lambda \nu {\delta _{ab}}{B_0}\left( {M;p,M} \right)
\end{align}

\begin{minipage}{\linewidth}
\centering
\begin{tikzpicture}[scale=1]
	\FtriangleSeagullT{15}{-2.5}{\Fhiggs}
\end{tikzpicture}
\end{minipage}
\begin{align}
V_9^H &= 8{\kappa ^2}\nu {\delta _{ab}}{B_0}\left( {m;m,{m_H}} \right)
\end{align}

\begin{minipage}{\linewidth}
\centering
\begin{tikzpicture}[scale=1]
	\FtriangleSeagullB{2}{-7}{\Fhiggs}
\end{tikzpicture}
\end{minipage}
\begin{align}
V_{10}^H &= 8{\kappa ^2}\nu {\delta _{ab}}{B_0}\left( {m;m,{m_H}} \right) \label{pionpionhiggsvertlast}
\end{align}

\subsection{Pion-Pion-Pion-Pion Vertex}

\begin{minipage}{\linewidth}
\centering
	\begin{tikzpicture}[scale=1]
	\FFourtoFour{2}{2}{\Frho}
	\end{tikzpicture}
\end{minipage}
\begin{align*}
V_1^\phi \left( s \right) &= 2n{e^4}\left( {2{\delta _{ab}}{\delta _{cd}} + {\delta _{ac}}{\delta _{bd}} + {\delta _{ad}}{\delta _{bc}}} \right){B_0}\left( {M;\sqrt{s},M} \right)
\end{align*}
Including the Mandlestam $t$ and $u$ channels
\begin{align*}
V_{abcd}^{1\phi }\left( {s,t,u} \right) &= 2n{e^4} \Biggl[ \left( {2{\delta _{ab}}{\delta _{cd}} + {\delta _{ac}}{\delta _{bd}} + {\delta _{ad}}{\delta _{bc}}} \right){B_0}\left( {M;\sqrt s ,M} \right) + \\
&+\left( {{\delta _{ab}}{\delta _{cd}} + 2{\delta _{ac}}{\delta _{bd}} + {\delta _{ad}}{\delta _{bc}}} \right){B_0}\left( {M;\sqrt t ,M} \right) + \left( {{\delta _{ab}}{\delta _{cd}} + {\delta _{ac}}{\delta _{bd}} + 2{\delta _{ad}}{\delta _{bc}}} \right){B_0}\left( {M;\sqrt u ,M} \right) \Biggr] \this \label{phi4first}
\end{align*}

\begin{minipage}{\linewidth}
\centering
	\begin{tikzpicture}[scale=1]
	\FFourtoFour{2}{2}{\Fhiggs}
	\end{tikzpicture}
\end{minipage}
\begin{align*}
V_2^\phi\left(s\right) = 4{\kappa ^2}{\delta _{ab}}{\delta _{cd}}{B_0}\left( {{m_H};\sqrt{s},{m_H}} \right)
\end{align*}
Including the $t$ and $u$ Mandlestam channels
\begin{align}
V_{abcd}^{2\phi }\left( {s,t,u} \right) = 4{\kappa ^2}\left[ {{\delta _{ab}}{\delta _{cd}}{B_0}\left( {{m_H};\sqrt s ,{m_H}} \right) + {\delta _{ac}}{\delta _{bd}}{B_0}\left( {{m_H};\sqrt t ,{m_H}} \right) + {\delta _{ad}}{\delta _{bc}}{B_0}\left( {{m_H};\sqrt u ,{m_H}} \right)} \right]
\end{align}

\begin{minipage}{\linewidth}
\centering
	\begin{tikzpicture}[scale=1]
	\FFourtoFour{2}{2}{\Fchi}
	\end{tikzpicture}
\end{minipage}
\begin{align*}
V_3^\phi\left(s\right)  = 12{\kappa ^2}{\delta _{ab}}{\delta _{cd}}{B_0}\left( {M;\sqrt{s},M} \right)
\end{align*}
with 
\begin{align}
V_{abcd}^{3\phi }\left( {s,t,u} \right) = 12{\kappa ^2}\left[ {{\delta _{ab}}{\delta _{cd}}{B_0}\left( {M;\sqrt s ,M} \right) + {\delta _{ac}}{\delta _{bd}}{B_0}\left( {M;\sqrt t ,M} \right) + {\delta _{ad}}{\delta _{bc}}{B_0}\left( {M;\sqrt u ,M} \right)} \right]
\end{align}

\begin{minipage}{\linewidth}
\centering
	\begin{tikzpicture}[scale=1]
	\FFourtoFour{2}{2}{\Fphi}
	\end{tikzpicture}
\end{minipage}
\begin{align*}
V_4^\phi\left(s\right)  = \lambda _4^2\left( {7{\delta _{ab}}{\delta _{cd}} + 2{\delta _{ac}}{\delta _{bd}} + 2{\delta _{ad}}{\delta _{bc}}} \right){B_0}\left( {m;\sqrt{s},m} \right)
\end{align*}
with 
\begin{align}
V_{abcd}^{4\phi }\left( {s,t,u} \right) &= \lambda _4^2 \Biggl[\left( {7{\delta _{ab}}{\delta _{cd}} + 2{\delta _{ac}}{\delta _{bd}} + 2{\delta _{ad}}{\delta _{bc}}} \right){B_0}\left( {m;\sqrt s ,m} \right) + \nonumber \\
 &+ \left( {2{\delta _{ab}}{\delta _{cd}} + 7{\delta _{ac}}{\delta _{bd}} + 2{\delta _{ad}}{\delta _{bc}}} \right){B_0}\left( {m;\sqrt t ,m} \right) + \left( {2{\delta _{ab}}{\delta _{cd}} + 2{\delta _{ac}}{\delta _{bd}} + 7{\delta _{ad}}{\delta _{bc}}} \right){B_0}\left( {m;\sqrt u ,m} \right) \Biggr]
\end{align}

\begin{minipage}{\linewidth}
\centering
	\begin{tikzpicture}[scale=1]
	\FThreetoFourtriangle{1}{1}{\Fphi}{\Frho}{\Fphi}
	\end{tikzpicture}
\end{minipage}
\begin{align*}
V_5^\phi\left(s\right)  &=  - {e^2}{\lambda _4}\left( { - 4{\delta _{ab}}{\delta _{cd}} + {\delta _{ac}}{\delta _{bd}} + {\delta _{ad}}{\delta _{bc}}} \right) \Biggl[ 2{B_0}\left( {m;m,M} \right) - {B_0}\left( {m;\sqrt{s},m} \right) +\\
&+  \left( {4{m^2} - {M^2} - 2{s}} \right){C_0}\left( {m;{p_2},M;\sqrt{s},m} \right) \Biggr]
\end{align*}
with 
\begin{align*}
&V_{abcd}^{5\phi}\left( {s,t,u} \right) =  - {e^2}{\lambda _4} \Biggl\{ \\
&\left( { - 4{\delta _{ab}}{\delta _{cd}} + {\delta _{ac}}{\delta _{bd}} + {\delta _{ad}}{\delta _{bc}}} \right)\left[ {2{B_0}\left( {m;m,M} \right) - {B_0}\left( {m;\sqrt s ,m} \right) + \left( {4{m^2} - {M^2} - 2s} \right){C_0}\left( {m;{p_2},M;\sqrt s ,m} \right)} \right] + \\
&\left( {{\delta _{ab}}{\delta _{cd}} - 4{\delta _{ac}}{\delta _{bd}} + {\delta _{ad}}{\delta _{bc}}} \right)\left[ {2{B_0}\left( {m;m,M} \right) - {B_0}\left( {m;\sqrt t ,m} \right) + \left( {4{m^2} - {M^2} - 2t} \right){C_0}\left( {m; - {p_3},M;\sqrt t ,m} \right)} \right] + \\
&\left( {{\delta _{ab}}{\delta _{cd}} + {\delta _{ac}}{\delta _{bd}} - 4{\delta _{ad}}{\delta _{bc}}} \right)\left[ {2{B_0}\left( {m;m,M} \right) - {B_0}\left( {m;\sqrt u ,m} \right) + \left( {4{m^2} - {M^2} - 2u} \right){C_0}\left( {m; - {p_4},M;\sqrt u ,m} \right)} \right]\Biggr\}  \this
\end{align*}

\begin{minipage}{\linewidth}
\centering
	\begin{tikzpicture}[scale=1]
	\FThreetoFourtriangle{1}{1}{\Frho}{\Fphi}{\Frho}
	\end{tikzpicture}
\end{minipage}
\begin{align*}
V_7^\phi \left( s \right) &= {e^4}\left( {2{\delta _{ab}}{\delta _{cd}} + {\delta _{ac}}{\delta _{bd}} + {\delta _{ad}}{\delta _{bc}}} \right) \biggl[ 2{B_0}\left( {M;\sqrt s ,M} \right) - {B_0}\left( {m,m,M} \right)\\
& + \frac{1}{2}\left( {8{m^2} - 2{M^2} - s} \right){C_0}\left( {m,{p_2},M,\sqrt s ,M} \right) \biggr]
\end{align*}
with 
\begin{align*}
\frac{1}{{{e^4}}}V_{abcd}^{7\phi }\left( {s,t,u} \right) &= \left( {2{\delta _{ab}}{\delta _{cd}} + {\delta _{ac}}{\delta _{bd}} + {\delta _{ad}}{\delta _{bc}}} \right) \biggl[ 2{B_0}\left( {M;\sqrt s ,M} \right) - {B_0}\left( {m,m,M} \right) + \\
 &+ \frac{1}{2}\left( {8{m^2} - 2{M^2} - s} \right){C_0}\left( {m,{p_2},M,\sqrt s ,M} \right) \biggr] + \\
 &+ \left( {{\delta _{ab}}{\delta _{cd}} + 2{\delta _{ac}}{\delta _{bd}} + {\delta _{ad}}{\delta _{bc}}} \right) \biggl[2{B_0}\left( {M;\sqrt t ,M} \right) - {B_0}\left( {m,m,M} \right) + \\
 &+ \frac{1}{2}\left( {8{m^2} - 2{M^2} - t} \right){C_0}\left( {m, - {p_1},M,\sqrt t ,M} \right) \biggr] + \\
 &+ \left( {{\delta _{ab}}{\delta _{cd}} + {\delta _{ac}}{\delta _{bd}} + 2{\delta _{ad}}{\delta _{bc}}} \right) \biggl[2{B_0}\left( {M;\sqrt u ,M} \right) - {B_0}\left( {m,m,M} \right) + \\
 &+ \frac{1}{2}\left( {8{m^2} - 2{M^2} - u} \right){C_0}\left( {m, - {p_1},M,\sqrt u ,M} \right) \biggr] \this
\end{align*}

\begin{minipage}{\linewidth}
\centering
	\begin{tikzpicture}[scale=1]
	\FThreetoFourtriangle{1}{1}{\Fhiggs}{\Fphi}{\Fhiggs}
	\end{tikzpicture}
\end{minipage}
\begin{align*}
V_6^\phi\left( s \right)   = 32{\kappa ^3}{\nu ^2}{\delta _{ab}}{\delta _{cd}}{C_0}\left( {m_H;{p_2},{m}; \sqrt{s},{m_H}} \right)
\end{align*}
with 
\begin{align*}
V_{abcd}^{6\phi }\left( {s,t,u} \right) &= 32{\kappa ^3}{\nu ^2} \biggl[{\delta _{ab}}{\delta _{cd}}{C_0}\left( {{m_H},{p_2},m,\sqrt s ,{m_H}} \right) + {\delta _{ac}}{\delta _{bd}}{C_0}\left( {{m_H}, - {p_3},m,\sqrt t ,{m_H}} \right) + \\
& + {\delta _{ad}}{\delta _{bc}}{C_0}\left( {{m_H}, - {p_4},m,\sqrt u ,{m_H}} \right) \biggr] \this
\end{align*}

\begin{minipage}{\linewidth}
\centering
	\begin{tikzpicture}[scale=1]
	\FThreetoFourtriangle{1}{1}{\Fphi}{\Fhiggs}{\Fphi}
	\end{tikzpicture}
\end{minipage}
\begin{align*}
V_8^\phi\left( s \right)  = 16{\kappa ^2}{\nu ^2}{\lambda _4}\left( {{\delta _{ab}}{\delta _{cd}} + {\delta _{ac}}{\delta _{bd}} + {\delta _{ad}}{\delta _{bc}}} \right){C_0}\left( {m;{p_2},{m_H};\sqrt{s},m} \right)
\end{align*}
with
\begin{align*}
V_{abcd}^{8\phi }\left( {s,t,u} \right) &= 16{\kappa ^2}{\nu ^2}{\lambda _4}\left( {{\delta _{ab}}{\delta _{cd}} + {\delta _{ac}}{\delta _{bd}} + {\delta _{ad}}{\delta _{bc}}} \right) \biggl[ {C_0}\left( {m;{p_2},{m_H};\sqrt s ,m} \right) + \\
&+{C_0}\left( {m; - {p_3},{m_H};\sqrt t ,m} \right) + {C_0}\left( {m; - {p_4},{m_H};\sqrt u ,m} \right) \biggr] \this
\end{align*}

\begin{minipage}{\linewidth}
\centering
	\begin{tikzpicture}[scale=1]
	\Fbox{2}{2}{\Fphi}{\Frho}{\Fphi}{\Frho}
	\end{tikzpicture}
\end{minipage}
\begin{align*}
V_9^\phi &= {e^4}\left( {{\delta _{ab}}{\delta _{cd}} + {\delta _{ac}}{\delta _{bd}}} \right) \Biggl[ {B_0}\left( {m,\sqrt s ,m} \right) + \frac{{4\left( {s + t - 4{m^2}} \right)}}{{t - 4{m^2}}}\left[ {{B_0}\left( {M,t,M} \right) - {B_0}\left( {m,m,M} \right)} \right] + \\
& + \frac{2}{{4{m^2} - t}}\left( {32{m^4} - 8{m^2}\left( {{M^2} + 2s + t} \right) + 2{M^2}\left( {s + t} \right) + 3st} \right){C_0}\left( {m,{p_1},M,{p_3},M} \right) + \\
+&2\left( { - 4{m^2} + {M^2} + 2s} \right){C_0}\left( {{m^2},{p_1},M,{p_3},M} \right) + {\left( { - 4{m^2} + {M^2} + 2s} \right)^2}{D_0}\left( {m,{p_2},M,\sqrt s ,m,{p_4},M} \right)\Biggr]
\end{align*}
with 

\begin{align*}
&\frac{1}{{{e^4}}}V_{abcd}^{9\phi }\left( {s,t,u} \right) = \left( {{\delta _{ab}}{\delta _{cd}} + {\delta _{ac}}{\delta _{bd}}} \right) \Biggl[ {B_0}\left( {m,\sqrt s ,m} \right) + \frac{{4\left( {s + t - 4{m^2}} \right)}}{{t - 4{m^2}}}\left[ {{B_0}\left( {M,\sqrt t ,M} \right) - {B_0}\left( {m,m,M} \right)} \right] + \\
 &+ \frac{2}{{4{m^2} - t}}\left( {32{m^4} - 8{m^2}\left( {{M^2} + 2s + t} \right) + 2{M^2}\left( {s + t} \right) + 3st} \right){C_0}\left( {m,{p_1},M,{p_3},M} \right) + \\
 &+ 2\left( {2s - 4{m^2} + {M^2}} \right){C_0}\left( {{m^2},{p_1},M,{p_3},M} \right) + {\left( {2s - 4{m^2} + {M^2}} \right)^2}{D_0}\left( {m,{p_2},M,\sqrt s ,m,{p_4},M} \right)\Biggr] + \\
 &+ \left( {{\delta _{ac}}{\delta _{bd}} + {\delta _{ad}}{\delta _{bc}}} \right) \Biggl[{B_0}\left( {m,\sqrt u ,m} \right) + \frac{{4\left( {u + t - 4{m^2}} \right)}}{{t - 4{m^2}}}\left[ {{B_0}\left( {M,\sqrt t ,M} \right) - {B_0}\left( {m,m,M} \right)} \right] + \\
 &+ \frac{2}{{4{m^2} - t}}\left( {32{m^4} - 8{m^2}\left( {{M^2} + 2u + t} \right) + 2{M^2}\left( {u + t} \right) + 3ut} \right){C_0}\left( {m, - {p_3},M, - {p_1},M} \right) + \\
 &+ 2\left( {2u - 4{m^2} + {M^2}} \right){C_0}\left( {{m^2}, - {p_3},M, - {p_1},M} \right) + {\left( {2u - 4{m^2} + {M^2}} \right)^2}{D_0}\left( {m,{p_2},M,\sqrt u ,m,{p_4},M} \right)\Biggr] + \\
 &+ \left( {{\delta _{ab}}{\delta _{cd}} + {\delta _{ad}}{\delta _{bc}}} \right) \Biggl[{B_0}\left( {m,\sqrt u ,m} \right) + \frac{{4\left( {u + s - 4{m^2}} \right)}}{{s - 4{m^2}}}\left[ {{B_0}\left( {M,\sqrt s ,M} \right) - {B_0}\left( {m,m,M} \right)} \right] + \\
 &+ \frac{2}{{4{m^2} - s}}\left( {32{m^4} - 8{m^2}\left( {{M^2} + 2u + s} \right) + 2{M^2}\left( {u + s} \right) + 3us} \right){C_0}\left( {m,{p_2},M, - {p_1},M} \right) + \\
 &+ 2\left( {2u - 4{m^2} + {M^2}} \right){C_0}\left( {{m^2},{p_2},M, - {p_1},M} \right) + {\left( {2u - 4{m^2} + {M^2}} \right)^2}{D_0}\left( {m, - {p_3},M,\sqrt u ,m,{p_4},M} \right) \Biggr]\this
\end{align*}

\begin{minipage}{\linewidth}
\centering
	\begin{tikzpicture}[scale=1]
	\Fbox{2}{2}{\Frho}{\Fphi}{\Frho}{\Fphi}
	\end{tikzpicture}
\end{minipage}
\begin{align*}
V_{10}^\phi  &= {e^4}\left( {{\delta _{ab}}{\delta _{cd}} + {\delta _{ac}}{\delta _{bd}}} \right) \Biggl[ {B_0}\left( {m,\sqrt s ,m} \right) + \frac{{4\left( {s + t - 4{m^2}} \right)}}{{t - 4{m^2}}}\left[ {{B_0}\left( {M,\sqrt{t},M} \right) - {B_0}\left( {m,m,M} \right)} \right] + \\
& + \frac{2}{{4{m^2} - t}}\left( {32{m^4} - 8{m^2}\left( {{M^2} + 2s + t} \right) + 2{M^2}\left( {s + t} \right) + 3st} \right){C_0}\left( {m,{p_1},M,{p_3},M} \right) + \\
+&2\left( { - 4{m^2} + {M^2} + 2s} \right){C_0}\left( {{m^2},{p_1},M,{p_3},M} \right) + {\left( { - 4{m^2} + {M^2} + 2s} \right)^2}{D_0}\left( {m,{p_2},M,\sqrt s ,m,{p_4},M} \right)\Biggr]
\end{align*}
with
\begin{align*}
&\frac{1}{{{e^4}}}V_{abcd}^{10\phi }\left( {s,t,u} \right) = \left( {{\delta _{ab}}{\delta _{cd}} + {\delta _{ac}}{\delta _{bd}}} \right) \Biggl[ {B_0}\left( {m,\sqrt t ,m} \right) + \frac{{4\left( {s + t - 4{m^2}} \right)}}{{s - 4{m^2}}}\left[ {{B_0}\left( {M,\sqrt t ,M} \right) - {B_0}\left( {m,m,M} \right)} \right] + \\
 &- 2\left( {2t - 4{m^2} + {M^2}} \right){C_0}\left( {M,\sqrt s ,M,{p_2},m} \right) + {\left( {2t - 4{m^2} + {M^2}} \right)^2}D\left( {M,{p_2},m,\sqrt s ,M,{p_4},m} \right) + \\
 &+ \frac{2}{{4{m^2} - s}}\left( {32{m^4} - 8{m^2}\left( {{M^2} + s + 2t} \right) + 2{M^2}\left( {s + t} \right) + 3st} \right){C_0}\left( {M,\sqrt s ,M,{p_2},m} \right) \Biggr] + \\
 &+ \left( {{\delta _{ac}}{\delta _{bd}} + {\delta _{ad}}{\delta _{bc}}} \right) \Biggl[{B_0}\left( {m,\sqrt t ,m} \right) + \frac{{4\left( {u + t - 4{m^2}} \right)}}{{u - 4{m^2}}}\left[ {{B_0}\left( {M,\sqrt t ,M} \right) - {B_0}\left( {m,m,M} \right)} \right] + \\
 &- 2\left( {2t - 4{m^2} + {M^2}} \right){C_0}\left( {M,\sqrt u ,M,{p_2},m} \right) + {\left( {2t - 4{m^2} + {M^2}} \right)^2}D\left( {M,{p_2},m,\sqrt u ,M,{p_4},m} \right) + \\
 &+ \frac{2}{{4{m^2} - u}}\left( {32{m^4} - 8{m^2}\left( {{M^2} + u + 2t} \right) + 2{M^2}\left( {u + t} \right) + 3ut} \right){C_0}\left( {M,\sqrt t ,M,{p_2},m} \right) \Biggr]] + \\
 &+ \left( {{\delta _{ab}}{\delta _{cd}} + {\delta _{ad}}{\delta _{bc}}} \right) \Biggl[{B_0}\left( {m,\sqrt s ,m} \right) + \frac{{4\left( {s + u - 4{m^2}} \right)}}{{u - 4{m^2}}}\left[ {{B_0}\left( {M,\sqrt s ,M} \right) - {B_0}\left( {m,m,M} \right)} \right] + \\
 &- 2\left( {2s - 4{m^2} + {M^2}} \right){C_0}\left( {M,\sqrt u ,M, - {p_3},m} \right) + {\left( {2s - 4{m^2} + {M^2}} \right)^2}D\left( {M, - {p_3},m,\sqrt u ,M,{p_4},m} \right) + \\
 &+ \frac{2}{{4{m^2} - s}}\left( {32{m^4} - 8{m^2}\left( {{M^2} + u + 2s} \right) + 2{M^2}\left( {u + s} \right) + 3us} \right){C_0}\left( {M,\sqrt t ,M, - {p_3},m} \right) \Biggr] \this
\end{align*}

\begin{minipage}{\linewidth}
\centering
	\begin{tikzpicture}[scale=1]
	\Fbox{2}{2}{\Fhiggs}{\Fphi}{\Fhiggs}{\Fphi}
	\end{tikzpicture}
\end{minipage}

\begin{align*}
V_{11}^\phi  = {\left( {4\kappa \nu } \right)^4}{\delta _{ab}}{\delta _{cd}}{D_0}\left( {{m_H};{p_2},m;p,{m_H};{p_4},m} \right)
\end{align*}
with
\begin{align*}
\frac{1}{{{{\left( {4\kappa \nu } \right)}^4}}}V_{abcd}^{11\phi }\left( {s,t,u} \right) &= {\delta _{ab}}{\delta _{cd}}{D_0}\left( {{m_H};{p_2},m;\sqrt s ,{m_H};{p_4},m} \right) + {\delta _{ac}}{\delta _{bd}}{D_0}\left( {{m_H}; - {p_3},m;\sqrt t ,{m_H};{p_4},m} \right) + \\
&+{\delta _{ad}}{\delta _{bc}}{D_0}\left( {{m_H};{p_2},m;\sqrt u ,{m_H};{p_4},m} \right) \this
\end{align*}

\begin{minipage}{\linewidth}
\centering
	\begin{tikzpicture}[scale=1]
	\Fbox{2}{2}{\Fphi}{\Fhiggs}{\Fphi}{\Fhiggs}
	\end{tikzpicture}
\end{minipage}
\begin{align*}
V_{12}^\phi  = {\left( {4\kappa \nu } \right)^4}{\delta _{ac}}{\delta _{bd}}{D_0}\left( {m;{p_2},{m_H};\sqrt s ,m;{p_4},{m_H}} \right)
\end{align*}
with
\begin{align*}
\frac{1}{{{{\left( {4\kappa \nu } \right)}^4}}}V_{abcd}^{12\phi }\left( {s,t,u} \right) &= {\delta _{ab}}{\delta _{cd}}{D_0}\left( {m; - {p_3},{m_H};\sqrt t ,m;{p_4},{m_H}} \right) + {\delta _{ac}}{\delta _{bd}}{D_0}\left( {m;{p_2},{m_H};\sqrt s ,m;{p_4},{m_H}} \right) + \\
 &+ {\delta _{ad}}{\delta _{bc}}{D_0}\left( {m; - {p_4},{m_H};\sqrt u ,m; - {p_2},{m_H}} \right) \this \label{phi4last}
\end{align*}


\section{Renormalization}
\allowdisplaybreaks
We shall use the On Shell Renormalization conditions to define the mass and residue of the pole which will fix the 
renormalization constants. The conditions are
\begin{align}
{\left. {\Pi \left( {{p^2}} \right)} \right|_{{p^2} = {m^2}}} &= 0 \label{OMS1}\\
{\left. {\frac{d}{{d{p^2}}}\Pi \left( {{p^2}} \right)} \right|_{{p^2} = {m^2}}} &= 0 \label{OMS2}
\end{align}
where $\Pi \left( {{p^2}} \right)$ represents the self energy, with the mass being set by condition \eqref{OMS1} and the residue set with \eqref{OMS2}.

\subsection{Rho Self Energy Renormalization}
The counter term for the rho propagator can be written in terms of the projectors as
\begin{align*}
\bar \Gamma _{ab{\rm{CT}}}^{\mu \nu } &= i{\delta ^{ab}}\left[ {\left( {{p^2}\delta {\bar{Z}_A} - \delta {\bar{M}^2}} \right){g^{\mu \nu }} + \left( {\delta \bar{\xi}  - \delta {\bar{Z}_A}} \right){k^\mu }{k^\nu }} \right]\\
 &=  i{\delta ^{ab}}\left[ {\left( {{p^2}\delta {\bar{Z}_A} - \delta {\bar{M}^2}} \right)P_T^{\mu \nu } + \left( {{p^2}\delta \bar{\xi}  - \delta {\bar{M}^2}} \right)P_L^{\mu \nu }} \right]\this
\end{align*}
The rho self energy is the sum of contributions including the symmetry factors from \eqref{rhoselfEfirst}-\eqref{rhoselfElast} and the counter term:
\begin{align}
 - i\pi _{ab}^{\mu \nu } &= \sum\limits_{j = 1}^{11} {\left[ { - i\pi _{abj}^{\mu \nu }} \right]} +  i{\delta ^{ab}}\left[ {\left( {{p^2}\delta {\bar{Z}_A} - \delta {\bar{M}^2}} \right)P_T^{\mu \nu } + \left( {{p^2}\delta \bar{\xi}  - \delta {\bar{M}^2}} \right)P_L^{\mu \nu }} \right]\nonumber \\
 &=  \frac{i}{\left(4\pi\right)^2}{\delta _{ab}}{\pi ^{\mu \nu }} +  \frac{i}{\left(4\pi\right)^2}{\delta ^{ab}}\left[ {\left( {{p^2}\delta {Z_A} - \delta {M^2}} \right)P_T^{\mu \nu } + \left( {{p^2}\delta \xi  - \delta {M^2}} \right)P_L^{\mu \nu }} \right]\\
  &:= \frac{i}{\left(4\pi\right)^2} {\delta _{ab}}\left( \pi_T P^{\mu \nu }_T +\pi_L P^{\mu \nu }_L\right) + \frac{i}{\left(4\pi\right)^2}{\delta ^{ab}}\left[ {\left( {{p^2}\delta {Z_A} - \delta {M^2}} \right)P_T^{\mu \nu } + \left( {{p^2}\delta \xi  - \delta {M^2}} \right)P_L^{\mu \nu }} \right]
\end{align}
with 
\begin{align*}
{\pi _{T }}&:=\frac{{65{e^2}{p^2}}}{{12\varepsilon }} + \frac{1}{\varepsilon }\left[ {{e^2}\left( {2{m^2} + \frac{{9{M^4}}}{{2m_H^2}} + 11{M^2} - \frac{{m_H^2}}{2}} \right) + \frac{{12\kappa {m^2}{M^2}}}{{m_H^2}} + \frac{{3{\lambda }{M^4}}}{{2m_H^2}} + \frac{{3{\lambda}{M^2}}}{2}} \right] + \\
& + \left( { - \frac{{2{e^2}}}{3} + \frac{{12{M^2}\kappa }}{{m_H^2}}} \right){A_F}\left[ m \right] + \left[ {{e^2}\left( {\frac{7}{6} + \frac{{9{M^2}}}{{2m_H^2}}} \right) + \frac{{3{M^2}{\lambda }}}{{2m_H^2}}} \right]{A_F}\left[ M \right] + \left( { - \frac{{{e^2}}}{3} + \frac{{3{M^2}{\lambda }}}{{2m_H^2}}} \right){A_F}\left[ {{m_H}} \right] + \\
& + \frac{8}{3}{e^2}{m^2}{B_F}\left[ {m,p,m} \right] + 12{e^2}{M^2}{B_F}\left[ {M,p,M} \right] + {e^2}\left( { - \frac{{13{M^2}}}{6} - \frac{{m_H^2}}{6}} \right){B_F}\left[ {M,p,{m_H}} \right] + \\
& + \frac{{{e^2}}}{{{p^2}}}\left( {\frac{1}{{12}}\left( {{M^2} - m_H^2} \right)\left( {{A_F}\left[ {{m_H}} \right] - {A_F}\left[ M \right]} \right) + \left( {\frac{{{M^4}}}{{12}} - \frac{{{M^2}m_H^2}}{6} + \frac{{m_H^4}}{{12}}} \right){B_F}\left[ {M,p,{m_H}} \right]} \right) + \\
& + \frac{{12{m^2}{M^2}\kappa {\gamma _e}}}{{m_H^2}} + \frac{3}{2}{M^2}\lambda {\gamma _e} + \frac{{3{M^4}\lambda {\gamma _e}}}{{2m_H^2}} + \\
& + {e^2}\left( {\frac{{8{m^2}}}{3} + \frac{{23{M^2}}}{6} - \frac{{3{M^4}}}{{m_H^2}} - \frac{{m_H^2}}{6} + 2{m^2}{\gamma _e} + 11{M^2}{\gamma _e} + \frac{{9{M^4}{\gamma _e}}}{{2m_H^2}} - \frac{{m_H^2{\gamma _e}}}{2}} \right) + \\
& + {e^2}{p^2}\left( { - \frac{2}{3}{B_F}\left[ {m,p,m} \right] + 6{B_F}\left[ {M,p,M} \right] + \frac{1}{{12}}{B_F}\left[ {M,p,{m_H}} \right] - \frac{7}{{18}} + \frac{{65{\gamma _e}}}{{12}}} \right) \this \label{rhoSEt}
\end{align*}
and
\begin{align*}
{\pi _L} &:=  - \frac{{{e^2}{p^2}}}{{2\varepsilon }} + \frac{1}{\varepsilon }\left[ {{e^2}\left( {2{m^2} + \frac{{9{M^4}}}{{2m_H^2}} + 11{M^2} - \frac{{m_H^2}}{2}} \right) + \frac{{12\kappa {m^2}{M^2}}}{{m_H^2}} + \frac{{3{\lambda }{M^4}}}{{2m_H^2}} + \frac{{3{\lambda}{M^2}}}{2}} \right] + \\
 &+ \left( {2{e^2} + \frac{{12{M^2}\kappa }}{{m_H^2}}} \right){A_F}\left[ m \right] + \left( {{e^2}\left( {9 + \frac{{9{M^2}}}{{2m_H^2}}} \right) + \frac{{3{M^2}{\lambda}}}{{2m_H^2}}} \right){A_F}\left[ M \right] + \left( { - \frac{{{e^2}}}{2} + \frac{{3{M^2}{\lambda }}}{{2m_H^2}}} \right)A_F\left[ {{m_H}} \right] + \\
 &+ 4{e^2}{M^2}{B_F}\left[ {M,p,M} \right] - 2{e^2}{M^2}{B_F}\left[ {M,p,{m_H}} \right] + \\
 &+ \frac{{{e^2}}}{{{p^2}}}\left( {\frac{1}{4}\left( {{M^2} - m_H^2} \right)\left( {{A_F}\left[ M \right] - {A_F}\left[ {{m_H}} \right]} \right) + \left( { - \frac{{{M^4}}}{4} + \frac{{{M^2}m_H^2}}{2} - \frac{{m_H^4}}{4}} \right){B_F}\left[ {M,p,{m_H}} \right]} \right) + \\
 &+ {e^2}{p^2}\left( { - \frac{1}{2}{B_F}\left[ {M,p,M} \right] - \frac{{{\gamma _e}}}{2}} \right) + \frac{{12{m^2}{M^2}\kappa {\gamma _e}}}{{m_H^2}} + \frac{3}{2}{M^2}{\lambda}{\gamma _e} + \frac{{3{M^4}{\lambda}{\gamma _e}}}{{2m_H^2}} + \\
 &+ {e^2}\left( { - 4{M^2} - \frac{{3{M^4}}}{{m_H^2}} + 2{m^2}{\gamma _e} + 11{M^2}{\gamma _e} + \frac{{9{M^4}{\gamma _e}}}{{2m_H^2}} - \frac{{m_H^2{\gamma _e}}}{2}} \right) \this \label{rhoSEl}
\end{align*}
where $\pi ^{\mu \nu }$ will in general have the form 
\begin{align}
{\pi ^{\mu \nu }} = i\left( {AP_T^{\mu \nu } + BP_L^{\mu \nu }} \right)
\end{align}
with
\begin{align}
A&: = \frac{1}{{{{\left( {4\pi } \right)}^2}}}\left( {{\pi _T} + {p^2}\delta {Z_A} - \delta {M^2}} \right) \label{tselfA}\\
B&: = \frac{1}{{{{\left( {4\pi } \right)}^2}}}\left( {{\pi _L} + {p^2}\delta \xi  - \delta {M^2}} \right)\label{lselfA}
\end{align}
Applying the On Shell Renormalization conditions \eqref{OMS1} and \eqref{OMS2} to \eqref{tselfA} and \eqref{lselfA} we obtain 
the renormalization constants:
\begin{align*}
 - \delta {Z_A} &= \frac{{65{e^2}}}{{12\varepsilon }} - \frac{2}{3}{e^2}{B_F}\left[ {m,M,m} \right] + 6{e^2}{B_F}\left[ {M,M,M} \right] + \frac{1}{{12}}{e^2}{B_F}\left[ {M,M,{m_H}} \right] + \\
 &- \frac{{{e^2}}}{{{M^4}}}\left[ {\frac{1}{{12}}\left( {{M^2} - m_H^2} \right)\left( {{A_F}\left[ {{m_H}} \right] - {A_F}\left[ M \right]} \right) + \left( {\frac{{{M^4}}}{{12}} - \frac{{{M^2}m_H^2}}{6} + \frac{{m_H^4}}{{12}}} \right){B_F}\left[ {M,M,{m_H}} \right]} \right] + \\
 &+ \left( {\frac{{8{e^2}{m^2}}}{3} - \frac{{2{e^2}{M^2}}}{3}} \right){\left. {\frac{\partial }{{\partial p}}B_F\left[ {m,p,m} \right]} \right|_{p = M}} + 18{e^2}{M^2}{\left. {\frac{\partial }{{\partial p}}B_F\left[ {M,p,M} \right]} \right|_{p = M}} + \\
 &+ \left( {\frac{{{e^2}{M^2}}}{{12}} + {e^2}\left( { - \frac{{13{M^2}}}{6} - \frac{{m_H^2}}{6}} \right) + \frac{{{e^2}}}{{{M^2}}}\left( {\frac{{{M^4}}}{{12}} - \frac{{{M^2}m_H^2}}{6} + \frac{{m_H^4}}{{12}}} \right)} \right){\left. {\frac{\partial }{{\partial p}}B_F\left[ {M,p,{m_H}} \right]} \right|_{p = M}} + \\
 &+ {e^2}\left( { - \frac{7}{{18}} + \frac{{65{\gamma _e}}}{{12}}} \right) \this\\ \\
\delta {M^2} &= \frac{1}{\varepsilon }\left[ {{e^2}\left( {2{m^2} + \frac{{9{M^4}}}{{2m_H^2}} + 11{M^2} - \frac{{m_H^2}}{2}} \right) + \frac{{12\kappa {m^2}{M^2}}}{{m_H^2}} + \frac{{3{\lambda}{M^4}}}{{2m_H^2}} + \frac{{3{\lambda}{M^2}}}{2}} \right] + \\
 &+ \left( { - \frac{{2{e^2}}}{3} + \frac{{12{M^2}\kappa }}{{m_H^2}}} \right){A_F}\left[ m \right] + \left( {{e^2}\left( {\frac{7}{6} + \frac{{9{M^2}}}{{2m_H^2}}} \right) + \frac{{3{M^2}{\lambda}}}{{2m_H^2}}} \right){A_F}\left[ M \right] + \\
 &+ \left( { - \frac{{{e^2}}}{3} + \frac{{3{M^2}{\lambda _4}}}{{2m_H^2}}} \right){A_F}\left[ {{m_H}} \right] + \frac{8}{3}{e^2}{m^2}{B_F}\left[ {m,M,m} \right] + \frac{2}{3}{e^2}{M^2}{B_F}\left[ {m,M,m} \right] + \\
 &+ 6{e^2}{M^2}{B_F}\left[ {M,M,M} \right] - \frac{1}{{12}}{e^2}{M^2}{B_F}\left[ {M,M,{m_H}} \right] + {e^2}\left( { - \frac{{13{M^2}}}{6} - \frac{{m_H^2}}{6}} \right){B_F}\left[ {M,M,{m_H}} \right] + \\
 &+ \frac{{2{e^2}}}{{{M^2}}}\left[ {\frac{1}{{12}}\left( {{M^2} - m_H^2} \right)\left( {{A_F}\left[ {{m_H}} \right] - {A_F}\left[ M \right]} \right) + \left( {\frac{{{M^4}}}{{12}} - \frac{{{M^2}m_H^2}}{6} + \frac{{m_H^4}}{{12}}} \right){B_F}\left[ {M,{M^2},{m_H}} \right]} \right] + \\
 &+ \left( { - \frac{8}{3}{e^2}{m^2}{M^2} + \frac{{2{e^2}{M^4}}}{3}} \right){\left. {\frac{\partial }{{\partial p}}B_F\left[ {m,p,m} \right]} \right|_{p = M}} - 18{e^2}{M^4}{\left. {\frac{\partial }{{\partial p}}B_F\left[ {M,p,M} \right]} \right|_{p = M}} + \\
 &+ \left( { - \frac{1}{{12}}{e^2}{M^4} - {e^2}{M^2}\left( { - \frac{{13{M^2}}}{6} - \frac{{m{h^2}}}{6}} \right) - {e^2}\left( {\frac{{{M^4}}}{{12}} - \frac{{{M^2}m_H^2}}{6} + \frac{{m_H^4}}{{12}}} \right)} \right){\left. {\frac{\partial }{{\partial p}}B_F\left[ {M,p,{m_H}} \right]} \right|_{p = M}} + \\
 &+ {M^2}\left( { - \frac{2}{3}{e^2}{B_F}\left[ {m,M,m} \right] + 6{e^2}{B_F}\left[ {M,M,M} \right] + \frac{1}{{12}}{e^2}{B_F}\left[ {M,M,{m_H}} \right]} \right)\\
 &+ {e^2}{M^2}\left( { - \frac{7}{{18}} + \frac{{65{\gamma _e}}}{{12}}} \right) + \frac{{12{m^2}{M^2}\kappa {\gamma _e}}}{{m_H^2}} + \frac{3}{2}{M^2}{\lambda}{\gamma _e} + \frac{{3{M^4}{\lambda}{\gamma _e}}}{{2m_H^2}} - {e^2}{M^2}\left( { - \frac{7}{{18}} + \frac{{65{\gamma _e}}}{{12}}} \right) + \\
 &+ {e^2}\left( {\frac{{8{m^2}}}{3} + \frac{{23{M^2}}}{6} - \frac{{3{M^4}}}{{m_H^2}} - \frac{{m_H^2}}{6} + 2{m^2}{\gamma _e} + 11{M^2}{\gamma _e} + \frac{{9{M^4}{\gamma _e}}}{{2m_H^2}} - \frac{{m_H^2{\gamma _e}}}{2}} \right) \this \\ \\
 - \delta \xi  &=  - \frac{{{e^2}}}{{2\varepsilon }} - \frac{2}{3}{e^2}{B_F}\left[ {m,M,m} \right] + 6{e^2}{B_F}\left[ {M,M,M} \right] + \frac{1}{{12}}{e^2}{B_F}\left[ {M,M,{m_H}} \right] + \\
 &- \frac{{{e^2}}}{{{M^4}}}\left( {\frac{1}{{12}}\left( {{M^2} - m_H^2} \right)\left( {A_F\left[ {{m_H}} \right] - A_F\left[ M \right]} \right) + \left( {\frac{{{M^4}}}{{12}} - \frac{{{M^2}m_H^2}}{6} + \frac{{m_H^4}}{{12}}} \right){B_F}\left[ {M,M,{m_H}} \right]} \right) + \\
 &+ \left( {\frac{{8{e^2}{m^2}}}{3} - \frac{{2{e^2}{M^2}}}{3}} \right){\left. {\frac{\partial }{{\partial p}}B_F\left[ {m,p,m} \right]} \right|_{p = M}} + 18{e^2}{M^2}{\left. {\frac{\partial }{{\partial p}}B_F\left[ {M,p,M} \right]} \right|_{p = M}} + \\
 &+ \left[ {\frac{{{e^2}{M^2}}}{{12}} + {e^2}\left( { - \frac{{13{M^2}}}{6} - \frac{{m_H^2}}{6}} \right) + \frac{{{e^2}}}{{{M^2}}}\left( {\frac{{{M^4}}}{{12}} - \frac{{{M^2}m_H^2}}{6} + \frac{{m_H^4}}{{12}}} \right)} \right]{\left. {\frac{\partial }{{\partial p}}B_F\left[ {M,p,{m_H}} \right]} \right|_{p = M}} + \\
 &+ {e^2}\left( { - \frac{7}{{18}} + \frac{{65{\gamma _e}}}{{12}}} \right) \this
\end{align*}

The contributions from infinitely many self energy insertions leads to a geometric series which can be summed with $D: = {p^2} - {M^2} + i\varepsilon $,
\begin{align*}
iD_{ab}^{(1)\mu \nu } &= \frac{{ - i}}{D}{\delta _{ab}}\left( {P_T^{\mu \nu } + P_L^{\mu \nu }} \right) \\ 
&+ \sum\limits_{n = 1}^\infty  {\frac{{ - i}}{D}{\delta _{ab}}\left( {P_T^{\mu {\alpha _1}} + P_L^{\mu {\alpha _1}}} \right)\prod\limits_{k = 1}^n {\left[ {i\pi _{{\alpha _{2k-1}}}^{{\alpha _{2k}}}\frac{{ - i}}{D}\left( {P_{T{\alpha _{2k}}}^{{\alpha _{2k+1}}} + P_{L{\alpha _{2k}}}^{{\alpha _{2k + 1}}}} \right)} \right]} } \left[ {i\pi _{{\alpha _{2n + 1}}}^{{\alpha _{2n + 2}}}\frac{{ - i}}{D}\left( {P_{T{\alpha _{2n + 2}}}^\nu  + P_{L{\alpha _{2n + 2}}}^\nu } \right)} \right]\\
&= \frac{{ - i}}{D}{\delta _{ab}}\left( {P_T^{\mu \nu } + P_L^{\mu \nu }} \right) + \frac{{ - i}}{D}{\delta _{ab}}\left( {P_T^{\mu {\alpha _1}} + P_L^{\mu {\alpha _1}}} \right) \cdot i\left( {AP_{T{\alpha _1}}^{{\alpha _2}} + BP_{L{\alpha _1}}^{{\alpha _2}}} \right) \cdot \frac{{ - i}}{D}\left( {P_{T{\alpha _2}}^\nu  + P_{L{\alpha _2}}^\nu } \right) +  \cdots \\
&= \frac{{ - i}}{D}{\delta _{ab}}\left( {P_T^{\mu \nu } + P_L^{\mu \nu }} \right) + \frac{{ - i}}{{{D^2}}}{\delta _{ab}}\left( {AP_T^{\mu \nu } + BP_L^{\mu \nu }} \right) + \frac{{ - i}}{{{D^3}}}{\delta _{ab}}\left( {{A^2}P_T^{\mu \nu } + {B^2}P_L^{\mu \nu }} \right) +  \cdots \\
&= \frac{{ - i{\delta _{ab}}}}{D}\left\{ {\left[ {1 + \frac{A}{D} + {{\left( {\frac{A}{D}} \right)}^2} +  \cdots } \right]P_T^{\mu \nu } + \left[ {1 + \frac{B}{D} + {{\left( {\frac{B}{D}} \right)}^2} +  \cdots } \right]P_L^{\mu \nu }} \right\}\\
&= \frac{{ - i{\delta _{ab}}}}{D}\left[ {\frac{1}{{1 - \frac{A}{D}}}P_T^{\mu \nu } + \frac{1}{{1 - \frac{B}{D}}}P_L^{\mu \nu }} \right]\\
&= \frac{{ - i{\delta _{ab}}}}{{D - A}}\left[ {P_T^{\mu \nu } + \frac{{D - A}}{{D - B}}P_L^{\mu \nu }} \right]\\
&= \frac{{ - i{\delta _{ab}}}}{{D - A}}\left[ {{g^{\mu \nu }} + \frac{{B - A}}{{D - B}}P_L^{\mu \nu }} \right] \\
&\cong \frac{{ - i{\delta _{ab}}}}{{D - A}} g^{\mu \nu }  \this
\end{align*}
Since the longitudinal projector is made up of on shell momenta which will always encounter the on shell momentum from the Feynman vertices, the longitudinal piece of the one loop propagator will not contribute to scattering amplitudes.

\subsection{Higgs Self Energy Renormalization}
The counter term for the Higgs propagator is 
\begin{align}
\overline \Gamma  _{\rm{CT}}= i\left( { {k^2}\delta {\bar{Z}_H} - \delta {\bar{m}_H ^2}} \right)
\end{align}
The of Higgs self energy diagrams from \eqref{higgsselfEfirst} to \eqref{higgsselfElast} including the symmetry factors and the counter term is
\begin{align*}
 - i{\Pi _H} &= \sum\limits_{j = 1}^{11} {\left[ { - i\pi _j^H} \right]}  + i\left( {{p^2}\delta {{\bar Z}_H} - \delta \bar m_H^2} \right)\\
 &= \frac{i}{{{{\left( {4\pi } \right)}^2}}}\left[ {{\pi _H} + {k^2}\delta {Z_H} - \delta m_H^2} \right] \this
\end{align*}
with 
\begin{align*}
{\pi _H} &: =  - \frac{{3{e^2}{p^2}}}{{4\varepsilon }} + \frac{1}{\varepsilon } \Biggl[\frac{{27{e^2}{M^2}}}{2} + 3{m^2}\kappa  + \frac{{3{M^2}\lambda }}{8} - \frac{{27{M^4}\lambda }}{{4m_H^2}} + \frac{{3m_H^2\lambda }}{4} + \\
&+ \frac{{{M^2}}}{{{e^2}}}\left( {48{\kappa ^2} - \frac{{18{m^2}\kappa \lambda }}{{m_H^2}} + \frac{{3{\lambda ^2}}}{4} - \frac{{9{M^2}{\lambda ^2}}}{{4m_H^2}}} \right) \Biggr] + \frac{{9{M^4}\lambda }}{{2m_H^2}} + 3{m^2}\kappa {\gamma _e} + \frac{3}{8}{M^2}\lambda {\gamma _e} + \\
 &- \frac{{27{M^4}\lambda {\gamma _e}}}{{4m_H^2}} + \frac{3}{4}m_H^2\lambda {\gamma _e} + 3\kappa {A_F}\left[ m \right] + \frac{3}{8}\lambda {A_F}\left[ M \right] - \frac{{27{M^2}\lambda {A_F}\left[ M \right]}}{{4m_H^2}} + \frac{3}{4}\lambda {A_F}\left[ {mh} \right] + \\
&+{e^2}{p^2}\left( { - \frac{{3{\gamma _e}}}{4} - \frac{3}{4}{B_F}\left[ {M,p,M} \right]} \right) + \\
 &+ {e^2}\left( { - \frac{{15{M^2}}}{2} + \frac{{27{M^2}{\gamma _e}}}{2} + \frac{{21{A_F}\left[ M \right]}}{8} + \frac{{87}}{8}{M^2}{B_F}\left[ {M,p,M} \right]} \right) + \\
 &+ \frac{1}{{{e^2}}} \Biggl[48{M^2}{\kappa ^2}{\gamma _e} - \frac{{18{m^2}{M^2}\kappa \lambda {\gamma _e}}}{{m_H^2}} + \frac{3}{4}{M^2}{\lambda ^2}{\gamma _e} - \frac{{9{M^4}{\lambda ^2}{\gamma _e}}}{{4m_H^2}} - \frac{{18{M^2}\kappa \lambda {A_F}\left[ m \right]}}{{m_H^2}} + \\
 &- \frac{{9{M^2}{\lambda ^2}}}{{4m_H^2}}\left( {{A_F}\left[ M \right] + {A_F}\left[ {{m_H}} \right]} \right) + 48{M^2}{\kappa ^2}{B_F}\left[ {m,p,m} \right] + \frac{3}{4}{M^2}{\lambda ^2}{B_F}\left[ {M,p,M} \right] + \\
 &+ \frac{9}{4}{M^2}{\lambda ^2}{B_F}\left[ {{m_H},p,{m_H}} \right]\Biggr] \this \label{higgsselfE}
\end{align*}

Applying the On Shell Renormalization conditions \eqref{OMS1} and \eqref{OMS2} to \eqref{higgsselfE} we obtain the renormalization constants:
\begin{align*}
 - \delta {Z_H} &=  - \frac{{3{e^2}}}{{4\varepsilon }} + \frac{{87}}{8}{e^2}{M^2}{\left. {\frac{\partial }{{\partial p}}{B_F}\left[ {M,p,M} \right]} \right|_{p = {m_H}}} - \frac{3}{4}{e^2}m_H^2{\left. {\frac{\partial }{{\partial p}}{B_F}\left[ {M,p,M} \right]} \right|_{p = {m_H}}} + \\
 &+ \frac{1}{{{e^2}}}\frac{\partial }{{\partial p}}{\left. {\left[ {48{M^2}{\kappa ^2}{B_F}\left[ {m,p,m} \right] + \frac{3}{4}{M^2}{\lambda ^2}{B_F}\left[ {M,p,M} \right] + \frac{9}{4}{M^2}{\lambda ^2}{B_F}\left[ {{m_H},p,{m_H}} \right]} \right]} \right|_{p = {m_H}}} + \\
 &+ {e^2}\left( { - \frac{3}{4}{B_F}\left[ {M,{m_H},M} \right] - \frac{{3{\gamma _e}}}{4}} \right) \this
\\ \\
\delta m_H^2 &= \Biggl[\frac{{27{e^2}{M^2}}}{2} + 3{m^2}\kappa  + \frac{{3{M^2}\lambda }}{8} - \frac{{27{M^4}\lambda }}{{4m_H^2}} + \frac{{3m_H^2\lambda }}{4} + \\
 &+ \frac{{{M^2}}}{{{e^2}}}\left( {48{\kappa ^2} - \frac{{18{m^2}\kappa \lambda }}{{m_H^2}} + \frac{{3{\lambda ^2}}}{4} - \frac{{9{M^2}{\lambda ^2}}}{{4m_H^2}}} \right) \Biggr]\frac{1}{\varepsilon } - \frac{{15}}{2}{e^2}{M^2} + \frac{{9{M^4}\lambda }}{{2m_H^2}} + \\
 &+ \left( {3\kappa  - \frac{{18{M^2}\kappa \lambda }}{{{e^2}m_H^2}}} \right){A_F}\left[ m \right] + \left( {\frac{{21{e^2}}}{8} + \frac{{3\lambda }}{8} - \frac{{27{M^2}\lambda }}{{4m_H^2}} - \frac{{9{M^2}{\lambda ^2}}}{{4{e^2}m_H^2}}} \right){A_F}\left[ M \right] + \\
 &+ \left( {\frac{{3\lambda }}{4} - \frac{{9{M^2}{\lambda ^2}}}{{4{e^2}m_H^2}}} \right){A_F}\left[ {{m_H}} \right] + \frac{{48{M^2}{\kappa ^2}}}{{{e^2}}}{B_F}\left[ {m,{m_H},m} \right] + \\
 &+ \left( {\frac{{87{e^2}{M^2}}}{8} + \frac{{3{M^2}{\lambda ^2}}}{{4{e^2}}}} \right){B_F}\left[ {M,{m_H},M} \right] + \frac{{9{M^2}{\lambda ^2}}}{{4{e^2}}}{B_F}\left[ {{m_H},{m_H},{m_H}} \right] + \\
 &- \frac{{48{M^2}{m_H}{\kappa ^2}}}{{{e^2}}}{\left. {\frac{\partial }{{\partial p}}{B_F}\left[ {m,p,m} \right]} \right|_{p = {m_H}}} + \\
 &+ \left( { - \frac{{87}}{8}{e^2}{M^2}m_H^2 + \frac{{3{e^2}m_H^4}}{4} - \frac{{3{M^2}m_H^2{\lambda ^2}}}{{4{e^2}}}} \right){\left. {\frac{\partial }{{\partial p}}{B_F}\left[ {M,p,M} \right]} \right|_{p = {m_H}}} + \\
 &- \frac{{9{M^2}m_H^2{\lambda ^2}}}{{4{e^2}}}{\left. {\frac{\partial }{{\partial p}}{B_F}\left[ {{m_H},p,{m_H}} \right]} \right|_{p = {m_H}}} + \frac{{27}}{2}{e^2}{M^2}{\gamma _e} + 3{m^2}\kappa {\gamma _e} + \frac{{48{M^2}{\kappa ^2}{\gamma _e}}}{{{e^2}}} + \frac{3}{8}{M^2}\lambda {\gamma _e} + \\
 &- \frac{{27{M^4}\lambda {\gamma _e}}}{{4m_H^2}} + \frac{3}{4}m_H^2\lambda {\gamma _e} - \frac{{18{m^2}{M^2}\kappa \lambda {\gamma _e}}}{{{e^2}m_H^2}} + \frac{{3{M^2}{\lambda ^2}{\gamma _e}}}{{4{e^2}}} - \frac{{9{M^4}{\lambda ^2}{\gamma _e}}}{{4{e^2}m_H^2}} \this
\end{align*}

\subsection{Pion-Pion-Rho Vertex Renormalization}
For the vertex renormalization, we shall resort to using the Minimal Subtraction scheme. A conventional condition is to define the coupling at zero transferred three momentum. This is difficult to impose since the three point scalar function $C_0$ is difficult to manipulate. Summing up the vertex corrections \eqref{pionpionrhovertfirst} to \eqref{pionpionrhovertlast}
\begin{align}
\sum\limits_{k = 1}^{10} {V_k^\mu } =  \frac{i}{{{{\left( {4\pi } \right)}^2}}}{\varepsilon _{cab}}{\left( {{p_1} - {p_2}} \right)^\mu }\left( {i{V_e}} \right)
\end{align}
where $V_e$ is defined as
\begin{align*}
{V_e}&: = \frac{{{e^3}}}{{2\varepsilon }} + {e^3}\left( { - \frac{{3{A_F}\left[ m \right]}}{{2{m^2}}} + \frac{{3{A_F}\left[ M \right]}}{{2{m^2}}} + \left( {\frac{9}{2} - \frac{{3{M^2}}}{{2{m^2}}}} \right){B_F}\left[ {m,m,M} \right] + 3{\gamma _e}} \right) + \frac{1}{{4{m^2} - {p^2}}} \times  \Biggl\{ \\
&e{\kappa ^2}{\nu ^2}\left( {32{B_F}\left[ {m,m,{m_H}} \right] - 32{B_F}\left[ {m,p,m} \right] + \left( {64{m^2} - 32m_H^2 - 16{p^2}} \right){C_0}\left[ {{m_H},{p_1},m, - {p_2},m} \right]} \right) + \\
& + {e^3} \biggl[ \left( {6 - \frac{{3{p^2}}}{{2{m^2}}}} \right)\left( {{A_F}\left[ m \right] - {A_F}\left[ M \right]} \right) + \left( { - 24{m^2} + 10{M^2} + 9{p^2} - \frac{{3{M^2}{p^2}}}{{2{m^2}}}} \right){B_F}\left[ {m,m,M} \right] + \\
& + \left( {8{m^2} - 2{M^2} - 4{p^2}} \right){B_F}\left[ {m,p,m} \right] + \left( { - 2{M^2} - {p^2}} \right){B_F}\left[ {M,p,M} \right] + \\
& + \left( { - 8{m^2}{M^2} + 2{M^4} - 4{m^2}{p^2} + 2{M^2}{p^2} + \frac{{{p^4}}}{2}} \right){C_0}\left[ {m,{p_1},M, - {p_2},M} \right] + \kappa {\nu ^2} \bigg( - 4{B_F}\left[ {m,m,M} \right] + \\
& - 4{B_F}\left[ {m,m,{m_H}} \right] + 8{B_F}\left[ {M,p,{m_H}} \right] + \left( {32{m^2} - 4{M^2} - 4m_H^2 - 4{p^2}} \right){C_0}\left[ {m,{p_1},M, - {p_2},{m_H}} \right] \bigg) + \\
&\left( { - 16{m^4} + 12{m^2}{M^2} - 2{M^4} + 12{m^2}{p^2} - 5{M^2}{p^2} - 2{p^4}} \right){C_0}\left[ {m,{p_2},M,p,m} \right] - 10{m^2}{\gamma _e} + \frac{{5{p^2}{\gamma _e}}}{2}  \biggr] \Biggr\} \this \label{Ve}
\end{align*}
The one loop vertex correction including the counter term for the pion-pion-rho vertex correction is
\begin{align*}
{e^{\left( 1 \right)}}{\varepsilon _{cab}}{\left( {{p_1} - {p_2}} \right)^\mu } &= \sum\limits_{k = 1}^{10} {V_k^\mu }  + \delta {\overline Z _e}{\varepsilon _{cab}}{\left( {{p_1} - {p_2}} \right)^\mu }\\
 &= \sum\limits_{k = 1}^{10} {V_k^\mu }  + \frac{1}{{{{\left( {4\pi } \right)}^2}}}\delta {Z_e}{\varepsilon _{cab}}{\left( {{p_1} - {p_2}} \right)^\mu } \\
 \implies  - {\left( {4\pi } \right)^2}{e^{\left( 1 \right)}} &= {V_e} - \delta {\overline Z _e} \this \label{pionpionrhovert1}
\end{align*}
with ${e^{\left( 1 \right)}}$ the finite one loop correction to the coupling. Comparing \eqref{pionpionrhovert1} with \eqref{Ve} we can read off the renormalization constant:
\begin{align}
\delta {Z_e}: = \frac{{{e^3}}}{{2\varepsilon }}
\end{align}

\subsection{Pion-Pion-Higgs Vertex Renormalization}
The one loop vertex corrections \eqref{pionpionhiggsvertfirst} to \eqref{pionpionhiggsvertlast} to the pion-pion-Higgs vertex is summed to be 
\begin{align}
\sum\limits_{k = 1}^{10} {V_k^H}  = \frac{i}{{{{\left( {4\pi } \right)}^2}}}{\delta _{ab}}{V_H}
\end{align}
where $V_H$ is defined to be
\begin{align*}
{V_H} &= \frac{1}{\varepsilon }\left( {16{e^4}\nu  - 16{e^2}\kappa \nu  + 16{\kappa ^2}\nu  + 6\kappa \lambda \nu  + 20\kappa \nu {\lambda _4}} \right) - 8{e^4}\nu  + 16{\kappa ^2}\nu {B_F}\left[ {m,m,{m_H}} \right] + \\
 &+ \left( {16{e^4}\nu  + 3\kappa \lambda \nu } \right){B_F}\left[ {M,p,M} \right] + 3\kappa \lambda \nu {B_F}\left[ {{m_H},p,{m_H}} \right] + 24{\kappa ^2}\lambda {\nu ^3}{C_0}\left[ {m,{p_1},{m_H}, - {p_2},{m_H}} \right]\\
 &+ 64{\kappa ^3}{\nu ^3}{C_0}\left[ {m,{p_2},{m_H},p,m} \right] + 16{e^4}\nu {\gamma _e} + 16{\kappa ^2}\nu {\gamma _e} + 6\kappa \lambda \nu {\gamma _e} + 20\kappa \nu {B_F}\left[ {m,p,m} \right]{\lambda _4} + 20\kappa \nu {\gamma _e}{\lambda _4} + \\
&{e^2}\kappa \nu \Biggl[ - 16{\gamma _e} - 8{B_F}\left[ {m,m,M} \right] + 8{B_F}\left[ {m,p,m} \right] - 16{B_F}\left[ {M,p,M} \right] + \\
&\left( { - 32{m^2} + 8{M^2} + 4{p^2}} \right){C_0}\left[ {m,{p_1},M, - {p_2},M} \right] + \left( { - 32{m^2} + 8{M^2} + 16{p^2}} \right){C_0}\left[ {M,{p_1},m, - {p_2},m} \right] \Biggr] \this \label{Vh}
\end{align*}
The one loop correction to the pion-pion-Higgs verterx with the counter term is 
\begin{align*}
iV_H^{\left( 1 \right)}{\delta _{ab}} &= \sum\limits_{k = 1}^{10} {V_k^\mu }  - i\delta {\overline Z _H}{\delta _{ab}}\\
 &= \frac{i}{{{{\left( {4\pi } \right)}^2}}}{\delta _{ab}}{V_H} - \frac{i}{{{{\left( {4\pi } \right)}^2}}}\delta { Z _H}{\delta _{ab}}\\
 \Rightarrow {\left( {4\pi } \right)^2}V_H^{\left( 1 \right)} &= {V_H} - \delta {Z _H} \this \label{pionpionhiggsvertcond}
\end{align*}
Comparing the expressions \eqref{Vh} with \eqref{pionpionhiggsvertcond} we can read off the renormalization constant
\begin{align}
\delta {Z_H} = \frac{1}{\varepsilon }\left( {16{e^4}\nu  - 16{e^2}\kappa \nu  + 16{\kappa ^2}\nu  + 6\kappa \lambda \nu  + 20\kappa \nu {\lambda _4}} \right)
\end{align}
where $V_H^{\left( 1 \right)}$ is the finite one loop correction to the pio-pion-Higgs coupling.

\subsection{Pion-Pion-Pion-Pion Vertex Renormalization}
The one loop vertex corrections \eqref{phi4first} to \eqref{phi4last} to the pion-pion-pion-pion vertex is summed to be 
\begin{align*}
iV^{\lambda}_{abcd}&:=\sum\limits_{k = 1}^{12} V_{abcd}^{k,\phi}  = \frac{i}{{{{\left( {4\pi } \right)}^2}}}V_{abcd}^{\phi} \\
&= \frac{i}{{{{\left( {4\pi } \right)}^2}}}\biggl[ \left( {40{e^4} + 16{\kappa ^2} + 2{e^2}\lambda 4 + 11\lambda {4^4}} \right)\left( {{\delta _{ab}}{\delta _{cd}} + {\delta _{ac}}{\delta _{bd}} + {\delta _{ad}}{\delta _{bc}}} \right) \biggr]\frac{1}{\varepsilon} +\text{UV finite terms} \this \label{phi4diverg}
\end{align*}
The counter term for this vertex is
\begin{align*}
\overline \Gamma  _{abcd\rm{CT}}&= - i{{\delta {\bar{Z}_{{\phi ^4}}}}}\left( {{\delta _{ab}}{\delta _{cd}} + {\delta _{bc}}{\delta _{da}} + {\delta _{bd}}{\delta _{ac}}} \right) \\
&= - \frac{i}{\left(4\pi\right)^2}{{\delta {Z_{{\phi ^4}}}}}\left( {{\delta _{ab}}{\delta _{cd}} + {\delta _{bc}}{\delta _{da}} + {\delta _{bd}}{\delta _{ac}}} \right) \this \label{phi4counterterm}
\end{align*}
Matching coefficients between \eqref{phi4diverg} and \eqref{phi4counterterm}, we see the value of the counter term is
\begin{align}
\delta Z_{\phi ^4}:=\left( {40{e^4} + 16{\kappa ^2} + 2{e^2}\lambda 4 + 11\lambda {4^4}} \right)\frac{1}{\varepsilon}
\end{align}
We have completed the renormalization of all the 1PI diagrams that can contribute to pion-pion scattering processes. 
%
\chapter{Conclusion and Outlook}
\lettrine[lines=4]{\initfamily\textcolor{darkblue}{W}}{e} set off on a journey to extend the Kroll-Lee-Zumino model which only had as its particle content the charged pions and neutral rho. This was done by using the larger gauge group SU(2). This group allowed for the inclusion of a  larger particle content with the charged and neutral pions and rhos. Though there was a price to be paid in the form of simplicity. Whereas for the Kroll-Lee-Zumino model, the neutral rho mass could be included externally by hand without breaking the U(1) gauge invariance, such a procedure in the SU(2) extension was not possible as it breaks the gauge invariance of the theory. The predictive power of gauge theories were too alluring to sacrifice and this forced us to generate the mass for the rho via Spontaneous Symmetry breaking using the Higgs mechanism while preserving the gauge invariance. 


We went on to calculate the pion-pion scattering amplitudes. The scattering lengths $a$ and $b$ were then computed with values summarized in the table \ref{treeresultstable}, listed below for convenience:

\begin{table}[H]
	\begin{center}
		\begin{tabular}{@{}cccccccc@{}}
			
			\toprule
Lengths&Weinberg$^d$&$\chi$PT$(1^{st}\mathcal{O})^a$&$\chi$PT$(2^{nd}\mathcal{O})^a$&NABKLZ$^\dagger$&Colangelo$^b$&Bijnens$^c$&Exp$^{abc}$\\
			\midrule
			\midrule
$a^0_0$ &$0.20$  	&$0.16$	 &$0.20$				&$0.21$ 					&$0.220$				&$0.219$& $0.220\pm0.005$ \\
$b^0_0$ &			&$0.18$	 &$0.26$				&$0.30$ 					&$0.276$				&$0.279$& $0.25\pm0.03$\\


$a^0_2\times 10^{3}$ & 			&$0$&$2$&$2.06$		& $1.75$	&$2.2$  & $1.7\pm 3$ \\
$b^0_2\times 10^{4}$ &		    &   &   & $-5.23$ 	& $-3.55$   &$-3.2$ &\\
			
			\midrule 
			

$a^1_1$ &			  	&$0.030$ &$0.036$				&$0.0528$				&$0.0379$	&$0.0378$ &$0.038 \pm 0.002 $\\
$b^1_1$ &				&$0$	 &$0.043$ 				&$0.0053$				&$0.0057$	&$0.0059$ &\\
	
			
			\midrule
			
$a^2_0$ & $-0.06$			 &$-0.045$&$-0.041$		&$-0.0456$				&$-0.0444$		&$-0.0420$& $-0.044\pm 0.001$ \\
$b^2_0$ &			 &  	  &  			&$-0.0225$				&$-0.0803$		&$-0.0756$&$-0.082\pm0.008$\\


$a^2_2\times 10^{4}$ & 			  	&$0$ 	&$3.5$ 		& $-2.03$	& $1.70$ 	&$2.90$		&$1.3\pm 3$ \\
$b^2_2\times 10^{4}$ &				&$-8.9$ &$-7$ 	& $-0.53$	& $-3.26$	&$-3.60$	&$-8.2$\\

			\bottomrule
		\end{tabular}
	\end{center}
	\medskip
		\raggedright
		{\footnotesize
		}
	\caption{Summary of predicted values and experimental data of the scattering length.}
\end{table}

We sought to increase the precision with the inclusion of the one loop corrections to the tree scattering lengths. This entailed computing the $\sim 85$ one-particle-irreducible diagrams. The calculations were done using dimensional regularization to parameterize the divergences. These divergences were absorbed into the counter terms using the On Shell renormalization conditions for the self energies and Minimal Subtraction for the vertices. All the pieces required to compute the one loop correction to the tree scattering lengths have been calculated but due to time constraints we could not complete the program of computing the one loop correction to the scattering lengths.

For the future, one possible path lies ahead with:
\begin{itemize}
\item Completing the program with the inclusion of the one loop corrections to the scattering lengths. At this stage we cannot estimate the size of the corrections, since the expansion parameter for the perturbative series is of the form $\left(\dfrac{e}{4\pi}\right)^2 = 0.2< 1$, this could lead to a reasonable perturbative series, though the size of the coefficients to the expansion parameter cannot be commented on without further analysis. 

\item Changing from the isospin basis to the physical basis gives direct access to the rho decay rates.

\item Computing the pion form factor using this quantum field theory as this has been a fruitful path of investigation for the U(1) Kroll-Lee-Zumino model.

\end{itemize}

For further development of the model, one could consider if the $\omega(782)$ could be included to this model? One could follow the path taken in the standard electroweak model and introduce an additional vector field and mix one component of the rho triplet with the new vector field to give the $\rho^0$ and $\omega$. This requires investigation whether this is possible. 
%
%
%
%
\cleardoublepage
\part*{Appendix}
\appendix

\chapter{Scalar Functions}\label{ScalarFunctions}
\numberwithin{equation}{chapter}
The scalar functions used in these calculations are defined below. In $n=4-2\varepsilon$ dimensions
\begin{align}
\Delta &:= \frac{1}{\varepsilon } + {\gamma _e}\\
{\gamma _e} &:=  - {\gamma _E} + \ln \left( {4\pi } \right)
\end{align}

The one point scalar function:
\begin{align}
{A_0}\left( m \right) &= \int {\frac{{{d^n}k}}{{{{\left( {2\pi } \right)}^n}}}\frac{1}{{{k^2} - {m^2}}}} \nonumber \\
 &= \frac{i}{{{{\left( {4\pi } \right)}^2}}}\left[ {{m^2}\Delta  + {A_F}\left( m \right) + O\left( \varepsilon  \right)} \right]
\end{align}
with the finite part defined as
\begin{align}
{A_F}\left( m \right): = {m^2}\left( {\ln \frac{{{m^2}}}{{{\mu ^2}}} + 1} \right)
\end{align}

The two point scalar function:
\begin{align}
{B_0}\left( {{m_0};p,{m_1}} \right)&: = \int {\frac{{{d^n}k}}{{{{\left( {2\pi } \right)}^n}}}\frac{1}{{\left( {{k^2} - m_0^2} \right)\left[ {{{\left( {k + p} \right)}^2} - m_1^2} \right]}}} \nonumber\\
 &= \frac{i}{{{{\left( {4\pi } \right)}^2}}}\left[ {\Delta  + {B_F}\left( {{m_0};p,{m_1}} \right) + O\left( \varepsilon  \right)} \right]
\end{align}
with the finite part
\begin{align}
{B_F}\left( {{m_0};p,{m_1}} \right): = 2 - \ln \frac{{{m_0}{m_1}}}{{{\mu ^2}}} + \frac{{m_0^2 - m_1^2}}{{{p^2}}}\ln \frac{{{m_1}}}{{{m_0}}} - \frac{{{m_0}{m_1}}}{{{p^2}}}\left( {\frac{1}{r} - r} \right)\ln \left( r \right)
\end{align}
the value of $r$ is found from the roots of the polynomial equation
\begin{align}
{x^2} + \frac{{m_0^2 + m_1^2 - {p^2} - i\varepsilon }}{{{m_0}{m_1}}}x + 1 = \left( {x + r} \right)\left( {x + \frac{1}{r}} \right)
\end{align}
with 
\begin{align}
r &= \frac{1}{{2{m_0}{m_1}}}\left( {m_0^2 + m_1^2 - {p^2} - i\varepsilon  \pm \sqrt {{{\left( {m_0^2 + m_1^2 - {p^2} - i\varepsilon } \right)}^2} - 4m_0^2m_1^2} } \right)\\
\frac{1}{r} - r &=  \mp \frac{1}{{{m_0}{m_1}}}\sqrt {{{\left( {m_0^2 + m_1^2 - {p^2} - i\varepsilon } \right)}^2} - 4m_0^2m_1^2} 
\end{align}

The three and four point scalar functions are finite in $4$ dimensions and are defined as
\begin{align}
{C_0}\left( {{m_0},{p_1},{m_1},{p_2},{m_2}} \right): = \int {\frac{{{d^n}k}}{{{{\left( {2\pi } \right)}^n}}}\frac{1}{{\left( {{k^2} - m_0^2} \right)\left[ {{{\left( {k + {p_1}} \right)}^2} - m_1^2} \right]\left[ {{{\left( {k + {p_2}} \right)}^2} - m_2^2} \right]}}} 
\end{align}
and 
\begin{align}
{D_0}\left( {{m_0},{p_1},{m_1},{p_2},{m_2},{p_3},{m_3}} \right): = \int {\frac{{{d^n}k}}{{{{\left( {2\pi } \right)}^n}}}\frac{1}{{\left( {{k^2} - m_0^2} \right)\left[ {{{\left( {k + {p_1}} \right)}^2} - m_1^2} \right]\left[ {{{\left( {k + {p_2}} \right)}^2} - m_2^2} \right]\left[ {{{\left( {k + {p_3}} \right)}^2} - m_3^2} \right]}}} 
\end{align}

%
%
\chapter{Self Energy}
\label{selfenergies}
For completeness, the reduction for the self energies are presented below.
\section{Vector Self Energy}

\subsection{$\phi$ bubble}
\begin{minipage}{\linewidth}
\begin{tikzpicture}[scale=1]
	\Fbubble{0}{2}{\Frho}{\Fphi}{\Fphi}{\Frho};
\end{tikzpicture}
\end{minipage}

\begin{align*}
{\Gamma _1} = e{\varepsilon _{adc}}{\left( {2k + p} \right)^\mu } \qquad {\Gamma _2} = e{\varepsilon _{bef}}{\left( {2k + p} \right)^\nu }
\end{align*}
\begin{align*}
 - i\pi _{ab1}^{\mu \nu } &= \int {\frac{{{d^n}k}}{{{{\left( {2\pi } \right)}^n}}}{\Gamma _1} \cdot {D^{df}}\left( k \right) \cdot {\Gamma _2} \cdot {D^{ce}}\left( {k + p} \right)} \\
 &= \int {\frac{{{d^n}k}}{{{{\left( {2\pi } \right)}^n}}}} \frac{{e{\varepsilon _{adc}}{{\left( {2k + p} \right)}^\mu } \cdot {\delta ^{df}} \cdot e{\varepsilon _{bef}}{{\left( {2k + p} \right)}^\nu }{\delta ^{ce}}}}{{\left( {{k^2} - {m^2}} \right)\left[ {{{\left( {k + p} \right)}^2} - {m^2}} \right]}}\\
 &= {e^2}{\varepsilon _{cda}}{\varepsilon _{cdb}}\int {\frac{{{d^n}k}}{{{{\left( {2\pi } \right)}^n}}}} \frac{{{{\left( {2k + p} \right)}^\mu }{{\left( {2k + p} \right)}^\nu }}}{{\left( {{k^2} - {m^2}} \right)\left[ {{{\left( {k + p} \right)}^2} - {m^2}} \right]}}\\
 &= {e^2}{\varepsilon _{cda}}{\varepsilon _{cdb}}I_1^{\mu \nu }
\end{align*}
Using ${\varepsilon _{cda}}{\varepsilon _{cdb}} = 2{\delta _{ab}}$
\begin{align*}
 - i\pi _{ab1}^{\mu \nu } = 2{e^2}{\delta _{ab}}I_1^{\mu \nu }
\end{align*}

Making some definitions
\begin{align*}
N_1^{\mu \nu }&:= {\left( {2k + p} \right)^\mu }{\left( {2k + p} \right)^\nu }\\
 &= 4{k^\mu }{k^\nu } + 2\left( {{k^\mu }{p^\nu } + {p^\mu }{k^\nu }} \right) + {p^\mu }{p^\nu }\\
{D_1}&:= {k^2} - {m^2}\\
{D_2}&:= {\left( {k + p} \right)^2} - {m^2}\\
 &= {k^2} + 2k \cdot p + {p^2} - {m^2}\\
 &= {D_1} + 2k \cdot p + {p^2}
\end{align*}
So we have the relations
\begin{align*}
{k^2} &= {D_1} + {m^2}\\
2k \cdot p &= {D_2} - {D_1} - {p^2}
\end{align*}
Splitting the integral over the transverse and longitudinal parts
\begin{align*}
I_1^{\mu \nu }&:= \int {\frac{{{d^n}k}}{{{{\left( {2\pi } \right)}^n}}}} \frac{{N_1^{\mu \nu }}}{{{D_1}{D_2}}}\\
 &= {f_{T1}}P_T^{\mu \nu } + {f_{L1}}P_L^{\mu \nu }
\end{align*}
and we can extract the coefficients using the transverse and longitudinal projectors.
\begin{align*}
{P_{L\mu \nu }}I_1^{\mu \nu } &= {f_{L1}}\\
{P_{T\mu \nu }}I_1^{\mu \nu } &= \left( {n - 1} \right){f_{T1}}
\end{align*}

\begin{align*}
{P_{L\mu \nu }}N_1^{\mu \nu } &= \frac{1}{{{p^2}}}{\left( {2k \cdot p + {p^2}} \right)^2}\\
 &= \frac{1}{{{p^2}}}\left( {D_2^2 - 2{D_1}{D_2} + D_1^2} \right)\\
\\
{P_{T\mu \nu }}N_1^{\mu \nu } &= {g_{\mu \nu }}N_1^{\mu \nu } - {P_{L\mu \nu }}N_1^{\mu \nu }\\
{g_{\mu \nu }}N_1^{\mu \nu } &= 4{k^2} + 2k \cdot p + {p^2}\\
 &= 2{D_1} + 2{D_2} + 4{m^2} - {p^2}
\end{align*}

\begin{align*}
{f_{L1}} &= \int {\frac{{{d^n}k}}{{{{\left( {2\pi } \right)}^n}}}} \frac{{{P_{L\mu \nu }}N_1^{\mu \nu }}}{{{D_1}{D_2}}}\\
 &= \frac{1}{{{p^2}}}\int {\frac{{{d^n}k}}{{{{\left( {2\pi } \right)}^n}}}} \frac{{D_2^2 - 2{D_1}{D_2} + D_1^2}}{{{D_1}{D_2}}}\\
 &= \frac{1}{{{p^2}}}\int {\frac{{{d^n}k}}{{{{\left( {2\pi } \right)}^n}}}\left( {\frac{{{D_2}}}{{{D_1}}} + \frac{{{D_1}}}{{{D_2}}} - 2} \right)} \\
 &= \frac{1}{{{p^2}}}\int {\frac{{{d^n}k}}{{{{\left( {2\pi } \right)}^n}}}\left( {\frac{{{D_2}}}{{{D_1}}} + \frac{{{D_1}}}{{{D_2}}}} \right)} 
\end{align*}

\begin{align*}
\int {\frac{{{d^n}k}}{{{{\left( {2\pi } \right)}^n}}}\frac{{{D_1}}}{{{D_2}}}}  = \int {\frac{{{d^n}k}}{{{{\left( {2\pi } \right)}^n}}}\frac{{{k^2} - {m^2}}}{{{{\left( {k + p} \right)}^2} - {m^2}}}} 
\end{align*}
Making a change of variables
${l^\mu } =  - {\left( {k + p} \right)^\mu } \Rightarrow {k^\mu } =  - {\left( {l + p} \right)^\mu } \Rightarrow \int {\dfrac{{{d^n}k}}{{{{\left( {2\pi } \right)}^n}}}}  = \int {\dfrac{{{d^n}l}}{{{{\left( {2\pi } \right)}^n}}}} $

\begin{align*}
\int {\frac{{{d^n}k}}{{{{\left( {2\pi } \right)}^n}}}\frac{{{D_1}}}{{{D_2}}}}  &= \int {\frac{{{d^n}l}}{{{{\left( {2\pi } \right)}^n}}}\frac{{{{\left( {l + p} \right)}^2} - {m^2}}}{{{l^2} - {m^2}}}} \\
 &= \int {\frac{{{d^n}k}}{{{{\left( {2\pi } \right)}^n}}}\frac{{{D_2}}}{{{D_1}}}} \\
 &= \int {\frac{{{d^n}k}}{{{{\left( {2\pi } \right)}^n}}}\left( {1 + \frac{{2k \cdot p}}{{{k^2} - {m^2}}} + \frac{{{p^2}}}{{{k^2} - {m^2}}}} \right)} \\
 &= {p^2}{A_0}\left( m \right)
\end{align*}

\begin{align*}
{f_{L1}} &= \frac{1}{{{p^2}}}\int {\frac{{{d^n}k}}{{{{\left( {2\pi } \right)}^n}}}\left( {\frac{{{D_2}}}{{{D_1}}} + \frac{{{D_1}}}{{{D_2}}}} \right)} \\
 &= 2{A_0}\left( m \right)
\end{align*}

\begin{align*}
\left( {n - 1} \right){f_{T1}} &= {P_{T\mu \nu }}I_1^{\mu \nu }\\
 &= \int {\frac{{{d^n}k}}{{{{\left( {2\pi } \right)}^n}}}} \frac{{{P_{T\mu \nu }}N_1^{\mu \nu }}}{{{D_1}{D_2}}}\\
 &= \int {\frac{{{d^n}k}}{{{{\left( {2\pi } \right)}^n}}}} \left( {\frac{{{g_{\mu \nu }}N_1^{\mu \nu }}}{{{D_1}{D_2}}} - \frac{{{P_{L\mu \nu }}N_1^{\mu \nu }}}{{{D_1}{D_2}}}} \right)\\
 &= \int {\frac{{{d^n}k}}{{{{\left( {2\pi } \right)}^n}}}} \frac{{2{D_1} + 2{D_2} + 4{m^2} - {p^2}}}{{{D_1}{D_2}}} - \int {\frac{{{d^n}k}}{{{{\left( {2\pi } \right)}^n}}}} \frac{{{P_{L\mu \nu }}N_1^{\mu \nu }}}{{{D_1}{D_2}}}\\
 &= \int {\frac{{{d^n}k}}{{{{\left( {2\pi } \right)}^n}}}} \left[ {\frac{2}{{{D_1}}} + \frac{2}{{{D_2}}} + \frac{{4{m^2} - {p^2}}}{{{D_1}{D_2}}}} \right] - {f_{L1}}\\
 &= 4{A_0}\left( m \right) + \left( {4{m^2} - {p^2}} \right){B_0}\left( {m;p,m} \right) - 2{A_0}\left( m \right)\\
 &= 2{A_0}\left( m \right) + \left( {4{m^2} - {p^2}} \right){B_0}\left( {m;p,m} \right)\\\\
\implies{f_{T1}} &= \frac{1}{{n - 1}}\left[ {2{A_0}\left( m \right) + \left( {4{m^2} - {p^2}} \right){B_0}\left( {m;p,m} \right)} \right]
\end{align*}

\begin{align*}
I_1^{\mu \nu } &= \frac{1}{{n - 1}}\left[ {2{A_0}\left( m \right) + \left( {4{m^2} - {p^2}} \right){B_0}\left( {m;p,m} \right)} \right]P_T^{\mu \nu } + 2{A_0}\left( m \right)P_L^{\mu \nu }
\end{align*}
\begin{align*}
 - i\pi _{ab1}^{\mu \nu } &= 2{e^2}{\delta _{ab}}I_1^{\mu \nu }\\
 &= \frac{{2{e^2}{\delta _{ab}}}}{{n - 1}}\left[ {2{A_0}\left( m \right) + \left( {4{m^2} - {p^2}} \right){B_0}\left( {m;p,m} \right)} \right]P_T^{\mu \nu } + 4{e^2}{\delta _{ab}}{A_0}\left( m \right)P_L^{\mu \nu }
\end{align*}

\subsection{$\phi$ tadpole}
\begin{minipage}{\linewidth}
\begin{tikzpicture}[scale=1]
	\Ftadpole{4.5}{2}{\Frho}{\Fphi};
\end{tikzpicture}
\end{minipage}
\begin{align*}
\Gamma  = i{e^2}{g^{\mu \nu }}\left( {{\varepsilon _{jac}}{\varepsilon _{jbd}} + {\varepsilon _{jad}}{\varepsilon _{jbc}}} \right)
\end{align*}
\begin{align*}
 - i\pi _{ab2}^{\mu \nu } &= \int {\frac{{{d^n}k}}{{{{\left( {2\pi } \right)}^n}}}\Gamma  \cdot {D^{cd}}\left( k \right)} \\
 &= \int {\frac{{{d^n}k}}{{{{\left( {2\pi } \right)}^n}}}i{e^2}{g^{\mu \nu }}\left( {{\varepsilon _{jac}}{\varepsilon _{jbd}} + {\varepsilon _{jad}}{\varepsilon _{jbc}}} \right) \cdot \frac{{i{\delta ^{cd}}}}{{{k^2} - {m^2}}}} \\
 &=  - {e^2}{g^{\mu \nu }}\left( {{\varepsilon _{jac}}{\varepsilon _{jbc}} + {\varepsilon _{jac}}{\varepsilon _{jbc}}} \right)\int {\frac{{{d^n}k}}{{{{\left( {2\pi } \right)}^n}}}\frac{1}{{{k^2} - {m^2}}}} \\
 &=  - 2{e^2}{g^{\mu \nu }}{\varepsilon _{jac}}{\varepsilon _{jbc}}{A_0}\left( m \right)\\
 &=  - 4{e^2}{\delta _{ab}}{g^{\mu \nu }}{A_0}\left( m \right)\\
 &=  - 4{e^2}{\delta _{ab}}\left( {P_T^{\mu \nu } + P_L^{\mu \nu }} \right){A_0}\left( m \right)\\
 &=  - 4{e^2}{\delta _{ab}}{A_0}\left( m \right)P_T^{\mu \nu } - 4{e^2}{\delta _{ab}}{A_0}\left( m \right)P_L^{\mu \nu }
\end{align*}

\subsection{H-$\rho$ bubble}
\begin{minipage}{\linewidth}
\begin{tikzpicture}[scale=1]
	\Fbubble{9}{-1}{\Frho}{\Frho}{\Fhiggs}{\Frho};
\end{tikzpicture}
\end{minipage}

\begin{align*}
{\Gamma _1} &= i{e^2}\nu {\delta ^{ac}}{g^{\alpha \mu }} \qquad {\Gamma _2} = i{e^2}\nu {\delta ^{bd}}{g^{\beta \nu }}\\
D_{\alpha \beta }^{cd}\left( k \right) &= \frac{{ - i{\delta _{cd}}{g_{\alpha \beta }}}}{{{k^2} - {M^2}}} \qquad D\left( {k + p} \right) = \frac{i}{{{{\left( {k + p} \right)}^2} - m_H^2}}
\end{align*}

\begin{align*}
 - i\pi _{ab3}^{\mu \nu } &= \int {\frac{{{d^n}k}}{{{{\left( {2\pi } \right)}^n}}}{\Gamma _1} \cdot D_{\alpha \beta }^{cd}\left( k \right) \cdot {\Gamma _2} \cdot D\left( {k + p} \right)} \\
 &=  - {\left( {{e^2}\nu } \right)^2}{\delta _{ab}}{g^{\mu \nu }}\int {\frac{{{d^n}k}}{{{{\left( {2\pi } \right)}^n}}}\frac{1}{{\left( {{k^2} - {M^2}} \right)\left[ {{{\left( {k + p} \right)}^2} - m_H^2} \right]}}} \\
 &=  - {e^2}{\left( {e\nu } \right)^2}{\delta _{ab}}\left( {P_T^{\mu \nu } + P_L^{\mu \nu }} \right){B_0}\left( {M;p,{m_H}} \right)\\
 &=  - {e^2}{\left( {e\nu } \right)^2}{\delta _{ab}}{B_0}\left( {M;p,{m_H}} \right)P_T^{\mu \nu } - {e^2}{\left( {e\nu } \right)^2}{\delta _{ab}}{B_0}\left( {M;p,{m_H}} \right)P_L^{\mu \nu }
\end{align*}
Recall the definition $M: = e\nu $
\begin{align*}
 - i\pi _{ab3}^{\mu \nu } =  - {e^2}{M^2}{\delta _{ab}}{B_0}\left( {M;p,{m_H}} \right)P_T^{\mu \nu } - {e^2}{M^2}{\delta _{ab}}{B_0}\left( {M;p,{m_H}} \right)P_L^{\mu \nu }
\end{align*}

\subsection{H tadpole}
\begin{minipage}{\linewidth}
\begin{tikzpicture}[scale=1]
	\Ftadpole{13.5}{-1}{\Frho}{\Fhiggs};
\end{tikzpicture}
\end{minipage}
\begin{align*}
\Gamma  = \frac{1}{2}i{e^2}{\delta _{ab}}{g^{\mu \nu }} \qquad D\left( k \right) = \frac{i}{{{k^2} - m_H^2}}
\end{align*}
\begin{align*}
 - i\pi _{ab4}^{\mu \nu } &= \int {\frac{{{d^n}k}}{{{{\left( {2\pi } \right)}^n}}}\Gamma  \cdot D\left( k \right)} \\
 &=  - \frac{1}{2}{e^2}{\delta _{ab}}{g^{\mu \nu }}\int {\frac{{{d^n}k}}{{{{\left( {2\pi } \right)}^n}}}\frac{1}{{{k^2} - m_H^2}}} \\
 &=  - \frac{1}{2}{e^2}{\delta _{ab}}\left( {P_T^{\mu \nu } + P_L^{\mu \nu }} \right){A_0}\left( {{m_H}} \right)\\
&=  - \frac{1}{2}{e^2}{\delta _{ab}}{A_0}\left( {{m_H}} \right)P_T^{\mu \nu } - \frac{1}{2}{e^2}{\delta _{ab}}{A_0}\left( {{m_H}} \right)P_L^{\mu \nu }
\end{align*}

\subsection{H-Goldstone bubble}
\begin{minipage}{\linewidth}
\begin{tikzpicture}[scale=1]
	\Fbubble{9}{-1}{\Frho}{\Fchi}{\Fhiggs}{\Frho};
\end{tikzpicture}
\end{minipage}
\begin{align*}
{\Gamma _1} &= \frac{1}{2}e{\delta _{ac}}{\left( {2k + p} \right)^\mu }\qquad {\Gamma _2} = \frac{1}{2}e{\delta _{bd}}{\left( {2k + p} \right)^\nu }\\
{D^{cd}}\left( k \right) &= \frac{{i{\delta ^{cd}}}}{{{k^2} - {M^2}}}\qquad D\left( {k + p} \right) = \frac{i}{{{{\left( {k + p} \right)}^2} - m_H^2}}
\end{align*}
\begin{align*}
 - i\pi _{ab5}^{\mu \nu } &= \int {\frac{{{d^n}k}}{{{{\left( {2\pi } \right)}^n}}}{\Gamma _1} \cdot {D^{cd}}\left( k \right) \cdot {\Gamma _2} \cdot D\left( {k + p} \right)} \\
 &=  - \frac{1}{4}{e^2}{\delta _{ab}}\int {\frac{{{d^n}k}}{{{{\left( {2\pi } \right)}^n}}}\frac{{{{\left( {2k + p} \right)}^\mu }{{\left( {2k + p} \right)}^\nu }}}{{\left( {{k^2} - {M^2}} \right)\left[ {{{\left( {k + p} \right)}^2} - m_H^2} \right]}}} \\
 &=  - \frac{1}{4}{e^2}{\delta _{ab}}I_5^{\mu \nu }
\end{align*}
Some definitions
\begin{align*}
N_5^{\mu \nu }&: = {\left( {2k + p} \right)^\mu }{\left( {2k + p} \right)^\nu }\\
 &= 4{k^\mu }{k^\nu } + 2\left( {{k^\mu }{p^\nu } + {p^\mu }{k^\nu }} \right) + {p^\mu }{p^\nu }\\\\
{D_1}&: = {k^2} - {M^2}\\
{D_2}&: = {\left( {k + p} \right)^2} - m_H^2\\
 &= {D_1} + 2k \cdot p + {M^2} + {p^2} - m_H^2
\end{align*}
Now we have the relations
\begin{align*}
{k^2} &= {D_1} + {M^2}\\
2k \cdot p &= {D_2} - {D_1} + m_H^2 - {M^2} - {p^2}
\end{align*}
\begin{align*}
I_5^{\mu \nu }&: = \int {\frac{{{d^n}k}}{{{{\left( {2\pi } \right)}^n}}}} \frac{{N_5^{\mu \nu }}}{{{D_1}{D_2}}}\\
 &= {f_{T5}}P_T^{\mu \nu } + {f_{L5}}P_L^{\mu \nu }
\end{align*}

\begin{align*}
{P_{L\mu \nu }}N_5^{\mu \nu } &= \frac{1}{{{p^2}}}{\left( {2k \cdot p + {p^2}} \right)^2}\\
 &= \frac{1}{{{p^2}}}\left[ {D_2^2 - 2{D_1}{D_2} + D_1^2 + 2{M^2}\left( {{D_1} - {D_2}} \right) + 2m_H^2\left( {{D_2} - {D_1}} \right) + {{\left( {m_H^2 - {M^2}} \right)}^2}} \right]\\
 &= \frac{1}{{{p^2}}}\left[ {D_2^2 - 2{D_1}{D_2} + D_1^2 + 2\left( {m_H^2 - {M^2}} \right)\left( {{D_2} - {D_1}} \right) + {{\left( {m_H^2 - {M^2}} \right)}^2}} \right]\\\\
{P_{T\mu \nu }}N_5^{\mu \nu } &= {g_{\mu \nu }}N_5^{\mu \nu } - {P_{L\mu \nu }}N_5^{\mu \nu }\\\\
{g_{\mu \nu }}N_5^{\mu \nu } &= 4{k^2} + 2k \cdot p + {p^2}\\
 &= 2{D_1} + 2{D_2} + 2m_H^2 + 2{M^2} - {p^2}
\end{align*}

\begin{align*}
{f_{L5}} &= \int {\frac{{{d^n}k}}{{{{\left( {2\pi } \right)}^n}}}} \frac{{{P_{L\mu \nu }}N_5^{\mu \nu }}}{{{D_1}{D_2}}}\\
 &= \frac{1}{{{p^2}}}\int {\frac{{{d^n}k}}{{{{\left( {2\pi } \right)}^n}}}} \frac{{D_2^2 - 2{D_1}{D_2} + D_1^2 + 2\left( {m_H^2 - {M^2}} \right)\left( {{D_2} - {D_1}} \right) + {{\left( {m_H^2 - {M^2}} \right)}^2}}}{{{D_1}{D_2}}}\\
 &= \frac{1}{{{p^2}}}\int {\frac{{{d^n}k}}{{{{\left( {2\pi } \right)}^n}}}\left[ {\frac{{{D_2}}}{{{D_1}}} + \frac{{{D_1}}}{{{D_2}}} - 2 + 2\left( {m_H^2 - {M^2}} \right)\left( {\frac{1}{{{D_1}}} - \frac{1}{{{D_2}}}} \right) + \frac{{{{\left( {m_H^2 - {M^2}} \right)}^2}}}{{{D_1}{D_2}}}} \right]} \\
 &= \frac{1}{{{p^2}}}\int {\frac{{{d^n}k}}{{{{\left( {2\pi } \right)}^n}}}\left[ {\frac{{{D_2}}}{{{D_1}}} + \frac{{{D_1}}}{{{D_2}}} + 2\left( {m_H^2 - {M^2}} \right)\left( {\frac{1}{{{D_1}}} - \frac{1}{{{D_2}}}} \right) + \frac{{{{\left( {m_H^2 - {M^2}} \right)}^2}}}{{{D_1}{D_2}}}} \right]} 
\end{align*}

\begin{align*}
\int {\frac{{{d^n}k}}{{{{\left( {2\pi } \right)}^n}}}\frac{{{D_1}}}{{{D_2}}}}  = \int {\frac{{{d^n}k}}{{{{\left( {2\pi } \right)}^n}}}\frac{{{k^2} - {M^2}}}{{{{\left( {k + p} \right)}^2} - m_H^2}}} 
\end{align*}
We make a change of variables ${l^\mu } =  - {\left( {k + p} \right)^\mu } \Rightarrow {k^\mu } =  - {\left( {l + p} \right)^\mu }\Rightarrow \int {\dfrac{{{d^n}k}}{{{{\left( {2\pi } \right)}^n}}}}  = \int {\dfrac{{{d^n}l}}{{{{\left( {2\pi } \right)}^n}}}} $
\begin{align*}
\int {\frac{{{d^n}k}}{{{{\left( {2\pi } \right)}^n}}}\frac{{{D_1}}}{{{D_2}}}} &= \int {\frac{{{d^n}l}}{{{{\left( {2\pi } \right)}^n}}}\frac{{{{\left( {l + p} \right)}^2} - {M^2}}}{{{l^2} - m_H^2}}} \\
 &= \int {\frac{{{d^n}k}}{{{{\left( {2\pi } \right)}^n}}}\frac{{{k^2} + 2k \cdot p + {p^2} - {M^2}}}{{{k^2} - m_H^2}}} \\
 &= \int {\frac{{{d^n}k}}{{{{\left( {2\pi } \right)}^n}}}\frac{{{k^2} - m_H^2 + 2k \cdot p + m_H^2 - {M^2} + {p^2}}}{{{k^2} - m_H^2}}} \\
 &= \int {\frac{{{d^n}k}}{{{{\left( {2\pi } \right)}^n}}}\left[ {1 + \frac{{2k \cdot p}}{{{k^2} - m_H^2}} + \frac{{m_H^2 - {M^2} + {p^2}}}{{{k^2} - m_H^2}}} \right]} \\
 &= \left( {m_H^2 - {M^2} + {p^2}} \right){A_0}\left( {{m_H}} \right)
\end{align*}

\begin{align*}
\int {\frac{{{d^n}k}}{{{{\left( {2\pi } \right)}^n}}}\frac{{{D_2}}}{{{D_1}}}} &= \int {\frac{{{d^n}k}}{{{{\left( {2\pi } \right)}^n}}}\frac{{{{\left( {k + p} \right)}^2} - m_H^2}}{{{k^2} - {M^2}}}} \\
 &= \int {\frac{{{d^n}k}}{{{{\left( {2\pi } \right)}^n}}}\frac{{{k^2} + 2k \cdot p + {p^2} - m_H^2}}{{{k^2} - {M^2}}}} \\
 &= \int {\frac{{{d^n}k}}{{{{\left( {2\pi } \right)}^n}}}\frac{{{k^2} - {M^2} + 2k \cdot p - m_H^2 + {M^2} + {p^2}}}{{{k^2} - {M^2}}}} \\
 &= \int {\frac{{{d^n}k}}{{{{\left( {2\pi } \right)}^n}}}\left[ {1 + \frac{{2k \cdot p}}{{{k^2} - {M^2}}} + \frac{{ - m_H^2 + {M^2} + {p^2}}}{{{k^2} - {M^2}}}} \right]} \\
 &= \left( { - m_H^2 + {M^2} + {p^2}} \right){A_0}\left( M \right)
\end{align*}

\begin{align*}
{f_{L5}} &= \frac{1}{{{p^2}}}\int {\frac{{{d^n}k}}{{{{\left( {2\pi } \right)}^n}}}\left[ {\frac{{{D_2}}}{{{D_1}}} + \frac{{{D_1}}}{{{D_2}}} + 2\left( {m_H^2 - {M^2}} \right)\left( {\frac{1}{{{D_1}}} - \frac{1}{{{D_2}}}} \right) + \frac{{{{\left( {m_H^2 - {M^2}} \right)}^2}}}{{{D_1}{D_2}}}} \right]} \\
 &= \frac{1}{{{p^2}}} \Bigg[ \left( { - m_H^2 + {M^2} + {p^2}} \right){A_0}\left( M \right) + \left( {m_H^2 - {M^2} + {p^2}} \right){A_0}\left( {{m_H}} \right) + 2\left( {m_H^2 - {M^2}} \right)\left( {{A_0}\left( M \right) - {A_0}\left( {{m_H}} \right)} \right) +\\ &+ {{\left( {m_H^2 - {M^2}} \right)}^2}{B_0}\left( {M;p,{m_H}} \right) \Bigg]\\
 &= \frac{1}{{{p^2}}}\Bigg[ \left( {m_H^2 - {M^2}} \right)\left( {{A_0}\left( {{m_H}} \right) - {A_0}\left( M \right)} \right) + {p^2}\left( {{A_0}\left( M \right) + {A_0}\left( {{m_H}} \right)} \right) + 2\left( {m_H^2 - {M^2}} \right)\left( {{A_0}\left( M \right) - {A_0}\left( {{m_H}} \right)} \right) +\\ &+ {{\left( {m_H^2 - {M^2}} \right)}^2}{B_0}\left( {M;p,{m_H}} \right) \Bigg]\\
 &= \frac{1}{{{p^2}}}\left[ { - \left( {m_H^2 - {M^2}} \right)\left( {{A_0}\left( {{m_H}} \right) - {A_0}\left( M \right)} \right) + {p^2}\left( {{A_0}\left( M \right) + {A_0}\left( {{m_H}} \right)} \right) + {{\left( {m_H^2 - {M^2}} \right)}^2}{B_0}\left( {M;p,{m_H}} \right)} \right]\\
 &= {A_0}\left( M \right) + {A_0}\left( {{m_H}} \right) + \frac{{m_H^2 - {M^2}}}{{{p^2}}}\left[ { - {A_0}\left( {{m_H}} \right) + {A_0}\left( M \right) + \left( {m_H^2 - {M^2}} \right){B_0}\left( {M;p,{m_H}} \right)} \right]
\end{align*}

\begin{align*}
\left( {n - 1} \right){f_{T5}} &= {P_{T\mu \nu }}I_5^{\mu \nu }\\
 &= \int {\frac{{{d^n}k}}{{{{\left( {2\pi } \right)}^n}}}} \frac{{{P_{T\mu \nu }}N_5^{\mu \nu }}}{{{D_1}{D_2}}}\\
 &= \int {\frac{{{d^n}k}}{{{{\left( {2\pi } \right)}^n}}}} \left( {\frac{{{g_{\mu \nu }}N_5^{\mu \nu }}}{{{D_1}{D_2}}} - \frac{{{P_{L\mu \nu }}N_5^{\mu \nu }}}{{{D_1}{D_2}}}} \right)\\
 &= \int {\frac{{{d^n}k}}{{{{\left( {2\pi } \right)}^n}}}} \frac{{2{D_1} + 2{D_2} + 2m_H^2 + 2{M^2} - {p^2}}}{{{D_1}{D_2}}} - \int {\frac{{{d^n}k}}{{{{\left( {2\pi } \right)}^n}}}} \frac{{{P_{L\mu \nu }}N_5^{\mu \nu }}}{{{D_1}{D_2}}}\\
 &= \int {\frac{{{d^n}k}}{{{{\left( {2\pi } \right)}^n}}}} \left[ {\frac{2}{{{D_1}}} + \frac{2}{{{D_2}}} + \frac{{2m_H^2 + 2{M^2} - {p^2}}}{{{D_1}{D_2}}}} \right] - {f_{L5}}\\
 &= 2{A_0}\left( M \right) + 2{A_0}\left( {{m_H}} \right) + \left( {2m_H^2 + 2{M^2} - {p^2}} \right){B_0}\left( {M;p,{m_H}} \right) - {A_0}\left( M \right) - {A_0}\left( {{m_H}} \right) +\\ &- \frac{{m_H^2 - {M^2}}}{{{p^2}}}\left[ { - {A_0}\left( {{m_H}} \right) + {A_0}\left( M \right) + \left( {m_H^2 - {M^2}} \right){B_0}\left( {M;p,{m_H}} \right)} \right]\\
 &= {A_0}\left( M \right) + {A_0}\left( {{m_H}} \right) + \left( {2m_H^2 + 2{M^2} - {p^2}} \right){B_0}\left( {M;p,{m_H}} \right) +\\ &- \frac{{m_H^2 - {M^2}}}{{{p^2}}}\left[ { - {A_0}\left( {{m_H}} \right) + {A_0}\left( M \right) + \left( {m_H^2 - {M^2}} \right){B_0}\left( {M;p,{m_H}} \right)} \right]
\end{align*}

\begin{align*}
I_5^{\mu \nu } &= {f_{T5}}P_T^{\mu \nu } + {f_{L5}}P_L^{\mu \nu }\\
 &= \frac{1}{{n - 1}}\Bigg\{ {A_0}\left( M \right) + {A_0}\left( {{m_H}} \right) + \left( {2m_H^2 + 2{M^2} - {p^2}} \right){B_0}\left( {M;p,{m_H}} \right)+ \\ &- \frac{{m_H^2 - {M^2}}}{{{p^2}}}\left[ { - {A_0}\left( {{m_H}} \right) + {A_0}\left( M \right) + \left( {m_H^2 - {M^2}} \right){B_0}\left( {M;p,{m_H}} \right)} \right] \Bigg\} P_T^{\mu \nu } + \\
 &+ \left\{ {{A_0}\left( M \right) + {A_0}\left( {{m_H}} \right) + \frac{{m_H^2 - {M^2}}}{{{p^2}}}\left[ { - {A_0}\left( {{m_H}} \right) + {A_0}\left( M \right) + \left( {m_H^2 - {M^2}} \right){B_0}\left( {M;p,{m_H}} \right)} \right]} \right\}P_L^{\mu \nu }
\end{align*}

\begin{align*}
 - i\pi _{ab5}^{\mu \nu } &=  - \frac{1}{4}{e^2}{\delta _{ab}}I_5^{\mu \nu }\\
 &=  - \frac{{{e^2}{\delta _{ab}}}}{{4\left( {n - 1} \right)}} \bigg\{ {A_0}\left( M \right) + {A_0}\left( {{m_H}} \right) + \left( {2m_H^2 + 2{M^2} - {p^2}} \right){B_0}\left( {M;p,{m_H}} \right) +\\ &- \frac{{m_H^2 - {M^2}}}{{{p^2}}}\left[ {{A_0}\left( M \right) - {A_0}\left( {{m_H}} \right) + \left( {m_H^2 - {M^2}} \right){B_0}\left( {M;p,{m_H}} \right)} \right] \bigg\} P_T^{\mu \nu } + \\
 &- \frac{1}{4}{e^2}{\delta _{ab}}\left\{ {{A_0}\left( M \right) + {A_0}\left( {{m_H}} \right) + \frac{{m_H^2 - {M^2}}}{{{p^2}}}\left[ { - {A_0}\left( {{m_H}} \right) + {A_0}\left( M \right) + \left( {m_H^2 - {M^2}} \right){B_0}\left( {M;p,{m_H}} \right)} \right]} \right\}P_L^{\mu \nu }
\end{align*}

\subsection{Goldstone bubble}
\begin{minipage}{\linewidth}
\begin{tikzpicture}[scale=1]
	\Fbubble{9}{-1}{\Frho}{\Fchi}{\Fchi}{\Frho};
\end{tikzpicture}
\end{minipage}
\begin{align*}
{\Gamma _1} &= \frac{1}{2}e{\varepsilon _{adc}}{\left( {2k + p} \right)^\mu } \qquad {\Gamma _2} = \frac{1}{2}e{\varepsilon _{bfg}}{\left( {2k + p} \right)^\nu }\\
{D^{dg}}\left( k \right) &= \frac{{i{\delta ^{dg}}}}{{{k^2} - {M^2}}} \qquad {D^{cf}}\left( {k + p} \right) = \frac{{i{\delta ^{cf}}}}{{{{\left( {k + p} \right)}^2} - {M^2}}}
\end{align*}

\begin{align*}
 - i\pi _{ab8}^{\mu \nu } &= \int {\frac{{{d^n}k}}{{{{\left( {2\pi } \right)}^n}}}{\Gamma _1} \cdot {D^{dg}}\left( k \right) \cdot {\Gamma _2} \cdot {D^{cf}}\left( {k + p} \right)} \\
 &=  - \frac{1}{4}{e^2}{\varepsilon _{adc}}{\varepsilon _{bcd}}\int {\frac{{{d^n}k}}{{{{\left( {2\pi } \right)}^n}}}\frac{{{{\left( {2k + p} \right)}^\mu }{{\left( {2k + p} \right)}^\nu }}}{{\left( {{k^2} - {M^2}} \right)\left[ {{{\left( {k + p} \right)}^2} - {M^2}} \right]}}} \\
 &=   \frac{1}{2}{e^2}{\delta _{ab}}\int {\frac{{{d^n}k}}{{{{\left( {2\pi } \right)}^n}}}\frac{{{{\left( {k + p} \right)}^\mu }{{\left( {k + p} \right)}^\nu }}}{{\left( {{k^2} - {M^2}} \right)\left[ {{{\left( {k + p} \right)}^2} - {M^2}} \right]}}} \\
 &=  \frac{1}{2}{e^2}{\delta _{ab}}I_8^{\mu \nu }
\end{align*}
Making some definitions
\begin{align*}
N_8^{\mu \nu }&:= {\left( {2k + p} \right)^\mu }{\left( {2k + p} \right)^\nu }\\
 &= 4{k^\mu }{k^\nu } + 2\left( {{k^\mu }{p^\nu } + {p^\mu }{k^\nu }} \right) + {p^\mu }{p^\nu }\\
{D_1}&:= {k^2} - {M^2}\\
{D_2}&:= {\left( {k + p} \right)^2} - {M^2}\\
 &= {k^2} + 2k \cdot p + {p^2} - {M^2}\\
 &= {D_1} + 2k \cdot p + {p^2}
\end{align*}
We have the relations
\begin{align*}
{k^2} &= {D_1} + {M^2}\\
2k \cdot p &= {D_2} - {D_1} - {p^2}
\end{align*}
\begin{align*}
I_8^{\mu \nu }&:= \int {\frac{{{d^n}k}}{{{{\left( {2\pi } \right)}^n}}}} \frac{{N_8^{\mu \nu }}}{{{D_1}{D_2}}}\\
 &= {f_{T8}}P_T^{\mu \nu } + {f_{L8}}P_L^{\mu \nu }
\end{align*}
Looking at $N_8 ^{\mu \nu}, D_1,D_2$ we see the this is the same integral as in the case of the $I_1^{\mu \nu }$ but with the replacement of $m \to M$.
\begin{align*}
{f_{T8}} &= {f_{T1}}\\
{f_{L8}} &= {f_{L1}}\\
I_8^{\mu \nu } &= \frac{1}{{n - 1}}\left[ {2{A_0}\left( M \right) + \left( {4{M^2} - {p^2}} \right){B_0}\left( {M;p,M} \right)} \right]P_T^{\mu \nu } + 2{A_0}\left( M \right)P_L^{\mu \nu }
\end{align*}

\begin{align*}
 - i\pi _{ab8}^{\mu \nu } &=   \frac{1}{2}{e^2}{\delta _{ab}}I_8^{\mu \nu }\\
 &=   \frac{{{e^2}{\delta _{ab}}}}{{2\left( {n - 1} \right)}}\left[ {2{A_0}\left( M \right) + \left( {4{M^2} - {p^2}} \right){B_0}\left( {M;p,M} \right)} \right]P_T^{\mu \nu } + {e^2}{\delta _{ab}}{A_0}\left( M \right)P_L^{\mu \nu }
\end{align*}

\subsection{Goldstone tadpole}
\begin{minipage}{\linewidth}
\begin{tikzpicture}[scale=1]
	\Ftadpole{13.5}{-1}{\Frho}{\Fchi};
\end{tikzpicture}
\end{minipage}
\begin{align*}
\Gamma  = \frac{1}{2}i{e^2}{\delta ^{ab}}{\delta ^{cd}}{g^{\mu \nu }} \qquad {D^{cd}}\left( k \right) = \frac{{i{\delta ^{cd}}}}{{{k^2} - {M^2}}}
\end{align*}

\begin{align*}
 - i\pi _{ab6}^{\mu \nu } &= \int {\frac{{{d^n}k}}{{{{\left( {2\pi } \right)}^n}}}\Gamma  \cdot {D^{cd}}\left( k \right)} \\
 &=  - \frac{1}{2}{e^2}{\delta ^{ab}}{\delta ^{cc}}{g^{\mu \nu }}\int {\frac{{{d^n}k}}{{{{\left( {2\pi } \right)}^n}}}\frac{1}{{{k^2} - {M^2}}}} \\
 &=  - \frac{3}{2}{e^2}{\delta ^{ab}}\left( {P_T^{\mu \nu } + P_L^{\mu \nu }} \right){A_0}\left( M \right)\\
 &=  - \frac{3}{2}{e^2}{\delta ^{ab}}{A_0}\left( M \right)P_T^{\mu \nu } - \frac{3}{2}{e^2}{\delta ^{ab}}{A_0}\left( M \right)P_L^{\mu \nu }
\end{align*}

\subsection{Ghost bubble}
\begin{minipage}{\linewidth}
\begin{tikzpicture}[scale=1]
	\Fbubble{9}{-1}{\Frho}{\Fgst}{\Fgst}{\Frho};
\end{tikzpicture}
\end{minipage}
\begin{align*}
{\Gamma _1} &=  - e{\varepsilon _{adc}}{\left( {k + p} \right)^\mu } \qquad {\Gamma _2} = e{\varepsilon _{bfg}}{\left( {k + p} \right)^\nu }\\
{D^{dg}}\left( k \right) &= \frac{{i{\delta ^{dg}}}}{{{k^2} - {M^2}}} \qquad {D^{cf}}\left( {k + p} \right) = \frac{{i{\delta ^{cf}}}}{{{{\left( {k + p} \right)}^2} - {M^2}}}
\end{align*}

\begin{align*}
 - i\pi _{ab7}^{\mu \nu } &= \left( { - 1} \right)\int {\frac{{{d^n}k}}{{{{\left( {2\pi } \right)}^n}}}{\Gamma _1} \cdot {D^{dg}}\left( k \right) \cdot {\Gamma _2} \cdot {D^{cf}}\left( {k + p} \right)} \\
 &=  - {e^2}{\varepsilon _{adc}}{\varepsilon _{bcd}}\int {\frac{{{d^n}k}}{{{{\left( {2\pi } \right)}^n}}}\frac{{{{\left( {k + p} \right)}^\mu }{{\left( {k + p} \right)}^\nu }}}{{\left( {{k^2} - {M^2}} \right)\left[ {{{\left( {k + p} \right)}^2} - {M^2}} \right]}}} \\
 &= 2{e^2}{\delta _{ab}}\int {\frac{{{d^n}k}}{{{{\left( {2\pi } \right)}^n}}}\frac{{{{\left( {k + p} \right)}^\mu }{{\left( {k + p} \right)}^\nu }}}{{\left( {{k^2} - {M^2}} \right)\left[ {{{\left( {k + p} \right)}^2} - {M^2}} \right]}}} \\
 &= 2{e^2}{\delta _{ab}}I_7^{\mu \nu }
\end{align*}

Some definitions
\begin{align*}
N_7^{\mu \nu }&:= {\left( {k + p} \right)^\mu }{\left( {k + p} \right)^\nu }\\
 &= {k^\mu }{k^\nu } + \left( {{k^\mu }{p^\nu } + {p^\mu }{k^\nu }} \right) + {p^\mu }{p^\nu }\\
{D_1}&:= {k^2} - {M^2}\\
{D_2}&:= {\left( {k + p} \right)^2} - {M^2}\\
 &= {k^2} + 2k \cdot p + {p^2} - {M^2}\\
 &= {D_1} + 2k \cdot p + {p^2}
\end{align*}
We have the relations
\begin{align*}
{k^2} &= {D_1} + {M^2}\\
2k \cdot p &= {D_2} - {D_1} - {p^2}
\end{align*}
\begin{align*}
I_7^{\mu \nu }&:= \int {\frac{{{d^n}k}}{{{{\left( {2\pi } \right)}^n}}}} \frac{{N_7^{\mu \nu }}}{{{D_1}{D_2}}}\\
 &= {f_{T7}}P_T^{\mu \nu } + {f_{L7}}P_L^{\mu \nu }
\end{align*}

\begin{align*}
{P_{L\mu \nu }}N_7^{\mu \nu } &= \frac{1}{{{p^2}}}\left[ {{{\left( {k \cdot p} \right)}^2} + {p^2}\left( {2k \cdot p} \right) + {p^4}} \right]\\
 &= \frac{1}{{4{p^2}}}\left[ {D_1^2 + D_2^2 - 2{D_1}{D_2} + 2{p^2}\left( {{D_2} - {D_1}} \right) + {p^4}} \right]\\\\
{P_{T\mu \nu }}N_7^{\mu \nu } &= {g_{\mu \nu }}N_7^{\mu \nu } - {P_{L\mu \nu }}N_7^{\mu \nu }\\\\
{g_{\mu \nu }}N_7^{\mu \nu } &= {k^2} + 2k \cdot p + {p^2}\\
 &= {D_2} + {M^2}
\end{align*}

\begin{align*}
{f_{L7}} &= \int {\frac{{{d^n}k}}{{{{\left( {2\pi } \right)}^n}}}} \frac{{{P_{L\mu \nu }}N_7^{\mu \nu }}}{{{D_1}{D_2}}}\\
 &= \frac{1}{{4{p^2}}}\int {\frac{{{d^n}k}}{{{{\left( {2\pi } \right)}^n}}}} \frac{{D_1^2 + D_2^2 - 2{D_1}{D_2} + 2{p^2}\left( {{D_2} - {D_1}} \right) + {p^4}}}{{{D_1}{D_2}}}\\
 &= \frac{1}{{4{p^2}}}\int {\frac{{{d^n}k}}{{{{\left( {2\pi } \right)}^n}}}} \left[ {\frac{{{D_1}}}{{{D_2}}} + \frac{{{D_2}}}{{{D_1}}} - 2 + 2{p^2}\left( {\frac{1}{{{D_1}}} - \frac{1}{{{D_2}}}} \right) + \frac{{{p^4}}}{{{D_1}{D_2}}}} \right]
\end{align*}
\begin{align*}
\int {\frac{{{d^n}k}}{{{{\left( {2\pi } \right)}^n}}}\frac{{{D_1}}}{{{D_2}}}}  = \int {\frac{{{d^n}k}}{{{{\left( {2\pi } \right)}^n}}}\frac{{{k^2} - {M^2}}}{{{{\left( {k + p} \right)}^2} - {M^2}}}} 
\end{align*}
Making a change of variables ${l^\mu } =  - {\left( {k + p} \right)^\mu } \Rightarrow {k^\mu } =  - {\left( {l + p} \right)^\mu } \Rightarrow \int {\dfrac{{{d^n}k}}{{{{\left( {2\pi } \right)}^n}}}}  = \int {\dfrac{{{d^n}l}}{{{{\left( {2\pi } \right)}^n}}}} $
\begin{align*}
\int {\frac{{{d^n}k}}{{{{\left( {2\pi } \right)}^n}}}\frac{{{D_1}}}{{{D_2}}}}  &= \int {\frac{{{d^n}l}}{{{{\left( {2\pi } \right)}^n}}}\frac{{{{\left( {l + p} \right)}^2} - {M^2}}}{{{l^2} - {M^2}}}} \\
 &= \int {\frac{{{d^n}k}}{{{{\left( {2\pi } \right)}^n}}}\frac{{{D_2}}}{{{D_1}}}} \\
 &= \int {\frac{{{d^n}k}}{{{{\left( {2\pi } \right)}^n}}}\left( {1 + \frac{{2k \cdot p}}{{{k^2} - {M^2}}} + \frac{{{p^2}}}{{{k^2} - {M^2}}}} \right)} \\
 &= {p^2}{A_0}\left( M \right)
\end{align*}

\begin{align*}
{f_{L7}} &= \frac{1}{{4{p^2}}}\left[ {2{p^2}{A_0}\left( M \right) + {p^4}{B_0}\left( {M;p,M} \right)} \right]\\
 &= \frac{1}{2}{A_0}\left( M \right) + \frac{1}{4}{p^2}{B_0}\left( {M;p,M} \right)
\end{align*}

\begin{align*}
\left( {n - 1} \right){f_{T7}} &= {P_{T\mu \nu }}I_7^{\mu \nu }\\
 &= \int {\frac{{{d^n}k}}{{{{\left( {2\pi } \right)}^n}}}} \frac{{{P_{T\mu \nu }}N_7^{\mu \nu }}}{{{D_1}{D_2}}}\\
 &= \int {\frac{{{d^n}k}}{{{{\left( {2\pi } \right)}^n}}}} \left( {\frac{{{g_{\mu \nu }}N_7^{\mu \nu }}}{{{D_1}{D_2}}} - \frac{{{P_{L\mu \nu }}N_7^{\mu \nu }}}{{{D_1}{D_2}}}} \right)\\
 &= \int {\frac{{{d^n}k}}{{{{\left( {2\pi } \right)}^n}}}} \frac{{{D_2} + {M^2}}}{{{D_1}{D_2}}} - \int {\frac{{{d^n}k}}{{{{\left( {2\pi } \right)}^n}}}} \frac{{{P_{L\mu \nu }}N_7^{\mu \nu }}}{{{D_1}{D_2}}}\\
 &= {A_0}\left( M \right) + {M^2}{B_0}\left( {M;p,M} \right) - {f_{L7}}\\
 &= {A_0}\left( M \right) + {M^2}{B_0}\left( {M;p,M} \right) - \frac{1}{2}{A_0}\left( M \right) - \frac{1}{4}{p^2}{B_0}\left( {M;p,M} \right)\\
 &= \frac{1}{2}{A_0}\left( M \right) + \left( {{M^2} - \frac{1}{4}{p^2}} \right){B_0}\left( {M;p,M} \right)
\end{align*}
\begin{align*}
I_7^{\mu \nu } &= {f_{T7}}P_T^{\mu \nu } + {f_{L7}}P_L^{\mu \nu }\\
 &= \frac{1}{{n - 1}}\left[ {\frac{1}{2}{A_0}\left( M \right) + \left( {{M^2} - \frac{1}{4}{p^2}} \right){B_0}\left( {M;p,M} \right)} \right]P_T^{\mu \nu } + \left[ {\frac{1}{2}{A_0}\left( M \right) + \frac{1}{4}{p^2}{B_0}\left( {M;p,M} \right)} \right]P_L^{\mu \nu }\\\\
 - i\pi _{ab7}^{\mu \nu } &= 2{e^2}{\delta _{ab}}I_7^{\mu \nu }\\
 &= \frac{{2{e^2}{\delta _{ab}}}}{{n - 1}}\left[ {\frac{1}{2}{A_0}\left( M \right) + \left( {{M^2} - \frac{1}{4}{p^2}} \right){B_0}\left( {M;p,M} \right)} \right]P_T^{\mu \nu } + 2{e^2}{\delta _{ab}}\left[ {\frac{1}{2}{A_0}\left( M \right) + \frac{1}{4}{p^2}{B_0}\left( {M;p,M} \right)} \right]P_L^{\mu \nu }
\end{align*}

\subsection{$\rho$ bubble}
\begin{minipage}{\linewidth}
\begin{tikzpicture}[scale=1]
	\Fbubble{9}{-1}{\Frho}{\Frho}{\Frho}{\Frho};
\end{tikzpicture}
\end{minipage}

\begin{align*}
{\Gamma _1} &= e{\varepsilon _{adc}}\left[ { - {{\left( {2k + p} \right)}^\mu }{g^{\alpha \beta }} + {{\left( {k + 2p} \right)}^\beta }{g^{\mu \alpha }} + {{\left( {k - p} \right)}^\alpha }{g^{\mu \beta }}} \right]\\
{\Gamma _2} &= e{\varepsilon _{fgb}}\left[ { - {{\left( {2k + p} \right)}^\nu }{g^{\kappa \lambda }} + {{\left( {k + 2p} \right)}^\lambda }{g^{\kappa \nu }} + {{\left( {k - p} \right)}^\kappa }{g^{\lambda \nu }}} \right]\\
D_{\beta \lambda }^{dg}\left( k \right) &= \frac{{ - i{\delta _{dg}}{g_{\beta \lambda }}}}{{{k^2} - {M^2}}}\qquad D_{\alpha \kappa }^{cf}\left( {k + p} \right) = \frac{{ - i{\delta _{cf}}{g_{\alpha \kappa }}}}{{{{\left( {k + p} \right)}^2} - {M^2}}}
\end{align*}

\begin{align*}
& - i\pi _{ab9}^{\mu \nu } = \int {\frac{{{d^n}k}}{{{{\left( {2\pi } \right)}^n}}}{\Gamma _1} \cdot D_{\beta \lambda }^{dg}\left( k \right) \cdot {\Gamma _2} \cdot D_{\alpha \kappa }^{cf}\left( {k + p} \right)} \\
 &=  - {e^2}{\varepsilon _{adc}}{\varepsilon _{bcd}}\int \dfrac{{{d^n}k}}{{{{\left( {2\pi } \right)}^n}}} \times\\ &\times\frac{{\left[ { - {{\left( {2k + p} \right)}^\mu }{g^{\alpha \beta }} + {{\left( {k + 2p} \right)}^\beta }{g^{\mu \alpha }} + {{\left( {k - p} \right)}^\alpha }{g^{\mu \beta }}} \right] \cdot {g_{\beta \lambda }}{g_{\alpha \kappa }} \cdot \left[ { - {{\left( {2k + p} \right)}^\nu }{g^{\kappa \lambda }} + {{\left( {k + 2p} \right)}^\lambda }{g^{\kappa \nu }} + {{\left( {k - p} \right)}^\kappa }{g^{\lambda \nu }}} \right]}}{{\left( {{k^2} - {M^2}} \right)\left[ {{{\left( {k + p} \right)}^2} - {M^2}} \right]}} \\
 &=   2{e^2}{\delta _{ab}} \times \\
 &\int {\frac{{{d^n}k}}{{{{\left( {2\pi } \right)}^n}}}\frac{{\left[ { - {{\left( {2k + p} \right)}^\mu }{g^{\alpha \beta }} + {{\left( {k + 2p} \right)}^\beta }{g^{\mu \alpha }} + {{\left( {k - p} \right)}^\alpha }{g^{\mu \beta }}} \right]\left[ { - {{\left( {2k + p} \right)}^\nu }{g_{\alpha \beta }} + {{\left( {k + 2p} \right)}_\beta }g_\alpha ^\nu  + {{\left( {k - p} \right)}_\alpha }g_\beta ^\nu } \right]}}{{\left( {{k^2} - {M^2}} \right)\left[ {{{\left( {k + p} \right)}^2} - {M^2}} \right]}}} \\
 &=   2{e^2}{\delta _{ab}}I_9^{\mu \nu }
\end{align*}
Some definitions
\begin{align*}
N_9^{\mu \nu }&:= \left[ { - {{\left( {2k + p} \right)}^\mu }{g^{\alpha \beta }} + {{\left( {k + 2p} \right)}^\beta }{g^{\mu \alpha }} + {{\left( {k - p} \right)}^\alpha }{g^{\mu \beta }}} \right]\left[ { - {{\left( {2k + p} \right)}^\nu }{g_{\alpha \beta }} + {{\left( {k + 2p} \right)}_\beta }g_\alpha ^\nu  + {{\left( {k - p} \right)}_\alpha }g_\beta ^\nu } \right]\\
 &= \left( {4n - 6} \right){k^\mu }{k^\nu } + \left( {2n - 3} \right)\left( {{k^\mu }{p^\nu } + {p^\mu }{k^\nu }} \right) + \left( {n - 6} \right){p^\mu }{p^\nu } + \left( {2{k^2} + 2k \cdot p + 5{p^2}} \right){g^{\mu \nu }}\\
{D_1}&:= {k^2} - {M^2}\\
{D_2}&:= {\left( {k + p} \right)^2} - {M^2}\\
 &= {k^2} + 2k \cdot p + {p^2} - {M^2}\\
 &= {D_1} + 2k \cdot p + {p^2}
\end{align*}
We have the relations
\begin{align*}
{k^2} &= {D_1} + {M^2}\\
2k \cdot p &= {D_2} - {D_1} - {p^2}
\end{align*}
We can simplify $N^{\mu\nu}_9$ with the above relations
\begin{align*}
N_9^{\mu \nu } = \left( {4n - 6} \right){k^\mu }{k^\nu } + \left( {2n - 3} \right)\left( {{k^\mu }{p^\nu } + {p^\mu }{k^\nu }} \right) + \left( {n - 6} \right){p^\mu }{p^\nu } + \left( {{D_1} + {D_2} + 2{M^2} + 4{p^2}} \right){g^{\mu \nu }}
\end{align*}
\begin{align*}
I_9^{\mu \nu }&:= \int {\frac{{{d^n}k}}{{{{\left( {2\pi } \right)}^n}}}} \frac{{N_9^{\mu \nu }}}{{{D_1}{D_2}}}\\
 &= {f_{T9}}P_T^{\mu \nu } + {f_{L9}}P_L^{\mu \nu }
\end{align*}

\begin{align*}
{P_{L\mu \nu }}N_9^{\mu \nu } &= \frac{1}{{{p^2}}}\left[ {\left( {4n - 6} \right){{\left( {k \cdot p} \right)}^2} + \left( {2n - 3} \right){p^2}\left( {2k \cdot p} \right) + \left( {n - 6} \right){p^4} + \left( {{D_1} + {D_2} + 2{M^2} + 4{p^2}} \right){p^2}} \right]\\
 &= \frac{1}{{{p^2}}}\left[ {\left( {n - \frac{3}{2}} \right)\left( {D_1^2 + D_2^2} \right) + \left( {3 - 2n} \right){D_1}{D_2} + {p^2}\left( {{D_1} + {D_2}} \right) + \frac{1}{2}{p^2}\left( {4{M^2} - {p^2}} \right)} \right]\\\\
{P_{T\mu \nu }}N_9^{\mu \nu } &= {g_{\mu \nu }}N_9^{\mu \nu } - {P_{L\mu \nu }}N_9^{\mu \nu }\\\\
{g_{\mu \nu }}N_9^{\mu \nu } &= \left( {4n - 6} \right){k^2} + \left( {2n - 3} \right)\left( {2k \cdot p} \right) + \left( {n - 6} \right){p^2} + n\left( {{D_1} + {D_2} + 2{M^2} + 4{p^2}} \right)\\
 &= 3\left( {n - 1} \right)\left( {{D_1} + {D_2}} \right) + 2 \cdot 3\left( {n - 1} \right){m^2} + 3\left( {n - 1} \right){p^2}\\
 &= 3\left( {n - 1} \right)\left[ {\left( {{D_1} + {D_2}} \right) + 2{m^2} + {p^2}} \right]
\end{align*}

\begin{align*}
{f_{L9}} &= \int {\frac{{{d^n}k}}{{{{\left( {2\pi } \right)}^n}}}} \frac{{{P_{L\mu \nu }}N_9^{\mu \nu }}}{{{D_1}{D_2}}}\\
 &= \frac{1}{{{p^2}}}\int {\frac{{{d^n}k}}{{{{\left( {2\pi } \right)}^n}}}} \frac{1}{{{D_1}{D_2}}}\left[ {\left( {n - \frac{3}{2}} \right)\left( {D_1^2 + D_2^2} \right) + \left( {3 - 2n} \right){D_1}{D_2} + {p^2}\left( {{D_1} + {D_2}} \right) + \frac{1}{2}{p^2}\left( {4{M^2} - {p^2}} \right)} \right]\\
 &= \frac{1}{{{p^2}}}\int {\frac{{{d^n}k}}{{{{\left( {2\pi } \right)}^n}}}} \left[ {\left( {n - \frac{3}{2}} \right)\left( {\frac{{{D_1}}}{{{D_2}}} + \frac{{{D_2}}}{{{D_1}}}} \right) + \left( {3 - 2n} \right) + {p^2}\left( {\frac{1}{{{D_1}}} + \frac{1}{{{D_2}}}} \right) + \frac{1}{2}{p^2}\left( {4{M^2} - {p^2}} \right)\frac{1}{{{D_1}{D_2}}}} \right]\\
 &= \frac{1}{{{p^2}}}\int {\frac{{{d^n}k}}{{{{\left( {2\pi } \right)}^n}}}} \left[ {\left( {n - \frac{3}{2}} \right)\left( {\frac{{{D_1}}}{{{D_2}}} + \frac{{{D_2}}}{{{D_1}}}} \right) + {p^2}\left( {\frac{1}{{{D_1}}} + \frac{1}{{{D_2}}}} \right) + \frac{1}{2}{p^2}\left( {4{M^2} - {p^2}} \right)\frac{1}{{{D_1}{D_2}}}} \right]
\end{align*}

\begin{align*}
\int {\frac{{{d^n}k}}{{{{\left( {2\pi } \right)}^n}}}\frac{{{D_1}}}{{{D_2}}}}  = \int {\frac{{{d^n}k}}{{{{\left( {2\pi } \right)}^n}}}\frac{{{k^2} - {M^2}}}{{{{\left( {k + p} \right)}^2} - {M^2}}}} 
\end{align*}
Making a change of variables ${l^\mu } =  - {\left( {k + p} \right)^\mu } \Rightarrow {k^\mu } =  - {\left( {l + p} \right)^\mu } \Rightarrow \int {\dfrac{{{d^n}k}}{{{{\left( {2\pi } \right)}^n}}}}  = \int {\dfrac{{{d^n}l}}{{{{\left( {2\pi } \right)}^n}}}} $
\begin{align*}
\int {\frac{{{d^n}k}}{{{{\left( {2\pi } \right)}^n}}}\frac{{{D_1}}}{{{D_2}}}}  &= \int {\frac{{{d^n}l}}{{{{\left( {2\pi } \right)}^n}}}\frac{{{{\left( {l + p} \right)}^2} - {M^2}}}{{{l^2} - {M^2}}}} \\
 &= \int {\frac{{{d^n}k}}{{{{\left( {2\pi } \right)}^n}}}\frac{{{D_2}}}{{{D_1}}}} \\
 &= \int {\frac{{{d^n}k}}{{{{\left( {2\pi } \right)}^n}}}\left( {1 + \frac{{2k \cdot p}}{{{k^2} - {M^2}}} + \frac{{{p^2}}}{{{k^2} - {M^2}}}} \right)} \\
 &= {p^2}{A_0}\left( M \right)
\end{align*}

\begin{align*}
{f_{L9}} &= \frac{1}{{{p^2}}}\int {\frac{{{d^n}k}}{{{{\left( {2\pi } \right)}^n}}}} \left[ {\left( {n - \frac{3}{2}} \right)\left( {\frac{{{D_1}}}{{{D_2}}} + \frac{{{D_2}}}{{{D_1}}}} \right) + {p^2}\left( {\frac{1}{{{D_1}}} + \frac{1}{{{D_2}}}} \right) + \frac{1}{2}{p^2}\left( {4{M^2} - {p^2}} \right)\frac{1}{{{D_1}{D_2}}}} \right]\\
 &= \frac{1}{{{p^2}}}\left[ {2\left( {n - \frac{3}{2}} \right){p^2}{A_0}\left( M \right) + 2{p^2}{A_0}\left( M \right) + \frac{1}{2}{p^2}\left( {4{M^2} - {p^2}} \right)B\left( {M;p,M} \right)} \right]\\
 &= \left( {2n - 1} \right){A_0}\left( M \right) + \frac{1}{2}\left( {4{M^2} - {p^2}} \right)B\left( {M;p,M} \right)
\end{align*}

\begin{align*}
\left( {n - 1} \right){f_{T9}} &= {P_{T\mu \nu }}I_9^{\mu \nu }\\
 &= \int {\frac{{{d^n}k}}{{{{\left( {2\pi } \right)}^n}}}} \frac{{{P_{T\mu \nu }}N_9^{\mu \nu }}}{{{D_1}{D_2}}}\\
 &= \int {\frac{{{d^n}k}}{{{{\left( {2\pi } \right)}^n}}}} \left( {\frac{{{g_{\mu \nu }}N_9^{\mu \nu }}}{{{D_1}{D_2}}} - \frac{{{P_{L\mu \nu }}N_9^{\mu \nu }}}{{{D_1}{D_2}}}} \right)\\
 &= \int {\frac{{{d^n}k}}{{{{\left( {2\pi } \right)}^n}}}} \frac{{3\left( {n - 1} \right)\left[ {\left( {{D_1} + {D_2}} \right) + 2{M^2} + {p^2}} \right]}}{{{D_1}{D_2}}} - \int {\frac{{{d^n}k}}{{{{\left( {2\pi } \right)}^n}}}} \frac{{{P_{L\mu \nu }}N_9^{\mu \nu }}}{{{D_1}{D_2}}}\\
 &= 3\left( {n - 1} \right)\int {\frac{{{d^n}k}}{{{{\left( {2\pi } \right)}^n}}}} \left[ {\frac{1}{{{D_1}}} + \frac{1}{{{D_2}}} + \frac{{2{M^2} + {p^2}}}{{{D_1}{D_2}}}} \right] - {f_{L9}}\\
 &= 3\left( {n - 1} \right)\left[ {2{A_0}\left( M \right) + \left( {2{M^2} + {p^2}} \right)B\left( {M;p,M} \right)} \right] - \left( {2n - 1} \right){A_0}\left( M \right) + \\
 & - \frac{1}{2}\left( {4{M^2} - {p^2}} \right)B\left( {M;p,M} \right)\\
\implies \left( {n - 1} \right){f_{T9}} &= \left( {4n - 5} \right){A_0}\left( M \right) + \frac{1}{2}\left[ 4{\left( {3n - 4} \right){M^2} + \left( {6n - 5} \right){p^2}} \right]B\left( {M;p,M} \right)
\end{align*}

\begin{align*}
I_9^{\mu \nu } &= {f_{T9}}P_T^{\mu \nu } + {f_{L9}}P_L^{\mu \nu }\\
 &= \left\{ {\frac{{4n - 5}}{{n - 1}}{A_0}\left( M \right) + \frac{1}{{2\left( {n - 1} \right)}}\left[ 4{\left( {3n - 4} \right){M^2} + \left( {6n - 5} \right){p^2}} \right]B\left( {M;p,M} \right)} \right\}P_T^{\mu \nu } +\\&+ \left[ {\left( {2n - 1} \right){A_0}\left( M \right) + \frac{1}{2}\left( {4{M^2} - {p^2}} \right)B\left( {M;p,M} \right)} \right]P_L^{\mu \nu }
\end{align*}

\begin{align*}
 - i\pi _{ab9}^{\mu \nu } &=   2{e^2}{\delta _{ab}}I_9^{\mu \nu }\\
 &=   2{e^2}{\delta _{ab}}\left\{ {\frac{{4n - 5}}{{n - 1}}{A_0}\left( M \right) + \frac{1}{{2\left( {n - 1} \right)}}\left[ 4{\left( {3n - 4} \right){M^2} + \left( {6n - 5} \right){p^2}} \right]B\left( {M;p,M} \right)} \right\}P_T^{\mu \nu } + \\
 &+ 2{e^2}{\delta _{ab}}\left[ {\left( {2n - 1} \right){A_0}\left( M \right) + \frac{1}{2}\left( {4{M^2} - {p^2}} \right)B\left( {M;p,M} \right)} \right]P_L^{\mu \nu }
\end{align*}

\subsection{$\rho$ tadpole}
\begin{minipage}{\linewidth}
\begin{tikzpicture}[scale=1]
	\Ftadpole{13.5}{-1}{\Frho}{\Frho};
\end{tikzpicture}
\end{minipage}
\begin{align*}
\Gamma  &=  - i{e^2}\left[ {{\varepsilon _{abe}}{\varepsilon _{cde}}\left( {{g^{\mu \lambda }}{g^{\nu \sigma }} - {g^{\mu \sigma }}{g^{\nu \lambda }}} \right) + {\varepsilon _{ace}}{\varepsilon _{bde}}\left( {{g^{\mu \nu }}{g^{\sigma \lambda }} - {g^{\mu \sigma }}{g^{\nu \lambda }}} \right) + {\varepsilon _{ade}}{\varepsilon _{bce}}\left( {{g^{\mu \nu }}{g^{\sigma \lambda }} - {g^{\mu \lambda }}{g^{\nu \sigma }}} \right)} \right]\\
D_{\sigma \lambda }^{cd}\left( k \right) &= \frac{{ - i{\delta _{cd}}{g_{\sigma \lambda }}}}{{{k^2} - {M^2}}}
\end{align*}

\begin{align*}
 - i\pi _{ab10}^{\mu \nu } &= \int {\frac{{{d^n}k}}{{{{\left( {2\pi } \right)}^n}}}\Gamma  \cdot D_{\sigma \lambda }^{cd}\left( k \right)} \\
 &=  - {e^2}\left[ {{\varepsilon _{abe}}{\varepsilon _{cde}}\left( {{g^{\mu \lambda }}{g^{\nu \sigma }} - {g^{\mu \sigma }}{g^{\nu \lambda }}} \right) + {\varepsilon _{ace}}{\varepsilon _{bde}}\left( {{g^{\mu \nu }}{g^{\sigma \lambda }} - {g^{\mu \sigma }}{g^{\nu \lambda }}} \right) + {\varepsilon _{ade}}{\varepsilon _{bce}}\left( {{g^{\mu \nu }}{g^{\sigma \lambda }} - {g^{\mu \lambda }}{g^{\nu \sigma }}} \right)} \right]{\delta _{cd}}{g_{\sigma \lambda }} \times \\ &\times \int {\frac{{{d^n}k}}{{{{\left( {2\pi } \right)}^n}}}\frac{1}{{{k^2} - {M^2}}}} \\
 &=  - {e^2}\left[ {{\varepsilon _{abe}}{\varepsilon _{cce}}\left( {{g^{\mu \lambda }}{g^{\nu \sigma }} - {g^{\mu \sigma }}{g^{\nu \lambda }}} \right){g_{\sigma \lambda }} + {\varepsilon _{ace}}{\varepsilon _{bce}}\left( {n{g^{\mu \nu }} - {g^{\mu \nu }}} \right) + {\varepsilon _{ace}}{\varepsilon _{bce}}\left( {n{g^{\mu \nu }} - {g^{\mu \nu }}} \right)} \right]{A_0}\left( M \right)\\
 &=  - 2{e^2}\left( {n - 1} \right){\varepsilon _{ace}}{\varepsilon _{bce}}{g^{\mu \nu }}{A_0}\left( M \right)\\
 &=  - 4{e^2}\left( {n - 1} \right){\delta _{ab}}\left( {P_T^{\mu \nu } + P_L^{\mu \nu }} \right){A_0}\left( M \right)\\
 &=  - 4{e^2}{\delta _{ab}}\left( {n - 1} \right){A_0}\left( M \right)P_T^{\mu \nu } - 4{e^2}{\delta _{ab}}\left( {n - 1} \right){A_0}\left( M \right)P_L^{\mu \nu }
\end{align*}

\subsection{lollipops}

\subsubsection{$\phi$}
\begin{minipage}{\linewidth}
\begin{tikzpicture}[scale=1]
	\Flollipop{9}{-5.5}{\Frho}{\Fphi};
\end{tikzpicture}
\end{minipage}
\begin{align*}
{\Gamma _2} = 2e{\varepsilon _{dab}}{k^\gamma } \qquad {D^{ab}}\left( k \right) = \frac{{i{\delta ^{ab}}}}{{{k^2} - {m^2}}}
\end{align*}
\begin{align*}
t_{A1}^{d\gamma } &= \int {\frac{{{d^n}k}}{{{{\left( {2\pi } \right)}^n}}}{\Gamma _2} \cdot {D^{ab}}\left( k \right)} \\
 &= 2ie{\varepsilon _{daa}}\int {\frac{{{d^n}k}}{{{{\left( {2\pi } \right)}^n}}}\frac{{{k^\gamma }}}{{{k^2} - {m^2}}}} \\
 &= 0
\end{align*}

\subsubsection{Goldstone}
\begin{minipage}{\linewidth}
\begin{tikzpicture}[scale=1]
	\Flollipop{9}{-5.5}{\Frho}{\Fchi};
\end{tikzpicture}
\end{minipage}
\begin{align*}
{\Gamma _2} =  - e{\varepsilon _{dab}}{k^\gamma } \qquad {D^{ab}}\left( k \right) = \frac{{i{\delta ^{ab}}}}{{{k^2} - {M^2}}}
\end{align*}
\begin{align*}
t_{A2}^{d\gamma } &= \int {\frac{{{d^n}k}}{{{{\left( {2\pi } \right)}^n}}}{\Gamma _2} \cdot {D^{ab}}\left( k \right)} \\
 &=  - ie{\varepsilon _{daa}}\int {\frac{{{d^n}k}}{{{{\left( {2\pi } \right)}^n}}}\frac{{{k^\gamma }}}{{{k^2} - {m^2}}}} \\
 &= 0
\end{align*}

\subsubsection{ghost}
\begin{minipage}{\linewidth}
\begin{tikzpicture}[scale=1]
	\Flollipop{9}{-5.5}{\Frho}{\Fgst};
\end{tikzpicture}
\end{minipage}
\begin{align*}
{\Gamma _2} = e{\varepsilon _{dab}}{k^\gamma } \qquad {D^{ab}}\left( k \right) = \frac{{i{\delta ^{ab}}}}{{{k^2} - {M^2}}}
\end{align*}
\begin{align*}
t_{A3}^{d\gamma } &= \int {\frac{{{d^n}k}}{{{{\left( {2\pi } \right)}^n}}}{\Gamma _2} \cdot {D^{ab}}\left( k \right)} \\
 &= ie{\varepsilon _{daa}}\int {\frac{{{d^n}k}}{{{{\left( {2\pi } \right)}^n}}}\frac{{{k^\gamma }}}{{{k^2} - {m^2}}}} \\
 &= 0
\end{align*}

\subsubsection{$\rho$}
\begin{minipage}{\linewidth}
\begin{tikzpicture}[scale=1]
	\Flollipop{9}{-5.5}{\Frho}{\Frho};
\end{tikzpicture}
\end{minipage}
\begin{align*}
{\Gamma _2} = e{\varepsilon _{dab}}\left[ {{k^\alpha }{g^{\beta \gamma }} + {k^\beta }{g^{\alpha \gamma }} - 2{k^\gamma }{g^{\alpha \beta }}} \right] \qquad D_{\alpha \beta }^{ab}\left( k \right) = \frac{{ - i{\delta ^{ab}}{g_{\alpha \beta }}}}{{{k^2} - {M^2}}}
\end{align*}
\begin{align*}
t_{A4}^{d\gamma } &= \int {\frac{{{d^n}k}}{{{{\left( {2\pi } \right)}^n}}}{\Gamma _2} \cdot {D^{ab}}\left( k \right)} \\
 &=  - ie{\varepsilon _{daa}}\int {\frac{{{d^n}k}}{{{{\left( {2\pi } \right)}^n}}}\frac{{{k^\alpha }{g^{\beta \gamma }} + {k^\beta }{g^{\alpha \gamma }} - 2{k^\gamma }{g^{\alpha \beta }}}}{{{k^2} - {m^2}}}} \\
 &= 0
\end{align*}

\subsubsection{Higgs}
\begin{minipage}{\linewidth}
\begin{tikzpicture}[scale=1]
	\Flollipop{9}{-5.5}{\Fhiggs}{\Fhiggs};
\end{tikzpicture}
\end{minipage}
\begin{align*}
{\Gamma _2} =  - \frac{3}{2}i\lambda \nu \qquad D\left( k \right) = \frac{i}{{{k^2} - m_H^2}}
\end{align*}
\begin{align*}
t_1^H &= \int {\frac{{{d^n}k}}{{{{\left( {2\pi } \right)}^n}}}{\Gamma _2} \cdot D\left( k \right)} \\
 &= \frac{3}{2}\lambda \nu \int {\frac{{{d^n}k}}{{{{\left( {2\pi } \right)}^n}}}\frac{1}{{{k^2} - m_H^2}}} \\
 &= \frac{3}{2}\lambda \nu {A_0}\left( {{m_H}} \right)
\end{align*}

\subsubsection{Goldstone}
\begin{minipage}{\linewidth}
\begin{tikzpicture}[scale=1]
	\Flollipop{9}{-5.5}{\Fhiggs}{\Fchi};
\end{tikzpicture}
\end{minipage}
\begin{align*}
{\Gamma _2} =  - \frac{1}{2}i\lambda \nu {\delta _{ab}}\qquad {D^{ab}}\left( k \right) = \frac{{i{\delta ^{ab}}}}{{{k^2} - {M^2}}}
\end{align*}
\begin{align*}
t_2^H &= \int {\frac{{{d^n}k}}{{{{\left( {2\pi } \right)}^n}}}{\Gamma _2} \cdot {D^{ab}}\left( k \right)} \\
 &= \frac{1}{2}\lambda \nu {\delta _{aa}}\int {\frac{{{d^n}k}}{{{{\left( {2\pi } \right)}^n}}}\frac{1}{{{k^2} - {M^2}}}} \\
 &= \frac{3}{2}\lambda \nu {A_0}\left( M \right)
\end{align*}

\subsubsection{$\rho$}
\begin{minipage}{\linewidth}
\begin{tikzpicture}[scale=1]
	\Flollipop{9}{-5.5}{\Fhiggs}{\Frho};
\end{tikzpicture}
\end{minipage}
\begin{align*}
{\Gamma _2} = \frac{1}{2}i{e^2}\nu {\delta _{ab}}{g^{\alpha \beta }} \qquad D_{\alpha \beta }^{ab}\left( k \right) = \frac{{ - i{\delta ^{ab}}{g_{\alpha \beta }}}}{{{k^2} - {M^2}}}
\end{align*}
\begin{align*}
t_3^H &= \int {\frac{{{d^n}k}}{{{{\left( {2\pi } \right)}^n}}}{\Gamma _2} \cdot D_{\alpha \beta }^{ab}\left( k \right)} \\
 &= \frac{1}{2}{e^2}\nu n{\delta ^{aa}}\int {\frac{{{d^n}k}}{{{{\left( {2\pi } \right)}^n}}}\frac{1}{{{k^2} - {M^2}}}} \\
 &= \frac{3}{2}{e^2}\nu n{A_0}\left( M \right)
\end{align*}

\subsubsection{$\phi$}
\begin{minipage}{\linewidth}
\begin{tikzpicture}[scale=1]
	\Flollipop{9}{-5.5}{\Fhiggs}{\Fphi};
\end{tikzpicture}
\end{minipage}
\begin{align*}
{\Gamma _2} =  - 4i\kappa \nu {\delta _{ab}} \qquad {D^{ab}}\left( k \right) = \frac{{i{\delta ^{ab}}}}{{{k^2} - {m^2}}}
\end{align*}
\begin{align*}
t_4^H &= \int {\frac{{{d^n}k}}{{{{\left( {2\pi } \right)}^n}}}{\Gamma _2} \cdot {D^{ab}}\left( k \right)} \\
 &= 4\kappa \nu {\delta ^{aa}}\int {\frac{{{d^n}k}}{{{{\left( {2\pi } \right)}^n}}}\frac{1}{{{k^2} - {m^2}}}} \\
 &= 12\kappa \nu {A_0}\left( m \right)
\end{align*}

\subsubsection{ghost}
\begin{minipage}{\linewidth}
\begin{tikzpicture}[scale=1]
	\Flollipop{9}{-5.5}{\Fhiggs}{\Fgst};
\end{tikzpicture}
\end{minipage}
\begin{align*}
{\Gamma _2} = \frac{1}{2}i{e^2}\nu {\delta _{ab}} \qquad {D^{ab}}\left( k \right) = \frac{{i{\delta ^{ab}}}}{{{k^2} - {M^2}}}
\end{align*}
\begin{align*}
t_5^H &= \int {\frac{{{d^n}k}}{{{{\left( {2\pi } \right)}^n}}}{\Gamma _2} \cdot {D^{ab}}\left( k \right)} \\
 &=  - \frac{1}{2}{e^2}\nu {\delta ^{aa}}\int {\frac{{{d^n}k}}{{{{\left( {2\pi } \right)}^n}}}\frac{1}{{{k^2} - {M^2}}}} \\
 &=  - \frac{3}{2}{e^2}\nu {A_0}\left( M \right)
\end{align*}

\begin{minipage}{\linewidth}
\begin{tikzpicture}[scale=1]
	\Flollipop{9}{-5.5}{\Fhiggs}{fill=gray!70};
\end{tikzpicture}
\end{minipage}
\begin{align*}
{T_H} &= \sum\limits_{k = 1}^5 {t_k^H} \\
 &= \frac{3}{2}\nu \left\{ {\lambda {A_0}\left( {{m_H}} \right) + \left[ {\lambda  + {e^2}\left( {n - 1} \right)} \right]{A_0}\left( M \right) + 8\kappa {A_0}\left( m \right)} \right\}
\end{align*}

\begin{minipage}{\linewidth}
\begin{tikzpicture}[scale=1]
	\FFlollipop{9}{-5.5}{\Frho}{\Fhiggs}{fill=gray!70};
\end{tikzpicture}
\end{minipage}
\begin{align*}
\Gamma  &= i{e^2}\nu {\delta _{ab}}{g^{\mu \nu }} \qquad D\left( 0 \right) = \frac{{ - i}}{{m_H^2}}\\
 &= i{e^2}\nu {\delta _{ab}}\left( {P_T^{\mu \nu } + P_L^{\mu \nu }} \right)
\end{align*}
\begin{align*}
 - i\pi _{ab11}^{\mu \nu } &= \Gamma  \cdot D\left( 0 \right) \cdot {T_H}\\
 &= i{e^2}\nu {\delta _{ab}}\left( {P_T^{\mu \nu } + P_L^{\mu \nu }} \right) \cdot \frac{{ - i}}{{m_H^2}} \cdot \frac{3}{2}\nu \left\{ {\lambda {A_0}\left( {{m_H}} \right) + \left[ {\lambda  + {e^2}\left( {n - 1} \right)} \right]{A_0}\left( M \right) + 8\kappa {A_0}\left( m \right)} \right\}\\
 &= \frac{3}{2}{e^2}{\nu ^2}{\delta _{ab}}\left( {P_T^{\mu \nu } + P_L^{\mu \nu }} \right)\left\{ {\lambda \frac{{{A_0}\left( {{m_H}} \right)}}{{m_H^2}} + \left[ {\lambda  + {e^2}\left( {n - 1} \right)} \right]\frac{{{A_0}\left( M \right)}}{{m_H^2}} + 8\kappa \frac{{{A_0}\left( m \right)}}{{m_H^2}}} \right\}\\
 &= \frac{3}{2}{M^2}{\delta _{ab}}\left\{ {\lambda \frac{{{A_0}\left( {{m_H}} \right)}}{{m_H^2}} + \left[ {\lambda  + {e^2}\left( {n - 1} \right)} \right]\frac{{{A_0}\left( M \right)}}{{m_H^2}} + 8\kappa \frac{{{A_0}\left( m \right)}}{{m_H^2}}} \right\}P_T^{\mu \nu } + \\
 &+ \frac{3}{2}{M^2}{\delta _{ab}}\left\{ {\lambda \frac{{{A_0}\left( {{m_H}} \right)}}{{m_H^2}} + \left[ {\lambda  + {e^2}\left( {n - 1} \right)} \right]\frac{{{A_0}\left( M \right)}}{{m_H^2}} + 8\kappa \frac{{{A_0}\left( m \right)}}{{m_H^2}}} \right\}P_L^{\mu \nu }
\end{align*}

\subsection{Summary}

\begin{minipage}{\linewidth}
\begin{tikzpicture}[scale=1]
	\Fbubble{0}{2}{\Frho}{\Fphi}{\Fphi}{\Frho};
\end{tikzpicture}
\end{minipage}
\begin{align*}
 - i\pi _{ab1}^{\mu \nu } &= 2{e^2}{\delta _{ab}}I_1^{\mu \nu }\\
 &= \frac{{2{e^2}{\delta _{ab}}}}{{n - 1}}\left[ {2{A_0}\left( m \right) + \left( {4{m^2} - {p^2}} \right){B_0}\left( {m;p,m} \right)} \right]P_T^{\mu \nu } + 4{e^2}{\delta _{ab}}{A_0}\left( m \right)P_L^{\mu \nu }
\end{align*}

\begin{minipage}{\linewidth}
\begin{tikzpicture}[scale=1]
	\Ftadpole{4.5}{2}{\Frho}{\Fphi};
\end{tikzpicture}
\end{minipage}
\begin{align*}
 - i\pi _{ab2}^{\mu \nu } &=  - 4{e^2}{\delta _{ab}}{A_0}\left( m \right)P_T^{\mu \nu } - 4{e^2}{\delta _{ab}}{A_0}\left( m \right)P_L^{\mu \nu }
\end{align*}

\begin{minipage}{\linewidth}
\begin{tikzpicture}[scale=1]
	\Fbubble{9}{-1}{\Frho}{\Frho}{\Fhiggs}{\Frho};
\end{tikzpicture}
\end{minipage}
\begin{align*}
 - i\pi _{ab3}^{\mu \nu } =  - {e^2}{M^2}{\delta _{ab}}{B_0}\left( {M;p,{m_H}} \right)P_T^{\mu \nu } - {e^2}{M^2}{\delta _{ab}}{B_0}\left( {M;p,{m_H}} \right)P_L^{\mu \nu }
\end{align*}

\begin{minipage}{\linewidth}
\begin{tikzpicture}[scale=1]
	\Ftadpole{13.5}{-1}{\Frho}{\Fhiggs};
\end{tikzpicture}
\end{minipage}
\begin{align*}
 - i\pi _{ab4}^{\mu \nu } &=  - \frac{1}{2}{e^2}{\delta _{ab}}{A_0}\left( {{m_H}} \right)P_T^{\mu \nu } - \frac{1}{2}{e^2}{\delta _{ab}}{A_0}\left( {{m_H}} \right)P_L^{\mu \nu }
\end{align*}

\begin{minipage}{\linewidth}
\begin{tikzpicture}[scale=1]
	\Fbubble{9}{-1}{\Frho}{\Fchi}{\Fhiggs}{\Frho};
\end{tikzpicture}
\end{minipage}
\begin{align*}
 - i\pi _{ab5}^{\mu \nu }&=  - \frac{{{e^2}{\delta _{ab}}}}{{4\left( {n - 1} \right)}} \bigg\{ {A_0}\left( M \right) + {A_0}\left( {{m_H}} \right) + \left( {2m_H^2 + 2{M^2} - {p^2}} \right){B_0}\left( {M;p,{m_H}} \right) +\\ &- \frac{{m_H^2 - {M^2}}}{{{p^2}}}\left[ {{A_0}\left( M \right) - {A_0}\left( {{m_H}} \right) + \left( {m_H^2 - {M^2}} \right){B_0}\left( {M;p,{m_H}} \right)} \right] \bigg\} P_T^{\mu \nu } + \\
 &- \frac{1}{4}{e^2}{\delta _{ab}}\left\{ {{A_0}\left( M \right) + {A_0}\left( {{m_H}} \right) + \frac{{m_H^2 - {M^2}}}{{{p^2}}}\left[ { - {A_0}\left( {{m_H}} \right) + {A_0}\left( M \right) + \left( {m_H^2 - {M^2}} \right){B_0}\left( {M;p,{m_H}} \right)} \right]} \right\}P_L^{\mu \nu }
\end{align*}

\begin{minipage}{\linewidth}
\begin{tikzpicture}[scale=1]
	\Fbubble{9}{-1}{\Frho}{\Fchi}{\Fchi}{\Frho};
\end{tikzpicture}
\end{minipage}
\begin{align*}
 - i\pi _{ab8}^{\mu \nu } &=   \frac{{{e^2}{\delta _{ab}}}}{{2\left( {n - 1} \right)}}\left[ {2{A_0}\left( M \right) + \left( {4{M^2} - {p^2}} \right){B_0}\left( {M;p,M} \right)} \right]P_T^{\mu \nu } + {e^2}{\delta _{ab}}{A_0}\left( M \right)P_L^{\mu \nu }
\end{align*}

\begin{minipage}{\linewidth}
\begin{tikzpicture}[scale=1]
	\Ftadpole{13.5}{-1}{\Frho}{\Fchi};
\end{tikzpicture}
\end{minipage}
\begin{align*}
 - i\pi _{ab6}^{\mu \nu } &=  - \frac{3}{2}{e^2}{\delta ^{ab}}{A_0}\left( M \right)P_T^{\mu \nu } - \frac{3}{2}{e^2}{\delta ^{ab}}{A_0}\left( M \right)P_L^{\mu \nu }
\end{align*}

\begin{minipage}{\linewidth}
\begin{tikzpicture}[scale=1]
	\Fbubble{9}{-1}{\Frho}{\Fgst}{\Fgst}{\Frho};
\end{tikzpicture}
\end{minipage}
\begin{align*}
- i\pi _{ab7}^{\mu \nu } &= \frac{{2{e^2}{\delta _{ab}}}}{{n - 1}}\left[ {\frac{1}{2}{A_0}\left( M \right) + \left( {{M^2} - \frac{1}{4}{p^2}} \right){B_0}\left( {M;p,M} \right)} \right]P_T^{\mu \nu } + 2{e^2}{\delta _{ab}}\left[ {\frac{1}{2}{A_0}\left( M \right) + \frac{1}{4}{p^2}{B_0}\left( {M;p,M} \right)} \right]P_L^{\mu \nu }
\end{align*}

\begin{minipage}{\linewidth}
\begin{tikzpicture}[scale=1]
	\Fbubble{9}{-1}{\Frho}{\Frho}{\Frho}{\Frho};
\end{tikzpicture}
\end{minipage}
\begin{align*}
 - i\pi _{ab9}^{\mu \nu } &=   2{e^2}{\delta _{ab}}\left\{ {\frac{{4n - 5}}{{n - 1}}{A_0}\left( M \right) + \frac{1}{{2\left( {n - 1} \right)}}\left[4 {\left( {3n - 4} \right){M^2} + \left( {6n - 5} \right){p^2}} \right]B\left( {M;p,M} \right)} \right\}P_T^{\mu \nu } + \\
 &+ 2{e^2}{\delta _{ab}}\left[ {\left( {2n - 1} \right){A_0}\left( M \right) + \frac{1}{2}\left( {4{M^2} - {p^2}} \right)B\left( {M;p,M} \right)} \right]P_L^{\mu \nu }
\end{align*}

\begin{minipage}{\linewidth}
\begin{tikzpicture}[scale=1]
	\Ftadpole{13.5}{-1}{\Frho}{\Frho};
\end{tikzpicture}
\end{minipage}
\begin{align*}
 - i\pi _{ab10}^{\mu \nu } &=  - 4{e^2}{\delta _{ab}}\left( {n - 1} \right){A_0}\left( M \right)P_T^{\mu \nu } - 4{e^2}{\delta _{ab}}\left( {n - 1} \right){A_0}\left( M \right)P_L^{\mu \nu }
\end{align*}

\begin{minipage}{\linewidth}
\begin{tikzpicture}[scale=1]
	\FFlollipop{9}{-5.5}{\Frho}{\Fhiggs}{fill=gray!70};
\end{tikzpicture}
\end{minipage}
\begin{align*}
 - i\pi _{ab11}^{\mu \nu } &= \frac{3}{2}{M^2}{\delta _{ab}}\left\{ {\lambda \frac{{{A_0}\left( {{m_H}} \right)}}{{m_H^2}} + \left[ {\lambda  + {e^2}\left( {n - 1} \right)} \right]\frac{{{A_0}\left( M \right)}}{{m_H^2}} + 8\kappa \frac{{{A_0}\left( m \right)}}{{m_H^2}}} \right\}P_T^{\mu \nu } + \\
 &+ \frac{3}{2}{M^2}{\delta _{ab}}\left\{ {\lambda \frac{{{A_0}\left( {{m_H}} \right)}}{{m_H^2}} + \left[ {\lambda  + {e^2}\left( {n - 1} \right)} \right]\frac{{{A_0}\left( M \right)}}{{m_H^2}} + 8\kappa \frac{{{A_0}\left( m \right)}}{{m_H^2}}} \right\}P_L^{\mu \nu }
\end{align*}

\section{Higgs Self Energy}

\subsection{$\phi$ bubble}
\begin{minipage}{\linewidth}
\begin{tikzpicture}[scale=1]
	\Fbubble{0}{2}{\Fhiggs}{\Fphi}{\Fphi}{\Fhiggs};
\end{tikzpicture}
\end{minipage}

\begin{align*}
{\Gamma _1} &=  - 4i\kappa \nu {\delta _{ab}} \qquad {\Gamma _2} =  - 4i\kappa \nu {\delta _{cd}} \\
{D^{bd}}\left( k \right) &= \frac{{i{\delta _{bd}}}}{{{k^2} - {m^2}}} \qquad {D^{ac}}\left( {k + p} \right) = \frac{{i{\delta _{ac}}}}{{{{\left( {k + p} \right)}^2} - {m^2}}}
\end{align*}
\begin{align*}
 - i\pi _1^H &= \int {\frac{{{d^n}k}}{{{{\left( {2\pi } \right)}^n}}}{\Gamma _2} \cdot {D^{bd}}\left( k \right) \cdot {\Gamma _1} \cdot {D^{ac}}\left( {k + p} \right)} \\
 &= {\left( {4\kappa \nu } \right)^2}{\delta _{aa}}\int {\frac{{{d^n}k}}{{{{\left( {2\pi } \right)}^n}}}\frac{1}{{\left( {{k^2} - {m^2}} \right)\left[ {{{\left( {k + p} \right)}^2} - {m^2}} \right]}}} \\
 &= 3{\left( {4\kappa \nu } \right)^2}{B_0}\left( {m;p,m} \right)\\
 &= 48{\kappa ^2}{\nu ^2}{B_0}\left( {m;p,m} \right)
\end{align*}

\subsection{$\phi$ tadpole}
\begin{minipage}{\linewidth}
\begin{tikzpicture}[scale=1]
	\Ftadpole{4.5}{2}{\Fhiggs}{\Fphi};
\end{tikzpicture}
\end{minipage}
\begin{align*}
\Gamma  =  - 2i\kappa {\delta _{ab}} \qquad {D^{ab}}\left( k \right) = \frac{{i{\delta _{ab}}}}{{{k^2} - {m^2}}}
\end{align*}
\begin{align*}
 - i\pi _2^H &= \int {\frac{{{d^n}k}}{{{{\left( {2\pi } \right)}^n}}}\Gamma  \cdot {D^{ab}}\left( k \right)} \\
 &= 2\kappa {\delta _{aa}}\int {\frac{{{d^n}k}}{{{{\left( {2\pi } \right)}^n}}}\frac{1}{{{k^2} - {m^2}}}} \\
 &= 6\kappa {A_0}\left( m \right)
\end{align*}

\subsection{$\rho$ bubble}
\begin{minipage}{\linewidth}
\begin{tikzpicture}[scale=1]
	\Fbubble{9}{-1}{\Fhiggs}{\Frho}{\Frho}{\Fhiggs};
\end{tikzpicture}
\end{minipage}

\begin{align*}
{\Gamma _1} &= i{e^2}\nu {\delta _{ab}}{g^{\alpha \beta }} \qquad {\Gamma _2} = i{e^2}\nu {\delta _{cd}}{g^{\gamma \delta }} \\
D_{\beta \delta }^{bd}\left( k \right) &= \frac{{ - i{\delta _{bd}}{g_{\beta \delta }}}}{{{k^2} - {M^2}}} \qquad D_{\alpha \gamma }^{ac}\left( {k + p} \right) = \frac{{ - i{\delta _{ac}}{g_{\alpha \gamma }}}}{{{{\left( {k + p} \right)}^2} - {M^2}}}
\end{align*}

\begin{align*}
 - i\pi _3^H &= \int {\dfrac{{{d^n}k}}{{{{\left( {2\pi } \right)}^n}}}{\Gamma _2} \cdot {D^{bd}}\left( k \right) \cdot {\Gamma _1} \cdot {D^{ac}}\left( {k + p} \right)} \\
 &= {e^4}{\nu ^2}n{\delta _{aa}}\int {\dfrac{{{d^n}k}}{{{{\left( {2\pi } \right)}^n}}}\frac{1}{{\left( {{k^2} - {M^2}} \right)\left[ {{{\left( {k + p} \right)}^2} - {M^2}} \right]}}} \\
 &= 3{e^4}{\nu ^2}n{B_0}\left( {M;p,M} \right)
\end{align*}

\subsection{$\rho$ tadpole}
\begin{minipage}{\linewidth}
\begin{tikzpicture}[scale=1]
	\Ftadpole{13.5}{-1}{\Fhiggs}{\Frho};
\end{tikzpicture}
\end{minipage}
\begin{align*}
\Gamma  = \frac{1}{2}i{e^2}{\delta _{ab}}{g^{\alpha \beta }} \qquad
D_{\alpha \beta }^{ab}\left( k \right) = \frac{{ - i{\delta _{ab}}{g_{\alpha \beta }}}}{{{k^2} - {M^2}}}
\end{align*}
\begin{align*}
 - i\pi _4^H &= \int {\frac{{{d^n}k}}{{{{\left( {2\pi } \right)}^n}}}\Gamma  \cdot D_{\alpha \beta }^{ab}\left( k \right)} \\
 &= \frac{1}{2}{e^2}n{\delta _{aa}}\int {\frac{{{d^n}k}}{{{{\left( {2\pi } \right)}^n}}}\frac{1}{{{k^2} - {M^2}}}} \\
 &= \frac{3}{2}{e^2}n{A_0}\left( M \right)
\end{align*}

\subsection{Goldstone bubble}
\begin{minipage}{\linewidth}
\begin{tikzpicture}[scale=1]
	\Fbubble{9}{-1}{\Fhiggs}{\Fchi}{\Fchi}{\Fhiggs};
\end{tikzpicture}
\end{minipage}
\begin{align*}
{\Gamma _1} &=  - \frac{1}{2}i\lambda \nu {\delta _{ab}} \qquad {\Gamma _2} =  - \frac{1}{2}i\lambda \nu {\delta _{cd}}\\
{D^{bd}}\left( k \right) &= \frac{{i{\delta _{bd}}}}{{{k^2} - {M^2}}} \qquad {D^{ac}}\left( {k + p} \right) = \frac{{i{\delta _{ac}}}}{{{{\left( {k + p} \right)}^2} - {M^2}}}
\end{align*}
\begin{align*}
 - i\pi _5^H &= \int {\frac{{{d^n}k}}{{{{\left( {2\pi } \right)}^n}}}{\Gamma _2} \cdot {D^{bd}}\left( k \right) \cdot {\Gamma _1} \cdot {D^{ac}}\left( {k + p} \right)} \\
 &= {\left( {\frac{1}{2}\lambda \nu } \right)^2}{\delta _{aa}}\int {\frac{{{d^n}k}}{{{{\left( {2\pi } \right)}^n}}}\frac{1}{{\left( {{k^2} - {M^2}} \right)\left[ {{{\left( {k + p} \right)}^2} - {M^2}} \right]}}} \\
 &= \frac{3}{4}{\lambda ^2}{\nu ^2}{B_0}\left( {M;p,M} \right)
\end{align*}

\subsection{Goldstone tadpole}
\begin{minipage}{\linewidth}
\begin{tikzpicture}[scale=1]
	\Ftadpole{13.5}{-1}{\Fhiggs}{\Fchi};
\end{tikzpicture}
\end{minipage}
\begin{align*}
\Gamma  =  - \frac{1}{4}i\lambda {\delta _{ab}}\qquad
{D^{ab}}\left( k \right) = \frac{{i{\delta _{ab}}}}{{{k^2} - {M^2}}}
\end{align*}
\begin{align*}
 - i\pi _6^H &= \int {\frac{{{d^n}k}}{{{{\left( {2\pi } \right)}^n}}}\Gamma  \cdot {D^{ab}}\left( k \right)} \\
 &= \frac{1}{4}\lambda {\delta _{aa}}\int {\frac{{{d^n}k}}{{{{\left( {2\pi } \right)}^n}}}\frac{1}{{{k^2} - {M^2}}}} \\
 &= \frac{3}{4}\lambda {A_0}\left( M \right)
\end{align*}

\subsection{Higgs bubble}
\begin{minipage}{\linewidth}
\begin{tikzpicture}[scale=1]
	\Fbubble{9}{-1}{\Fhiggs}{\Fhiggs}{\Fhiggs}{\Fhiggs};
\end{tikzpicture}
\end{minipage}
\begin{align*}
{\Gamma _1} &=  - \frac{3}{2}i\lambda \nu \qquad {\Gamma _2} =  - \frac{3}{2}i\lambda \nu \\
D\left( k \right) &= \frac{i}{{{k^2} - m_H^2}} \qquad D\left( {k + p} \right) = \frac{i}{{{{\left( {k + p} \right)}^2} - m_H^2}}
\end{align*}

\begin{align*}
 - i\pi _7^H &= \int {\frac{{{d^n}k}}{{{{\left( {2\pi } \right)}^n}}}{\Gamma _2} \cdot D\left( k \right) \cdot {\Gamma _1} \cdot D\left( {k + p} \right)} \\
 &= {\left( {\frac{3}{2}\lambda \nu } \right)^2}\int {\frac{{{d^n}k}}{{{{\left( {2\pi } \right)}^n}}}\frac{1}{{\left( {{k^2} - m_H^2} \right)\left[ {{{\left( {k + p} \right)}^2} - m_H^2} \right]}}} \\
 &= \frac{9}{4}{\lambda ^2}{\nu ^2}{B_0}\left( {{m_H};p,{m_H}} \right)
\end{align*}

\subsection{Higgs tadpole}
\begin{minipage}{\linewidth}
\begin{tikzpicture}[scale=1]
	\Ftadpole{13.5}{-1}{\Fhiggs}{\Fhiggs};
\end{tikzpicture}
\end{minipage}

\begin{align*}
\Gamma  =  - \frac{3}{2}i\lambda \qquad D\left( k \right) = \frac{i}{{{k^2} - m_H^2}}
\end{align*}

\begin{align*}
 - i\pi _8^H &= \int {\frac{{{d^n}k}}{{{{\left( {2\pi } \right)}^n}}}\Gamma  \cdot D\left( k \right)} \\
 &= \frac{3}{2}\lambda \int {\frac{{{d^n}k}}{{{{\left( {2\pi } \right)}^n}}}\frac{1}{{{k^2} - m_H^2}}} \\
 &= \frac{3}{2}\lambda {A_0}\left( {{m_H}} \right)
\end{align*}

\subsection{Goldstone-$\rho$ bubble}
\begin{minipage}{\linewidth}
\begin{tikzpicture}[scale=1]
	\Fbubble{9}{-1}{\Fhiggs}{\Frho}{\Fchi}{\Fhiggs};
\end{tikzpicture}
\end{minipage}
\begin{align*}
{\Gamma _1} &=  - \frac{1}{2}e{\delta _{ac}}{\left( {k + 2p} \right)^\alpha } \qquad {\Gamma _2} = \frac{1}{2}e{\delta _{bd}}{\left( {k + 2p} \right)^\beta }\\
D_{\alpha \beta }^{ac}\left( k \right) &= \frac{{ - i{\delta _{ac}}{g_{\alpha \beta }}}}{{{k^2} - {M^2}}} \qquad {D^{cd}}\left( {k + p} \right) = \frac{{i{\delta _{cd}}}}{{{{\left( {k + p} \right)}^2} - {M^2}}}
\end{align*}

\begin{align*}
 - i\pi _9^H &= \int {\frac{{{d^n}k}}{{{{\left( {2\pi } \right)}^n}}}{\Gamma _2} \cdot D_{\alpha \beta }^{ac}\left( k \right) \cdot {\Gamma _1} \cdot {D^{cd}}\left( {k + p} \right)} \\
 &=  - \frac{1}{4}{e^2}{\delta _{aa}}\int {\frac{{{d^n}k}}{{{{\left( {2\pi } \right)}^n}}}\frac{{{{\left( {k + 2p} \right)}^2}}}{{\left( {{k^2} - {M^2}} \right)\left[ {{{\left( {k + p} \right)}^2} - {M^2}} \right]}}} \\
 &=  - \frac{3}{4}{e^2}{I_9}
\end{align*}
We make some definitions
\begin{align*}
{D_1} &= {k^2} - {M^2}\\
{D_2} &= {\left( {k + p} \right)^2} - {M^2}\\
 &= {k^2} + 2k \cdot p + {p^2} - {M^2}\\
 &= {D_1} + 2k \cdot p + {p^2}
\end{align*}
Which gives us the relations
\begin{align*}
{k^2} &= {D_1} + {M^2}\\
2k \cdot p &= {D_2} - {D_1} - {p^2}
\end{align*}
We can write the numerator as
\begin{align*}
{N_9} &= {\left( {k + 2p} \right)^2}\\
 &= {k^2} + 2\left( {2k \cdot p} \right) + 4{p^2}\\
 &= 2{D_2} - {D_1} + {M^2} + 2{p^2}
\end{align*}
\begin{align*}
{I_9} &= \int {\frac{{{d^n}k}}{{{{\left( {2\pi } \right)}^n}}}\frac{{{N_9}}}{{{D_1}{D_2}}}} \\
 &= \int {\frac{{{d^n}k}}{{{{\left( {2\pi } \right)}^n}}}\frac{{2{D_2} - {D_1} + {M^2} + 2{p^2}}}{{{D_1}{D_2}}}} \\
 &= 2{A_0}\left( M \right) - {A_0}\left( M \right) + \left( {{M^2} + 2{p^2}} \right){B_0}\left( {M;p,M} \right)\\
 &= {A_0}\left( M \right) + \left( {{M^2} + 2{p^2}} \right){B_0}\left( {M;p,M} \right)
\end{align*}
\begin{align*}
 - i\pi _9^H &=  - \frac{3}{4}{e^2}{I_9}\\
 &=  - \frac{3}{4}{e^2}\left[ {{A_0}\left( M \right) + \left( {{M^2} + 2{p^2}} \right){B_0}\left( {M;p,M} \right)} \right]
\end{align*}

\subsection{Ghost bubble}
\begin{minipage}{\linewidth}
\begin{tikzpicture}[scale=1]
	\Fbubble{9}{-1}{\Fhiggs}{\Fgst}{\Fgst}{\Fhiggs};
\end{tikzpicture}
\end{minipage}
\begin{align*}
{\Gamma _1} &= \frac{1}{2}i{e^2}\nu {\delta _{ab}};{\Gamma _2} = \frac{1}{2}i{e^2}\nu {\delta _{cd}}\\
{D^{bd}}\left( k \right) &= \frac{{i{\delta _{bd}}}}{{{k^2} - {M^2}}} \qquad {D^{ac}}\left( {k + p} \right) = \frac{{i{\delta _{ac}}}}{{{{\left( {k + p} \right)}^2} - {M^2}}}
\end{align*}

\begin{align*}
 - i\pi _{10}^H &= \int {\frac{{{d^n}k}}{{{{\left( {2\pi } \right)}^n}}}{\Gamma _2} \cdot {D^{bd}}\left( k \right) \cdot {\Gamma _1} \cdot {D^{ac}}\left( {k + p} \right)} \\
 &= {\left( {\frac{1}{2}{e^2}\nu } \right)^2}{\delta _{aa}}\int {\frac{{{d^n}k}}{{{{\left( {2\pi } \right)}^n}}}\frac{1}{{\left( {{k^2} - {M^2}} \right)\left[ {{{\left( {k + p} \right)}^2} - {M^2}} \right]}}} \\
 &= \frac{3}{4}{e^4}{\nu ^2}{B_0}\left( {M;p,M} \right)\\
 &= \frac{3}{4}{e^2}{M^2}{B_0}\left( {M;p,M} \right)
\end{align*}

\subsection{Lollipops}
\begin{minipage}{\linewidth}
\begin{tikzpicture}[scale=1]
	\Flollipop{9}{-5.5}{\Fhiggs}{fill=gray!70};
\end{tikzpicture}
\end{minipage}
\begin{align*}
{T_H} = \frac{3}{2}\nu \left\{ {\lambda {A_0}\left( {{m_H}} \right) + \left[ {\lambda  + {e^2}\left( {n - 1} \right)} \right]{A_0}\left( M \right) + 8\kappa {A_0}\left( m \right)} \right\}
\end{align*}

\begin{minipage}{\linewidth}
\begin{tikzpicture}[scale=1]
	\FFlollipop{9}{-5.5}{\Fhiggs}{\Fhiggs}{fill=gray!70};
\end{tikzpicture}
\end{minipage}
\begin{align*}
\Gamma  =  - \frac{3}{2}i\lambda \nu \qquad D\left( 0 \right) = \frac{{ - i}}{{m_H^2}}
\end{align*}
\begin{align*}
 - \pi _{11}^H &= \Gamma  \cdot D\left( 0 \right) \cdot {T_H}\\
 &=  - \frac{3}{2}i\lambda \nu  \cdot \frac{{ - i}}{{m_H^2}} \cdot \frac{3}{2}\nu \left\{ {\lambda {A_0}\left( {{m_H}} \right) + \left[ {\lambda  + {e^2}\left( {n - 1} \right)} \right]{A_0}\left( M \right) + 8\kappa {A_0}\left( m \right)} \right\}\\
 &=  - \frac{9}{4}\lambda {\nu ^2}\left\{ {\lambda \frac{{{A_0}\left( {{m_H}} \right)}}{{m_H^2}} + \left[ {\lambda  + {e^2}\left( {n - 1} \right)} \right]\frac{{{A_0}\left( M \right)}}{{m_H^2}} + 8\kappa \frac{{{A_0}\left( m \right)}}{{m_H^2}}} \right\}
\end{align*}

\subsection{Summary}

\begin{minipage}{\linewidth}
\begin{tikzpicture}[scale=1]
	\Fbubble{0}{2}{\Fhiggs}{\Fphi}{\Fphi}{\Fhiggs};
\end{tikzpicture}
\end{minipage}
\begin{align*}
 - i\pi _1^H = 48{\kappa ^2}{\nu ^2}{B_0}\left( {m;p,m} \right)
\end{align*}

\begin{minipage}{\linewidth}
\begin{tikzpicture}[scale=1]
	\Ftadpole{4.5}{2}{\Fhiggs}{\Fphi};
\end{tikzpicture}
\end{minipage}
\begin{align*}
 - i\pi _2^H = 6\kappa {A_0}\left( m \right)
\end{align*}

\begin{minipage}{\linewidth}
\begin{tikzpicture}[scale=1]
	\Fbubble{9}{-1}{\Fhiggs}{\Frho}{\Frho}{\Fhiggs};
\end{tikzpicture}
\end{minipage}
\begin{align*}
 - i\pi _3^H = 3{e^4}{\nu ^2}n{B_0}\left( {M;p,M} \right)
\end{align*}

\begin{minipage}{\linewidth}
\begin{tikzpicture}[scale=1]
	\Ftadpole{13.5}{-1}{\Fhiggs}{\Frho};
\end{tikzpicture}
\end{minipage}
\begin{align*}
 - i\pi _4^H = \frac{3}{2}{e^2}n{A_0}\left( M \right)
\end{align*}

\begin{minipage}{\linewidth}
\begin{tikzpicture}[scale=1]
	\Fbubble{9}{-1}{\Fhiggs}{\Fchi}{\Fchi}{\Fhiggs};
\end{tikzpicture}
\end{minipage}
\begin{align*}
 - i\pi _5^H = \frac{3}{4}{\lambda ^2}{\nu ^2}{B_0}\left( {M;p,M} \right)
\end{align*}

\begin{minipage}{\linewidth}
\begin{tikzpicture}[scale=1]
	\Ftadpole{13.5}{-1}{\Fhiggs}{\Fchi};
\end{tikzpicture}
\end{minipage}
\begin{align*}
 - i\pi _6^H = \frac{3}{4}\lambda {A_0}\left( M \right)
\end{align*}

\begin{minipage}{\linewidth}
\begin{tikzpicture}[scale=1]
	\Fbubble{9}{-1}{\Fhiggs}{\Fhiggs}{\Fhiggs}{\Fhiggs};
\end{tikzpicture}
\end{minipage}
\begin{align*}
 - i\pi _7^H = \frac{9}{4}{\lambda ^2}{\nu ^2}{B_0}\left( {{m_H};p,{m_H}} \right)
\end{align*}

\begin{minipage}{\linewidth}
\begin{tikzpicture}[scale=1]
	\Ftadpole{13.5}{-1}{\Fhiggs}{\Fhiggs};
\end{tikzpicture}
\end{minipage}

\begin{align*}
 - i\pi _8^H = \frac{3}{2}\lambda {A_0}\left( {{m_H}} \right)
\end{align*}

\begin{minipage}{\linewidth}
\begin{tikzpicture}[scale=1]
	\Fbubble{9}{-1}{\Fhiggs}{\Frho}{\Fchi}{\Fhiggs};
\end{tikzpicture}
\end{minipage}
\begin{align*}
 - i\pi _9^H =  - \frac{3}{4}{e^2}\left[ {{A_0}\left( M \right) + \left( {{M^2} + 2{p^2}} \right){B_0}\left( {M;p,M} \right)} \right]
\end{align*}

\begin{minipage}{\linewidth}
\begin{tikzpicture}[scale=1]
	\Fbubble{9}{-1}{\Fhiggs}{\Fgst}{\Fgst}{\Fhiggs};
\end{tikzpicture}
\end{minipage}
\begin{align*}
 - i\pi _{10}^H = \frac{3}{4}{e^2}{M^2}{B_0}\left( {M;p,M} \right)
\end{align*}

\begin{minipage}{\linewidth}
\begin{tikzpicture}[scale=1]
	\FFlollipop{9}{-5.5}{\Fhiggs}{\Fhiggs}{fill=gray!70};
\end{tikzpicture}
\end{minipage}
\begin{align*}
 - i\pi _{11}^H =  - \frac{9}{4}\lambda {\nu ^2}\left\{ {\lambda \frac{{{A_0}\left( {{m_H}} \right)}}{{m_H^2}} + \left[ {\lambda  + {e^2}\left( {n - 1} \right)} \right]\frac{{{A_0}\left( M \right)}}{{m_H^2}} + 8\kappa \frac{{{A_0}\left( m \right)}}{{m_H^2}}} \right\}
\end{align*}

\addcontentsline{toc}{chapter}{Bibliography}

\bibliographystyle{unsrt}
\bibliography{phdcitelist}

\begin{thebibliography}{10}

\bibitem{Yukawa1935}
Yukawa H.
\newblock On the interaction of elementary particles.
\newblock {\em Proc.Phys.Math.Soc.Jap. 17 (1935) 48;}, 1935.

\bibitem{GELLMANN1964a}
M.~Gell-Mann.
\newblock A schematic model of baryons and mesons.
\newblock {\em Physics Letters}, 8(3):214 -- 215, 1964.

\bibitem{Zweig1964a}
G.~Zweig.
\newblock An su(3) model for strong interaction symmetry and its breaking.
\newblock Technical Report CERN Report No.8182/TH.401, CERN, 1964.

\bibitem{Zweig1964b}
G.~Zweig.
\newblock An su(3) model for strong interaction symmetry and its breaking:{II}.
\newblock Technical Report CERN Report No.8419/TH.412, CERN, 1964.

\bibitem{Feynman1969a}
Feynman R.~P.
\newblock The behavior of hadron collisions at extreme energies.
\newblock In {\em High Energy Collisions: Third International Conference at
  Stony Brook, N.Y.}, 1969.

\bibitem{Feynman1969b}
Feynman R.~P.
\newblock Very high-energy collisions of hadrons.
\newblock {\em Phys. Rev. Lett.}, 23:1415--1417, Dec 1969.

\bibitem{Bjorken1975}
J.~D. Bjorken and E.~A. Paschos.
\newblock Inelastic electron-proton and $\ensuremath{\gamma}$-proton scattering
  and the structure of the nucleon.
\newblock {\em Phys. Rev.}, 185:1975--1982, Sep 1969.

\bibitem{Politzer1973}
H.~David Politzer.
\newblock Reliable perturbative results for strong interactions?
\newblock {\em Phys. Rev. Lett.}, 30:1346--1349, Jun 1973.

\bibitem{GrossWilczek1973}
David~J. Gross and Frank Wilczek.
\newblock Ultraviolet behavior of non-abelian gauge theories.
\newblock {\em Phys. Rev. Lett.}, 30:1343--1346, Jun 1973.

\bibitem{Boyle2009}
Boyle P.A.
\newblock Kaon physics from lattice.
\newblock In {\em PoS(KAON09)002}, 2009.

\bibitem{Yeh2002}
Tsung-Wen Yeh.
\newblock Applicability of perturbative qcd and nlo power corrections for the
  pion form factor.
\newblock {\em Phys. Rev. D}, 65:074016, Mar 2002.

\bibitem{Nambu1957}
Yoichiro Nambu.
\newblock Possible existence of a heavy neutral meson.
\newblock {\em Phys. Rev.}, 106:1366--1367, Jun 1957.

\bibitem{SAKURAI1960}
J.J Sakurai.
\newblock Theory of strong interactions.
\newblock {\em Annals of Physics}, 11(1):1 -- 48, 1960.

\bibitem{SAKURAI1969}
Jun~John Sakurai.
\newblock {\em Currents and mesons}.
\newblock University of Chicago press, 1969.

\bibitem{Bauer1978}
T.~H. Bauer, R.~D. Spital, D.~R. Yennie, and F.~M. Pipkin.
\newblock The hadronic properties of the photon in high-energy interactions.
\newblock {\em Rev. Mod. Phys.}, 50:261--436, Apr 1978.

\bibitem{engel1996}
R.~Engel and J.~Ranft.
\newblock Hadronic photon-photon interactions at high energies.
\newblock {\em Phys. Rev. D}, 54:4244--4262, Oct 1996.

\bibitem{KLZ1967}
Norman~M. Kroll, T.~D. Lee, and Bruno Zumino.
\newblock Neutral vector mesons and the hadronic electromagnetic current.
\newblock {\em Phys. Rev.}, 157:1376--1399, May 1967.

\bibitem{Li2012}
Ling-Fong Li.
\newblock {Introduction to Renormalization in Field Theory}.
\newblock In {\em 100 Years of Subatomic Physics}, pages 465--491. WORLD
  SCIENTIFIC, 2013.

\bibitem{GHOSE1972}
P.~Ghose and A.~Das.
\newblock A variant of the stuckelberg formalism for massive gauge fields and
  applications.
\newblock {\em Nuclear Physics B}, 41(1):299 -- 316, 1972.

\bibitem{GALE1991}
Charles Gale and Joseph~I. Kapusta.
\newblock Vector dominance model at finite temperature.
\newblock {\em Nuclear Physics B}, 357(1):65 -- 89, 1991.

\bibitem{CADWillers2007}
Cesareo~A. Dominguez, Juan~I. Jottar, Marcelo Loewe, and Bernard Willers.
\newblock Pion form factor in the {Kroll-Lee-Zumino} model.
\newblock {\em Phys. Rev. D}, 76:095002, Nov 2007.

\bibitem{CADWillers2008}
C.~A. Dominguez, M.~Loewe, and B.~Willers.
\newblock Scalar radius of the pion in the {Kroll-Lee-Zumino} renormalizable
  theory.
\newblock {\em Phys. Rev. D}, 78:057901, Sep 2008.

\bibitem{Pelaez2005}
J.~R. Pel\'aez and F.~J. Yndur\'ain.
\newblock Pion-pion scattering amplitude.
\newblock {\em Phys. Rev. D}, 71:074016, Apr 2005.

\bibitem{Pelaez2006}
R.~Kami\ifmmode~\acute{n}\else \'{n}\fi{}ski, J.~R. Pel\'aez, and F.~J.
  Yndur\'ain.
\newblock Pion-pion scattering amplitude. {II}. {I}mproved analysis above
  $\overline{K}k$ threshold.
\newblock {\em Phys. Rev. D}, 74:014001, Jul 2006.

\bibitem{Pelaez2008}
R.~Kami\ifmmode~\acute{n}\else \'{n}\fi{}ski, J.~R. Pel\'aez, and F.~J.
  Yndur\'ain.
\newblock Pion-pion scattering amplitude. {III}. {I}mproving the analysis with
  forward dispersion relations and roy equations.
\newblock {\em Phys. Rev. D}, 77:054015, Mar 2008.

\bibitem{Pelaez2011}
R.~Garc\'{\i}a-Mart\'{\i}n, R.~Kami\ifmmode~\acute{n}\else \'{n}\fi{}ski, J.~R.
  Pel\'aez, J.~Ruiz~de Elvira, and F.~J. Yndur\'ain.
\newblock Pion-pion scattering amplitude. {IV}. {I}mproved analysis with once
  subtracted roy-like equations up to 1100 mev.
\newblock {\em Phys. Rev. D}, 83:074004, Apr 2011.

\bibitem{Batley2006}
J.R.~Batley et~al.
\newblock Observation of a cusp like structure in the invariant mass
  distribution from decay and determination of the $\pi \pi$ scattering
  lengths.
\newblock {\em Physics Letters B}, 633(2-3):173 -- 182, 2006.

\bibitem{Adeva2009}
B.~Adeva et~al.
\newblock Evidence for $\pi${K}-atoms with {DIRAC}.
\newblock {\em Physics Letters B}, 674(1):11 -- 16, 2009.

\bibitem{Patrignani}
C.~et~al. Patrignani.
\newblock {Review of Particle Physics}.
\newblock {\em Chin. Phys.}, C40(10):100001, 2016.

\bibitem{Bando1985}
M.~Bando, T.~Kugo, S.~Uehara, K.~Yamawaki, and T.~Yanagida.
\newblock Is the $\ensuremath{\rho}$ meson a dynamical gauge boson of hidden
  local symmetry?
\newblock {\em Phys. Rev. Lett.}, 54:1215--1218, Mar 1985.

\bibitem{Bohmbook}
{A}nsgar~{D}enner Manfred~{B}{\"o}hm and {H}ans {J}oos.
\newblock {\em Gauge {T}heories of the {S}trong and {E}lectroweak
  {I}nteraction}.
\newblock Vieweg {T}eubner {V}erlag, Stuttgart, Leipzig, Wiesbaden, 3rd
  edition, 2001.

\bibitem{Quigg2007}
Chris Quigg.
\newblock Spontaneous symmetry breaking as a basis of particle mass.
\newblock {\em Reports on Progress in Physics}, 70(7):1019, 2007.

\bibitem{tHooft1971167}
G.~'t~Hooft.
\newblock Renormalizable lagrangians for massive {Y}ang-{M}ills fields.
\newblock {\em Nuclear Physics B}, 35(1):167 -- 188, 1971.

\bibitem{FADDEEV1967}
L.D. Faddeev and V.N. Popov.
\newblock Feynman diagrams for the yang-mills field.
\newblock {\em Physics Letters B}, 25(1):29 -- 30, 1967.

\bibitem{ESPOSITO}
GIAMPIERO ESPOSITO, DIEGO~N. PELLICCIA, and FRANCESCO ZACCARIA.
\newblock Gribov problem for gauge theories: A pedagogical introduction.
\newblock {\em International Journal of Geometric Methods in Modern Physics},
  01(04):423--441, 2004.

\bibitem{Ghiotti2007}
Marco Ghiotti.
\newblock {\em {Gauge fixing and BRST formalism in non-Abelian gauge
  theories}}.
\newblock PhD thesis, Adelaide U., 2007.

\bibitem{FujikawaLee}
Kazuo Fujikawa, Benjamin~W. Lee, and A.~I. Sanda.
\newblock Generalized renormalizable gauge formulation of spontaneously broken
  gauge theories.
\newblock {\em Phys. Rev. D}, 6:2923--2943, Nov 1972.

\bibitem{Abers19731}
Ernest~S. Abers and Benjamin~W. Lee.
\newblock Gauge theories.
\newblock {\em Physics Reports}, 9(1):1 -- 2, 1973.

\bibitem{ChengLi}
T.P. Cheng and L.F. Li.
\newblock {\em Gauge {T}heory of {E}lementary {P}article {P}hysics}.
\newblock Oxford science publications. Clarendon Press, 1984.

\bibitem{Weinberg1996QFT2}
Steven Weinberg.
\newblock {\em {The quantum theory of fields. Vol. 2: Modern applications}}.
\newblock Cambridge University Press, 2013.

\bibitem{BerezinbookSuperAnalysis}
F.A. Berezin, A.A. Kirillov, J.~Niederle, and R.~Koteck{\`y}.
\newblock {\em Introduction to Superanalysis}.
\newblock Mathematical Physics and Applied Mathematics. Springer, 1987.

\bibitem{Spiesberger1986}
M.~Bohm, H.~Spiesberger, and W.~Hollik.
\newblock On the 1-loop renormalization of the electroweak standard model and
  its application to leptonic processes.
\newblock {\em Fortschritte der Physik}, 34(11):687--751, 1986.

\bibitem{Hioki1986}
Zenro Hioki.
\newblock Introduction to higher order analysis of the electroweak interaction.
\newblock {\em Acta Physica Polonica, Series B}, 17(12):1037--1084, 1986.

\bibitem{BardinbookSM}
D.~Bardin and G.~Passarino.
\newblock {\em The Standard Model in the Making: Precision Study of the
  Electroweak Interactions}.
\newblock International series of monographs on physics. Clarendon Press, 1999.

\bibitem{Bjorken1965}
James~D. Bjorken and Sidney~D. Drell.
\newblock {\em {Relativistic quantum fields}}.
\newblock Dover Publications, Incorporated, 1965.

\bibitem{Colangelo2001125}
G.~Colangelo, J.~Gasser, and H.~Leutwyler.
\newblock $\pi\pi$ scattering.
\newblock {\em Nuclear Physics B}, 603(1-2):125--179, 2001.

\bibitem{CADLushozi2015}
CA~Dominguez, M~Loewe, and M~Lushozi.
\newblock Scalar form factor of the pion in the kroll-lee-zumino field theory.
\newblock {\em Advances in High Energy Physics}, 2015, 2015.

\bibitem{Borsanyi2013}
Szabolcs Bors\'anyi, Stephan D\"urr, Zolt\'an Fodor, Stefan Krieg, Andreas
  Sch\"afer, Enno~E. Scholz, and K\'alm\'an~K. Szab\'o.
\newblock Su(2) chiral perturbation theory low-energy constants from
  $2\mathbf{+}1$ flavor staggered lattice simulations.
\newblock {\em Phys. Rev. D}, 88:014513, Jul 2013.

\bibitem{Scherer2003}
Stefan Scherer.
\newblock {\em Introduction to Chiral Perturbation Theory}, pages 277--538.
\newblock Springer US, Boston, MA, 2003.

\bibitem{Pelaez2016}
Jose~R. Pelaez.
\newblock From controversy to precision on the sigma meson: A review on the
  status of the non-ordinary resonance.
\newblock {\em Physics Reports}, 658:1 -- 111, 2016.

\bibitem{donoghue2014dynamics}
John~F Donoghue, Eugene Golowich, and Barry~R Holstein.
\newblock {\em Dynamics of the standard model}, volume~35.
\newblock Cambridge university press, 2014.

\bibitem{Bijnens2000}
G.~Amoros, J.~Bijnens, and P.~Talavera.
\newblock K$_{l4}$ form-factors and $\pi-\pi$ scattering.
\newblock {\em Nuclear Physics B}, 585(1):293 -- 352, 2000.

\bibitem{Weinberg1966}
Steven Weinberg.
\newblock Pion scattering lengths.
\newblock {\em Phys. Rev. Lett.}, 17:616--621, Sep 1966.

\bibitem{DasbookQFT}
A.~Das.
\newblock {\em Lectures on Quantum Field Theory}.
\newblock World Scientific, 2008.

\bibitem{tHooft1972189}
G.~'t~Hooft and M.~Veltman.
\newblock Regularization and renormalization of gauge fields.
\newblock {\em Nuclear Physics B}, 44(1):189 -- 213, 1972.

\bibitem{CollinsbookRenorm}
J.C. Collins.
\newblock {\em Renormalization: An Introduction to Renormalization, the
  Renormalization Group and the Operator-Product Expansion}.
\newblock Cambridge University Press, 1984.

\bibitem{Passarino1979151}
G.~Passarino and M.~Veltman.
\newblock One-loop corrections for $e^{+} e^{-}$ annihilation into
  $\mu^{+}\mu^-$ in the {W}einberg model.
\newblock {\em Nuclear Physics B}, 160(1):151 -- 207, 1979.

\bibitem{tHooft1979365}
G.~'t~Hooft and M.~Veltman.
\newblock Scalar one-loop integrals.
\newblock {\em Nuclear Physics B}, 153:365 -- 401, 1979.

\bibitem{Denner1991}
Ansgar Denner.
\newblock {Techniques for calculation of electroweak radiative corrections at
  the one loop level and results for W physics at LEP-200}.
\newblock {\em Fortsch. Phys.}, 41:307--420, 1993.

\end{thebibliography}

\end{document}